\documentclass[ reprint,showpacs,twocolumn,superscriptaddress,nofootinbib,aps]{revtex4-2}
\usepackage{tabu}

\usepackage{amsmath,amsthm,amssymb}
\usepackage{graphicx}  
\usepackage{xcolor}
\usepackage{mathtools}
\usepackage{bbm}
\usepackage{subfigure}
\usepackage{dcolumn}
\usepackage{braket}
\usepackage{lipsum}
\usepackage{bm}
\usepackage{comment}
\usepackage{wasysym}
\usepackage{pifont}
\usepackage{soul}
\usepackage{dsfont}
\usepackage{physics}
\usepackage[normalem]{ulem}

\newcommand{\CZ}{\textrm{CZ}}
\newcommand{\kket}[1]{| #1\rangle \! \rangle}
\newcommand{\bbra}[1]{\langle \! \langle #1|}

\newcommand{\avg}[1]{\left\langle #1 \right\rangle}
\newcommand{\R}[1]{{\color{red} #1}}
\definecolor{DarkGreen}{RGB}{0,128,0}
\newcommand{\G}[1]{{\color{DarkGreen} #1}}
\newcommand{\B}[1]{{\color{blue} #1}}

\makeatletter
\newcommand\mathcircled[1]{%
  \mathpalette\@mathcircled{#1}%
}
\newcommand\@mathcircled[2]{%
  \tikz[baseline=(math.base)] \node[draw,ellipse,inner sep=1pt] (math) {$\m@th#1#2$};%
}

\makeatother

\usepackage{tikz}

\definecolor{dodgerblue}{HTML}{1E90FF}
\usepackage[colorlinks=true,citecolor=dodgerblue,linkcolor=dodgerblue,urlcolor=dodgerblue,pdftitle={decohere D4}]{hyperref}

\tikzset{>=latex}

\begin{document}

\title{Decoherence and wavefunction deformation of $D_4$ non-Abelian  topological order}

\author{Pablo Sala}
\email{psala@caltech.edu}
\affiliation{Department of Physics and Institute for Quantum Information and Matter, California Institute of Technology, Pasadena, CA 91125, USA}
\affiliation{Walter Burke Institute for Theoretical Physics, California Institute of Technology, Pasadena, CA 91125, USA}

\author{Jason Alicea}
 \affiliation{Department of Physics and Institute for Quantum Information and Matter, California Institute of Technology, Pasadena, CA 91125, USA}
 \affiliation{Walter Burke Institute for Theoretical Physics, California Institute of Technology, Pasadena, CA 91125, USA}
 
\author{Ruben Verresen}
\email{verresen@uchicago.edu}
\affiliation{Pritzker School of Molecular Engineering, University of Chicago, Chicago, IL 60637, USA}
\affiliation{Department of Physics, Harvard University, Cambridge, MA 02138, USA}
\affiliation{Department of Physics, Massachusetts Institute of Technology, Cambridge, MA 02139, USA}

\date{\today}

\begin{abstract}
The effect of decoherence on topological order (TO) has been most deeply understood for the toric code, the paragon of Abelian TOs.
We show that certain non-Abelian TOs can be analyzed and understood to a similar degree, despite being significantly richer.
We consider both wavefunction deformations and quantum channels acting on $D_4$ TO, which has recently been realized on a quantum processor.
By identifying the corresponding local statistical mechanical spin or rotor model with $D_4$ symmetry, we find a remarkable stability against proliferating non-Abelian anyons.
This is shown by leveraging a reformulation in terms of the tractable O$(2)$ loop model in the pure state case, and $n$ coupled O$(2)$ loop models 
for Rényi-$n$ quantities in the decoherence case---corresponding to worldlines of the proliferating anyon with quantum dimension $2$.
In particular, we find that the purity ($n=2$) remains deep in the $D_4$ TO for any decoherence strength, while the $n \to \infty$ limit becomes critical upon maximally decohering a particular anyon type, similar to our wavefunction deformation result.
The information-theoretic threshold ($n\to 1$) appears to be controlled by a disordered version of these statistical mechanical models, akin to the toric code case although significantly more robust.
We furthermore use Monte Carlo simulations to explore the phase diagrams when multiple anyon types proliferate at the same time, leading to a continued stability of the $D_4$ TO in addition to critical phases with emergent $U(1)$ symmetry.
Instead of loop models, these are now described by net models corresponding to different anyon types coupled together according to fusion rules.
This opens up the exploration of statistical mechanical models for decohered non-Abelian TO, which can inform optimal decoders, and which in an ungauged formulation provides examples of non-Abelian strong-to-weak symmetry breaking.

 \end{abstract} 

\maketitle
\tableofcontents

\section{Introduction}

Exploring and defining mixed-state quantum phases of matter is receiving ever-growing interest~\cite{Hastings_11, Coser_2019, Tarun_finiteT,
 de_Groot_2022,bao2023mixedstate, fan2023diagnostics,LeeYouXu2022, Renorm_QECC_23,chen2023separability, wang2023intrinsic,Ma_23,Lu23,lee2022symmetry,Zhu23,lyons24,Mong_24,sohal_24,ellison2024classificationmixedstatetopologicalorders,tapestry_24,chen2024unconventional,Hauser_24,2024_sala_SSSB,Markov_length_24,TshungCheng_24,lee2024exactcalculationscoherentinformation,ma2024topological,rakovszky2024definingstablephasesopen, ma2024symmetry, hsin2023anomaliesaveragesymmetriesentanglement, su_2024, wang2024anomalyopenquantumsystems, Kawabata_24, guo_24, zhang2024quantumcommunicationmixedstateorder, lavasani2024stabilitygappedquantummatter, Lessa_24, lessa2024mixedstatequantumanomalymultipartite, li_lee_Yoshida_24, Leo_24, Lieu_24} 
at the intersection of condensed matter and quantum information. The motivation descends in part from the rapid development of highly controlled quantum platforms, whose sources of decoherence are more accurately characterized via effective (e.g., Pauli) noise models rather than by more traditional thermal ensembles~\cite{Nielsen_Chuang_2010}. This endeavor is also of intrinsic theoretical interest, as pursuing stable quantum coherent phenomena under imperfect conditions requires generalizing the conventional pure ‘ground state’ paradigm ~\cite{Wen_book,Sachdev_2023} to open quantum systems. 

One of the most striking emergent quantum phenomena is that of topological order (TO), for which a deep understanding has been achieved under ideal conditions~\cite{Wen_book,Sachdev_2023}.
Characterized by emergent anyons \cite{Leinaas_77, Goldin_81, Wilczek_82} and ground state degeneracy \cite{Einarsson90}, these systems can be exploited to robustly store and manipulate quantum information \cite{Kitaev_2003,Freedman_2000gwh, Freedman_2006,Nayak_08,Terhal_15}. The simplest TO corresponds to a two-dimensional $\mathbb{Z}_2$ deconfined gauge theory \cite{read_91,WEN1991}, which is microscopically realized by the celebrated toric code model on a square lattice~\cite{Kitaev_2003}. Correspondingly, much literature~\cite{WANG200331,fan2023diagnostics,bao2023mixedstate,LeeYouXu2022, Renorm_QECC_23,chen2023separability, wang2023intrinsic,Mong_24,lyons24,tapestry_24,chen2024unconventional,Hauser_24,2024_sala_SSSB,Markov_length_24,TshungCheng_24,lee2024exactcalculationscoherentinformation, sohal_24, ellison2024classificationmixedstatetopologicalorders} inspired by the seminal work of Dennis et al.~\cite{Dennis_2002}, has focused on extending this TO to mixed states, appearing as a result of local decohering processes. While this quantum memory breaks down at any finite temperature~\cite{Nussinov_08, Bravyi_2009, Hastings_11,Poulin_13,Brown_16,Tarun_finiteT}, it has been found---originally in the context of quantum error correction---that quantum information encoded on the ground state subspace is stable if decoherence does not exceed a certain strength~\cite{Dennis_2002}. It turns out that this error correction threshold $p_c$ (corresponding to an ideal decoder) is set by the finite temperature phase transition of the 2$D$ random bond Ising model (RBIM), where the disorder is sourced by different error configurations~\footnote{The mapping to the $2$-dimensional RBIM only holds under the assumption of perfect syndrome measurements. If one incorporates imperfect measurements as part of the physical process, then the underlying model becomes the $3$-dimensional random-plaquette model~\cite{Dennis_2002,WANG200331}}. 
More recently, Refs.~\onlinecite{bao2023mixedstate,fan2023diagnostics,LeeYouXu2022} followed a different, more agnostic approach to specific error correction schemes and proposed to directly characterize the decohered density matrix $\rho$ by evaluating various information-theoretic quantities. However, since such quantities are generally difficult to compute, the authors proposed to look at their $n$-Rényi generalizations associated with different moments $\textrm{tr}(\rho^n)$ to detect the non-trivial effects of decoherence. Even though these critical thresholds $p_c^{(n)}$ do in general depend on the Rényi index $n$, such generalizations simplify the calculations and connect to simpler statistical mechanical (``stat-mech''). The ``intrinsic'' (or ``information-theoretic'') error threshold $p_c$ can be obtained by taking an $n\to 1$ replica limit, although finite $p_c^{(n)}$'s already inform about singularities in the spectrum of $\rho$.

However, the rich landscape of TOs consists mostly of non-Abelian states \cite{Goldin_85,Wen_91_FQH,MOORE1991362, Moore1989}, whose anyons have internal structure---forming the hardware of a topological quantum computer \cite{Kitaev_2003,Nayak_08}.
Previous works considering the effect of decoherence on non-Abelian TO have either focused on error-correcting schemes for certain classes of TOs (without identifying stat-mech models) \cite{Wootton_14,Brell_14,Wootton_16,Burton_2017,Dauphinais_2017,Schotte_22a,Schotte_22b}, or explored what type of novel intrinsically-mixed phases can in principle emerge for maximal decoherence rate \cite{Mong_24, ellison2024classificationmixedstatetopologicalorders,sohal_24}. 
However, it is a priori not clear whether one can hope to understand the effects of decoherence of any non-Abelian TO to a degree that is akin to our understanding of the structurally much simpler Abelian case; it is similarly unclear whether one should expect the physics to be significantly different and more interesting. 
Here we answer both these open questions in the \emph{affirmative}.
In particular, we study $D_4$ TO, which has recently been experimentally realized in a trapped-ion processor \cite{iqbal2023creation}. The fact that this was the first non-Abelian TO to be realized using a quantum processor is due to it, in some sense, being the simplest non-Abelian TO---thereby making it an ideal candidate to play a paradigmatic role akin to the toric code for Abelian TO. Indeed, although it has as many as $22$ anyons\footnote{More than for $D(S_3)$ TO~\cite{Kitaev_2003} which counts $6$ anyons, or double Fibonacci (dFib)~\cite{Levin_Wen_05} which has only $4$.}, it is rather minimal since all non-Abelian anyons fuse to Abelian ones. This is known as acyclic or nilpotent property and was crucial to the theory proposal\footnote{The fact this TO can be obtained by a single round of measurements is believed to be impossible~\cite{Tantivasadakarn_2023} for smaller TOs such as $D(S_3)$ and dFib TO.} of its realization. This is related to $D_4$ TO being achingly close to an Abelian TO: it can be thought as a `twisted' Abelian gauge theory~\cite{Yoshida_2016,Dijkgraaf1990,Dijkgraaf1991,Hu13}. Although the (doubled) Ising anyon TO shares a similar property, $D_4$ TO has an emergent (non-fermionic) gauge group symmetry, similar to the toric code.

With this motivation, the present work is a detailed study of decohering $D_4$ TO.
We consider both \emph{pure wavefunction deformations}---realized as imaginary time evolutions for a time $\beta$ acting on the $D_4$ topologically order ground state---as well as \emph{local quantum channels} acting on an initial pure density matrix with some error rate $p$. In the following, we refer to both kinds of scenarios as errors. Pure wavefunction deformations have been previously considered in the literature (see, e.g., Refs.~\onlinecite{Haegeman_15,GuoYi_19} for deformations of the toric code wavefunction and Refs.~\cite{Mari_n_2017, Xu_2021, Xu_2022, Schotte_2019, Fendley_2008} when considering non-Abelian topological order), exploiting both numerical and analytical tractability in particular using tensor network formalism. However, our motivations to consider such wavefunction deformations as a prelude to understanding decohered mixed states are the following. First, this scenario provides a conceptually simple way to perturb the pure state while retaining a corresponding local parent Hamiltonian for the modified wavefunction, and involves similar algebraic calculations. Second, the combination of such deformations on top of zero-correlation length wavefunctions together with local quantum channels, allows us to controllably induce a finite-correlation length in the system, which will turn out to provide useful insights. And finally, as we will find, the deformed wavefunctions turn out to be relevant for understanding the behavior of the stat-mech model appearing in $\textrm{tr}(\rho^n)$ in the $n\to\infty$ limit.

A key finding of this work is that while Abelian errors lead, as for the toric code case, to Ising-like stat-mech models, we find that the proliferation of non-Abelian anyons in $D_4$ TO leads to a $D_4$-symmetric spin model. 
This larger symmetry gives rise to a field theory description at the critical threshold which showcases an emergent U$(1)$ symmetry, and hence vastly stabilizes the system against both types of errors. We discuss this formulation hand in hand with an exact high-temperature expansion of this model, which is given by an O$(2)$ loop model on a honeycomb lattice~\cite{Nienhuis_81, Nienhuis_82, peled2019lectures} (indeed, the usual Ising model is an O$(1)$ loop model). It is no coincidence that $d=2$ is the quantum dimension of the non-Abelian anyon proliferated by the underlying error.

First,  when subjecting $D_4$ TO to the proliferation of Abelian anyons via Pauli noise, we recover the same phenomenology as found for the decohered toric code ground state~\cite{fan2023diagnostics}. Specifically, $D_4$ TO breaks down beyond a certain error strength.
However, when deforming the wavefunction by creating non-Abelian anyons pairs, we find that the system is robust to any finite deformation. This stability relates to the fact that the norm of the deformed wavefunction, signaling ground state phase transitions, maps to an O$(2)$ loop model (see review~\cite{peled2019lectures}). Here, loop configurations correspond to the worldlines of non-Abelian anyons with quantum dimension $d=2$, contributing with a topological tension on top of a local string tension given by the error strength. This is then able to give rise to a critical state, which in this case corresponds to a Berezinskii-Kosterlitz-Thouless (BKT) critical point and that can be pushed to an extended gapless phase.

Similarly, when subjecting the system to decoherence, we find that $\textrm{tr}(\rho^n)$ leads to $n$ coupled O$(2)$ loop models. For example, when computing the purity $(\textrm{tr}(\rho^2)$, both copies are strongly coupled leading to a O$(4)$ loop model.  This translates into the robustness of $D_4$ TO even at the maximum error rate (under Pauli noise). While the phase diagram of higher Rényis with $n>2$ is more subtle, we can also argue that in $n\to \infty$ is very stable, and is likely to attain a threshold $p_c^{(\infty)}$ only at the maximum error rate. On more rigorous footing, we have shown that unlike what has been found for the toric code~\cite{fan2023diagnostics}, the thresholds $p_c^{(n)}$ for different Rényi index $n$ are \emph{not a monotonic increasing function} of $n$. However, the intrinsic threshold $p_c$ (as discussed in Ref.~\cite{Dennis_2002}) is controlled by the ``replica limit'' $n\to 1$, or rather by a transition in information theoretical quantities that involve the von Neumann entanglement of $\rho$. We are able to identify the local stat mech model determining this threshold (analogous to the RBIM for the toric code) for the decohered density matrix at the maximal error rate, where we can perform a full diagonalization. In particular, we find that the robustness of this non-Abelian quantum memory can be characterized by the free energy cost of inserting a symmetry defect line on a disordered $D_4$ rotor model.

Finally, we combine both ``Abelian'' and ``non-Abelian'' errors and characterize the resulting phase diagram for both pure wavefunction deformations, and for the decohered mixed state as signaled by a singular behavior in the purity $\textrm{tr}(\rho^2)$. We find a rich phase diagram that displays several short-range correlated phases, corresponding to various types of TOs, as well as an extended critical phase in the former case which is now stabilized by the presence of Abelian anyons (see Sec.~\ref{sec:D4_XpZ}). Such characterization is possible due to the alternative and equivalent formulations of the resulting loop models using explicitly local variables, which permits an efficient implementation of classical Monte Carlo simulations and the use of local order parameters.

We organize the remainder of the paper as follows. In Sec.~\ref{sec:D4_main} we start by reviewing a microscopic model realizing $D_4$ TO, and its connection to a symmetry enriched $\mathbb{Z}_2\times \mathbb{Z}_2$ toric code by via a (un)gauging unitary transformation. We then consider the effect of local phase $Z$ errors in Sec.~\ref{sec:Z_error}, which proliferate pairs of Abelian charges. First, we characterize the resulting phase diagram of the corresponding deformed wavefunction (Sec.~\ref{sec:Z_purewf} ), and then the fate of the decohered density matrix as measured by the purity (Sec.~\ref{sec:Z_dech}). Section~\ref{sec:D4_X} then follows a similar structure, first discussing $X$ pure wavefunction deformations in subsection~\ref{sec:X_puredef}, which proliferates pairs of non-Abelian fluxes. We show that the worldlines of these non-Abelian anyons with quantum dimension $d=2$, lead to O$(2)$ loop models when computing the norm of the deformed wavefunction, which is in turn reformulated in terms of local Ising-like interactions and as a four-state clock model. In the next subsection~\ref{subsubsec:X_dech}, we then follow a similar discussion for the purity. However, this section includes three additional results: First in subsection~\ref{sec:high_trrho} we compute higher-moments $\textrm{tr}(\rho^n)$ of the decohered density matrix where the non-Abelian nature of the proliferating anyon establishes a clear difference with respect to the toric code case. Second, in subsection~\ref{sec:n_to_1} we consider the limit $n\to \infty$, and prove that the phase diagram as characterized by $\textrm{tr}(\rho^\infty)$ matches that of deformed wavefunction. In Sec.~\ref{sec:D4_XpZ} we then combine both $Z$ Abelian and $X$ non-Abelian errors on different sublattices, and study the phase diagram of both the resulting deformed wavefunction and decohered density matrix. We also include the statistical models characterizing the effect of proliferating more than one type of non-Abelian anyons in Sec.~\ref{sec:var_NA}. Finally, we employ the ungauging maps to relate the decoherence transition of $D_4$ TO to the phenomenon of strong-to-weak spontaneous $D_4$ symmetry breaking in subsection~\ref{sec:D4_SSB_X}. We conclude in Sec.~\ref{sec:conclusions} discussing the main conclusions of this work as well as various open questions.

{In a companion work~\cite{short_paper}, we explore and confirm some of these results for a broad range of TOs, including the quantum double construction, and the Kiteav honeycomb model. The latter provides an example of a non-fixed point wave function with the noise model creating non-Abelian anyons which are neither self-bosons nor have an integer quantum dimension. Moreover, this companion work shows the dependence of the results on the underlying microscopic lattice.}

\section{Review of $D_4$ topological order}
\label{sec:D4_main}

\begin{figure}
    \centering
    \includegraphics[width=\linewidth]{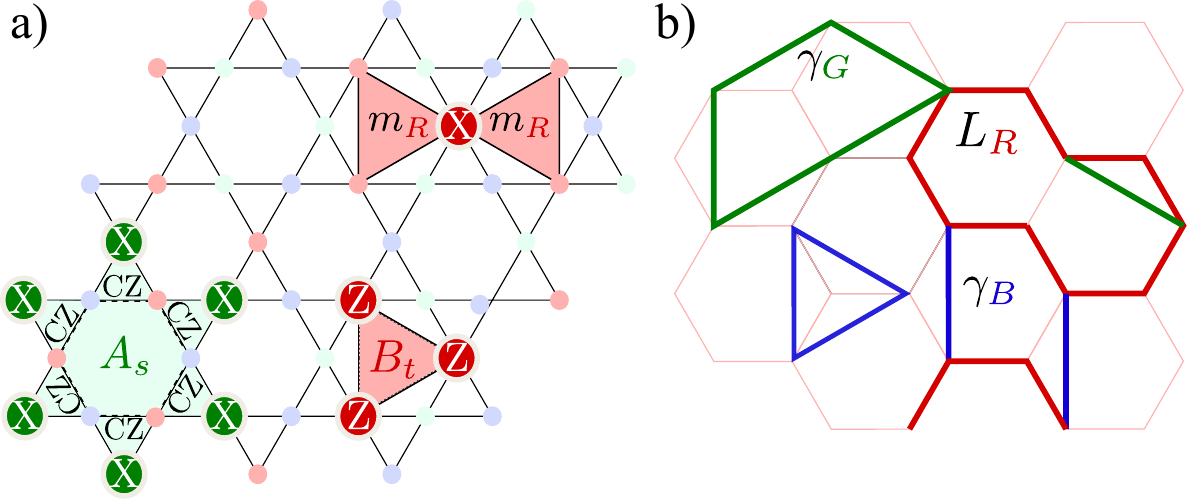}
    \caption{\textbf{Microscopic realization of $D_4$ topological order on the kagome lattice.} (a) Local Hamiltonian terms $A_s, B_t$ (specifically we show $\G{A_s}$ and $\R{B_t}$) in Eq.~\eqref{eq:H_D4}. Pairs of non-Abelian (Abelian) $m_\R{R}$ ($e_\B{B}$) fluxes (charges) are created by the action of local $\R{X_r}$ ($\B{Z_b}$) Paulis acting on the red (blue) sublattice $\R{\mathcal{R}_R}$ ($\B{\mathcal{R}_B}$).  (b) The worldlines of non-Abelian $m_\R{R}$ anyons form closed loops ($L_\R{R}$, red) on the honeycomb lattice (whose vertices lie at the center of triangle terms $\R{B_t}$). On the other hand, Abelian charges $e_\G{G}$ and $e_\B{B}$ either from closed loops on a triangular lattice (whose vertices lie at the center of star terms $\B{A_s},\G{A_s}$), \emph{or} they can end on the red loop, due to the non-trivial fusion $m_\R{R} \times m_\R{R} = 1 + e_\B{B} + e_\G{G} + e_\B{B} \times e_\G{G}$. This configuration spaces define a \emph{net model}.}
    \label{fig:D4_TO}
\end{figure}

We consider a microscopic model with $D_4$ non-Abelian topological order (TO) following Ref.~\onlinecite{Yoshida_2016}. It is a solvable spin-1/2 model which is in the same phase of matter as the quantum double $\mathcal{D}(D_4)$ \cite{propitius,Kitaev_2006,Lootens22}, which can be thought of as the deconfined phase of an (emergent) $D_4$ gauge theory. Since the dihedral group $D_4 \cong \mathbb Z_4 \rtimes \mathbb Z_2$ is the symmetry group of the square and hence non-Abelian, this deconfined phase has non-Abelian anyon excitations, as we will discuss. However, the model introduced in Ref.~\onlinecite{Yoshida_2016} can also be thought of as a $\mathbb Z_2 \times \mathbb Z_2 \times \mathbb Z_2$ gauge theory with a subtle kind of `twist' which is responsible for its non-Abelian nature\footnote{More concretely, it is obtained by gauging a $\mathbb{Z}_2\times\mathbb{Z}_2\times\mathbb{Z}_2$ symmetry protected topological (SPT) phase on a triangular lattice~\cite{Yoshida_2016}.} \cite{Dijkgraaf1990,Dijkgraaf1991,Hu13}. In Sec.~\ref{sec:mapping_main} we will use an alternative perspective, where this non-Abelian phase of matter is obtained by gauging the $\mathbb Z_2 \times \mathbb Z_2$ symmetry of a symmetry-enriched toric code \cite{Barkeshli19,Chen17,BenZion16,Stephen20}.

Our Hilbert space consists of spin-$1/2$ (qubits) on the vertices of a kagome lattice, and for convenience we will work with periodic boundary conditions. Since this lattice is tripartite, we distinguish among {\color{red} red} ($\R{\mathcal{R}_R}$), {\G{green} ($\G{\mathcal{R}_G}$), and {\color{blue} blue} ($\B{\mathcal{R}_B}$)  sublattices; see Fig.~\ref{fig:D4_TO}a. Equivalently, one can think of qubits lying on the edges of three interleaved honeycomb lattices colored  {\color{red} R}, {\G{G}}, and {\color{blue} B} (the red honeycomb lattice corresponding to $\R{\mathcal{R}_R}$ is shown in Fig.~\ref{fig:D4_TO}b). 

\subsection{Microscopic Hamiltonian and anyon content}
In this work we denote by $X,Y,Z$ the three Pauli matrices corresponding to $\sigma^x,\sigma^y$ and $\sigma^z$ respectively. We consider a spin-1/2 Hamiltonian realizing $D_4$ topological order on the sites of the kagome lattice \cite{Yoshida_2016} (following the notation of Ref.~\onlinecite{iqbal2023creation}):
\begin{equation} \label{eq:H_D4}
    H=-\sum_{s\in \{\textrm{\ding{65}}\}}A_s - \sum_{t\in \{\triangleright, \triangleleft\}} B_t
\end{equation}
with the $12$-body operators $A_s=\prod_{i_{\textrm{in}}=1}^6 \textrm{CZ}_{i_{\textrm{in}},i_{\textrm{in}}+1}\prod_{i_{\textrm{out}}=1}^6X_{i_{\textrm{out}}}$ and the  $3$-body triangle operators $B_t=\prod_{j\in \triangleright, \triangleleft}Z_j$. For the former, $\textrm{CZ}_{i,j}=\frac{1}{2}(1+Z_i + Z_j - Z_iZ_j) = e^{i \pi \frac{1-Z_i}{2} \frac{1-Z_j}{2}}$ denotes the controlled-$Z$ gate which assigns the phase $-1$ if and only if the two spins sitting at sites $i,j$ are in the $\ket{\downarrow}$ state. The first product runs over the $i_{\textrm{in}}$ sites lying within the internal hexagon of a star $\ding{65}$, while the second product runs over the exterior $i_{\rm out}$ sites. 
See Fig.~\ref{fig:D4_TO}a for the graphical definition of $A_s$ and $B_t$.

Analogously to the $2$D toric code, $A_s$'s can be viewed as plaquette operators defined on the intertwined colored honeycomb lattices, while the $B_t$'s correspond to star operators imposing Gauss's law on each honeycomb-lattice vertex. Each of these operators have an associated color inherited from the 3-colorable lattice.   
Although $[A_s,B_t]=0$ and $[B_t, B_{t'}]=0$, the $A_s$'s fail to commute with each other (due to the $\CZ$ rings lying within each plaquette term).  The latter property is important, because otherwise the Hamiltonian would reduce to three decoupled copies of the 2D toric code.  
Fortunately, the ground state manifold corresponds to $B_t=+1$ in every triangle. In this sector, $\left.[A_s,A_{s'}]\right|_{B_t=+1}=0$. Consequently, $A_s=B_t=+1$ holds in the ground-state subspace. 

This topological order hosts a total of 22 anyons (for a detailed description see the Appendices of Refs.~\onlinecite{iqbal2023creation, Shortest_NA}).  Of these anyons, $8$ are Abelian (i.e., have quantum dimension $d=1$), and the remaining are non-Abelian with quantum dimension $d=2$. All Abelian anyons, modulo the trivial particle, correspond to self-bosonic electric charges~\footnote{Calling certain anyons `charges' and others `fluxes' depends on how one describes it as a gauge theory, which is a matter of choice. Here we follow the convention of this state as a twisted $\mathbb{Z}_2^3$ gauge theory as in Ref.~\cite{Dijkgraaf1990,Dijkgraaf1991,Hu13,Yoshida_2016,iqbal2023creation}.} that we label $e_{\R{R}}, e_\G{G}, e_\B{B}, e_{\R{R}\G{G}}, e_{\R{R}\B{B}}, e_{\B{B}\G{G}}$, and $ e_{\R{R}\G{G}\B{B}}$. The first three charges ---which generate the others via fusion (i.e., for example $e_{\R{R}\G{G}}=e_\R{R}\times e_\G{G}$)---arise from violation of a kagome star operator $A_s$. Namely, a kagome star with $A_s=-1$ corresponds to a charge $e_{\R{R}}, e_{\G{G}}$, or $e_\B{B}$ depending on the color of the flipped operator. These anyons can be created in pairs by the action of local Pauli $Z$ operators. 
For example, a pair of $e_\B{B}$'s are created at the stars $\B{s}_i, \B{s}_f$ by the action of the Wilson operator $\B{\mathcal{Z}}_{s_i}^{s_f}\equiv\prod_{\B{b}\in \B{\gamma}_{x,y}}{Z}_{\B{b}}$, where $\B{\gamma}_{x,y}$ is any open string with support only on the blue sublattice $\B{\mathcal{R}_B}$ and with $x,y$ vertices lying in $s_i,s_f$, respectively. Hence,  for closed contractible loops with $s_i = s_f$, $\B{\mathcal{Z}}_{s_i}^{s_i}$ acts as the identity on the $D_4$ ground state, since such an operator creates and then fuses Abelian anyons. Logical operators $\mathcal{Z}_{\R{R}, \G{G}, \B{B}}$---which can act nontrivially in the ground state manifold---correspond to the product of $Z$ operators along closed \emph{non-contractible} (both horizontal and vertical) loops with $s_i=s_f$ and with support only on the corresponding sublattice $\mathcal{R}_{\R{R}, \G{G}, \B{B}}$. 

The remaining $14$ non-Abelian anyons have either bosonic, fermionic, or semionic self-statistics depending on their topological spin.  
The fundamental fluxes $m_{\R{R}},  m_{\G{G}}, m_{\B{B}}$---which together with the electric charges above generate all anyons via fusion---can also be created in pairs and correspond to violations of the triangle operator $B_t=-1$. For example,  a pair of fluxes on nearby triangles can be generated by a single Pauli $X$ operator acting on the common vertex (see the two coinciding red triangles in Fig.~\ref{fig:D4_TO}a). Creation of a pair of distant fluxes $m_{\R{R}}$  at triangles $\R{t}_i, \R{t}_f$, however, requires a string of $X$'s supplemented by a unitary circuit whose depth is linear (in the distance between the two triangles) \cite{iqbal2023creation}:
\begin{equation} \label{eq:X_if}
    \R{\mathcal{X}}_{t_i}^{t_f} = \prod_{\R{r}\in \R{\gamma}_{x,y}}X_\R{r} \prod_{\B{b}< \G{g}\in \R{\gamma}_{x,y}} \CZ_{\B{b}\G{g}},
\end{equation}
where $\R{\gamma}_{x,y}$ is an open string connecting sites $x,y$ lying on triangles $t_i, t_f$ respectively, and where `$\B{b}< \G{g}\in \R{\gamma}_{x,y}$' means we go over all blue and green sites which are passed upon traversing the red string and consider all pairs where blue is on the left of green\footnote{The orientation of the string is set by the fact that the initial triangle $\R{t}_i$ lives in either a blue or green plaquette.}.
The additional $\CZ$'s along the path appear due to the non-Abelian nature of the $m_{\R{R}}$'s:   
Unlike for Abelian anyons, fusion of two non-Abelian fluxes leads to more than one possible outcome, e.g., $m_\R{R}\times m_\R{R}=1+e_\G{G}+e_\B{B}+e_\G{G}e_\B{B}$.  Consequently, with only a product of $X$'s, non-Abelian anyons created by adjacent $X$ operators would fuse nontrivially---creating a superposition of allowed anyon fusion products along the entire length of the $X$ string.  (This point will be very important in later sections!) 
The CZ's ``clean up'' these excitations such that the operator $\R{\mathcal{X}}_{t_i}^{t_f}$ creates a pair of fluxes in the trivial fusion channel, \emph{without} excess anyons in the intervening region. See additional details in Ref.~\onlinecite{iqbal2023creation}. A graphical example of this operator is shown in Fig.~\ref{fig:ungauging_X}.
We note that Eq.~\eqref{eq:X_if} also affects the $A_s$ operators with $t_i,t_f$ lying at their center, such that $A_s$ takes an indefinite value with $\langle A_s\rangle=0$, which is related to the higher-dimensional internal space associated with the non-Abelian anyons. Similarly to the logical operators $\mathcal{Z}_{\R{R}, \G{G}, \B{B}}$ discussed above, a second set of logical operators $\mathcal{X}_{\R{R}, \G{G}, \B{B}}$ analogously correspond to closed non-contractible loop configurations with $t_i=t_f$.

Hereafter we take $\ket{D_4}$ to correspond to the ground state where all ${\mathcal{Z}}$ logical operators of the theory act as the identity.

\subsection{Ungauging and disentangling maps} \label{sec:mapping_main}

Here we highlight a particular change of variables that will prove useful later on. It involves dualizing some of the sublattices, which can be interpreted as \emph{ungauging}, thereby mapping it to a simpler theory.

In Sec.~\ref{sec:D4_X}, we will study the proliferation of fundamental non-Abelian fluxes. Since all three sublattices are manifestly on the same footing in the above $D_4$ TO model, without loss of generality we will choose to proliferate fluxes $m_\R{R}$ associated to the red sublattice. For this reason, the following change of variables treats the red sublattice on a different footing than blue and green. In particular, we will interpret the red sites of the kagome lattice as living on the \emph{links} of a honeycomb lattice. We will then dualize the green and blue sublattices of the kagome lattice in the sector $\B{B_t}=\G{B_t}=1$, such that they map to spin-1/2's living on the \emph{vertices} of the aforementioned honeycomb lattice (with blue and green corresponding to the two sublattices of the honeycomb).

In conclusion, we map the kagome lattice to a `heavy-hexagonal' lattice, with spin-1/2's on the vertices and bonds of the honeycomb lattice. 
Operators (within the $\B{B_t}=\G{B_t}=1$ sector) map to the new Hilbert space as follows: while $\R{Z} \to \R{Z}$ is unchanged, we have
\begin{align} 
&\raisebox{-20pt}{
\includegraphics[width=0.7\linewidth]{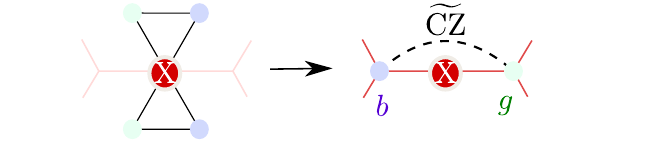}
} \label{eq:ungauge_X} \\
&\raisebox{-35pt}{
\includegraphics[width=0.7\linewidth]{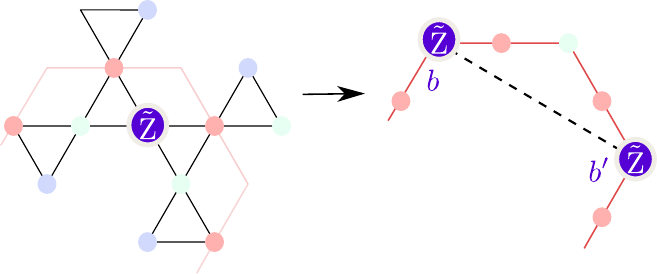}
} \label{eq:ungauge_Z}
\end{align}
and similarly upon exchanging the blue and green sublattices. We have not shown how $\B{A_s}$ and $\G{A_s}$ map since it is somewhat more complicated~\footnote{For completeness, $\B{A_s} \to \B{X_v} \prod_{\langle v,v' \rangle} \left(\G{C}\R{B}\right)_{\G{v'},\R{t}}$, where we are controlling the action of the three-body $\R{B_t}$ term on the state of the three neighboring green vertices on the honeycomb lattice. Clearly if $\R{B_t}=1$, one can forget about this controlled-term.} and will not explicitly be used in the main text, however, when acting on states where $\R{B_t} = 1$, they will simply map to single-site $\B{X}$ and $\G{X}$, respectively. We will use a tilde $\tilde{\cdot}$ to refer to the ungauged degrees of freedom on the green and blue sublattices. Here and in the following section, we make a slight abuse of notation and use $\mathcal{R}_{\G{G},\B{B}}$ to denote both green and blue sublattices on the kagome lattice, as well a sublattices of the honeycomb lattice respectively.

One can straightforwardly see that the above mapping indeed preserves the Pauli algebra. The precise structure of the mapping was chosen such that it maps the $D_4$ TO to a much simpler state:
\begin{equation} \label{eq:ung_state}
\ket{D_4}_{\text{kagome}} \to \ket{\R{\textrm{T.C.}}}_{\R{\mathcal{R}_R}}\otimes \underbrace{\ket{\G{\tilde{+}}}^{\otimes |\G{\mathcal{R}_G}|}_{\G{g}}\otimes \ket{\B{\tilde{+}}}^{\otimes |\B{\mathcal{R}_B}|}_{\B{b}}}_{=\ket{\tilde{+}}^{\otimes N_\text{sites}}}
\end{equation}
Indeed, it can be checked that $\R{A_s}$ maps to the usual toric code stabilizer $\prod_{j\in \R{\hexagon}} X_j$. This can be interpreted as mapping the $D_4$ TO to a toric code which is symmetry-enriched with a $\mathbb Z_2 \times \mathbb Z_2$ symmetry corresponding to spin-flip symmetries of the blue and green sublattices \cite{BenZion16,Stephen20}, where we have chosen to put the symmetry-enrichment in the operator algebra rather than the state~\footnote{If we had instead put it in the state, then Eq.~~\eqref{eq:ung_state} would have an additional finite-depth `CCZ' circuit along the bonds of the honeycomb lattice, and, e.g., Eq.~\eqref{eq:ungauge_X} would simplify to just mapping $\R{X} \to \R{X}$.}. The result of mapping the Wilson operator $\R{\mathcal X}_{t_i}^{t_f}$ in Eq.~\eqref{eq:X_if} is
\begin{equation} \label{eq:X_log_ung}
 \R{\mathcal{X}}_{t_i}^{t_f} \to \bigg( \prod_{\R{r}\in \R{\gamma}_{x,y}}X_\R{r} \bigg) \times \CZ_{\B{x}\G{y}},
\end{equation}
which is just the usual toric code anyon string, dressed with a CZ connecting its two endpoints.
We provide a visual derivation of this in Fig.~\ref{fig:ungauging_X} of App.~\ref{app:mappings}.

\vskip 0.1in
In the next two sections we consider Pauli errors of a single type, leaving the combination of several types of errors for later sections. For these, the previous mapping will be of great use to simplify the calculations and map the system to local stat-mech models.  We first consider errors that generate Abelian charges, and hence are parallel to previous studies of the toric code ground state under decoherence~\cite{Dennis_2002,WANG200331,fan2023diagnostics,bao2023mixedstate, LeeYouXu2022, Renorm_QECC_23,chen2023separability, wang2023intrinsic, Mong_24, tapestry_24,Hauser_24,2024_sala_SSSB,Markov_length_24,TshungCheng_24,lee2024exactcalculationscoherentinformation}. We then consider the proliferation of non-Abelian anyons. For each case we will examine pure wavefunction deformation and then treat decohered mixed states.  

\section{ $Z$ phase error: Abelian anyon proliferation}  \label{sec:Z_error}

In this section, we first study the effects of proliferating the \emph{Abelian} anyons $e_{\R{R}},e_\G{G},e_\B{B}$ of $D_4$ topological order. This will in large part reproduce the phenomenology known for the toric code, but it allows us to introduce useful concepts and provides a point of comparison for when we consider the effect of non-Abelian anyons in Sec.~\ref{sec:D4_X}. Moreover, in Sec.~\ref{sec:D4_XpZ} we will study the interplay of proliferating both Abelian and non-Abelian anyons at the same time.

\subsection{Pure wavefunction deformation\label{sec:Z_purewf}} 
We start by considering pure wavefunction deformations of the form 
\begin{equation} \label{eq:psi_abel}
    |\psi(\beta^z_\B{B},\beta^z_\G{G},\beta^z_\R{R} )\rangle=\prod_{\footnotesize c \in \{ \B B , \G G ,\R R  \}}e^{ \frac{\beta_c^z}{2} \sum_{j\in \mathcal{R}_c} Z_j}\ket{D_4} .
\end{equation}
Here, we allow for different deformation strengths $\beta^z_c$ in each sublattice.
The exponential represents a non-unitary operator that restructures the weight on the configurations present in $\ket{D_4}$.   
As we saw in Sec.~\ref{sec:D4_main}, $Z$ operators generate pairs of Abelian charges $e_c$, where their color $c=$ \R{R}, \G{G}, \B{B} depends on the sublattice on which the $Z$'s act. 
Since this ``Abelian'' local deformation does not couple different sublattices, it will turn out that the resulting phase transitions are decoupled for each of the three colors. Hence, we start by focusing on a single color---\B{B} for concreteness. 

To characterize the phase diagram of $ |\psi(\beta^z_\B{B})\rangle$ (where $\beta^z_\G{G}=\beta^z_\R{R}=0$), it is sufficient to look at the wavefunction overlap
\begin{equation}
\mathcal{Z}_{\ket{\psi}}(\beta^z_\B{B}) = \langle \psi(\beta^z_\B{B}) | \psi(\beta^z_\B{B}) \rangle.
\end{equation}
If we expand out $| \psi(\beta^z_\B{B}) \rangle$ into a basis which is diagonal in the non-unitary perturbation (in this case $e^{ \frac{\beta^z_\B{b}}{2} \sum_{j\in \mathcal{R}_\B{b}} Z_j}$), we can directly interpret $\mathcal{Z}_{\ket{\psi}}(\beta^z_\B{B})$ as a two-dimensional classical partition function. All its classical correlation functions directly capture observables which are diagonal in that basis. In fact, we prove in Appendix~\ref{app:pure_def_stat} that \emph{all} local observables (even off-diagonal ones) of the quantum state are captured by correlation functions in the classical model (due to our perturbation only involving local interactions~\cite{Verstraete_06}). We have thus reduced this two-dimensional quantum problem to studying a two-dimensional classical problem.

To calculate this partition function, we use the fact that $e^{ \alpha \vec n \cdot \vec \sigma} = \cosh(\alpha) + \sinh(\alpha) \vec n \cdot \vec \sigma$ where $\vec \sigma = (X,Y,Z)$ and $\vec n$ is any unit vector:
\begin{equation}
\begin{aligned} \label{eq:Z_B}
    &\mathcal{Z}_{\ket{\psi}}(\beta^z_\B{B})= \langle D_4|e^{ \beta^z_\B{B} \sum_{j \in \B{\mathcal{R}_B}} Z_j}|D_4\rangle
    \\ &\propto 
    \langle D_4|\prod_{j \in \B{\mathcal{R}_B}}\left[1+ \tanh(\beta^z_\B{B}) Z_j \right] |D_4\rangle ,
\end{aligned}
\end{equation}
where we dropped the analytic prefactor

$\cosh(\beta^z_\B{B})^{|\B{\mathcal{R}}_{\B{B}}|}$.  Expanding out the product yields
\begin{align} \label{eq:Z_B2}
    \mathcal{Z}_{\ket{\psi}}(\beta^z_\B{B})&= \sum_{n_{\B{B}}=0}^{|\B{\mathcal{R}}_{\B{B}}|}  \tanh(\beta^z_\B{B})^{n_{\B{B}}}
 \sum_{\{r_j\}_{j=1}^{n_{\B{B}}}}\langle D_4|\prod_{j=1}^{n_{\B{B}}} Z_{r_j} |D_4\rangle.
\end{align}
up to an overall prefactor that in the following we will also ignore. Here, $n_{\B{B}}$ counts the number of Pauli $Z$'s that act on the $D_4$ ground state at vertices $\{r_j\}$.   To evaluate $ \langle D_4|\prod_{j=1}^{n_{\B{B}}} Z_{r_j}  |D_4\rangle$, recall that $Z_{r_j}$ creates pairs of Abelian anyons $e_{\B{B}}$ at the center of the kagome star operators $\B{A}_s$.  If $\prod_{j=1}^n Z_{r_j}$ acting on the ket $\ket{D_4}$ includes at least one open string that generates unfused $e_\B{B}$ anyons, then the overlap with $\bra{D_4}$ vanishes.   Conversely, if $\prod_{j=1}^n Z_{r_j}$ forms a set of closed loops $L_{\B{B}}$---either contractible or non-contractible---then it acts on $\ket{D_4}$ as the identity as discussed in the previous section.   (Contractible closed loops act trivially because they create and then fuse Abelian anyons in a way that smoothly connects to doing nothing; non-contractible loops are trivial because we consider the ground state where $\mathcal{Z}$-type logical operators act trivially on $\ket{D_4}$.)  Closed-loop configurations $L_{\B{B}}$ correspondingly yield $\langle D_4|\prod_{j=1}^{n_{\B{B}}} Z_{r_j}  |D_4\rangle = 1$.

In summary, we find that (up to an inconsequential prefactor) Eq.~\eqref{eq:Z_B2} reduces to the partition function of an $O(1)$ loop model on the triangular lattice,
\begin{equation} \label{eq:Z1_betaB}
 \mathcal{Z}_{N=1}( t_\B{B})= \sum_{L_{\B{B}}} t_\B{B}^{|L_\B{B}|},
\end{equation}
defined on a triangular lattice with tension $t_\B{B}\equiv \tanh(\beta^z_\B{B})$. An example of a closed loop is shown in Fig.~\ref{fig:Abelian_triangle}. Panel a shows a closed loop configuration $L_\B{B}$ on the original kagome lattice, while panel b, shows the same loop configuration in the effective triangular lattice. In the exponent, $|L_{\B{B}}|$ denotes the total loop length in a given closed-loop configuration $L_{\B{B}}$. The derivation illustrates that we can think of these closed loops as worldlines of the Abelian anyons. As the name suggests, the $O(1)$ loop model is a special case of the $O(N)$ loop model, where each loop component\footnote{For loop models with intersections, one has to take care with how to define components; for $N=1$ this issue does not arise.} is weighted by an additional topological factor $N$~\cite{peled2019lectures}. We argue that it is not a coincidence that here we obtain $N=1$ for an Abelian anyon. As we will see in Sec.~\ref{sec:D4_X}, for non-Abelian anyons with non-trivial quantum dimension $d$, a loop model with $N=d$ can arise. We explore this for a broad range of non-Abelian TOs in companion paper \cite{short_paper}.

\begin{figure}
    \centering
    \includegraphics[width=\linewidth]{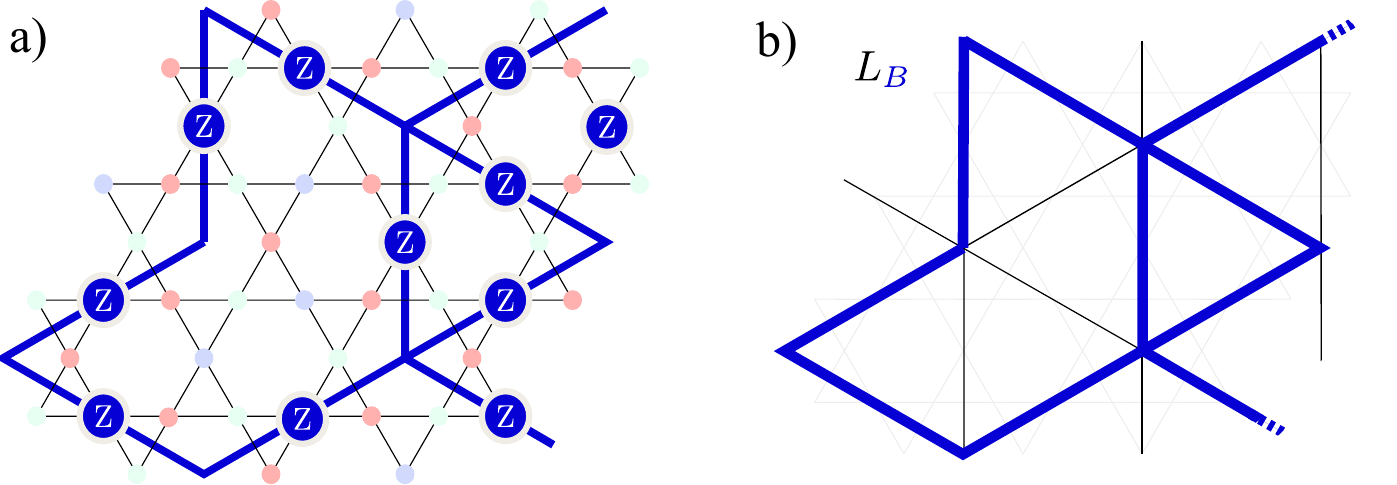}
    \caption{\textbf{Abelian anyons moving on triangular lattice.} Abelian charges $e_{\B{B}}$ move on the blue triangular lattice corresponding to the center of the star operators $\B{A_s}$. Panel a shows a closed loop configuration formed by the action of local Pauli $Z$ acting on $\B{\mathcal{R}_B}$. These form a closed (intersecting) loop configuration on the triangular lattice (panel b).}
    \label{fig:Abelian_triangle}
\end{figure}

Let us explore the phase diagram of Eq.~\eqref{eq:Z1_betaB}. When the perturbation is small and positive ($0 \leq \beta_\B{B}^z \ll 1$), we see that the string tension $t_\B{B} \ll 1$ penalizes loops. This small-loop phase corresponds to the initial $D_4$ TO. As we increase the perturbation, we will eventually cross a critical tension $t_c$, beyond which loops proliferate at all scales. Physically, this corresponds to a condensate of $e_\B{B}$. To infer the resulting phase of matter, let us consider the $\beta^z_\B{B}\to \infty$ limit and examine the local terms $A_s$, $B_t$ in the Hamiltonian (Eq.~\eqref{eq:H_D4}) for each of the three colored sublattices. In this limit, all qubits in the blue sublattice $\B{\mathcal{R}_B}$ are projected into the $\ket{\uparrow}$ state, and hence all $\B{B_t}$'s become trivial. Physically, this property reflects the confinement of $m_\B{B}$ fluxes by condensing $e_\B{B}$ charges. Of the remaining four types of $A_s, B_t$ ``stabilizers'', the red $\R{B_t}$ and green $\G{B_t}$ triangle operators are left untouched, while in the limit $\beta^z_\B{B}\to \infty$ all controlled-$Z$ gates in $\R{A_s}$, $\G{A_s}$ become identity\footnote{This implies the trivialization of the linear-depth circuit necessary for the the Wilson operators $\R{\mathcal{X}}_{t_i}^{t_f}$, $\G{\mathcal{X}}_{t_i}^{t_f}$ to produce a pair of non-Abelian fluxes in the undeformed $D_4$ state, since they now become Abelian anyons.}. We therefore obtain two decoupled sets of (commuting) stabilizers $\R{A_s}, \R{B_t}$ and $\G{A_s}, \G{B_t}$ for the red and green sublattices, respectively, corresponding to \emph{two decoupled copies of the toric code}. In particular, the resulting phase of matter\footnote{\label{foot:subtlety}In the present work we do not focus on the details of these beyond-treshold phases, but see Ref.~\onlinecite{sahay_2025} for subtleties that arise.} is thus Abelian.

It turns out the critical value of the string tension is $t_c = 2-\sqrt{3} \approx 0.268 $, i.e., $\beta_c = \frac{\ln 3}{4}$, and at this point the system is described by the Ising universality class. To see this, it is useful to note that the $O(1)$ loop model Eq.~\eqref{eq:Z1_betaB} can be rewritten as an \emph{exact} high-temperature expansion of the Ising model on the triangular lattice with inverse temperature $\beta^z_\B{B}$, for which the critical temperature is known via the Kramers-Wannier duality and star-triangle relation \cite{HOUTAPPEL1950425}. In fact, we can directly derive the Ising model by using the \emph{ungauging} transformation in Eq.~\eqref{eq:ungauge_Z}:
\begin{align} 
&\mathcal{Z}_{\ket{\psi}}(\beta^z_\B{B})=\langle D_4 | e^{\beta^z_\B{B} \sum_{j \in \B{\mathcal{R}_B}} Z_j}| D_4\rangle \\& = \frac{1}{2^{\mathcal{N}_{\B{B}}}}\sum_{\tilde{\sigma} }e^{\beta^z_\B{B} \sum_{ \langle b,b' \rangle_{\triangle}}\tilde{\sigma}_b \tilde{\sigma}_{b'}}.
\label{eq:ZB_Ising}
\end{align}
The phase where $e_{\B{B}}$ anyons condense corresponds to the ordered phase. Indeed, the two-point correlator $\avg{{\color{blue} \tilde{Z}}_b {\color{blue} \tilde{Z}}_{b'}}$ on the blue sublattice which diagnoses long-range magnetic order maps to the Wilson operator $\avg{\B{\mathcal{Z}}_{s_i}^{s_f}}$ under the (un)gauging map.

While one benefit of the loop model representation \eqref{eq:Z1_betaB} is that it makes manifest the physics of the model (i.e., proliferation of anyon worldlines), one advantage of the Ising model representation \eqref{eq:ZB_Ising} is that the partition function remains manifestly positive even if $\beta^z_\B{B} <0$. This introduces \emph{frustration} to the triangular lattice Ising model. This can also be seen in the loop model picture: different worldlines can now destructively interfere, such that anyon condensation is more difficult to achieve\footnote{We note this physical mechanism has also recently been explored from a Hamiltonian perspective in Ref.~\cite{unifyingQSL}, where it was found to lead to an infinitely robust spin liquid.}. The consequence is that the $D_4$ phase is robust for any \emph{finite} negative perturbation strength. In the limit $\beta_{\B{B}}^z \to -\infty$, the system becomes critical, described by a conformal field theory (CFT) with central charge $c = 1$~\cite{triang_Ising,blote_93}. Indeed, in this limit, it is well-known that the ground state manifold of the triangular lattice Ising model can be mapped onto a dimer model on the honeycomb lattice, since each triangle has exactly one frustrated bond \cite{Nienhuis84b,Moessner00,RK88,Read91,Fisher63}. Since the honeycomb lattice is bipartite, this gives rise to a $U(1)$ gauge theory with algebraic correlations. Interestingly, similarly to the ferromagnetic case, it can be shown that the remaining (green and red) sublattices define two decoupled toric codes, coexisting with the gapless degrees of freedom residing on the blue sublattice.

\vspace{50pt}

Lastly, we note that very similar results follow when considering Abelian deformations on two different sublattices---say, blue and green. For instance, in the strongly-ferromagnetic ($t_\B{B},t_\G{G}\gg 1$) case, the system becomes a single copy of the toric code on the honeycomb lattice with stabilizers given by $\R{A_s}$ and $\R{B_t}$ after setting $\B{Z_b}=\G{Z_g}=+1$. 

\subsection{Decohered mixed state} \label{sec:Z_dech}

 Next we consider deformations implemented by a composition of local $Z$ quantum channels (i.e., completely positive trace preserving maps) $\mathcal{E}_j(\rho_0)=(1-p_{\B{B}})\rho_0 + p_{\B{B}}Z_j \rho_0 Z_j$, with $p_{\B{B}}$ the error rate and $\rho_0=\ket{D_4}\bra{D_4}$ the pure-state density matrix for $D_4$ topological order.  Allowing for such errors on all sites $j$ on the blue sublattice $\B{\mathcal{R}_B}$, the density matrix evolves according to
 \begin{equation}
 \begin{aligned}
    &\rho_0 \to \rho=\mathcal{E}(\rho_0) = \prod_j \mathcal{E}_j(\rho_0).
    \end{aligned}
\end{equation}

 Unlike for ground states, the meaning of a mixed state phase---even an operational definition---is an active area of research. From the perspective of quantum error correction, one cares about the amount of quantum information preserved under the action of a quantum channel (as measured by the coherent information)~\cite{Dennis_2002, fan2023diagnostics,LeeYouXu2022}. Alternatively, one might characterize the decohered density matrix $\rho$ according to the existence, or lack thereof, of an unraveling in terms of short-range entangled states~\cite{chen2023separability,Chen_2024}. These two characterizations correspond to the more ``intrinsic'' properties of $\rho$. Yet, such quantities are generally difficult to calculate; see, e.g., Ref.~\onlinecite{lee2024exactcalculationscoherentinformation}. Recently, various works~\cite{bao2023mixedstate,ashida2023systemenvironment,Chen_2024,LeeYouXu2022,lee2022symmetry} have considered looking for singularities of the moments $\textrm{tr}(\rho^n)$ of the density matrix (of which the purity $\textrm{tr}(\rho^2)$ is typically the simplest moment), since these moments determine the full spectrum\footnote{Specht's theorem says that two hermitian matrices $A$ and $B$ are unitarily equivalent if and only if $\textrm{tr}(A^n) = \textrm{tr}(B^n)$ for all $n \in \mathbb N$.} of $\rho$. This approach provides valuable insight, although it may not suffice for defining a mixed state phase~\cite{Lessa_24}. 
 (In the context of strong-to-weak spontaneous symmetry breaking, Ref.~\onlinecite{Lessa_24} showed that mixed state phases can instead be characterized by the behavior of the quantum fidelity.) In fact, both the universality of the transition as well as its location will in general depend on $n$. This property stands to reason: For example, when computing $\textrm{tr}(\rho^n)$ for a thermal density matrix, the inverse temperature is amplified by a factor of $n$. Following the above references, in this subsection we characterize the effect of $Z$ errors on $D_4$ topological order via the purity $\textrm{tr}(\rho^2)$, delaying a more detailed discussion of the ``intrisic'' threshold to later sections where we will also focus on the more interesting case of proliferating non-Abelian anyons.
 
 Although it is not essential and an alternative derivation will be discussed in later sections, a similar approach to the deformed wavefunction (Sec.~\ref{sec:Z_purewf}) can be applied to a decohered density matrix $\rho$.  To this end, we vectorize $\rho$ by mapping it to a vector $\kket{\rho}\in \mathcal{H}\otimes\mathcal{H}$ in a doubled Hilbert space (a detailed explanation can be e.g., found in Appendix B.2 of Ref.~\onlinecite{Chen_2024}).  The quantum channel then becomes a non-negative (and non-unitary) operator acting in the doubled Hilbert space.
 Upon vectorizing the pure-state density matrix~\cite{Chen_2024} as $\kket{\rho_0}=\ket{D_4}\otimes\ket{D_4}$, the vectorized decohered density matrix reads
\begin{equation} 
    \kket{\rho(p_{\B{B}})}=\kket{\mathcal{E}(\rho_0)}=\prod_{j \in \B{\mathcal{R}_B}} \left(1-p_{\B{B}} + p_{\B{B}} Z_j\otimes Z_j\right)\kket{\rho_0}.
\end{equation}
Equivalently, we can write
\begin{equation} 
    \kket{\rho(p_{\B{B}})} \propto e^{\mu_{\B{B}} \sum_{j \in \B{\mathcal{R}_B}} Z_j\otimes Z_j}\kket{\rho_0}
\end{equation}
with $\tanh(\mu_{\B{B}})=\frac{p_{\B{B}}}{1-p_{\B{B}}}$, which resembles the form of the deformed pure state $\ket{\psi(\beta^z_{\B{B}})}$. 

The norm of $\kket{\rho(p_{\B{B}})}$ corresponds to the purity $\textrm{tr}(\rho^2)$ and can be written as:
\begin{equation}
\begin{aligned}
    &{\rm tr} ( \rho^2)=\langle \! \langle \rho(p_{\B{B}})\kket{\rho(p_{\B{B}})}\\
    & \propto \bbra{\rho_0}\prod_{j \in \B{\mathcal{R}_B}} \left(1 + r_{\B{B}} Z_j\otimes Z_j\right)\kket{\rho_0}.
    \end{aligned}
\end{equation}
Here

$r_{\B{B}}\equiv \frac{2p_{\B{B}}(1-p_{\B{B}})}{(1-p_{\B{B}})^2 +p_{\B{B}}^2}$, which is invariant under $p_{\B{B}} \to 1-p_{\B{B}}$. Adopting the same steps as used to derive Eq.~\eqref{eq:Z1_betaB}, we again find the partition function of a $O(1)$ loop model on a triangular lattice, 
\begin{equation}
    \textrm{tr}\left(\rho^2\right)
    \;
    \overset{\rm decohere}{\underset{e_\B{B}}{\propto}}
    \;
    \sum_{L_{\B{B}}} r_{\B{B}}^{|L_{\B{B}}|},
\end{equation}
with tension given by $r_{\B{B}}\in [0,1]$. Notice that in this case $\mu_\B{B}$ ($r_\B{B}$) is always non-negative, and hence there is no analog of the antiferromagnetic case discussed in the previous version. Similar to the deformed wavefunction, this decohered mixed state showcases a finite-temperature phase transition separating a disordered from a ferromagnetically ordered phase at a finite $p_\B{B}<1/2$. This transition indicates that beyond a finite error threshold $p_c^{(n=2)}=\frac{1}{2} \left( 1 - 3^{-\frac{1}{4}}\right) \approx 0.12$, the $D_4$ topological order is lost as measured by the purity.

For completeness, we note that following the usual computations demonstrated for the square lattice \cite{chen2023separability,lee2024exactcalculationscoherentinformation}, one can show that the spectrum of $\rho$ is given by the random-bond Ising model (RBIM)~\cite{SCHWARTZ1980115,Harris_1974,Nishimori_81,Ozeki_1993,doussal_88,Honecker_2001,Gruzberg_2001,deQueiroz06} on the honeycomb lattice with inverse temperature $\beta =\frac{1}{2} \ln \frac{1-p}{p}$, or equivalently, $p=\frac{1}{1+e^{2\beta}}$. More precisely, different eigenvalues are labeled by different disorder realizations of the RBIM, each of which can undergo a critical point. In fact, the largest eigenvalue of this ensemble corresponds to the \emph{clean} Ising model, which on the honeycomb lattice has a critical $\beta_c = \frac{\ln(2+\sqrt{3})}{2}$. This informs us that $\lim_{n \to \infty} {\rm tr}(\rho^n)$ has a transition at $p_c^{(n=\infty)} = \frac{1}{3+\sqrt{3}} \approx 0.21$. If one instead takes the whole ensemble into consideration, it turns out the weighting is such that the disorder probability $p$ coincides with the decoherence rate $p$. This means that the von Neumann entropy then corresponds to the quenched average free energy of the 2D RBIM along the Nishimori line~\cite{Nishimori_81}. On the honeycomb lattice this is known to occur at $p_c^{(n=1)} \approx 0.068$ \cite{deQueiroz06}. Note that the critical values grow monotonically with the above Rényi indices ($n=1,2,\infty$), which has also been observed in earlier work~\cite{fan2023diagnostics}. However, we will argue this does not need to hold for non-Abelian cases.

\section{ $X$ flip error: non-Abelian anyon proliferation}
\label{sec:D4_X}
We now consider deforming the $D_4$ topological order by applying local $X$ Paulis. We first consider a single sublattice (e.g., $\R{\mathcal{R}_R}$), since together with the previous section, this analysis will give us the main insights to obtain the corresponding stat-mech models for general (commuting) deformations. As in the Abelian case, we will first consider pure wavefunction deformation (Sec.~\ref{sec:X_puredef}) as a warm-up before advancing to the mixed state set-up (Sec.~\ref{subsubsec:X_dech}). 

\subsection{Pure wavefunction deformation} \label{sec:X_puredef}

As we saw in Sec.~\ref{sec:D4_main}, $X$ Paulis createss pairs of non-Abelian $m_{\R{R}}$ charges with quantum dimension $d=2$. The red sublattice $\R{\mathcal{R}_R}$ itself forms a (super)kagome lattice. It will be useful to interpret this as the medial lattice of a honeycomb lattice, such that the red qubits live on its bonds (see Fig.~\ref{fig:D4_TO}b). Then the non-Abelian $m_\R{R}$'s lie on the vertices of this honeycomb lattice.
We consider a non-unitarily deformed unnormalized wavefunction
\begin{equation} \label{eq:X_puredef}
    |\psi(\beta^x_{\R{R}} )\rangle  = e^{\frac{ \beta^x_{\R{R}}}{2} \sum_{j\in \R{\mathcal{R}_R}} X_j}\ket{D_4}.
\end{equation}
Following precisely the same steps as for $Z$ errors in the previous section, the norm can be expressed as 
\begin{equation}
\begin{aligned} \label{eq:X_r}
    &\mathcal{Z}_{\ket{\psi}}(\beta^x_{\R{R}})= \langle D_4|e^{ \beta^x_{\R{R}} \sum_{j \in \R{\mathcal{R}_R}} X_j}|D_4\rangle
    \\
    &\propto \langle D_4|\prod_{j \in \R{\mathcal{R}_R}}\left[1+ \tanh(\beta^x_{\R{R}}) X_j \right] |D_4\rangle 
    \\ &= \sum_{n_{\R{R}}=0}^{|\R{\mathcal{R}_R}|}  \tanh(\beta^x_{\R{R}})^{n_{\R{R}}} \sum_{\{r_j\}_{j=1}^{n_\R{R}}}\langle D_4|\prod_{j=1}^{n_{\R{R}}} X_{r_j} |D_4\rangle.
\end{aligned}
\end{equation}

\begin{figure}
    \centering
    \includegraphics[width=0.85\linewidth]{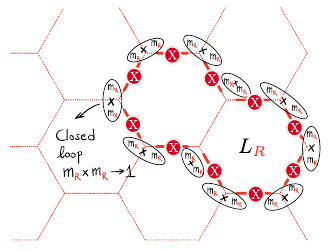}
    \caption{\textbf{Closed loop condition.} Every local $\R{X_r}$ acting on the red sublattice creates a pair of $m_\R{R}$ fluxes. Finite contributions in the partition function \eqref{eq:X_r} correspond to fusing pairs of nearby fluxes $
    {m_\R{R}\times m_\R{R}} 
$ into the identity channel $m_\R{R}\times m_\R{R}\to 1$. For anyons belonging to different pairs, this fusion process occurs with probability $1/d^2_{m_{\R{R}}}$, giving rise to a finite tension in the loop model. Every closed string moreover contributes with a topological factor given by the quantum dimension to the number of connected components of the loop configuration $2^{C_{L_{\R{R}}}}$. }
    \label{fig:mR_schematic}
\end{figure}

To proceed we need to evaluate $ \langle D_4|\prod_{j=1}^{n_{\R{R}}} X_{r_j}  |D_4\rangle$ with $\{r_j\}$ a collection of $n_{\R{R}}$ vertices.  
Let us first intuitively understand which configurations yield a nonzero contribution.  Similar to the case of $Z$ errors, the expectation value vanishes unless $\prod_{j=1}^n X_{r_j}$ forms a closed-loop configuration $L_{\R{R}}$. Indeed, any open strings yield unfused non-Abelian anyons in the $D_4$ ket, leading to orthogonality with the $D_4$ bra. We thus have:
\begin{equation} \label{eq:eq_21}
\mathcal{Z}_{\ket{\psi}}(\beta^x_{\R{R}}) \propto \sum_{L_{\R{R}}} \tanh(\beta^x_{\R{R}})^{|L_{\R{R}}|} f(L_\R{R}),
\end{equation}
with $L_{\R{R}}$ any contractible\footnote{{Only contractible loops will contribute, as otherwise the eigenvalue of the logical $\R{\mathcal{Z}}$ operator will be toggled when considering the initial state $\ket{D_4}$. This condition is relaxed when one instead considers e.g., the superposition $\ket{D_4} + \R{\mathcal{X}}\ket{D_4}$.}} closed-loop configuration, and $f(L_\R{R})\equiv\langle D_4|\prod_{j\in L_{\R{R}}} X_{j} |D_4\rangle$.  

Similarly to the Abelian case, we again find a loop model\footnote{The reason we find a loop model instead of a trivalent graph is that $m_\R{R}$ does not appear in the fusion $m_\R{R}\times m_\R{R}$.} (albeit now on the honeycomb lattice), and $\tanh(\beta^x_{\R{R}})$ contributes to the string tension. However, unlike the Abelian case, we now have the remaining expectation value $ \langle D_4|\prod_{j\in L_{\R{R}}} X_{j} |D_4\rangle$. In the Abelian case, this would simply be unity since it is the expectation value of a (product of) ground state stabilizer(s). This is not the case now: only the expectation value of the true non-Abelian anyon string operator would give a unity value in the ground state, but that is a complicated linear-depth circuit (see Eq.~\eqref{eq:X_if}). Instead we can think of $f(L_\R{R})$ as the overlap between the ground state $\bra{D_4}$ and a state $\prod_{j\in L_{\R{R}}} X_{j} |D_4\rangle$ where we have first created $|L_{\R{R}}|/2$ pairs of adjacent $m_{\R{R}}$ fluxes which are then pair-wise fused as sketched in Fig.~\ref{fig:mR_schematic}. However, this fusion can contain non-trivial anyons:
\begin{equation}
m_{\R{R}} \times m_{\R{R}} = 1 + e_{\B{B}} + e_{\G{G}} + e_{\B{B}} \times e_{\G{G}} \; .
\end{equation}
Only the `1' term will give a non-zero overlap with the ground state. We thus expect $f(L_{\R{R}}) < 1$.

We will momentarily give a physical way of deriving $f(L_\R{R})$ using the fusion-based interpretation, but let us first show the algebraic way. We exploit the ungauging and disentangling maps introduced in Sec.~\ref{sec:mapping_main}, which tells us:
\begin{align} \label{eq:f_LR_pre}
f(L_\R{R}) 
&= \bra{\G{\tilde{+}}}^{\otimes |\G{\mathcal{R}_G}|}_{\G{g}}\bra{\B{\tilde{+}}}^{\otimes |\B{\mathcal{R}_B}|}_{\B{b}} \!\!\! \!\prod_{\footnotesize \langle \G{g},\B{b} \rangle \in L_\R{R}} \!\!\! \widetilde{\CZ}_{\G{g}\B{b}} \ket{\G{\tilde{+}}}^{\otimes |\G{\mathcal{R}_G}|}_{\G{g}}\ket{\B{\tilde{+}}}^{\otimes |\B{\mathcal{R}_B}|}_{\B{b}} \\
&= \prod_{\ell_\R{R} \in L_\R{R}} \frac{1}{2^{|\ell_\R{R}|}}\; \textrm{tr}\left( \prod_{n=1}^{|\ell_{\R{R}}|} \widetilde{\CZ}_{n,n+1} \right), \label{eq:trCZ}
\end{align}
where we have decomposed $L_\R{R} = \bigoplus \ell_{\R{R}}$ into its connected components, and where the trace is over a circuit of $\CZ$'s on a ring with periodic boundary conditions. We note this expression was also obtained in a study of symmetry-protected topological phases~\cite{zhang2024strangecorrelationfunctionaverage}. Focusing on a single component and by writing Eq.~\eqref{eq:trCZ} as as tensor network (see Appendix~\ref{app:X_error}), one can straightforwardly show that
\begin{equation} \label{eq:fL}
f(L_\R{R}) = \frac{1}{\sqrt{2}^{|L_\R{R}|}}\;\textrm{tr}(H^{|L_\R{R}|}) = \frac{\textrm{tr}(\mathds 1)}{\sqrt{2}^{|L_\R{R}|}} = \frac{2}{\sqrt{2}^{|L_\R{R}|}},
\end{equation}
where $H = \frac{X+Z}{\sqrt{2}}$ is the Hadamard matrix and we used that $|L_\R{R}|$ is automatically even on a honeycomb lattice.

Putting everything together, we have obtained that the wavefunction norm is described by the following honeycomb loop model:
\begin{equation} \label{eq:main_ZX}
\mathcal{Z}_{\ket{\psi}}(\beta_{\R{R}}^x)
\propto \sum_{L_{\R{R}}} t_\R{R}^{|L_{\R{R}}|} \; 2^{C_{L_{\R{R}}}},
\end{equation}
with a string tension $t_\R{R}=\frac{\tanh(\beta^x_{\R{R}})}{\sqrt{2}}$ and where $C_{L_{\R{R}}}$ counts the number of loops (or equivalently, connected components) in the closed-loop configuration $L_{\R{R}}$. Consider for example the loop configuration in Fig.~\ref{fig:mR_schematic} with only one loop, and hence $C_{L_{\R{R}}}=1$.

A more physical way of deriving the above expression for $f(L_\R{R})$ is as follows. As we discussed, the (normalized) wavefunction $|L_{\R{R}}\rangle \equiv \prod_{j\in L_{\R{R}}} X_{j} |D_4\rangle$ includes all possible outcomes resulting from the fusion of two nearby fluxes $m_\R{R}$, one of which leads to a term proportional to the ground state $\ket{D_4}$. To obtain its numerical prefactor, we can use the fact that different anyon pairs are created separately, and then the fusion outcomes of two (causally disconnected) anyons belonging to two different pairs occur with a probability weighted by the quantum dimension of the fusion product~\cite{Shi_2020,Preskill_LN}, i.e., $p(m_\R{R}\times m_{\R{R}}\to a)=d_a/d^2_{m_{\R{R}}}$. Moreover, for every connected component of the closed loop configuration $L_\R{R}$, an additional $d_{m_\R{R}}$ appears. This is a consequence of the fact that once $|L_{\R{R}}|-1$ such pairs have fused into the vacuum, then the remaining pair is necessarily in the trivial channel, since all pairs were created from the vacuum (indeed, now the remaining anyons are no longer causally disconnected). Hence, we find
\begin{equation}
    |L_{\R{R}}\rangle = \frac{2^{C_{L_{\R{R}}}}}{\sqrt{2}^{|L_{\R{R}}|}}\ket{D_4} + \cdots
\end{equation}
where $\cdots$ is orthogonal to $\ket{D_4}$. Clearly $f(L_\R{R}) = \langle D_4 | L_\R{R} \rangle$ exactly picks up this prefactor, arriving at Eq.~\eqref{eq:fL}. The reason why here we find $\sqrt{d_{m_{\R{R}}}}^{|L_{\R{R}|}}$ rather than $d_{m_{\R{R}}}^{|L_{\R{R}}|}$ is because for this microscopic model a pair of nearest $m_\R{R}$ anyons can only fuse into either the trivial or an Abelian charge $e_c$ whose color $c=\G{G}, \B{B}$ depends on the lattice site. Indeed, more generally the length-dependent factor can depend on microscopic details, but the topological piece of $d_{m_{\R{R}}}^{C_\R{R}}$ which depends on the number of components should be universal.

Equation~\eqref{eq:main_ZX} is the partition function for an $O(2)$ loop model with tension $t_\R{R}$ defined on the $\R{\mathcal{R}_R}$ honeycomb lattice~\cite{Nienhuis_81,Nienhuis_82,duminilcopin2020macroscopicloopsloopon}. Since all loop configurations have even length $|L_{\R{R}}|$, the result does not depend on the sign of $\beta_{\R{R}}^x$.
It is useful to contrast the O$(2)$ loop model in Eq.~\eqref{eq:main_ZX}  to the O$(1)$ loop model we obtained in Eq.~\eqref{eq:Z1_betaB} when proliferating Abelian $e_\B{B}$ charges via $\B{Z}$ deformation. Unlike for the latter, proliferating non-Abelian fluxes with quantum dimension $d=2$ gives rise to an additional topological factor $2^{C_{L_{\R{R}}}}$ that enhances the probability of having a larger number of disconnected loops than in the Abelian case. In fact, this observation is key to understand the robustness of the $\ket{D_4}$ ground state to the proliferation of $m_{\R{R}}$ fluxes.

In general, O$(N)$ loop models with \emph{loop weight} $N \in [-2,2]$ showcase two different phases separated by a critical point at the critical tension $t_c(N)=(2+\sqrt{2-N})^{-1/2}$ \cite{Nienhuis_82}: A dilute (or small loop) phase for $t<t_c(N)$, and a dense phase for $t>t_c(N)$. For $N=2$ the critical point at $t_c(2)=1/\sqrt{2}$ is described by a BKT transition which extends into an extended gapless phase described by a Luttinger liquid with central charge $c=1$ (see also Ref.~\cite{peled2019lectures}). Since $t_\R{R}\leq 1/\sqrt{2}$ for all $\beta^x_{\R{R}}\in [0,\infty)$, we find that \emph{$\ket{D_4}$ is robust to arbitrary large deformations of the type given in Eq.~\eqref{eq:X_puredef}}, and turns into a critical state in the projective limit $\beta^x_{\R{R}} = \infty$ (where each red qubit is projected into $\R{X} = 1$). In summary, its phase diagram is:
\begin{center}
\begin{tikzpicture}
\draw[->] (0,0) -- (6.5,0) node[right] {$\beta_{\R{R}}^x$};
\draw[-,line width=3,color=white] (5.4,-0.1) -- (5.5,0.1);
\draw[-] (5.45,-0.1) -- (5.55,0.1);
\draw[-] (5.35,-0.1) -- (5.45,0.1);
\filldraw[black] (0,0) circle (2pt) node[below] {$0$};
\filldraw[red] (6,0) circle (2pt) node[below] {$\infty$} node[above] {\small BKT};
\node[above] at (3,0) {$D_4$ topological order};
\end{tikzpicture}
\end{center}

\subsection{From loop to local spin models} \label{sec:Ising_OP}

\begin{figure}
    \centering
    \includegraphics[width=\linewidth]{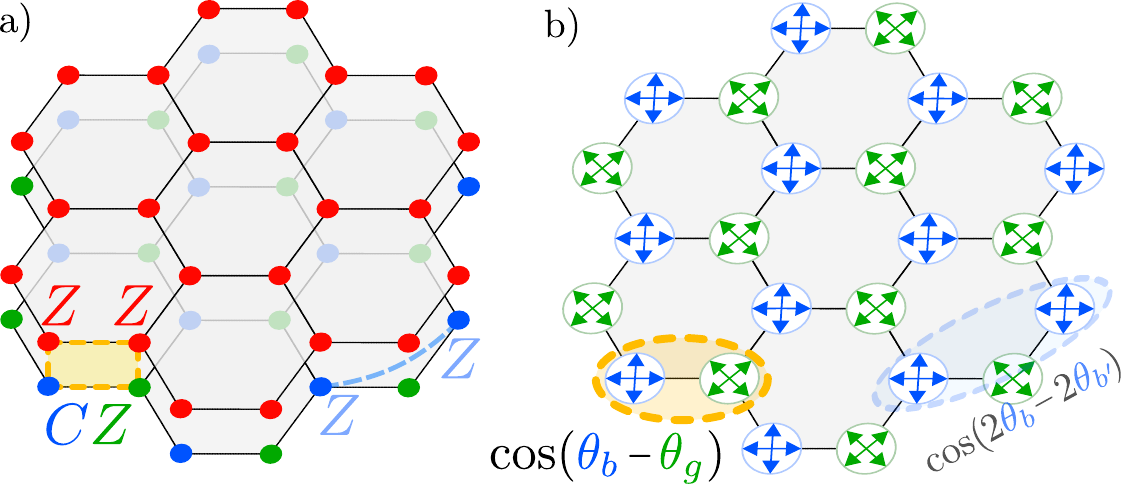}
    \caption{\textbf{Explicitly local stat-mech models for deformed wavefunction.} Formulation of the explicitly local stat-mech models in Secs.~\ref{sec:Ising_OP} and \ref{sec:comb_purewf} to characterize the phase diagram of the deformed $D_4$ wavefunction with non-Abelian $m_\R{R}$ anyons. Panel (a) corresponds to the formulation in terms of Ising variables on two honeycomb layers, while panel (b) is given in terms of a single-layer four-state clock model.}
    \label{fig:bil_honey}
\end{figure}

As in the Abelian case, it is useful to characterize the system by a local statistical mechanics model, as it can reveal hidden symmetries of the system and will also provide insight into the decohered case. As before, such a model can be derived by applying the ungauging maps directly at the level of the wavefunction.
In Sec.~\ref{sec:mapping_main} we saw how the $D_4$ state maps to a honeycomb toric code on the red lattice. Using the constraint $\R{A}_s=\prod_{j\in \R{\hexagon}}X_j=+1$ on each plaquette, we can moreover dualize this toric code to a trivial paramagnet on the vertices of the honeycomb lattice. This procedure leads to (compare to Eq.~\eqref{eq:ungauge_X}):
\begin{equation} \label{eq:map_X_red}
\begin{aligned}
     &\ket{\R{\textrm{T.C.}}}_{\R{\mathcal{R}_R}}\to \bigotimes_v\ket{\R{+}}_v,\\
    &\R{X}_\ell \widetilde{\CZ}_{\B{b},\G{g}} \to \R{Z}_b \R{Z}_g \widetilde{\CZ}_{\B{b},\G{g}}.
\end{aligned}
\end{equation} 
Hence, through these mappings, the original kagome lattice is mapped to a bilayer honeycomb lattice with red $\R{\sigma}=\pm 1$ Ising variables on the top layer, and $\B{\sigma}_b, \G{\sigma}_g$ on the bottom one with $g\in \G{G}$ and $b\in \B{B}$ denoting the two different sublattices (see Fig.~\ref{fig:bil_honey}a).

In summary, we can write the partition function obtained from the self-overlap of the deformed $\ket{D_4}$ wavefunction \eqref{eq:X_puredef}, as an \emph{Ising-like} partition function:
\begin{equation}
\begin{aligned}
\label{eq:main_tildeZ_x}
    {\mathcal{Z}}_{\ket{\psi}}(\beta^x_{\R{R}})&=\sum_{\{\sigma, \tilde{\sigma}\} }e^{-\beta_{\R{R}}^x H_{\ket{\psi}}(\sigma,\tilde{\sigma})},
\end{aligned}
\end{equation}
with the bilayer honeycomb Hamiltonian 
\begin{equation} \label{eq:main_H_X_I}
    H_{\ket{\psi}}=-\sum_{\langle i,j \rangle_{\hexagon}}\R{\sigma}_i \R{\sigma}_j\widetilde{\CZ}_{i,j},
\end{equation}
where $\widetilde{\CZ}_{ij} = \frac{1}{2} \left( 1+\tilde \sigma_i+\tilde \sigma_j-\tilde \sigma_i\tilde \sigma_j\right) \in \{-1,1\}$; the interaction is shown in Fig.~\ref{fig:bil_honey}a. It can be shown that an exact high-temperature expansion gives a closed-loop model on the honeycomb model~\footnote{{Notice that this high-temperature expansion includes both contractible and non-contractible loop configurations.}}---similar to the Ising case---but the dressing on the bottom layer $\widetilde{\CZ}_{ij}$ leads to an additional loop factor given by Eq.~\eqref{eq:trCZ}, which thus gives an $O(2)$ loop model in agreement with Eq.~\eqref{eq:main_ZX}.

Crucially, the spin Hamiltonian \eqref{eq:main_H_X_I} has an internal $D_4$ symmetry. To make this symmetry manifest, it is useful to repackage this honeycomb \emph{bilayer} of \emph{Ising} spins as a \emph{single} honeycomb lattice with \emph{two-component} spins\footnote{Intriguingly, the first equality can be interpreted as a mixed model where blue sites have `face-cubic' spins and green sites have `corner-cubic' spins. It is known that such spins lead to local spin models for $O(N)$ honeycomb loop models~\cite{Nienhuis_81,Chayes_2000}, but usually this only covers the range of tension $t \leq \frac{1}{n}$. Our mixed model gives a manifestly positive stat-mech model for $t \leq \frac{1}{\sqrt{n}}.$}:
\begin{equation} \label{eq:main_4clock}
H_{\ket{\psi}}=-\sqrt{2} \sum_{\langle b,g\rangle} \B{\boldsymbol n}_b \cdot \G{\boldsymbol n}_g =-\sqrt{2} \sum_{\langle b,g\rangle}\cos(\B{\theta_b}-\G{\theta_g})\; ,
\end{equation}
where
\begin{equation} \label{eq:nbng}
\B{\boldsymbol n}_b = \frac{1}{2} \left( \begin{array}{c} 
\R\sigma_b + \R\sigma_b \B{\tilde \sigma}_b \\
\R\sigma_b - \R\sigma_b \B{\tilde \sigma}_b 
\end{array} \right)
\textrm{ and }
\G{\boldsymbol n}_g = \frac{1}{\sqrt{2}} \left( \begin{array}{c} 
\R \sigma_g \\
\R \sigma_g \G{\tilde \sigma}_g
\end{array} \right),
\end{equation}
has been used in the first equality.
Moreover, since these vectors are normalized (i.e., $|\B{\boldsymbol n}_b|= |\G{\boldsymbol n}_g| = 1$), we can interpret them as taking values on the unit circle, although their discrete values depend on whether we are on the blue or green sublattice:
\begin{equation} \label{eq:theta_gb}
\B{\boldsymbol n}_b \in \; 
\raisebox{-11pt}{\begin{tikzpicture}
\draw[opacity=0.5] (0,0) circle (0.5);
\draw[->,blue] (0.05,0) -- (0.5,0);
\draw[->,blue] (0,0.05) -- (0,0.5);
\draw[->,blue] (-0.05,0) -- (-0.5,0);
\draw[->,blue] (0,-0.05) -- (0,-0.5);
\end{tikzpicture}}
\qquad \textrm{and} \qquad
\G{\boldsymbol n}_g \in \; 
\raisebox{-11pt}{\begin{tikzpicture}
\draw[opacity=0.5] (0,0) circle (0.5);
\draw[->,DarkGreen] (0.04,0.04) -- (0.353,0.353);
\draw[->,DarkGreen] (0.04,-0.04) -- (0.353,-0.353);
\draw[->,DarkGreen] (-0.04,0.04) -- (-0.353,0.353);
\draw[->,DarkGreen] (-0.04,-0.04) -- (-0.353,-0.353);
\end{tikzpicture}}
\; \; .
\end{equation}
Hence, we used the angle variables
\begin{equation} \label{eq:n_theta}
\B{n_b}=\left( \begin{array}{c}\cos(\B{\theta_b})\\\sin(\B{\theta_b})\end{array} \right) \textrm{ and }\G{n_g}=\left( \begin{array}{c}\cos(\G{\theta_g})\\ \sin(\G{\theta_g})\end{array} \right) 
\end{equation}
on the second equality of Eq.~\eqref{eq:main_4clock}.
These representations make clear that $H_{\ket{\psi}}$ is symmetric under an internal $D_4 \cong \mathbb Z_4 \rtimes \mathbb Z_2 \subset O(2)$ symmetry. More precisely, $\B{\boldsymbol n}_b \cdot \G{\boldsymbol n}_g$ is invariant if each $\bm n_j$ transforms under the symmetry group of the square, $D_4 \cong \langle R,S | R^4=S^2 =1, SRS = R^{-1} \rangle$, where $R$ rotates the vector counterclockwise by $90^\circ$ and $S$ mirrors top and bottom:

\begin{center}
\begin{tikzpicture}
\node at (0,0){
    \begin{tikzpicture}
    \draw[opacity=0.5] (0,0) circle (0.5);
    \draw[->] (0.6,0) to[out=90,in=-10] (0,0.6);
    \node at (0.7,0.4) {$R$};
    \draw[-,dashed,opacity=0.25] (0,0) -- (0.7,0);
    \draw[-,dashed,opacity=0.25] (0,0) -- (0,0.7);
    \end{tikzpicture}
};
\node at (3,0){
    \begin{tikzpicture}
    \draw[opacity=0.5] (0,0) circle (0.5);
    \draw[-,dashed,opacity=0.25] (-0.7,0) -- (0.7,0);
    \draw[<->] (0.6,-0.3) -- (0.6,0.3);
    \node at (0.85,0) {$S$};
    \end{tikzpicture}
};
\end{tikzpicture}
\end{center}
On the two-component vector, $R$ is represented by $-i \sigma^y$, and $S$ by $\sigma^z$, defining a two-dimensional irrep of $D_4$. From Eq.~\eqref{eq:nbng}, we infer that the $D_4$ symmetry acts as follows in the original variables of Eq.~\eqref{eq:main_H_X_I}:
\begin{equation} \label{eq:SR}
\begin{aligned}
    S:\, \left\{ \begin{array}{l}  \G{\tilde{\sigma}}_g \to -\G{\tilde{\sigma}}_g,\\ 
    \R{{\sigma}}_b\to \R{{\sigma}}_b \B{\tilde{\sigma}}_b ,\end{array} \right. ,\hspace{20pt}
    RS:\, \left\{ \begin{array}{l}  \B{\tilde{\sigma}}_b\to -\B{\tilde{\sigma}}_b,\\ 
    \R{{\sigma}}_g\to \R{{\sigma}}_g \G{\tilde{\sigma}}_g ,
    \end{array} \right.
\end{aligned}
\end{equation}
where $R=(RS) \times S$.

In addition to this internal $D_4$ symmetry, the model has a spatial $\mathbb Z_2$ symmetry, denoted $M$, that swaps the blue and green sublattices in Eq.~\eqref{eq:main_H_X_I}. This operation acts as a Hadamard matrix on the two-component vector, which geometrically corresponds to a mirror across an axis at $22.5^\circ$. Interestingly, $MS$ acts as an effective $45^\circ$ rotation. Hence the internal and external symmetries combine into a $D_8 \cong \mathbb Z_8 \rtimes \mathbb Z_2$ symmetry. This enlarged symmetry will be important for understanding the stability of a critical phase upon perturbing this model.

Lastly, we note that  Eq.~\eqref{eq:nbng} also makes apparent that we can write the model as a two-body Ising spin model by performing a change of basis $\R{\sigma}_i\to \R{\sigma}_i $, and $\tilde{\sigma}_i \to \R{\sigma}_i \tilde{\sigma}_i$ on every site. We then obtain
\begin{equation} \label{eq:main_H_X_II}
    H_{\ket{\psi}}=-\frac{1}{2} \sum_{\langle i,j \rangle_{\hexagon}}(\R{\sigma}_i \R{\sigma}_j+\R{\sigma}_i \tilde{\sigma}_j + \tilde{\sigma}_i \R{\sigma}_j -\tilde{\sigma}_i \tilde{\sigma}_j).
\end{equation}
We have found that this representation is especially useful for Monte-Carlo simulations of (perturbations of this) model (see Appendix~\ref{app:MC}), as we explore later. On the other hand, the rotor representation \eqref{eq:main_4clock} will prove rather useful for more analytic arguments.

\subsection{Decohered mixed state from purity} \label{subsubsec:X_dech}

The above detailed study of the wavefunction-deformed case will be useful for understanding the effect of decohering $D_4$ TO with non-Abelian anyons, which we turn to now. The composition of local quantum channels of the form $\mathcal{E}^{\R{X}}_j(\rho_0)=(1-p_{\R{R}})\rho_0 + p_{\R{R}}X_j \rho_0 X_j$ acting on the red sublattice $\R{\mathcal{R}_R}$ leads to the proliferation of incoherent $m_{\R{R}}$ fluxes. Similarly to the Abelian case (Sec.~\ref{sec:Z_error}), we will first study the purity $\textrm{tr}(\rho^2)$ of the decohered density matrix $\rho = \prod_j \mathcal E^{\R{X}}_j(\rho_0)$. For this, we again turn to the vectorized representation of $\rho$:
\begin{equation}
\begin{aligned}
     \kket{\rho}&=\prod_{j\in \R{\mathcal{R}_R}}\left(1-p_{\R{R}} +p_{\R{R}}X_j\otimes X_j\right)\kket{\rho_0}\\
     & \propto e^{\mu_{\R{R}}\sum_{j\in \mathcal{R}} X_j\otimes X_j}\kket{\rho_0}
\end{aligned}
\end{equation}
with $\tanh(\mu_\R{R})=p_\R{R}/(1-p_{\R{R}})$ and $\kket{\rho_0} = \ket{D_4} \otimes \ket{D_4}$. Since $\textrm{tr}\left(\rho^2\right) = \langle \! \langle \rho | \rho \rangle \! \rangle$, we can repeat the derivation in the previous subsection, which again leads to a honeycomb loop model, but now $f(L_\R{R})^2$ appears, rather than just $f(L_\R{R})$ as given in Eq.~\eqref{eq:fL}. As a result, we now obtain an O$(4)$ honeycomb loop model, rather than an O(2) loop model:
\begin{equation} \label{eq:X_trrho2}
\textrm{tr}\left(\rho^2\right)
\;
    \overset{\rm decohere}{\underset{m_\R{R}}{\propto}}
    \;
\sum_{L_{\R{R}}}\left(\frac{r_\R{R}}{2}\right)^{|L_{\R{R}}|}4^{C_{L_{\R{R}}}},
\end{equation}
with $r_\R{R}=\frac{2p_\R{R}(1-p_\R{R})}{p_\R{R}^2 +(1-p_\R{R})^2} \in [0,1]$. The fact that the loop weight $N$ is given by the square of the quantum dimension of $m_{\R{R}}$, i.e., $N=d_a^2$, instead of by $N=d_a$, is understood from the fact that such a local error creates pairs of fluxes $m_{\R{R}}\overline{m_{\R{R}}}$ combining the bra and ket subspaces together, similar to the toric code case~\cite{bao2023mixedstate}. This dependence will be clarified when we consider higher moments $\textrm{tr}(\rho^n)$ of the decohered density matrix.

Unlike the O(2) loop model, the O($N$) honeycomb loop model does not proliferate to a large loop phase\footnote{Interestingly, there is phase transition involving crystalline symmetry-breaking for very large string tension \cite{Chayes_2000,Guo_2000}, which is far outside the physical regime under consideration here.} for $N>2$. Hence, as far as the purity is concerned, decohering $D_4$ topological order with $m_\R{R}$ anyons does \emph{not} lead to a transition, even if $p=\frac{1}{2}$.
It is important to notice that the decohered density matrix $\rho_{\frac{1}{2}}=\mathcal{E}^{\R{X}}(\rho_0)$ when taking the error rate $p_\R{R}=\frac{1}{2}$ is a fixed point of the quantum channel $\mathcal{E}^{\R{X}}$ for any $p_\R{R}$, i.e., $\mathcal{E}^{\R{X}}(\rho_{\frac{1}{2}})=\rho_{\frac{1}{2}}$ for any $p_\R{R}$~\footnote{Notice that the quantum channel resulting from applying the channel $\mathcal{E}^{\R{X}}$ with the error strength $p_\R{R}$ $m$ times, is a quantum channel with the same form but error strength $p_{\R{R}}^{(m)}= \frac{1}{2} - \left(\frac{1}{2} - p_\R{R} \right)(1-2p_\R{R})^m \in [0,\frac{1}{2}]$. }. Hence, a single application of the channel with $p_\R{R}=\frac{1}{2}$ is sufficient to reach its fixed point. This fact in combination with the lack of singularity in $\textrm{tr}(\rho^2)$ indicates that $D_4$ topological order is infinitely robust under an $X$-decoherence channel acting on one of the sublattices, \emph{as far as the purity is concerned}. However, as we mentioned in the introduction, the critical threshold $p_c^{(n)}$ depends on the moment $\textrm{tr}(\rho^n)$ of the density matrix that is considered, with the `intrinsic' (i.e., quantum-information-theoretic) threshold $p_c$ appearing when evaluating the von Neumann entropy (formally, $n \to 1$). In the following sections we will find indications of robustness also for other values of $n$, although we will find a transition for $n= \infty$.

Similarly to the previous section for the pure wave function deformation, one can rewrite the resulting loop model in terms of an explicitly local Ising-like Hamiltonian:
\begin{equation} \label{eq:Z_XR}
   \textrm{tr}\left(\rho^2\right)
   \;
    \overset{\rm decohere}{\underset{m_\R{R}}{\propto}}
    \;
   \sum e^{-2\mu_{\R{R}}H_\rho}.
\end{equation}
In this case, due to the doubling of degrees of freedom, we find a \emph{tetralayer} (rather than bilayer) honeycomb model, governed by a Hamiltonian of the form
\begin{equation} 
    H_{\rho}=-\sum_{\langle i,j \rangle_{\hexagon}}\R{\sigma}^1_i \R{\sigma}^1_j\widetilde{\CZ}^{(2)}_{ij} \R{\sigma}^3_i \R{\sigma}^3_j\widetilde{\CZ}^{(4)}_{ij},
\end{equation}
with $\tanh(\mu_{\R{R}})=\frac{p_{\R{R}}}{1-p_{\R{R}}}$.
Here $\R{\sigma}^1$ and $\G{\tilde{\sigma}}^2_i, \B{\tilde{\sigma}}^2_i$ live on the first and second layer respectively (with the second set living on the two different sublattices as in Eq.~\eqref{eq:main_H_X_I}), while $\R{\sigma}^3$ and $\G{\tilde{\sigma}}^4_i, \B{\tilde{\sigma}}^4_i$ live on layers $3$ and $4$ (see Fig.~\ref{fig:trhon} expressed in terms of rotor variables). However, one can effectively reduce the number of layers from four to three by defining a new variable $\sigma^1_i\to \sigma_i = \sigma^1_i \sigma^3_i$.  The Hamiltonian then simplifies to a \emph{trilayer} honeycomb model with interactions
\begin{equation} \label{eq:main_rho_H_X}
    H_{\rho}=-\sum_{\langle i,j \rangle_{\hexagon}}\R{\sigma}_i \R{\sigma}_j\widetilde{\CZ}^{(2)}_{ij} \widetilde{\CZ}^{(3)}_{ij}.
\end{equation}
This has at least two independent $D_4$ symmetries involving the pair of layers $(1,2)$ and $(1,3)$. 

We can use this statistical mechanical model to evaluate specific information-theoretic diagnostics---or rather simpler-to-compute Rényi versions---of the stability of this topological order, as Ref.~\onlinecite{fan2023diagnostics} did for the toric code. In particular, let us consider the Rényi-$2$  quantum relative entropy measuring the distinguishability of two states $\rho$ (namely the decohered $D_4$ TO state) and $\rho_{xy}$ (to be specified later on). This quantity is defined as
\begin{equation} \label{eq:D2}
D^{(2)}(\rho||\rho_{xy})\equiv -\log\left( \frac{\textrm{tr}(\rho \rho_{xy})}{\textrm{tr}(\rho^2)}\right),
\end{equation}
and diverges when $\rho$ and $\sigma$ are orthogonal (hence distinguishable), while it saturates to a finite value when the two corrupted states cannot be distinguished. Since anyonic excitations are a defining characteristic of TO, checking whether they are still well-defined (i.e., orthogonal to the vacuum) for the decohered $D_4$ TO is relevant. Hence, we take $\rho_{xy}$ to be the decohered state after applying the quantum channel $\mathcal{E}^{\R{X}}$ to the initial state $\R{\mathcal{X}}_{x}^{y}\ket{D_4}$, hosting two non-Abelian anyons $m_{\R{R}}$ located on triangles $x$ and $y$; see Eq.~\eqref{eq:X_if}. Using the same ungauging maps previously employed in this section (in particular App.~\ref{app:mappings}), the Rényi-2 quantum relative entropy becomes the thermal correlation function
\begin{equation}
    D^{(2)}(\rho||\rho_{xy})=-\ln\left( \langle \R{\sigma}_{\G{x}} \R{\sigma}_{\B{y}}\widetilde{\CZ}^{(2)}_{{\G{x}}{\B{y}}} \widetilde{\CZ}^{(3)}_{{\G{x}}{\B{y}}} \rangle \right)
\end{equation}
evaluated on the stat-mech model of Eq.\eqref{eq:main_rho_H_X}. Since for all error rate the resulting O$(4)$ loop model lies within the small loop phase, $\langle \R{\sigma}_{\G{x}} \R{\sigma}_{\B{y}}\widetilde{\CZ}^{(2)}_{{\G{x}}{\B{y}}} \widetilde{\CZ}^{(3)}_{{\G{x}}{\B{y}}} \rangle$ decays exponentially with the distance and hence $D^{(2)}(\rho||\rho_{xy})\sim|\B{y}-\G{x}|$ diverges with the distance between the two $m_\R{R}$ anyons. 

The previous discussion, in particular the content of Eq.~\eqref{eq:X_trrho2}, appears to be at odds with the fact that fusion of $m_\R{R}$ anyons can also lead to Abelian charges $e_{\G{G}, \B{B}}$.  In particular, one might wonder why additional contributions coming from such anyons are absent from the partition function $\mathcal{Z}_{\rho}(p_{\R{R}})$. More plainly put, why do we obtain clean loop models where $m_\R{R} \times m_\R{R}$ always fuses trivially? This property relates to the particular choice of quantum channel which only proliferates $m_\R{R}$ (which will be discussed more in later sections), but it also reflects the fixed-point nature of the initial $D_4$ wavefunction we are considering.
In Sec.~\ref{sec:D4_XpZ}, we will study the deformation and decoherence proliferating both Abelian and non-Abelian anyons. Let us state a key result here, in the context of $\R{X}$-error on top of the deformed wavefunction we already introduced in Sec.~\ref{sec:Z_purewf} for perturbing with Abelian anyons. First, the relevant stat-mech model corresponds to a \emph{net} (rather than loop) model, with closed red loops $L_\R{R}$ together with blue $\gamma_{\B{B}}$ and green $\gamma_{\G{G}}$ closed loops or strings ending on $L_\R{R}$ loop configurations (see Fig.~\ref{fig:D4_TO}b). For the pure wavefunction deformation, we again derive Eq.~\eqref{eq:eq_21}, however now with 
\begin{equation}  \label{eq:over_EE}
f(L_\R{R};t_\G{G},t_\B{B}) = f(L_\R{R}) \times \mathcal{Z}_{L_\R{R}}(t_\G{G},t_\B{B})
\end{equation}
where $f(L_\R{R})$ was given in Eq.~\eqref{eq:fL}. Here, $t_{\G{G}, \B{B}}=\tanh(\beta^z_{\G{G}, \B{B}})$ and $\mathcal{Z}_{L_\R{R}}(t_\G{G},t_\B{B})$ is given by
\begin{equation} \label{eq:Z_gammaR}
    \mathcal{Z}_{L_{\R{R}}}(t_\G{G},t_\B{B})=\sum_{\gamma_{\G{G}}, \gamma_{\B{B}}}\sigma_{L_{\R{R}}}(\gamma_{\B{B}},\gamma_{\G{G}})t_\G{G}^{|\gamma_{\G{G}}|}t_\B{B}^{|\gamma_{\B{B}}|}.
\end{equation}
The symbol $\sigma_{L_{\R{R}}}(\gamma_{\B{B}},\gamma_{\G{G}})=\pm 1$ is a sign assignment which is consistent with anyon braiding properties and is specified in Appendix~\ref{app:comb_ZpX}. Hence, Abelian charges  $e_{\G{G}, \B{B}}$ will be generically generated when detuning from the fixed-point wavefunction. Notice that when either $t_\G{G}$ or $t_\B{B}$ vanish, $ \mathcal{Z}_{L_{\R{R}}}(t_\G{G},t_\B{B})$ becomes positive, and hence it corresponds to a classical partition function for every $L_{\R{R}}$.

Computing the purity $\textrm{tr}(\rho^2)$ involves a similar calculation where the main difference is replacing the topological factor in Eq.~\eqref{eq:over_EE}, by its square: $f(L_\R{R};t_\G{G},t_\B{B}) \to f(L_\R{R};t_\G{G},t_\B{B})^2$. Hence, the purity of the decohered density matrix reads
\begin{equation}
\textrm{tr}(\rho^2) \;
    \overset{\rm decohere}{\underset{m_\R{R}}{\propto}}
    \; \sum_{L_{\R{R}}} \left(\frac{r_\R{R}}{2}\right)^{|L_{\R{R}}|}4^{C_{L_{\R{R}}}}\mathcal{Z}^2_{L_{\R{R}}}(t_\G{G},t_\B{B}).
\end{equation}
This expression simplifies when both $t_{\G{G}},\, t_{\B{B}}\ll 1$.  First-order contributions in these tensions are given by the shortest configurations $\gamma_{\G{G}}, \gamma_{\B{B}}$. These correspond to either $|\gamma_{\G{G}}|=| \gamma_{\B{B}}|=0$, or to a string with the shortest possible length of either blue or green errors (namely with $|\gamma_{\G{G}}|,| \gamma_{\B{B}}|=1$) whose endpoints terminate on a closed loop configuration $L_{\R{R}}$. An example of such configuration is shown in Fig.~\ref{fig:D4_TO}b. For a given $L_{\R{R}}$ the partition function 
 $ \mathcal{Z}_{L_{\R{R}}}(t_\G{G},t_\B{B})$  then becomes $\mathcal{Z}_{L_{\R{R}}}(t_\G{G},t_\B{B})\approx 1 + \frac{|L_{\R{R}}|}{2}(t_\G{G}+t_\B{B})\approx ( 1 + (t_\G{G}+t_\B{B})/2)^{|L_{\R{R}}|} $, which can be interpreted as dressing the loop $L_\R{R}$. Hence, we find that the purity
\begin{equation}
\textrm{tr}(\rho^2) \;
    \overset{\rm decohere}{\underset{m_\R{R}}{\propto}}
    \; \sum_{L_{\R{R}}} \left(\frac{r_\R{R}( 2 + t_\G{G}+t_\B{B})}{4}\right)^{|L_{\R{R}}|}4^{C_{L_{\R{R}}}},
\end{equation}
is again given by an O$(4)$ honeycomb loop model where the tension is tuned by $r_\R{R}$ as well as by the tensions $t_\G{G},t_\B{B}$. This will prove useful later for detuning away from accidentally fine-tuned points.

\subsection{Higher moments $\textrm{tr}(\rho^n)$} \label{sec:high_trrho}

So far we characterized the decohered mixed state $\rho$ in terms of quantities that can be computed from its second moments (i.e., evaluating the expectation value of the density matrix $\rho^2/\textrm{tr}(\rho^2)$). We can similarly derive loop model descriptions of the higher moments $\textrm{tr}(\rho^n)$. This is most convenient to do in terms of the so-called error picture \cite{Dennis_2002,fan2023diagnostics}, where we observe we can write $\rho$ as a classical mixture of corrupted states $\prod_{j} \R{X}_{r_j} \ket{D_4}$. Then $\textrm{tr}(\rho^n)$ is naturally expressed in terms of overlaps of such corrupted wavefunctions, similarly to what we saw in Sec.~\ref{sec:X_puredef} and Sec.~\ref{subsubsec:X_dech}. Since we have already obtained a general expression for this wavefunction overlap in Eq.~\eqref{eq:fL}, we straightforwardly obtain stat-mech models for these higher moments. In particular, since Eq.~\eqref{eq:fL} took the form of the weight of an O(2) loop model, we naturally write $\textrm{tr}(\rho^n)$ as $\sim n$ coupled O(2) loop models. Indeed, the O(4) loop model we obtained for $n=2$ can be thought of as two tightly-bound O(2) loop models.
Similarly, $\textrm{tr}(\rho^3)$ admits a particularly simple expression as a coupled O$(2)$ loop model:
\begin{equation} 
\begin{aligned}
    \textrm{tr}(\rho^3)&\propto  \sum_{L^{(1)}, L^{(2)}}\prod_{\small L\in \{L^{(1)},L^{(2)}, 
L^{(1)}\oplus L^{(2)} \}}
{\tilde t}^{|L|} \;
2^{C_{L}},
\end{aligned}
\end{equation}
with $\tilde t^2 = \frac{p - p^2}{2 - 6 p + 6 p^2}$,
and where the sums are over contractible closed loop configurations $L^{(1)}, L^{(2)}$ on the honeycomb lattice. 
This can be thought of as three O(2) loop models, with a strong coupling that enforces the symmetric difference of all three loops to vanish. The coupled loop models for higher $n$ are summarized in Appendix~\ref{app:trrhon_local}, although they take a more complicated form.
In a companion work, we show how such coupled O$(N)$ loop models arise more generally for other types of topological order, with the loop weight $N$ corresponding to the quantum dimension of the proliferating anyon \cite{short_paper}.

However, to the best of our knowledge, the physics of these stat-mech loop models has not yet been explored in the literature, in contrast to the O(2) and O(4) loop models discussed above. For this reason, we focus on the local spin model representation of these stat-mech models. In particular, similar to the previous section, we can show that $\textrm{tr}(\rho^n)$ can be written in terms of $n$ coupled ``$ZZ\CZ$'' models:
\begin{equation} \label{eq:trrhon_local}
\textrm{tr} \left( \rho^n \right) \propto \sum_{ \{\sigma_j^{(s)},\tilde \sigma_j^{(s)} \}_{s=1}^n } \exp\left( \beta \sum_{\langle i,j\rangle_{\hexagon}} \sum_{s=1}^{n} h_{i,j}^{(s)} h_{i,j}^{(s+1)} 
\right)
\end{equation}
where $\tanh(\beta)=\frac{p_\R{R}}{1-p_\R{R}}\in [0,1]$ and each replica is a Ising honeycomb bilayer with
\begin{equation} \label{eq:hij}
h_{i,j}^{(s)} = \sigma_i^{(s)} \sigma_j^{(s)} \widetilde{\CZ}_{ij}^{(s)} \quad \textrm{and} \quad  h_{i,j}^{(n+1)} =h_{i,j}^{(1)} . 
\end{equation}
This is direct generalization of the $n=2$ case in Eq.~\eqref{eq:Z_XR}. This is derived in Appendix~\ref{app:trrhon_local} using the error picture. There we also show an equivalent rewriting, which reduces to the known result for the toric code case \cite{fan2023diagnostics} if we were to set $\widetilde{\CZ}_{ij} \to 1$. We note that $h_{i,j}^{(s)}$ is the same interaction we encountered in the section on wavefunction deformation (Secs.~\ref{sec:X_puredef} and \ref{sec:Ising_OP}, in particular Eq.~\eqref{eq:main_H_X_I}). One can thus also express it in terms of rotors as in Eq.~\eqref{eq:main_4clock}, to make the internal $D_4$ symmetry manifest. In other words, $\textrm{tr}(\rho^n)$ can be regarded as a local coupling of $n$ layers of rotors, each with $D_4$ symmetry. We visually represent the model in Fig.~\ref{fig:trhon}.

\begin{figure}
    \centering
    \includegraphics[width=0.95\linewidth]{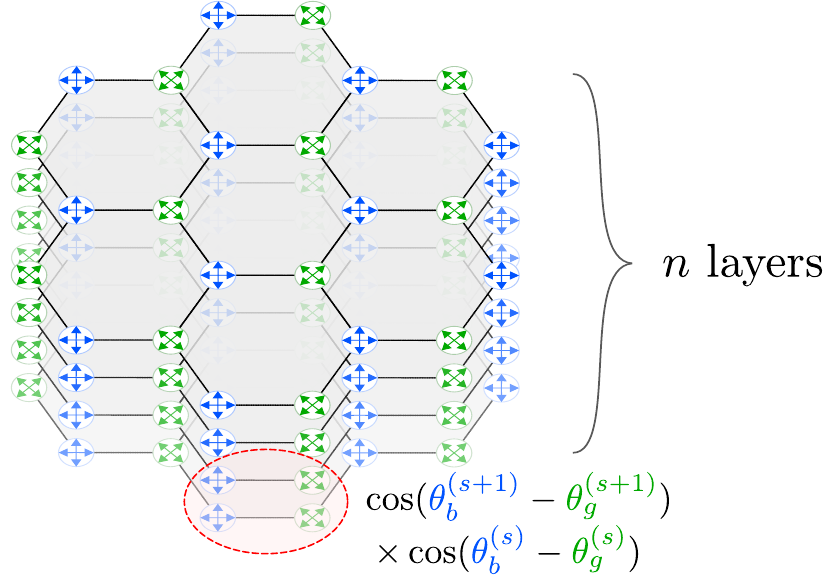}
    \caption{\textbf{Stat-mech model for $\textrm{tr}(\rho^n)$.} The $n$'th moment of the density matrix obtained by decohering $D_4$ TO with incoherent non-Abelian $m_\R{R}$ anyons leads to an $n$-fold stack of honeycomb rotor models, each layer with a discrete $D_4$ symmetry acting on the four-state rotors. The Hamiltonian of the stat-mech model has a nearest-layer interaction \eqref{eq:trrhon_local}, as shown in the figure. Alternatively, instead of four-state rotors, one can also express each layer as an Ising bilayer as in Eq.~\eqref{eq:hij}. These interactions resemble those encountered in the wavefunction deformation in Secs.~\ref{sec:X_puredef} and \ref{sec:Ising_OP}. The case of purity ($n=2$) is studied in detail in Sec.~\ref{subsubsec:X_dech}.}
    \label{fig:trhon}
\end{figure}

We have already seen that for $n=2$ (i.e., purity), the local stat mech model fails to order, even at zero temperature. This tells us that the purity does not detect any transition out of $D_4$ TO. If we increase $n$ and thereby add more layers to our stat mech model \eqref{eq:trrhon_local}, can this increase the tendency to order? To answer this question, it is instructive to consider the zero-temperature limit $\beta \to \infty$, more precisely setting $p_\R{R} = \frac{1}{2}$, where the model has the highest chance of ordering. Note that in this limit, Eq.~\eqref{eq:trrhon_local} only has nonzero (and equal) weight for configurations where $h_{i,j}^{(s)} h_{i,j}^{(s+1)} =1$. In other words, for the maximal decoherence rate $p_\R{R} = \frac{1}{2}$, the resulting stat mech model is the infinite-temperature ensemble in the constrained configuration space where $h_{i,j}^{(s)} = h_{i,j}^{(s')} \equiv \eta_{ij} \in \{ \pm 1 \}$. Hence, 
\begin{equation} \label{eq:trrhon_phalf}
\textrm{tr} \left( \rho^n \right) \;
\underset{p_\R{R} \to \frac{1}{2} }{\propto} \; 
\sum_{\{ \eta_{ij} \} } \left( \sum_{\{ \sigma_j, \tilde \sigma_j\} } \prod_{\langle i,j\rangle} \left[ 1 + \eta_{ij} \sigma_i \sigma_j \widetilde{\CZ}_{ij} \right] \right)^n.
\end{equation}

Considerable insight can be gained from Eq.~\eqref{eq:trrhon_phalf}. For instance, it indicates the fate of the $n \to \infty$ limit, which is where where the spin model \eqref{eq:trrhon_local} becomes a full-fledged 3d stat-mech model. Taking this limit in Eq.~\eqref{eq:trrhon_phalf} means only the $\{ \eta_{ij} \}$ configuration leading to the largest weight will survive. This turns out to be\footnote{More precisely, it holds for any `flux-free' configuration where $\prod_{\langle i,j \rangle \in \hexagon} \eta_{ij}=1$ for each hexagon, which means we can write $\eta_{ij} = \tau_i \tau_j$; by a change of variables $\sigma_i \to \sigma_i \tau_i$ this leads to the same partition function as $\eta_{ij} =1$.} $\eta_{ij} = 1$, as we will prove in the next subsection. Hence,
\begin{equation}  \label{eq:rhoinfty}
\lim_{n \to \infty} \sqrt[n]{\textrm{tr} \left( \rho^n \right)} \;
\underset{p_\R{R} \to \frac{1}{2} }{\propto} \; 
 \sum_{\{ \sigma_j, \tilde \sigma_j\} } \prod_{\langle i,j\rangle} \left[ 1 +\sigma_i \sigma_j \widetilde{\CZ}_{ij} \right].
\end{equation}
This is \emph{exactly} the zero-temperature limit of the stat-mech model we encountered when studying the pure wavefunction deformation in Secs.~\ref{sec:X_puredef} and \ref{sec:Ising_OP}. There we found that this model is perched at BKT criticality, with algebraic correlations. This shows that unlike for $n=2$, there exists a critical threshold $p_c^{(\infty)}$ for the $n\to\infty$ replica limit whose value is~\footnote{If $p_c^{(\infty)} < \frac{1}{2}$ it implies we have entered a phase with algebraic correlations.} $p_c^{(\infty)} \leq \frac{1}{2}$ (in fact in the next subsection we will argue $p_c^{(\infty)} = \frac{1}{2}$). In particular, this shows that $p_c^{(n)}$ is not an increasing function of $n$, unlike for the decohered toric code case~\cite{fan2023diagnostics}. This distinct behavior can be linked to the topological weights in the loop model representation: for $n=2$ the strong coupling of the two layers led to an O(4) loop model which is unable to proliferate, but for $n \to \infty$ the effects of the coupling are suppressed, revealing the critical point of the O(2) loop model, of which the local spin model in Eq.~\eqref{eq:rhoinfty} is an equivalent representation.

The result in Eq.~\eqref{eq:trrhon_phalf} even allows us to infer properties of the `true' threshold. Indeed, the form is amenable to taking the replica limit $n \to 1$. From this one can infer that the von Neumann entropy (at $p_\R{R} = \frac{1}{2}$) is dictated by a random-bond version of the `$ZZCZ$' model, which we discuss more rigorously and in detail in the next subsection. This suggests that the fate of $D_4$ as a quantum memory subjected to decoherence is tied up with whether or not this disordered spin model orders---and we have already established that even the clean model only reaches quasi-long-range order at zero temperature. However, we can avoid having to take the replica limit altogether: we now show how one can fully diagonalize $\rho$ for $p_\R{R} = \frac{1}{2}$, which means we have access not only to the full spectrum but also to all eigenstates, which in turn gives insight to how these results inform about $p_\R{R} < \frac{1}{2}$.

\subsection{Solving maximal decoherence $p_\R{R}=1/2$} \label{sec:n_to_1}

We have seen that unlike decohering with Abelian anyons, TO can be remarkably stable to decohering with non-Abelian anyons. For instance, the purity sees no transition at all upon maximally decohering $m_\R{R}$ in $D_4$. This raises the question: can the TO be stable even at maximal decoherence to the proliferation of a given (set of) anyons? Here we show how one can shed light on this question by exactly diagonalizing the decohered density matrix $\rho_{\frac{1}{2}}$ at maximum error rate $p_\R{R}=\frac{1}{2}$ in terms of a disordered spin model.

To diagonalize $\rho_{\frac{1}{2}}$ we notice that the local quantum channel $\mathcal{E}^{\R{X}}_{\R{r}}$ (for $p_{\R{R}} = \frac{1}{2}$) acting on a site ${\R{r}}$ of $\R{\mathcal{R}_R}$ can be written as a random projector channel:
\begin{equation} \label{eq:E_proj}
    \mathcal{E}^{\R{X}}_{\R{r}}(\cdot)=\sum_{\eta_\R{r}=\pm 1}P_{\eta_\R{r}}(\cdot ) P_{\eta_\R{r}},
\end{equation}
with $P_{\eta_\R{r}}=\frac{1}{2}(\mathds{1}+\eta_\R{r}X_{\R{r}})$. Hence, when acting on $\ket{D_4}$ and on all sites of the red sublattice leads to the maximally decohered density matrix 
\begin{equation} \label{eq:rho_diagonal}
    \rho_{\frac{1}{2}}=\frac{1}{4^{|\R{\mathcal{R}_R}|}} \sum_{\eta}\ket{\eta}\bra{\eta}
\end{equation}
 with $\eta=\{\eta_\R{r}=\pm 1\}$, and where the (un-normalized) states $\ket{\eta}$ associated to a given configuration $\eta$ are given by
\begin{equation} \label{eq:eta}
    \ket{\eta}=\prod_{\R{r} \in \R{\mathcal{R}_R}} (1+\eta_\R{r} X_\R{r})\ket{D_4}.
\end{equation}  
The orthogonality of the projectors $P_{s_\R{r}}$ directly leads to the orthogonality condition $\avg{\eta|\eta'}= \avg{\eta|\eta}\prod_{\R{r}\in \R{\mathcal{R}_R}}\delta_{\eta_\R{r},\eta'_\R{r}}$, hence the states $\ket{\eta}$ correspond to eigenvectors of $\rho_{\frac{1}{2}}$. In other words, Eq.~\eqref{eq:rho_diagonal} constitutes a diagonalized density matrix. The corresponding (non-negative) eigenvalues are then given by $P(\eta)=\avg{\eta|\eta}/4^{|\R{\mathcal{R}_R}|}$ which more explicitly reads
\begin{align} 
    P(\eta)=\frac{1}{4^{|\R{\mathcal{R}_R}|}}\sum_{ L_{\R{R}}} \left(\prod_{\R{e} \in L_{\R{R}}}\frac{\eta_\R{e}}{\sqrt{2}} \right) 2^{C_{L_{\R{R}}}},
\end{align}
where we have explicitly used that $f(L_{\R{R}})= 2^{C_{L_{\R{R}}}}/\sqrt{2}^{|L_\R{R}|}$ (see Eq.~\eqref{eq:fL}). Therefore, the eigenvalues $P(\eta)$ correspond to the partition function of a random O$(2)$ loop model with a signed-disorder tension given by $\eta_{\R{e}}/\sqrt{2}$ on a bond $\R{e}$ of the honeycomb lattice.

Let us now consider the stat-mech model appearing in the limit $\lim_{n\to\infty}\textrm{tr}(\rho^n)^{1/n}$ which equals to $\max_{\eta}P(\eta)$ (up to a system-size-dependent constant factor related to its degeneracy). Since $f(L_{\R{R}})\geq 0$, one finds the largest eigenvalue to be
\begin{equation} \label{eq:max_Peta}
    \max_{\eta}P(\eta)=\frac{1}{4^{|\R{\mathcal{R}_R}|}}\sum_{ L_{\R{R}}}\frac{2^{C_{L_{\R{R}}}}}{\sqrt{2}^{L_{\R{R}}}},
\end{equation}
which is attained for all eigenstates for which $\prod_{\R{e}\in \R{\hexagon}} \eta_{\R{r}}=+1$ for all plaquettes $\R{\hexagon}$. This provides a derivation of Eq.~\eqref{eq:rhoinfty}, which is written as the local stat-mech model formulation of the O(2) loop model at the BKT critical point. As discussed in the previous subsection, this shows that unlike the case for purity ($n=2$), there is a $p_c^{(\infty)} \leq \frac{1}{2}$. In fact, we will now argue $p_c^{(\infty)}=\frac{1}{2}$.

The previous calculation can be repeated when considering the deformed wavefunction $|\psi(\beta^z_{\G{G}}, \beta^z_{\B{B}})\rangle$ instead of $\ket{D_4}$ (see Sec.~\ref{subsubsec:X_dech} where we discussed the case of purity). In this case the eigenvalues become
\begin{align} \label{eq:lambda_eta}
    P(\eta)=\frac{1}{4^{|\R{\mathcal{R}_R}|}}\sum_{ L_{\R{R}}} \left(\prod_{\R{e} \in L_{\R{R}}}\frac{\eta_\R{e}}{\sqrt{2}} \right) 2^{C_{L_{\R{R}}}} \frac{\mathcal{Z}_{L_{\R{R}}}(t_\G{G},t_\B{B})}{\mathcal{Z}_{\emptyset}(\G{t_G},\B{t_B})}.
\end{align}
where $\mathcal{Z}_{\emptyset}(\G{t_G},\B{t_B}) \geq 0$ is the norm of $|\psi(\beta^z_{\G{G}}, \beta^z_{\B{B}})\rangle$, which agrees with Eq.~\eqref{eq:Z_gammaR} for the trivial loop configuration $L_{\R{R}}=\emptyset$.
Moreover, whenever $\mathcal{Z}_{L_{\R{R}}}(t_\G{G},t_\B{B})\geq 0$, one finds the largest eigenvalue is given by
\begin{equation} \label{eq:ninfty_beta}
\begin{aligned}
    &\max_{\eta}P(\eta)=\frac{1}{4^{|\R{\mathcal{R}_R}|}}\sum_{L_{\R{R}}} \frac{2^{C_{L_{\R{R}}}}}{\sqrt{2}^{|L_{\R{R}}|}}\frac{\mathcal{Z}_{L_{\R{R}}}(t_\G{G},t_\B{B})}{\mathcal{Z}_{\emptyset}(\G{t_G},\B{t_B})}.
\end{aligned}
\end{equation}
The condition $\mathcal{Z}_{L_{\R{R}}}(t_\G{G},t_\B{B})\geq 0$ for all $L_{\R{R}}$ is clearly satisfied when either $t_\G{G}$ or $t_\B{B}$ vanish (since in this case $\sigma_{L_{\R{R}}}(\gamma_\G{G},\gamma_\B{B})$ in Eq.~\eqref{eq:Z_gammaR} is positive), or when both of them are positive but sufficiently small as shown at the end of Sec.~\ref{subsubsec:X_dech}. In this case we then find that 
\begin{equation} \label{eq:Z2_tgtb}
\lim_{n\to\infty}\textrm{tr}(\rho^n)^{1/n}\approx \sum_{L_{\R{R}}} 2^{C_{L_{\R{R}}}} \left(\frac{t_{\R{R}}}{\sqrt{2}}\right)^{^{|L_{\R{R}}|}}
\end{equation}
corresponds to the partition function of the O$(2)$ loop model on the honeycomb lattice with (tunnable) tension $t_{\R{R}}\approx 1 + \frac{t_\B{B} + t_\G{G}}{2}$. 
Therefore, for $t_\B{B}, t_\G{G}>0 $ the system lies within the extended gapless phase; attains a critical point for $t_\B{B}= t_\G{G}=0$; and finally lies within the shortly correlated phase for negative $t_\B{B}, t_\G{G} $!
This means that any infinitesimal negative value of $t_\B{B}$ or $t_\G{G}$ is sufficient to prevent there being a transition for $n=\infty$, similar to the purity case ($n=2$). If we presume $p_c^{(\infty)}$ is a continuous function of $t_\B{B},t_\G{G}$, this suggests that for $t_\B{B}=0=t_\G{G}$, we have $p_c^{(\infty)} = \frac{1}{2}$. However, this is based on the aforementioned perturbative expansion, the validity of which would be interesting to explore in future work.

Knowledge of the full spectrum of $\rho_{\frac{1}{2}}$ allows us to compute various information-theoretic quantities which relate to the underlying random O$(2)$ loop model stat-mech model. To do so, it is desirable to express the eigenvalues $P(\eta)$ in Eq.~\eqref{eq:lambda_eta} as the partition function of a local stat-mech model with \emph{non-negative Boltzmann weights}. In particular, (up to $\eta$-independent overall factors) Eq.~\eqref{eq:lambda_eta} can be rewritten as
\begin{equation} \label{eq:Peta_local}
\begin{aligned}
    P(\eta)&\propto \sum_{\{\sigma, \tilde{\sigma}\}} \prod_{\langle i,j\rangle}\left(1+\eta_{ij} \sigma_i\sigma_j\widetilde{\CZ}_{ij}\right)\\
    &\times e^{\beta^z_{\B{B}} \sum_{\langle \langle \B{b},\B{b'}\rangle \rangle} \tilde{\sigma}_\B{b}\tilde{\sigma}_\B{b'}+\beta^z_{\G{G}} \sum_{\langle \langle \G{g},\G{g'}\rangle \rangle} \tilde{\sigma}_\G{g}\tilde{\sigma}_\G{g'}},
\end{aligned}
\end{equation}
defined on a bilayer honeycomb lattice as in Fig.~\ref{fig:bil_honey}a (recall that $t_\G{G}=\tanh(\beta^z_{\G{G}})$ and $t_\B{B}=\tanh(\beta^z_{\B{B}})$ ). Note that this agrees with Eq.~\eqref{eq:trrhon_phalf} for $\beta_{\B{B}}^z = \beta_{\G{G}}^z=0$. The von Neumann entropy of $\rho_{\frac{1}{2}}$ then corresponds to the quenched disorder average of the free energy of the partition function in Eq.~\eqref{eq:Peta_local}:
\begin{equation}
S_{\text{vN}}(\rho_{\frac{1}{2}})=-\sum_\eta P(\eta) \ln[P(\eta)]. \label{eq:SvN}
\end{equation}
The fact that the probability of a given disorder configuation $\eta$ is proportional to the partition function $P(\eta)$ itself, is an example of a Nishimori condition~\cite{Nishimori_81}.

Eq.~\eqref{eq:SvN} suggests that determining whether the $D_4$ topological quantum memory persists to the maximal-decoherence limit, is equivalent to asking whether this disordered `$ZZ\CZ$' model remains in the paramagnetic phase~\footnote{Notice the difference with the analysis of the toric code~\cite{Dennis_2002,fan2023diagnostics}, where the paramagnetic phase of the RBIM corresponds to breakdown of the quantum memory. }. Indeed, since we have full access to the \emph{eigenstates} of $\rho_\frac{1}{2}$, one can confirm that the fidelity between the decohered copies of two (initially) distinct logical states\footnote{In particular, the initial $\ket{D_4}$ and the state obtained by applying the logical $\mathcal X$ string operator (Eq.~\ref{eq:X_if}) around the torus.} of the $D_4$ can be related to a thermodynamic quantity of the disordered spin model. Writing $\rho = \mathcal{E}^{\R{X}}( \ket{D_4}\bra{D_4})$ and $\sigma = \mathcal{E}^{\R{X}}( \R{\mathcal{X}}\ket{D_4}\bra{D_4}\R{\mathcal{X}})$ with $\R{\mathcal{X}}$ the logical operator as defined below Eq.~\eqref{eq:X_if}, acting on a non-contractible loop (e.g., the horizontal direction) around the torus, the quantum fidelity between these two quantities reads 
\begin{equation}
F(\rho, \sigma)\equiv \textrm{tr}\left(\sqrt{\sqrt{\rho}\sigma\sqrt{\rho}}\right) \propto {\frac{1}{2}\sum_{\{\eta\}}  P(\eta)\frac{\left|1-e^{-\Delta F_\mathcal{C}}\right|}{1+e^{-\Delta F_\mathcal{C}}} }\label{eq:fidelity}
\end{equation}
where $\Delta F_{\mathcal{C}}$, is the free energy difference of inserting a line of flipped (antiferromagnetic) bonds along a non-contractible loop $\mathcal{C}$ perpendicular to $\R{\mathcal{X}}$, i.e., with a symmetry defect line~\footnote{We define $\Delta F_{\mathcal{C}}=-\ln(Z_-(\eta)/Z(\eta))$, where $\mathcal{Z}(\eta)\propto P(\eta)$, and $\mathcal Z_-(\eta)$ is the partition function with a line $\mathcal{C}$ of flipped antiferromagnetic bonds.}. See details of the derivation in App.~\ref{ref:fidelity}. 
In the paramagnetic phase, this symmetry defect line is invisible such that $F(\rho, \sigma)\approx 0$ in the thermodynamic limit, whereas in the ordered phase the defect would frustrate the order $\Delta F_{\mathcal{C}}\approx |\mathcal{C}|$, making the fidelity nonzero, and hence signaling the breakdown of the (non-Abelian) quantum memory. Notice that the same condition, namely the free energy cost of inserting a domain wall, is usually employed (although in the opposite direction with the ferromagnetic phase corresponding to a good quantum memory) to diagnose the breakdown of an Abelian quantum memory~\cite{Dennis_2002,WANG200331,Kubica_2018}.

This then raises the question: what is the fate of the disordered spin model? We already discussed that in the clean case, i.e., for $\eta_{\G{g}\B{b}} =+1$ on every link, the system showcases quasi-long range order only for $\beta^x_{\R{R}}=\infty$. It is then suggestive that if this ``ferromagnetic'' model fails to develop long-range order, the disordered one will follow the same fate. Hence, we expect that disorder will hinder even more the system from ordering, leading to lower ordering temperatures and in turn to the absence of a transition all together. This intuition appears to be supported by the observation that 
\begin{equation} \label{eq:P_eta_loop}
P(\eta)\approx \sum_{L_{\R{R}}} 2^{C_{L_{\R{R}}}} \left(\prod_{\R{e} \in L_{\R{R}}}\frac{\eta_\R{e}t_\R{R}}{\sqrt{2}} \right),
\end{equation}
appearing in the regime $t_\G{G},t_\B{B}\ll 1$ for any disorder configuration $\eta$, resembles disordered XY models previously studied in the literature~\cite{Shimada_2014,Alba_2009}. Notice that the clean versions of the XY model and that of the O$(2)$ loop model in Eq.~\eqref{eq:P_eta_loop} share the same critical behavior. It has been found that for the former, various forms of disorder cause the critical temperature separating a quasi-long-range order from a paramagnetic phase to decrease with increasing disorder strength. Hence, it is suggestive that a similar outcome holds for Eq.~\eqref{eq:P_eta_loop}, in which case either $p_c^{(1)} = \frac{1}{2}$ or the $D_4$ quantum memory would be stable even in the maximal-decoherence limit (which seems especially likely if we allow ourselves to tune to negative $t_\G{G}$ or $t_\B{B}$ in the initial wavefunction, as discussed above). Nonetheless, we notice that unlike the XY model, the system in Eq.~\eqref{eq:P_eta_loop} appears to be highly frustrated, so we cannot exclude that disorder increases the critical temperature by partially alleviating the frustration (which would in turn imply a finite $p_c^{(1)}$).
We leave a more careful analysis for future work.

\section{Combined Abelian and non-Abelian errors}
\label{sec:D4_XpZ}
We now combine both Abelian $Z$ and non-Abelian errors $X$ acting on different sublattices\footnote{Notice that $X$ and $Z$ pure wavefunction deformations acting on the same lattice do not commute. One can show that the resulting stat-mech model for a given ordering of the deformations, corresponds to coupled self-avoiding O$(2)$ and O$(1)$ loop models.} on the initial wavefunction $\ket{D_4}$. As the previous section has made explicit, the evaluation of the expectation values of the corresponding errors on the $\ket{D_4}$ TO ground state will give rise to a non-trivial coupling among different errors. 
In particular, we consider $X$ acting on $\R{\mathcal{R}_R}$, and $Z$ acting on $\B{\mathcal{R}_B}$ and $\G{\mathcal{R}_G}$ with different tunable strengths.

\begin{figure}
    \centering
    
    \includegraphics[width=\linewidth]{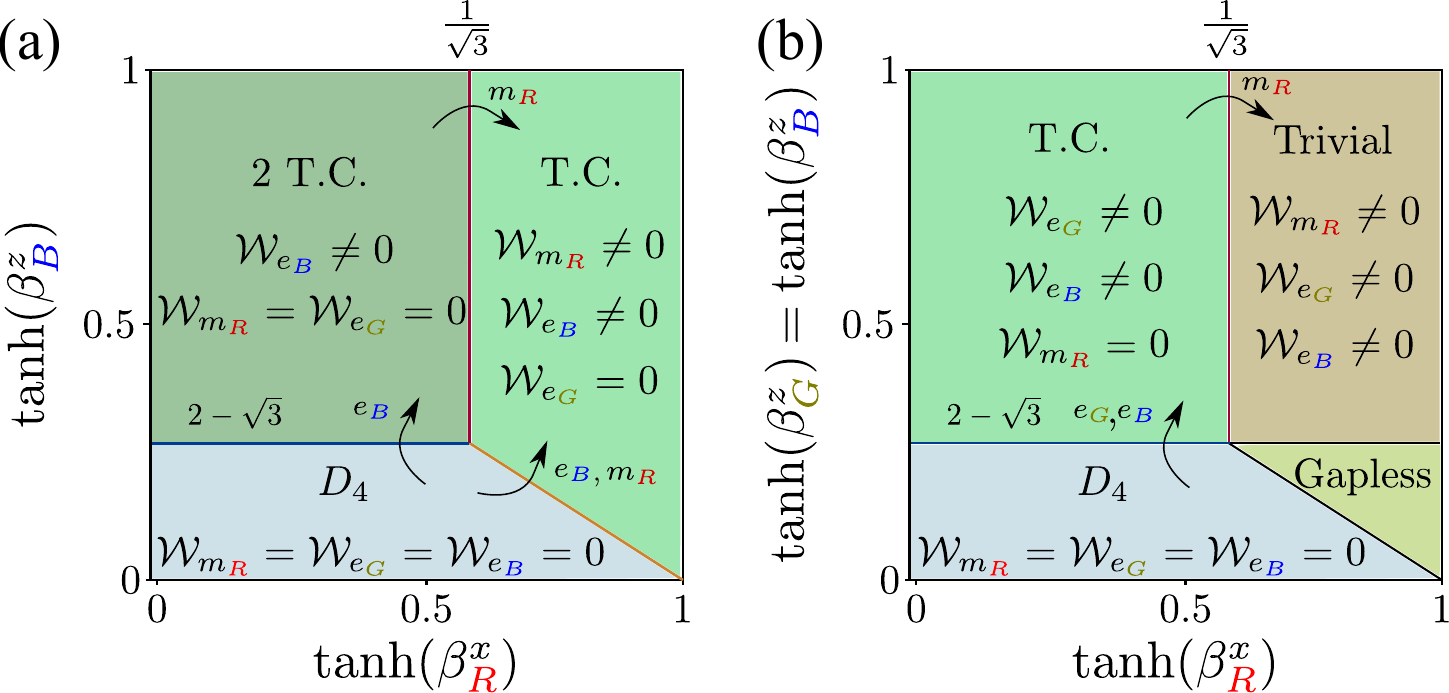}
    
    \caption{\textbf{Schematic phase diagram for pure wavefunction deformations.} Both panels include the value of the relevant order parameters indicating condensation of anyons as well as the resulting phases \R{(see Footnote~\ref{foot:subtlety})} labeled as: $D_4\to$ $D_4$ topological order, $2 \textrm{T.C.}\to$ topological order equivalent to two copies of the toric code phase, $\textrm{T.C.}\to$ toric code phase. Panel (a) shows the phase diagram as a function of  $\beta_{\R{R}}^x$ and  $\beta_{\B{B}}^z$ for $\beta_{\G{G}}^z=0$.  Panel (b) displays the phase diagram as a function of  $\beta_{\R{R}}^x$ and $\beta_{\G{G}}^z=\beta_{\B{B}}^z$. This shows an additional ``gapless'' phase diagnosed by the power-law decay of two-point correlation functions as shown in Fig.~\ref{fig:Correlations_gapless}. Transitions among phases can be understood as condensation of anyons indicated by an arrow. The underlying stat-mech model in the Ising-like representation is displayed in Eq.~\eqref{eq:main_H_ZZCZ_ZZ}. }
    \label{fig:Phase_diag_D4}
\end{figure}

\subsection{Pure wavefunction deformation} \label{sec:comb_purewf}
 Let us consider the unnormalized wavefunction 
 \begin{equation} \label{eq:psi_XZ}
 \begin{aligned}
     &|{\psi}(\beta^x_{\R{R}},\beta^z_{\G{G}}=0, \beta^z_{\B{B}})\rangle=e^{\frac{\beta^x_{\R{R}}}{2} \sum_{j\in \R{\mathcal{R}_R}} X_j} e^{\frac{ \beta^z_{\B{B}}}{2} \sum_{j\in \B{\mathcal{R}_B}} Z_j} \ket{D_4}.
\end{aligned}
 \end{equation}
 One finds that its norm reads (again  neglecting overall analytic prefactors)
 \begin{equation} \label{eq:main_purewf_bega0}
 \begin{aligned}
     &\mathcal{Z}_{\ket{\psi}}(\beta^x_{\R{R}}, t_\B{B})= \sum_{L_{\R{R}}}\sum_{\gamma_{\B{B}}}\tanh(\beta^x_{\R{R}})^{|L_{\R{R}}|} \\ &\times t_\B{B}^{|\gamma_{\B{B}}|}\langle{D_4|\prod_{j\in L_{\R{R}}}X_j \prod_{i\in \gamma_{\B{B}}} Z_i|D_4}\rangle,
 \end{aligned}
 \end{equation}
 with $t_\B{B}=\tanh(\beta^z_{\B{B}})$, and where now the errors are coupled through the last factor. As before, only closed red loop configurations $L_{\R{R}}$ on the honeycomb lattice lead to a non-vanishing contribution (since $\R{\mathcal{R}_R}$ and $\B{\mathcal{R}_B}$ do not overlap). Deferring detailed calculations to Appendix~\ref{app:comb_ZpX}, we now explain the constraint on the possible configurations of $\gamma_{\B{B}}$ (since it will no longer be simply closed loops, we will use $\gamma_\B{B}$ rather than $L_\B{B}$ to emphasize this). First, recall that while the O$(2)$ loop model describing the proliferation of non-Abelian fluxes is defined on the $\R{\mathcal{R}_R}$ honeycomb lattice, the analogous formulation for Abelian $e_{\B{B}}$  charges is given in terms of O$(1)$ loop model on a triangular lattice (see Fig.~\ref{fig:Abelian_triangle}). This corresponds to the dual triangular lattice of the $\B{\mathcal{R}_B}$ honeycomb. The case where $L_{\R{R}}$ and $\gamma_{\B{B}}$ do not intersect reduces to the discussion in the previous section, with $\gamma_{\B{B}}$ required to be a closed loop configuration. However, since an $L_{\R{R}}$ loop contains $e_{\B{B}}$ Abelian anyons coming from the fusion channel $m_{\R{R}}\times m_{\R{R}}=1+e_{\B{B}} + e_{\G{G}}+ e_{\B{B}}e_{\G{G}}$, a $\gamma_{\B{B}}$ configuration can end on a $L_{\R{R}}$ loop. Putting it all together we find
 \begin{equation} \label{eq:main_XR_ZB}
 \begin{aligned}
     &\mathcal{Z}_{\ket{\psi}}(\beta^x_{\R{R}}, t_\B{B})= \\&\sum_{L_{\R{R}}}\sum_{\gamma_{\B{B}}} t_\B{B}^{|\gamma_{\B{B}}|} \left(\frac{\tanh(\beta_{\R{R}}^x)}{\sqrt{2}}\right)^{|L_{\R{R}}|}2^{C_{L_{\R{R}}}},
    \end{aligned}
 \end{equation}
i.e., coupled O$(2)$ and O$(1)$ loop models where $\sum_{\gamma_{\B{B}}}$ corresponds to the sum over all configurations in the triangular \B{blue} sublattice that either form closed loops, or strings whose endpoints terminate on a $L_{\R{R}}$ loop (see Fig.~\ref{fig:D4_TO}b).
 
 As a next step, we add Abelian $Z$ errors on the remaining $\G{\mathcal{R}_G}$ sublattice. A similar calculation to the one above leads to the classical partition function
\begin{equation} \label{eq:main_X_R_Z_BG}
\begin{aligned}
     &\mathcal{Z}_{\ket{\psi}}(\beta^x_{\R{R}}, t_\G{G},t_\B{B})=\sum_{L_{\R{R}}}\sum_{\gamma_{\B{B}}}\sum_{\gamma_{\G{G}}}\sigma_{L_{\R{R}}}(\gamma_{\B{B}},\gamma_{\G{G}})\\
     &\times t_\B{B}^{|\gamma_{\B{B}}|}t_\G{G}^{|\gamma_{\G{G}}|} \left(\frac{\tanh(\beta_{\R{R}}^x)}{\sqrt{2}}\right)^{|L_{\R{R}}|}2^{C_{L_{\R{R}}}},
\end{aligned}
 \end{equation}
 where now both $\gamma_{\B{B}}$ and $\gamma_{\G{G}}$ are loop configurations that can be either closed or end up on a $L_{\R{R}}$ loop configuration on the blue and green triangular sublattices, respectively. This corresponds to two O$(1)$ loop models (one per excited Abelian anyon type) coupled to a O$(2)$ loop model describing the proliferation of non-Abelian charges. Moreover, unlike for vanishing $\beta_{\G{G}}^z$, there is an additional sign contribution $ \sigma_{L_{\R{R}}}(\gamma_{\B{B}},\gamma_{\G{G}})=\pm 1$. Since in the following this sign will turn out not to be necessary, we defer a more detailed discussion to Appendix~\ref{app:comb_ZpX}.

 Similarly to Eq.~\eqref{eq:main_tildeZ_x} we can alternatively derive an Ising-like Hamiltonian which now reads (including the inverse temperature dependence) 
 \begin{equation} \label{eq:main_H_ZZCZ_ZZ}
 \begin{aligned}
    H_{\ket{\psi}}&=-\beta^x_{\R{R}}\sum_{\langle i,j \rangle_{\hexagon} } {\color{red} \sigma}_i {\color{red} \sigma}_j \widetilde{\CZ}_{i,j} - \beta^z_{\B{B}}  \sum_{\langle\langle b,b^\prime \rangle \rangle_{\hexagon} } \tilde{{\B{\sigma}}}_b \tilde{\B{\sigma}}_{b^\prime} \\ &- \beta^z_{\G{G}} \sum_{\langle\langle g,g^\prime \rangle \rangle_{\hexagon} } \tilde{\G{{\sigma}}}_g \tilde{\G{{\sigma}}}_{g^\prime}, 
    \end{aligned}
\end{equation}
with $b,b^\prime; g,g^\prime$ sites on the blue $\B{B}$ and green $\G{G}$ sublattices of the bottom layer, respectively, as shown in Fig.~\ref{fig:bil_honey}a. Notice that $D_4$ symmetry is enlarged to a $D_8$ symmetry group when  $\beta^z_{\G{G}}=\beta^z_{\B{B}}$, due to the additional $M$ mirror symmetry defined below Eq.~\eqref{eq:SR}.

\subsubsection{Order parameters}

The $D_4$ symmetry group has eight different proper non-trivial subgroups which correspond to either $\mathbb{Z}_2$ (appearing $5$ times), $\mathbb{Z}_4$ (appearing once), or $\mathbb{Z}_2\times \mathbb{Z}_2$ (appearing twice). These label the different possible thermal phases that can appear as a result of spontaneous symmetry breaking in the corresponding stat-mech model.
The advantage of the explicitly local formulation in terms of Ising variables as in Eq.~\eqref{eq:main_H_ZZCZ_ZZ}, relative to that in Eq.~\eqref{eq:main_X_R_Z_BG}, is to permit an efficient characterization of the system using Monte-Carlo methods. In particular, one can numerically investigate the phase diagram via the three order parameters discussed in the following.

First, and as discussed in Sec.~\ref{sec:Z_purewf}, condensation of $e_{\B{B}}$ charges is measured by long-range order of the two-point correlation $\mathcal{W}_{e_{\B{B}}}(\B{b},\B{b'})\equiv\avg{{\color{blue} \tilde{\sigma}}_\B{b} {\color{blue} \tilde{\sigma}}_\B{b'}}$ in the limit $|x-y|\to\infty$.
It is useful to repackage this into a single number which is nonzero only in the condensed phase:
\begin{equation} \label{eq:Web_OP}
\begin{aligned}
    \mathcal{W}_{e_{\B{B}}}
    &\equiv \frac{ \sum_{\B{b},\B{b'}} \mathcal W_{e_\B{B}}(\B{b},\B{b'}) }{ \sum_{\B{b},\B{b'}} 1  } = \frac{1}{|\B{\mathcal{R}_B}|^2}
    \avg{ \sum_{b,b'\in \B{\mathcal{R}_B}}{\color{blue} \tilde{\sigma}}_b {\color{blue}\tilde{\sigma}}_{b'} }
\end{aligned}
\end{equation}
with a similar definition of $\mathcal{W}_{e_{\G{G}}}$ for the green sublattice. 
 As usual, $\mathcal{W}_{e_\B{B}}\neq 0$ implies condensation of the order parameter $\langle \sum_b \tilde \sigma_b\rangle$. Using Eq.~\eqref{eq:n_theta}, we can also rewrite $\mathcal{W}_{e_{\B{B}}}$ as follows
 \begin{equation}
     \mathcal{W}_{e_{\B{B}}}\propto \avg{ \sum_{b,b'\in \B{\mathcal{R}_B}}\cos(2\B{\theta}_b-2\B{\theta}_{b'})}
 \end{equation}
If in the thermodynamic limit $\mathcal{W}_{e_{\B{B}}} \neq 0$ ($ \mathcal{W}_{e_{\G{G}}} \neq 0$), the resulting state can be invariant only under the $\mathbb{Z}_2\times \mathbb{Z}_2$ group generated by $S, R^2$ ($SR,R^2$) defined in Eq.~\eqref{eq:SR}. If both $\mathcal{W}_{e_{\B{B}}}$ and $\mathcal{W}_{e_{\G{G}}}$ are nonzero, only the $\mathbb{Z}_2$ subgroup generated by $R^2$ remains as a potential symmetry.

Analogous to the Abelian case, an order parameter can be defined for the ``condensation'' of non-Abelian anyons $m_{\R{R}}$. Apart from the application of a string of $\R{X}$ on $\R{\mathcal{R}_R}$, one requires a linear-depth circuit to fix the $m_{\R{R}}\times m_{\R{R}}$ fusion channel into the identity one as described in Eq.~\eqref{eq:X_if}. Under the ungauging map specified in Sec.~\ref{sec:mapping_main}, the Wilson operator maps to the (local) thermal correlation $\mathcal{W}_{m_{\R{R}}}(\G{g},\B{b})\equiv\avg{\R{\sigma}_\G{g}\R{\sigma}_\B{b}\widetilde{\CZ}_{\G{g},\B{b}}}$  (see derivation at the end of Appendix~\ref{app:mappings}). Similarly to the Abelian case, we can repackage this into a single symmetric order parameter that is efficiently computable via Monte-Carlo methods: 
\begin{equation} \label{eq:WmR_OP}
    \mathcal{W}_{m_{\R{R}}}\equiv \frac{1}{|\G{\mathcal{R}_G}||\B{\mathcal{R}_B}|} \avg{\sum_{\G{g}\in \G{\mathcal{R}_G}}\sum_{\B{b}\in \B{\mathcal{R}_B}}{\color{red} \sigma}_\G{g} {\color{red}\sigma}_\B{b}\widetilde{\CZ}_{\G{g},\B{b}}}.
\end{equation}

In terms of the four-state clock formulation discussed in Sec.~\ref{sec:Ising_OP} (in particular around Eq.~\eqref{eq:main_4clock}), this correlation becomes 
\begin{equation} 
\begin{aligned}
    \mathcal{W}_{m_{\R{R}}}&\propto \avg{\sum_{\G{g}\in \G{\mathcal{R}_G}}\sum_{\B{b}\in \B{\mathcal{R}_B}}\G{\bm{n}}_\G{g}\cdot \B{\bm{n}}_\B{b}} \\ &\propto \avg{\sum_{\G{g}\in \G{\mathcal{R}_G}}\sum_{\B{b}\in \B{\mathcal{R}_B}}\cos(\G{\theta}_g-\B{\theta}_b)},
\end{aligned}
\end{equation}
a form that we will exploit to understand the remaining symmetry after different symmetry breaking patterns. Notice that the variable $\G{\theta}_g-\B{\theta}_b$ is invariant under the symmetry transformation $R$, i.e., $\frac{\pi}{2}$ rotations. In the following we will find that $\mathcal{W}_{m_{\R{R}}}$ only takes a finite value when in addition either $\mathcal{W}_{e_{\G{G}}}$ or $\mathcal{W}_{e_{\B{B}}}$ are also non-vanishing. As detailed in Table~\ref{tab:OP_table}, for a finite value of $\mathcal{W}_{m_{\R{R}}}$ the $D_4$ symmetry is either broken down to a $\mathbb{Z}_2$ subgroup generated by either $SR$ or $S$, or fully broken. Other symmetry-breaking patterns can potentially also arise, and would be interesting to explore, but do not appear to be relevant for the present study.

\subsubsection{Phase diagram} \label{sec:phadag_wf}

\begin{table}
    \centering
    \begin{tabular}{c|ccc|c}
        \shortstack{Remaining \\ Symmetry } & $\mathcal{W}_{e_{\G{G}}}$  & $\mathcal{W}_{e_{\B{B}}}$ &  $\mathcal{W}_{m_{\R{R}}}$ & Phase \\ [0.5ex]  \hline \hline 
        $\mathbb{Z}_2\times \mathbb{Z}_2\cong\langle SR, R^2 \rangle$ & $\neq 0$  & $0$  & $0$ & $\ket{\R{\textrm{T.C.}}}\ket{\B{\textrm{T.C.}}}$ \\
       $\mathbb{Z}_2\times \mathbb{Z}_2 \cong\langle S, R^2 \rangle$ & $0$   & $\neq 0$ & $0$ & $\ket{\R{\textrm{T.C.}}}\ket{\G{\textrm{T.C.}}}$   \\
        $\mathbb{Z}_2\ \cong\langle SR \rangle$  & $\neq 0$  & $0$  & $\neq 0$ &$\ket{\B{\textrm{T.C.}}}$  \\
        
        $\mathbb{Z}_2 \cong\langle S \rangle$ & $0$   & $\neq 0$ & $\neq 0$ & $\ket{\G{\textrm{T.C.}}}$ \\ 
        
        $\mathbb{Z}_2\ \cong\langle R^2 \rangle$ & $\neq 0$   & $\neq 0$ & $0$ & $\ket{\R{\textrm{T.C.}}}$  \\
        $\emptyset$ & $\neq 0$   & $\neq 0$ & $\neq 0$ & Trivial  \\
    \end{tabular}
    \caption{\textbf{Order parameters and remaining symmetry group for deformed pure wavefunction.} Phase diagram as characterized by the order parameters $\mathcal{W}_{e_{\G{G}}}$, $\mathcal{W}_{e_{\B{B}}}$,  $\mathcal{W}_{m_{\R{R}}}$. These signal spontaneous symmetry breaking of the $D_4$ symmetry down to a subgroup specified in the first column. The resulting quantum many-body phase is specified in the right-most column. When $\beta^z_\G{G}=\beta_{\B{B}}^z$, an additional mirror sublattice symmetry $M$ exists. }
    \label{tab:OP_table}
\end{table}

\begin{figure*}
    \centering
    \includegraphics[width=0.88\linewidth]{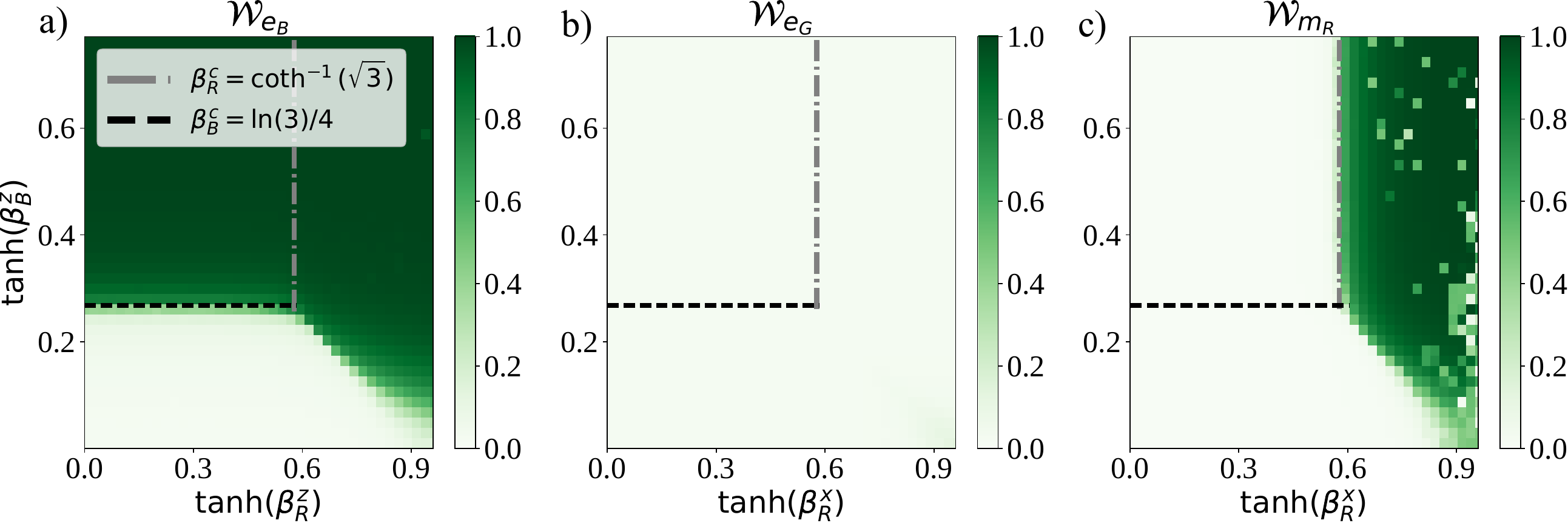}
    \caption{ \textbf{Numerical phase diagram for pure wavefunction deformations with $\beta^z_\G{G}=0$ using Monte-Carlo. } Numerical Monte-Carlo simulations with linear sizes $\mathcal{N}_x=80, \mathcal{N}_y=40$ and periodic boundary conditions. Dashed black horizontal lines indicates the value of $\beta^z_\B{B}=\frac{\ln 3}{4}$ at which an Ising transition takes place for $\beta^x_\R{R}=0$. Similarly, dotted-dashed grey vertical line marks the Ising transition occuring in the limit $\beta^z_{\B{B}}=\infty$ attained at $\beta^x_\R{R}=\coth^{-1}(\sqrt{3})$. Panels a and b show the behavior of the order parameters $\mathcal{W}_{e_\B{B}}$, $\mathcal{W}_{e_\G{G}}$ defined in Eq.~\eqref{eq:Web_OP} signaling the condensation of Abelian $e_\B{B}$ and $e_\G{G}$ charges respectively. Panel c shows instead the behavior of $\mathcal{W}_{m_\R{R}}$ signaling the condensation of non-Abelian fluxes $m_\R{R}$.  Results are averaged over $70000$ metropolis steps. Additional details of the simulations are provided in Appendix~\ref{app:MC}. }
    \label{fig:Purewf_phaseDiag_bG=0}
\end{figure*}

In the following, we will combine the knowledge of the phase diagram of decoupled O$(1)$ and O$(2)$ loop models, together with Monte-Carlo simulations, to characterize the phase diagram of the deformed wavefunction in Eq.~\eqref{eq:psi_XZ}. Recall that at the end of Sec.~\ref{sec:n_to_1}, we found that this phase diagram corresponds to that of $\textrm{tr}(\rho^n)$ for the decohered density matrix in the limit $n\to \infty$. Sketches of the resulting phase diagrams for $\beta^z_\G{G}=0$, and for $\beta^z_\G{G}=\beta^z_\B{B}$, respectively appear in Figs.~\ref{fig:Phase_diag_D4}a and b. Numerical results obtained via Monte-Carlo simulations are additionally shown in Figs.~\ref{fig:Purewf_phaseDiag_bG=0} and \ref{fig:Purewf_phaseDiag_bG=bB}, respectively. These were obtained using a bilayer honeycomb lattice with linear sizes  $\mathcal{N}_x=80, \mathcal{N}_y=40$ and periodic boundary conditions in both directions. Final data were then obtained averaging over $\mathcal{N}_x\times \mathcal{N}_y \times 7\cdot 10^4$ realizations. Additional details regarding Monte-Carlo simulations are contained in Appendix~\ref{app:MC}.

Let us first consider small deformations. We note that combining the ungauging and disentangling mappings in Sec.~\ref{sec:mapping_main} with Witten's conjugation method~\cite{Witten982,Wouters_2021}, one can find a local Hamiltonian for which $|{\psi}(\beta^x_{\R{R}},\beta^z_{\G{G}}, \beta^z_{\B{B}})\rangle$ is a ground state. In particular, small deformations correspond to small perturbations at the Hamiltonian level. Since $\ket{D_4}$ is the ground state of a gapped Hamiltonian, $|{\psi}(\beta^x_{\R{R}},\beta^z_{\G{G}}, \beta^z_{\B{B}})\rangle$ remains within the $D_4$ TO phase for sufficiently small $\beta^x_{\R{R}},\beta^z_{\G{G}}, \beta^z_{\B{B}}$. Within this phase, none of the order parameters introduced in the previous section picks up a finite expectation value---consistent with the presence of a fully $D_4$ symmetric phase. This phase corresponds to the lower left corners in panels a and b of Fig.~\ref{fig:Phase_diag_D4}. 

Now, by tuning $\beta^x_{\R{R}},\beta^z_{\G{G}}, \beta^z_{\B{B}}$ to larger values, a ground state phase transition associated to the condensation of different types of anyons can occur. To simplify matters, let us first consider the scenario with $\beta_\G{G}=0$. Weights appearing in the loop model partition function $\mathcal{Z}_{\ket{\psi}}(\beta^x_\R{R},\beta^z_\B{B})$ then become positive, leading to the simplified expression in Eq.~\eqref{eq:main_XR_ZB}. The cases where either $\beta^x_\R{R}$ or $\beta^z_\B{B}$ are set to zero respectively correspond to O$(1)$ and O$(2)$ loop models that we already encountered in Secs.~\ref{sec:Z_purewf} and \ref{sec:X_puredef}. The former admits a finite-temperature phase transition at $\tanh(\beta^z_\B{B})=2-\sqrt{3}$ beyond which the $D_4$ TO becomes an Abelian TO equivalent to two copies of the TC, as already discussed in Sec.~\ref{sec:Z_purewf}, {including the subtlety mentioned in Footnote~\ref{foot:subtlety}}. We label this phase by the state $\ket{2\,\textrm{T.C.}}$ (upper left corner in Fig.~\ref{fig:Phase_diag_D4}a). This phase transition can be diagnosed by the quantity $\mathcal{W}_{e_\B{B}}$ acquiring a finite value for sufficiently large $\beta^z_\B{B}$. This corresponds to the spontaneous symmetry breaking of $D_4$ down to the $\mathbb{Z}_2\times \mathbb{Z}_2$ subgroup generated by $\langle R^2, S\rangle$. This characterization is consistent with our numerical results in Fig.~\ref{fig:Purewf_phaseDiag_bG=0}a, where indeed $\mathcal{W}_{e_\B{B}}$ acquires a finite value at $\tanh(\beta^z_{\B{B}})>2-\sqrt{3}$ (dashed black horizontal line). Moreover, panels b and c show that in this phase $\mathcal{W}_{m_\R{R}}=\mathcal{W}_{e_\B{B}}=0$. In Fig.~\ref{fig:Purewf_phaseDiag_bG=bB}, we also characterize the phase diagram when tuning $\beta^z_\G{G}=\beta^z_\B{B}$. While the loop model for $\beta^z_\B{B}, \beta^z_\G{G}$ finite has negative Boltzmann weights, its Ising-like formulation in Eqs.~\eqref{eq:main_H_X_I}, \eqref{eq:main_H_X_II} does not, allowing to address them using Monte-Carlo. Numerical results are shown in Fig.~\ref{fig:Purewf_phaseDiag_bG=bB}a where $\mathcal{W}_{e_\B{B}}, \mathcal{W}_{e_\G{G}}$ take the same numerical values due to the $M$ mirror symmetry. Only the $\mathbb{Z}_2$ symmetry group generated by $R^2$ then remains, and the system reduces to the TC phase $\ket{\textrm{T.C.}}$. 

Our numerical results also show that this critical point [corresponding to one (two decoupled) Ising critical points when $\beta^z_\G{G}=0$ ($\beta^z_\G{G}=\beta^z_\B{B}$)] extends into a line when tuning $\beta^x_\R{R}$ to finite values (see dashed black horizontal line in panel a in Figs.~\ref{fig:Purewf_phaseDiag_bG=0} and  ~\ref{fig:Purewf_phaseDiag_bG=bB}). But how large can $\beta^x_\R{R}$ become before a different transition occurs?
Let us consider the regime with $\beta^z_\B{B}\to \infty$, i.e., $\tanh(\beta^z_\B{B})\to 1$ (upper horizontal boundaries in Fig.~\ref{fig:Phase_diag_D4}). To understand this limit, we set again $\beta^z_\G{G}=0$ and exactly rewrite the partition function in Eq.~\eqref{eq:main_XR_ZB} as 
 \begin{equation} 
 \begin{aligned}
     &\mathcal{Z}_{\ket{\psi}}(\beta^x_{\R{R}}, \beta^z_{\B{B}}) \\
     &=\sum_{L_{\R{R}}} \left(\frac{\tanh(\beta_{\R{R}}^x)}{\sqrt{2}}\right)^{|L_{\R{R}}|}\prod_{\ell_{\R{R}} \in L_{\R{R}}}\mathcal{Z}^{\textrm{Ising}}_{\ell_{\R{R}}}(\tilde{\beta}^z_\B{B}),
    \end{aligned}
 \end{equation}
 where $\mathcal{Z}^{\textrm{Ising}}_{\ell_{\R{R}}}(\tilde{\beta}^z_\B{B})$ is the partition function of the $2$D Ising model on the honeycomb lattice within the interior of a connected component $\ell_{\R{R}}$ of the loop configuration $L_{\R{R}}=\oplus \ell_{\R{R}}$, at inverse temperature $e^{-2\tilde{\beta}^z_\B{B}}=\tanh({\beta}^z_\B{B})$. Here the factor $2^{C_{L_{\R{R}}}}$ is reabsorbed into the Ising partition function $\mathcal{Z}^{\textrm{Ising}}_{\ell_{\R{R}}}(\tilde{\beta}^z_\B{B})$ in a given component $\ell_{\R{R}}$, using the exact relation $\mathcal{Z}^{\textrm{Ising}}_{\ell_{\R{R}}}(\tilde{\beta}^z_\B{B})=2\mathcal{Z}^{\textrm{O}(1)}_{\ell_{\R{R}}}({\beta}^z_\B{B})$~\footnote{Here the factor of $2$ relates to the degeneracy of domain wall configurations leading to the same loop configuration.}.  To perform this mapping, one can imagine there are two blue Ising spins on every vertex of the honeycomb lattice, which are decoupled if a red loop runs through and which are infinitely coupled otherwise. In this formulation, we can see that in the limit $\beta^z_\B{B}\to \infty$ (i.e., $\tilde{\beta}^z_\B{B}\to 0$), $\mathcal{Z}^{\textrm{Ising}}_{\ell_{\R{R}}}(\tilde{\beta}^z_\B{B})=2^{N(\ell_{\R{R}})}$, where $N(\ell_{\R{R}})$ corresponds to the number of spins in a component $\ell_{\R{R}}$. Hence,
\begin{equation} 
 \begin{aligned}
     &\mathcal{Z}_{\ket{\psi}}(\beta^x_{\R{R}}, \beta^z_{\B{B}}\to\infty) =\sum_{L_{\R{R}}} \left(\frac{\tanh(\beta_{\R{R}}^x)}{\sqrt{2}}\right)^{|L_{\R{R}}|}\sqrt{2}^{|L_{\R{R}}|}\\ &=\sum_{L_{\R{R}}} \tanh(\beta_{\R{R}}^x)^{|L_{\R{R}}|},
    \end{aligned}
 \end{equation}
 namely, we recover an O$(1)$ loop model on the honeycomb lattice which has an Ising critical point at $\tanh(\beta^x_\R{R})=1/\sqrt{3}$~\cite{HOUTAPPEL1950425}. Physically, we can interpret this result as showing that condensing $e_\B{B}$ anyons dresses up the non-Abelian $m_\R{R}$ anyons---downgrading them to Abelian, i.e., $d_{m_\R{R}}=2\to 1$.  In fact, this is consistent with the condensation picture: Upon condensing $e_\B{B}$, the non-Abelian flux $m_\R{R}$ (and similarly for $m_{\G{G}}$) splits up into quantum-dimension-one $m_\R{R}^{(1)}$ anyons, $m_\R{R}=m_\R{R}^{(1)}\oplus m_\R{R}^{(1)}e_\G{G}$. These correspond to the Abelian anyons of the $\ket{2, \textrm{T.C.}}$ phase with stabilizers $\R{A^{\textrm{TC}}_s}, \R{B_t}$  and $\G{A^{\textrm{TC}}_s}, \G{B_t}$, where $\R{A^{\textrm{TC}}_s}$ is defined via $\R{A^{\textrm{TC}}_s}=\prod_{j\in \R{\hexagon}} \R{X}_j$, and analogously for $\G{A^{\textrm{TC}}_s}$. This is analogous to splitting of the boson $\sigma\bar{\sigma}$ into charge and flux when condensing $\psi \bar{\psi}$ in an Ising$\times\overline{\text{Ising}}$ TO~\cite{Burnell_2018}. 

\begin{figure}
    \centering
    \includegraphics[width=1.05\linewidth]{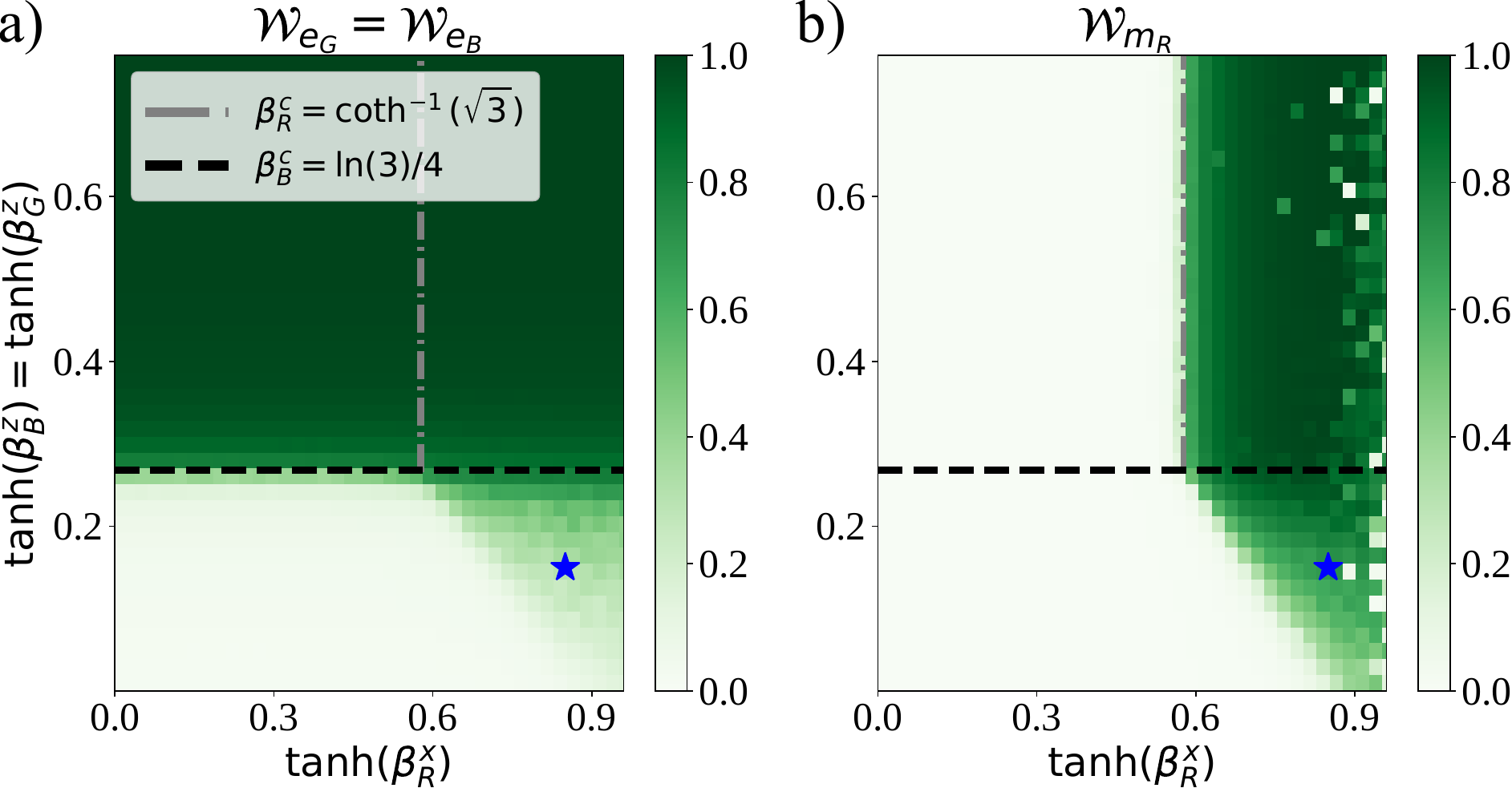}
    \caption{ \textbf{Numerical phase diagram for pure wavefunction deformations with $\beta^z_\G{G}=\beta^z_\B{B}$ using Monte-Carlo. } Numerical Monte-Carlo simulations with linear sizes $\mathcal{N}_x=80, \mathcal{N}_y=40$ and periodic boundary conditions. Because of the choice $\beta^z_\G{G}=\beta^z_\B{B}$, the order parameters $\mathcal{W}_{e_\B{B}}$, $\mathcal{W}_{e_\G{G}}$ shown in panel a numerically agree throughout the phase diagram. Panel b shows instead the behavior of $\mathcal{W}_{m_\R{R}}$ signaling the condensation of non-Abelian fluxes $m_\R{R}$. The lightly shaded region in both panels corresponds to a gapless phase. For any point in this region the three order parameters decay with system size size. Moreover, in Fig.~\ref{fig:Correlations_gapless} we show the power-law behavior of two-point correlations evaluated for the parameters indicated by $\B{\star}$. Additional details of numerical simulations are the same as in Fig.~\ref{fig:Purewf_phaseDiag_bG=0}. }
    \label{fig:Purewf_phaseDiag_bG=bB}
\end{figure}

Suppose now that the system lies deep within the $\ket{2\,\textrm{T.C.}}$ phase (namely $\beta^z_\B{B}\to\infty$ and $\beta^x_\R{R}\ll 1$), and that we increase the value of $\beta^x_\R{R}$. We just found that a phase transition can occur at a finite $\beta^x_\R{R}$, where the $\mathbb{Z}_2\times \mathbb{Z}_2$ can break down to either the trivial or to a $\mathbb{Z}_2$ subgroup. The latter can be generated by either $S$ or $SR$. As specified in Table~\ref{tab:OP_table}, the former path $(\mathbb{Z}_2\times \mathbb{Z}_2\to \langle S\rangle)$ is signaled by a non-zero value of the quantity $\mathcal{W}_{m_\R{R}}$.
Fig.~\ref{fig:Purewf_phaseDiag_bG=0}c shows that $\mathcal{W}_{m_\R{R}}$ acquires a finite value roughly beyond $\tanh(\beta^x_\R{R})=1/\sqrt{3}$, as also happens in the limit $\beta^z_\B{B}\to \infty$. 
Nonetheless, we can follow a similar discussion as in Sec.~\ref{sec:Z_purewf} and understand the fate of $\ket{2\, \textrm{T.C.}}$ by looking at its stabilizers in the limit $\beta^x_\R{R} \to \infty$. In this limit, $m_\R{R}$ anyons---now corresponding to Abelian excitations of $\R{B_t}$---condense, leading to the confinement of $e_\R{R}$ charges, with which they braid non-trivially. Algebraically, this makes the stabilizers $\R{A^{\textrm{TC}}_s}$ to become trivial $\R{A^{\textrm{TC}}_s}\to \mathds{1}$, and hence a single TC copy is left. This is stabilized by the commuting operators $\G{A^{\textrm{TC}}_s}, \G{B_t}$. 

Instead, when taking $\beta^z_\G{G}=\beta^z_\B{B}$, and assuming that the system lies deeply within the $\ket{\textrm{T.C.}}$ phase (upper left phase in Fig.~\ref{fig:Purewf_phaseDiag_bG=bB}b), the system is a common eigenstate of the stabilizers $\R{A^{\textrm{TC}}_s}, \R{B_t}$. In the limit $\beta^x_{\R{R}}\to \infty$, one then predicts a condensation transition of the (now Abelian) fluxes $m_{\R{R}}$ leading to a topologically trivial phase. This is captured by $\mathcal{W}_{m_{\R{R}}}$ acquiring a finite value as shown in Fig.~\ref{fig:Purewf_phaseDiag_bG=bB}b.

 Finally, let us explore the regime with $\beta^x_\R{R}\to \infty$, but finite $\beta^z_\B{B},\beta^z_\G{G}$. Following our discussion at the end of Sec.~\ref{sec:n_to_1}, we find that in the limit $\beta^x_\R{R}\to \infty$ (vertical right boundary in panels a and b in Fig.~\ref{fig:Phase_diag_D4}), the system is described by the stat-mech model $\sum_{L_{\R{R}}} \frac{2^{C_{L_{\R{R}}}}}{\sqrt{2}^{|L_{\R{R}}|}}\mathcal{Z}_{L_{\R{R}}}(t_\G{G},t_\B{B})$, which corresponds a O$(2)$ loop model for $t_\G{G}=t_\B{B}=0$ at the BKT critical point separating an extended gapless phase from one with exponentially decaying correlations~\cite{peled2019lectures}. As we already found, in the limit of small (and positive) $t_\G{G},t_\B{B}$, one can then explore its extended gapless phase. In the next section, we argue that while only a critical line separating the $D_4$ TO from the TC phase remains when setting $t_\G{G}=0$ (see Fig.~\ref{fig:Phase_diag_D4}a), an extended gapless phase appears for small but finite $t_\G{G}=t_\B{B}$ (see lower right triangular region in Fig.~\ref{fig:Phase_diag_D4}b), whose stability is related to the enhanced $D_4 \to D_8$ symmetry for this choice of parameters. See next subsection. In fact, a closer look at the numerical results in Fig.~\ref{fig:Purewf_phaseDiag_bG=bB} shows that for a point (blue star $\B{\star}$ in panel a) within the lightly shaded triangular region below the dashed black horizontal critical line, all order parameters $\mathcal{W}_{m_{\R{R}}},\mathcal{W}_{e_{\G{G}}},\mathcal{W}_{e_{\B{B}}}$ scale down with increasing system size---consistent with a gapless, symmetric phase with power-law-decaying correlations (see Fig.~\ref{fig:Correlations_gapless}) emerging in those regions.

\subsubsection{Field theory description}
In this section we aim to provide a consistent field theory interpretation of the results we discussed above. For this is useful to rewrite the local Hamiltonian in Eq.~\eqref{eq:main_H_ZZCZ_ZZ} using four-state clock variables in Eq.~\eqref{eq:nbng}. Recalling that $\B{n_b}=(\cos(\theta_\B{b}), \sin(\theta_\B{b}))^T$ and $\G{n_g}=(\cos(\theta_\G{g}), \sin(\theta_\G{g}))^T$ we find that the Hamiltonian \eqref{eq:main_H_ZZCZ_ZZ} takes the form

  \begin{equation} \label{eq:main_ZZCZ_ZZ_clock}
  \begin{aligned}
    H&=-\sqrt{2}\beta^x_\R{R}\sum_{\langle b,g\rangle_{\hexagon} }\cos(\B{\theta}_b-\G{\theta}_g)  - \beta^z_{\B{B}}\sum_{\langle\langle b,b^\prime \rangle \rangle_{\hexagon} } \cos(2\B{ \theta}_b -2\B{ \theta}_{b^\prime}) \\
   &-\beta^z_{\G{G}}\sum_{\langle\langle g,g^\prime \rangle \rangle_{\hexagon} } \cos(2\G{\theta}_g -2\G{ \theta}_{g^\prime}).
    \end{aligned}
\end{equation}
Here we have used that $\tilde{\sigma}_j\tilde{\sigma}_{j'}=\cos(2\theta_j-2\theta_{j'})$ when $j,j'$ lie on the same sublattice.
Recall that for all values of the parameters, the model has a $D_4=\langle R,S\rangle$ symmetry.

Let us first set $\beta^z_{\G{G}}=\beta^z_{\B{B}}=0$. In this case, the system is described by an O$(2)$ loop model which at the critical point---attained at $\beta^x_\R{R}=\infty$---corresponds to a compact boson at the BKT point.  
With this continuum formulation in mind, we define a $2\pi$-periodic, real-valued function $\theta(x,y)$ living on lattice sites now labeled $(x,y)$, constrained such that  $\theta(x,y)=\G{\theta}_g $ if $(x,y)\in \G{G}$ and $\theta(x,y)=\B{\theta}_b $ if $(x,y)\in \B{B}$. 
On $\theta(x,y)$, symmetry transformations act via $R:\theta \to \theta + \pi/2$, while $S:\theta \to -\theta$. Moreover, the mirror symmetry $M$ now acts as $M:\theta(x,y) \to \pi/4 -\theta(-x,y)$, namely $\G{\theta}_g \to \pi/4 -\B{\theta}_b$ and viceversa. 
To relax the four-state constraint in Eq.~\eqref{eq:nbng} we can include a softening potential with minima at the physical values of $\theta(x,y)$; in general this potential takes the form $+g_\B{B}\cos(4\B{\theta}_b)-g_\G{G}\cos(4\G{\theta}_g) + \cdots$, where the ellipsis denotes higher harmonics. Importantly, when $M$ symmetry is enforced, as is the case here since $\beta^z_{\G{G}}=\beta^z_{\B{B}}$, we must have $g_\B{B} = g_\G{G}$. 
Taking now the continuum limit by assuming that slowly varying configurations dominate\footnote{While this is a non-trivial assumption, notice that it correctly reproduces the known BKT critical point of the O$(2)$ loop model.}, the leading symmetry-allowed harmonic has the form $-g\cos(8\theta)$ due to cancellation of the $\cos(4\theta)$ contributions by $M$ symmetry.  One can view the surviving $-g\cos(8\theta)$ piece as implementing a common field constraint on the two sublattices.  

We then arrive at the Hamiltonian density
 \begin{equation}
     \mathcal{H} = \frac{K}{2\pi}\cos(\nabla \theta) -g\cos(8\theta).
 \end{equation}
One can recognize $\mathcal{H}$ as the XY model \cite{chaikin_lubensky_1995,dasgupta1981phase}, whose low-temperature physics can be recovered (provided the $\cos(8\theta)$ is irrelevant) by the $(2+0)$-dimensional quadratic theory $ \mathcal{H} = \frac{K}{2\pi}(\nabla \theta)^2$, or as the $(1+1)$-dimensional quantum field theory 
\begin{equation} \label{eq:quadratic}
     \mathcal{H} = \frac{1}{2\pi}\left(K(\partial_x \theta)^2 +\frac{1}{4K}(\partial_x \varphi)^2 \right) -g\cos(8\theta) - \lambda \cos(\varphi),
 \end{equation}
namely a Luttinger liquid with Luttinger parameter $K$ in the presence of additional potential terms. Here, $\varphi $ is a dual $2\pi$-periodic field satisfying\footnote{Our convention for the normalization of the fields connect to that in Ref.~\cite{fisher1996transportonedimensionalluttingerliquid} by sending $\varphi \to 2\varphi$. In this convention, the canonical commutator becomes $[\varphi(x), \partial_y \theta(y) ]=i\pi \delta(x-y)$, and the local Hamiltonian density $\mathcal{H} = \frac{1}{2\pi}\left(K(\partial_x \theta)^2 +\frac{1}{K}(\partial_x \varphi)^2 \right)$. } $[\varphi(x), \partial_y \theta(y) ]=i2\pi \delta(x-y)$, symmetric under $R$ and transforming as $\varphi\to -\varphi$ under $S$. Hence, $\cos(\varphi)$ is a symmetry-allowed local term with the smallest scaling dimension. The scaling dimension of $\cos(m\theta)$ is given by $m^2/(4K)$, and that of $\cos(n\varphi)$ by $n^2K$. 
The resulting compact boson field theory formulation when additional cosine terms are irrelevant, exactly matches the fact that the O$(2)$ loop model can be rewritten as a restricted Solid-on-Solid height-field model (discretized version of $\int(\nabla \varphi)^2$), whose critical point is described by a BKT transition~\cite{Nienhuis_84,domb2000phase}. 
It turns out that when $\beta^z_{\G{G}}=\beta^z_{\B{B}}=0$, the O$(2)$ loop model is parked\footnote{We thank Sagar Vijay for an insightful discussion on this point.} at $K=2$~\cite{Affleck_O2_K2}. For this value of the Luttinger parameter the $\cos(\varphi)$ term is marginally irrelevant for one choice of the sign of $\lambda$ and marginally relevant for the other, and is then responsible for gapping out the system into the `short loop phase'. In this work $\lambda$ is tuned by $\beta^x_{\R{R}}$ such that $\lambda\to 0$ when $\beta^x_{\R{R}}\to \infty$ (i.e., only one sign of $\lambda$ can be explored), and the short loop phase corresponds to $D_4$ TO.

So far we provided a field theory description along the horizontal axis in Fig.~\ref{fig:Phase_diag_D4} and close to the critical point $\beta^x_\R{R}=\infty$. Let us now include weak $\beta^z_{\G{G}},\beta^z_{\B{B}}\neq 0$.  
One important effect of this perturbation is to modify the value of $K$, and hence we expect additional transitions.
First, when $\beta^z_{\G{G}},\beta^z_{\B{B}}$ differ, the $M$ symmetry is broken---allowing a $\cos(4\theta)$ perturbation which is marginal for $K = 2$, coexisting with the $\cos(\varphi)$ term.  When the former becomes relevant (i.e., for $K>2$), the BKT point will be gapped out due to the pinning of $\theta$ by $\cos(4\theta)$. This scenario takes place when setting $\beta^z_\G{G}=0$ while keeping $\beta^z_{\B{B}}$ finite, leading to the toric code phase in Fig.~\ref{fig:Phase_diag_D4}a where $\B{\theta}_b$ is pinned, and hence $\mathcal{W}_{e_{\B{B}}}$ takes a finite value. On the other hand, for $K<2$, $\cos(\varphi)$ becomes relevant, driving a transition into the gapped $D_4$ TO phase. Finally, $\beta^z_{\G{G}}=\beta^z_{\B{B}}\neq 0$ (for which the $\cos(4\theta)$ term is again forbidden by symmetry) can lead to an extended gapless phase (see lower right triangle labeled `gapless' in Fig.~\ref{fig:Phase_diag_D4}b) if $2<K<8$ since then both $\cos(\varphi)$ and $\cos(8\theta)$ become irrelevant. 

We validate the consistency of this hypothesis by numerically evaluating the correlations $\mathcal{W}_{e_\B{B}}(x,y), \mathcal{W}_{m_\R{R}}(x,y)$ on a point belonging to this phase with parameters $\tanh(\beta^x_\R{R})=0.85$, and $\tanh(\beta^z_\G{G})=\tanh(\beta^z_\B{B})=0.15$. Using the previous mapping and assuming the existence of this gapless phase one finds that
\begin{equation}
    \mathcal{W}_{m_\R{R}}(x,y)=\langle \cos(\theta(x)-\theta(y))\rangle \sim |x-y|^{-\frac{1}{2K}},
\end{equation}
and similarly,
\begin{equation} \label{eq:WeB_xy}
    \mathcal{W}_{e_\B{B}}(x,y)=\langle \cos(2(\theta(x)-\theta(y))) \rangle \sim |x-y|^{-\frac{2}{K}}.
\end{equation}
Hence, both correlations are expected to decay as a power law with the distance, where the power-law exponent of $\mathcal{W}_{e_\B{B}}(x,y)$ is four times larger than that of $\mathcal{W}_{m_\R{R}}(x,y)$ regardless the value of $K$.  Fig.~\ref{fig:Correlations_gapless} shows that (i) both correlations indeed decay as a power law with $|x-y|$, and (ii) that the data approximately satisfies the relation $\mathcal{W}_{e_\B{B}}(x,y)\sim (\mathcal{W}_{m_\R{R}}(x,y))^4$ at long distances. Moreover, assuming Eq.~\eqref{eq:WeB_xy}, we extract that the Luttinger parameter is given by $K\approx 4 >2$. 

\begin{figure}
    \centering
    \includegraphics[width=0.71\linewidth]{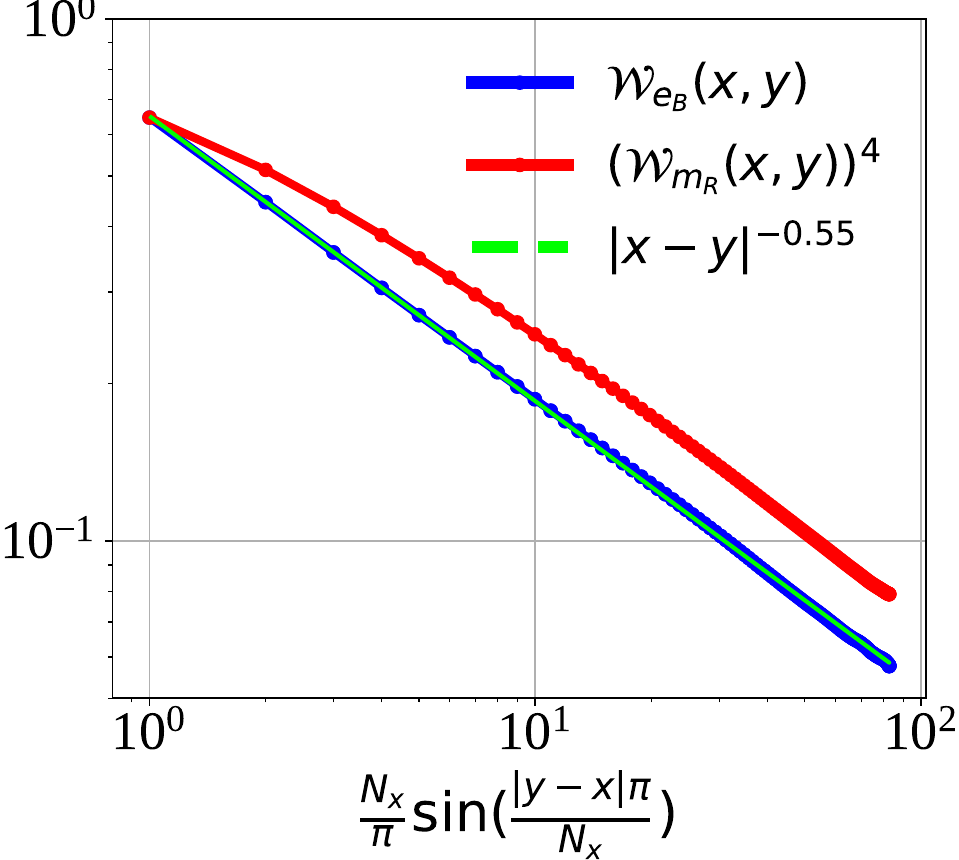}
    \caption{ \textbf{Power-law decay of correlations for $\beta_{\G{G}}^z=\beta_{\B{B}}^z$. } Numerical evaluation of correlation functions $\mathcal{W}_{e_{\B{B}}}(\B{b},\B{b'})\equiv\avg{{\color{blue} \tilde{\sigma}}_{\B{b}} {\color{blue} \tilde{\sigma}}_{\B{b}}}$ and $ \mathcal{W}_{m_{\R{R}}}(\B{b},\G{g})\equiv\avg{\R{\sigma}_{\B{b}}\R{\sigma}_{\G{g}}\widetilde{\CZ}_{\B{b},\G{g}}}$ within the extended gapless phase for parameters $\tanh(\beta_{\B{B}}^z)=0.15, \tanh(\beta_{\B{B}}^z)=0.85 $ indicated with $\B{\star}$ in Fig.~\ref{fig:Purewf_phaseDiag_bG=bB}a. For the former $x,y$ belong to the blue sublattice, while for the latter $x\in \G{G}$ and $y\in \B{B}$. The average $\langle \cdot \rangle$ corresponds to the thermal expectation value evaluated with respect the classical Hamiltonian in Eq.~\eqref{eq:main_H_ZZCZ_ZZ}. Numerical data corresponds to system size $\mathcal{N}_x=520, \mathcal{N}_y=260$ with periodic boundary conditions, $7\cdot 10^4$ equilibration time steps and $12\cdot 10^4$ Metropolis steps to obtain the plotted data.}
    \label{fig:Correlations_gapless}
\end{figure}

Finally, if $\beta^z_{\G{G}},\beta^z_{\B{B}}$ are strong, one sees from the lattice Hamiltonian that we should be able to condense, e.g.,
$\cos(2\theta)$ without condensing $\cos(\theta)$. This can be accomplished by e.g., introducing new `Ising' variables $\sigma$ that have the same symmetry charges as $\cos(2\theta)$. We provide a consistent field theory formulation in App.~\ref{app:field_theory}.

\subsection{Local quantum channels} \label{sec:LQC_comb}

We now describe the fate of the decohered density matrix as measured by the purity $\textrm{tr}(\rho^2)$, when subjecting $D_4$ TO to the composition of the following commuting local quantum channels
\begin{equation}
\begin{aligned} \label{eq:EX_D4_comb}
    &\mathcal{E}^{\R{X}}_{r}(\rho_0) = (1-p_\R{R})\rho_0 + p_\R{R} \R{X}_r\rho_0 \R{X}_r, \\ 
    &\mathcal{E}^{\B{Z}}_{b}(\rho_0) = (1-p_{\B{B}})\rho_0 + p_{\B{B}} \B{Z}_b \rho_0 \B{Z}_b,\\ 
    &\mathcal{E}^{\G{Z}}_{g}(\rho_0) = (1-p_\G{G})\rho_0 + p_\G{G} \G{Z}_g\rho_0 \G{Z}_g,
\end{aligned}
\end{equation}
with tunable error rates $p_\R{R}, p_\G{G}, p_\B{B} \in [0,1/2]$. 

\subsubsection{Phase diagram from $\textrm{tr}(\rho^\infty)$}

As discussed at the end of Sec.~\ref{sec:n_to_1}, the limit $\lim_{n\to \infty}\textrm{tr}(\rho^n)^{1/n}$ at $p_\R{R}=\frac{1}{2}$ is given by the norm of the deformed wavefunction $ |{\psi}(\beta^x_{\R{R}},\beta^z_{\G{G}}, \beta^z_{\B{B}})\rangle$ (up to possibly overall factors) when $\beta^x_{\R{R}}=\infty$, as long as the weight $\mathcal{Z}_{L_{\R{R}}}(t_\G{G},t_\B{B})\geq 0$ in Eq.~\eqref{eq:Z_gammaR} is non-negative. First, we showed that this is always true when $\beta^z_\G{G}=0$, and hence the phase diagram of $\textrm{tr}(\rho^\infty)$ is given by the right vertical boundary of the phase diagram in Fig.~\ref{fig:Phase_diag_D4}a. As explicitly shown in Fig.~\ref{fig:trrho_Inf}a, this showcases an unstable BKT critical point at $\beta^z_\B{B}=0$ beyond which $D_4$ TO is completely lost. 

On the other hand, we also argued that even when $\beta^z_\G{G}, \beta^z_\B{B}$ are finite, $\mathcal{Z}_{L_{\R{R}}}(t_\G{G},t_\B{B})$ remains non-negative. We use this fact to conjecture that the phase diagram of $\lim_{n\to \infty}\textrm{tr}(\rho^n)^{1/n}$ as a function of $\beta^z_\G{G}, \beta^z_\B{B}$ corresponds to the right vertical boundary in Fig.~\ref{fig:Phase_diag_D4}b. Unlike the previous case, this contains an extended gapless phase described by a compact boson for values of $\beta^z_\G{G}, \beta^z_\B{B}$ below a threshold.  The corresponding phase diagram is shown in Fig.~\ref{fig:trrho_Inf}b. Recall that as argued in Sec.~\ref{sec:n_to_1} the critical threshold for the $n\to \infty$ limit, $ p_c^{(\infty)}$, lower bounds the intrinsic threshold beyong which the capacity of $D_4$ TO as a good quantum memory decreases.

\subsubsection{Phase diagram from purity}

\begin{figure}
    \centering
    \includegraphics[width=\linewidth]{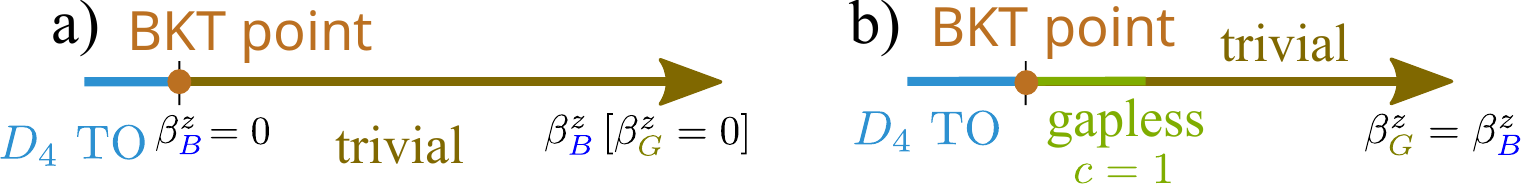}
    \caption{ \textbf{Phase diagram of $\textrm{tr}(\rho^\infty)$ for initial state $|\psi(\beta^z_{\G{G}}, \beta^z_{\B{B}},\beta^z_{\R{R}}=0)\rangle $ in Eq.~\eqref{eq:psi_abel} at $p_\R{R}=\frac{1}{2}$.} {The phase diagrams include the $D_4$ topological order phase, a gapless phase, and a ``trivial'' phase\textsuperscript{\ref{foot:subtlety}}. }
    \label{fig:trrho_Inf}}
\end{figure}

\begin{figure*}
    {\centering
    \includegraphics[width=0.8\linewidth]{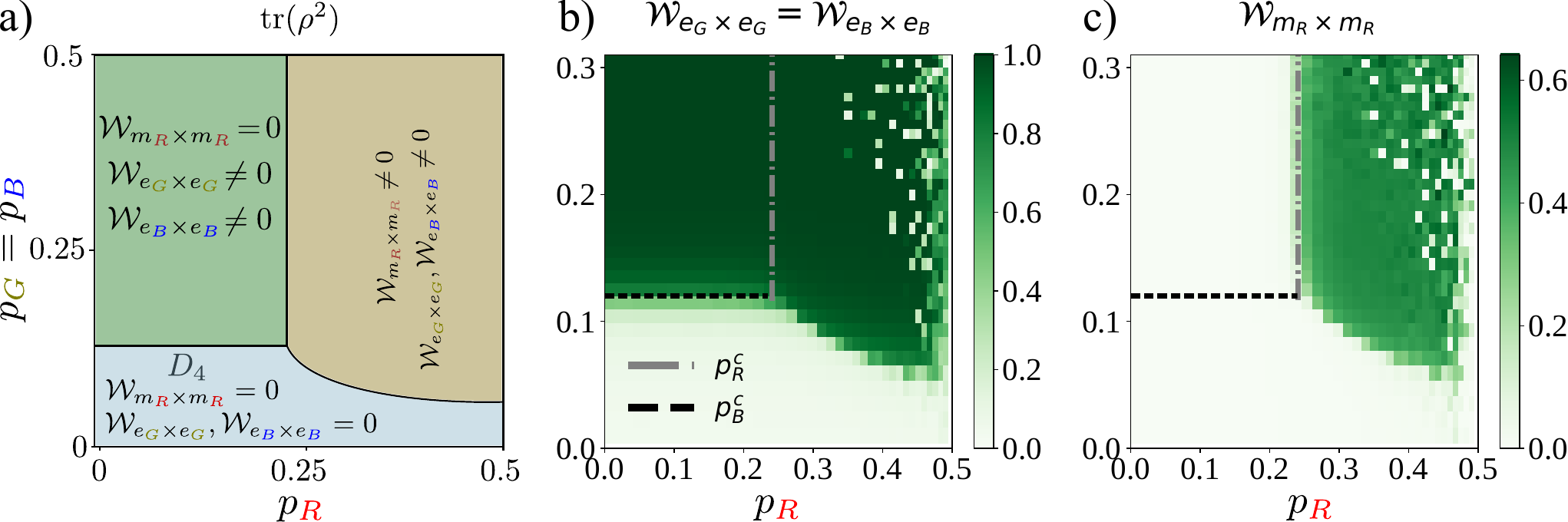}}
    \caption{\textbf{Phase diagram for decohered mixed state with $ p_\G{G}=p_\B{B}$ via $\textrm{tr}(\rho^2)$.} (a) Schematic phase diagram including the values of the relevant order parameters. $D_4$ TO is robust to finite error rates (light blue region). This is signaled by the order parameters acquiring a finite value for sufficiently large error rates.  (b-c) Numerical phase diagram with linear system sizes $\mathcal{N}_x=40$ and $\mathcal{N}_y=20$ and with periodic boundary conditions.  Dashed black horizontal and grey dotted-dashed lines indicate the critical values $p_\G{G}=p_\B{B}=(1-3^{-1/4})/2\approx 0.12$ and $p_\R{R}=(1-\sqrt{2-\sqrt{3}})/2\approx 0.24$ respectively.  Panel b and c show numerical results for the order parameters $\mathcal{W}_{e_\G{G}\times e_\G{G}}=\mathcal{W}_{e_\B{B}\times e_\B{B}}$ (Eq.~\eqref{eq:W_gg}) and $\mathcal{W}_{m_\R{R}\times m_\R{R}}$ (Eq.~\eqref{eq:W_rr}) respectively.} 
    \label{fig:mixed_pG=PB}
\end{figure*}

Following a similar derivation as in the previous sections one finds that the purity of the decohered density matrix takes the general form
 \begin{equation}
\begin{aligned}
     &\textrm{tr}(\rho^2) \;
    \overset{\rm decohere}{\underset{m_\R{R}, e_{\G{G}}, e_{\B{B}}}{\propto}}
    \; \sum_{L_{\R{R}}}\sum_{\gamma_{\B{B}}} \sum_{\gamma_{\G{G}}} \\ &\times (r_{\B{B}}^z)^{|\gamma_{\B{B}}|}(r_{\G{G}}^z)^{|\gamma_{\G{G}}|} \left(\frac{r_{\R{R}}^x}{2}\right)^{|L_{\R{R}}|}4^{C_{L_{\R{R}}}},
\end{aligned}
 \end{equation}
 i.e., three coupled $O(N)$ loop models with $N=1,4$. Notice the absence of the sign function $\sigma_{L_{\R{R}}}$ when comparing to Eq.~\eqref{eq:main_X_R_Z_BG}, the reason being that now every contribution for the pure wavefunction deformation is squared. Alternatively, one can write these coupled loop models in an explicitly local way similar to Eq.~\eqref{eq:main_rho_H_X} and given by (including temperature factors)
 \begin{equation} \label{eq:main_rho_H_XpZ}
 \begin{aligned}
    H_{\rho}&=-\beta^x_\R{R}\sum_{\langle i,j \rangle_{\hexagon}}\R{\sigma}_i \R{\sigma}_j\widetilde{\CZ}^{(2)}_{ij} \widetilde{\CZ}^{(3)}_{ij} \\  &- \beta^z_\B{B} \sum_{\langle \langle b,b'\rangle \rangle_{\hexagon} } \B{\tilde{\sigma}}_b^{(2)}\B{\tilde{\sigma}}_{b'}^{(2)} \B{\tilde{\sigma}}_b^{(3)}\B{\tilde{\sigma}}_{b'}^{(3)} \\ &- \beta^z_\G{G} \sum_{\langle \langle g,g'\rangle \rangle_{\hexagon } }\G{\tilde{\sigma}}_g^{(2)}\G{\tilde{\sigma}}_{g'}^{(2)} \G{\tilde{\sigma}}_g^{(3)}\G{\tilde{\sigma}}_{g'}^{(3)},
\end{aligned}
\end{equation}
defined on three honeycomb layers (one more than the stat-mech model associated to the deformed wavefunction). As discussed in previous sections, this model has an enhanced symmetry group, since $H_\rho$ is invariant under $D_4$ symmetry transformations involving either layers $1$ and $2$, or $1$ and $3$.

While a more detailed characterization of the resulting phase diagram as in Ref.~\onlinecite{Mong_24} is left for future work, we now briefly discuss the thresholds beyond which $D_4$ TO is lost in the presence of decoherence assuming $p_\G{G}=p_\B{B}$. As for the pure wavefunction deformation, the previous formulation permits a numerical evaluation of the phase diagram of the decohered density matrix using standard Monte-Carlo methods. 
We directly simulate $H_\rho$ in Eq.~\eqref{eq:main_rho_H_XpZ} on a trilayer honeycomb lattice with a brick-wall configuration with linear system sizes $\mathcal{N}_x=80$ and $\mathcal{N}_y=40$. 
Numerical results together with a sketch of the resulting phase diagram (panel a) are shown in Fig.~\ref{fig:mixed_pG=PB}. 

From a physics perspective, the main difference between decohering a density matrix and then computing its purity, and analyzing a pure wavefunction deformation, is that anyons are simultaneously created on both the bra and ket in the former case. Indeed, let us, e.g., consider the effect of the quantum channel $\mathcal{E}^{\B{Z}}_{b}$ in Eq.~\eqref{eq:EX_D4_comb}, and recall that the purity corresponds (up to an overall constant) to the norm of the unnormalized state 
\begin{equation}
\begin{aligned}
     \kket{\rho}&= e^{\mu_{\B{B}}\sum_{b\in \B{\mathcal{R}_B}} Z_j\otimes Z_j}\ket{D_4}\otimes \ket{D_4},
\end{aligned}
\end{equation}
with $\tanh(\mu_\B{B})=p_\B{B}/(1-p_{\B{B}})$. Hence, pairs of fluxes $e_\B{B}\times {\bar e_\B{B}}$ are simultaneously created on the first and second copy~\cite{bao2023mixedstate}. The corresponding order parameter capturing the condensation of these pairs is given by
\begin{equation} \label{eq:W_gg}
    \mathcal{W}_{e_{\B{B}}\times e_{\B{B}}}
    \equiv \frac{1}{|\B{\mathcal{R}_B}|^2}
    \avg{ \sum_{b,b'\in \B{\mathcal{R}_B}}{\color{blue} \tilde{\sigma}}^{(2)}_b {\color{blue} \tilde{\sigma}}^{(3)}_b  {\color{blue}\tilde{\sigma}}^{(2)}_{b'}{\color{blue}\tilde{\sigma}}^{(3)}_{b'} },
\end{equation}
and analogously defined for $\mathcal{W}_{e_{\G{G}}\times e_{\G{G}}}$. On the other hand, the condensation of $m_\R{R}\times \bar{m_\R{R}}$ is signaled by 
\begin{equation} \label{eq:W_rr}
    \mathcal{W}_{m_{\R{R}}\times m_{\R{R}}}\equiv \frac{1}{|\G{\mathcal{R}_G}||\B{\mathcal{R}_B}|} \avg{\sum_{\G{g}\in \G{\mathcal{R}_G}}\sum_{\B{b}\in \B{\mathcal{R}_B}}{\color{red} \sigma}_\G{g} {\color{red}\sigma}_\B{b}\widetilde{\CZ}^{(2)}_{\G{g},\B{b}} \widetilde{\CZ}^{(3)}_{\G{g},\B{b}}}.
\end{equation}

Panels b and c in Fig.~\ref{fig:mixed_pG=PB} show the dependence of these order parameters as a function of $p_\R{R}$ and $p_\G{G}=p_\B{B}$, resulting in the sketched diagram appearing in panel a. For this numerical analysis we considered a smaller system size $\mathcal{N}_x=40, \mathcal{N}_y=20$, since we found that closed to the zero temperature regime ($p_{\R{R}}\to 1/2$) the algorithm takes longer to converge. We believe the reason is the use of single site updates while interactions in $H_\rho$ involve local $6$-spins terms. Similar to the pure wavefunction deformation, an Ising phase transition occurs when $p_\R{R}=0$ and for $p_\G{G}=p_\B{B}=(1-3^{-1/4})/2$. In this case, the partition function with Hamiltonian $H_\rho$ in Eq.~\eqref{eq:main_rho_H_XpZ} corresponds to two decoupled $2$D ferromagnetic Ising model lying on a triangular lattice. Figure~\ref{fig:mixed_pG=PB}b shows that in fact, this point extends into a critical line (dashed black horizontal line) for finite values of $p_\R{R}$. Also, similarly to our analysis of the deformed wavefunction, we find an Ising critical point in the limit $p_\G{G}=p_\B{B}\to\infty$ for $p_\R{R}=(1-\sqrt{2-\sqrt{3}})/2$, which also extends into a critical line (dotted dashed grey vertical line) as shown in Fig.~\ref{fig:mixed_pG=PB}c. However, unlike for the deformed wavefunction, attaining a critical point at $\beta^x_\R{R}\to \infty$ for $\beta^z_\G{G}=\beta^z_\B{B}=0$, the horizontal axis with vanishing $p_\G{G},p_\B{B}$ is described by an O$(4)$ loop model---which shows no critical behavior for any value of $p_\R{R}$. Finally, our numerical results indicate that even for $p_\R{R}=1/2$, $D_4$ topological order is robust to finite error rates $p_{\G{G}}=p_{\B{B}}$. This seems natural since when $p_{\G{G}}=p_{\B{B}}=0$, the system is well within the shortly-correlated small loop phase of the O$(4)$ loop model, and so by infinitesimally increasing the coupling to $\gamma_\G{G},\gamma_\B{B}$, large red loop configurations $L_\R{R}$ (indicative of the dense loop phase) are not expected to be suddenly generated. Hence, a finite threshold is expected.

\subsection{Competing non-Abelian errors} \label{sec:var_NA}

At this point we have explored scenarios involving either a single non-Abelian error, or a combination of Abelian and non-Abelian errors acting on two different sublattices. In this last subsection we consider the scenario of non-Abelian $X$ errors acting on either two or all three sublattices. Suppose that $X$ errors act on both $\R{\mathcal{R}_R}$ and $\B{\mathcal{R}_B}$ sublattices, hence proliferating pairs of $m_\R{R}$ and $m_\B{B}$ fluxes, respectively. Unlike in our previous analysis, two non-Abelian anyons of different color have non-trivial braiding statistics. In fact, even when two pairs of $m_\R{R}$ and $m_\B{B}$ are generated out of the vacuum, their braiding toggles the fusion channel of both pairs from $1$ to $e_\G{G}$. Hence, when annihilated, a pair of $e_\G{G}$ are left behind~\cite{iqbal2023creation}. This scenario corresponds to either the deformed wave-function
 \begin{equation} 
 \begin{aligned}
     &|{\psi}(\beta^x_{\R{R}}, \beta^x_{\B{B}})\rangle=e^{\frac{\beta^x_{\R{R}}}{2} \sum_{r\in \R{\mathcal{R}_R}} X_r} e^{\frac{ \beta^z_{\B{B}}}{2} \sum_{b\in \B{\mathcal{R}_B}} X_b} \ket{D_4},
\end{aligned}
 \end{equation}
or the composition of the local quantum channels
\begin{equation}
\begin{aligned}
    &\mathcal{E}^{\R{X}}_{r}(\rho_0) = (1-p_\R{R})\rho_0 + p_\R{R} \R{X}_r\rho_0 \R{X}_r,\\
    &\mathcal{E}^{\B{X}}_{b}(\rho_0) = (1-p_\B{B})\rho_0 + p_\B{B} \B{X}_b\rho_0 \B{X}_b,
\end{aligned}
\end{equation}
with $r\in \R{\mathcal{R}_R}$ and $b\in \B{\mathcal{R}_B}$. We find the classical partition function (see Appendix~\ref{app:2_NA}) 
  \begin{equation}
  \begin{aligned}
     \mathcal{Z}_{\ket{\psi}}(\beta^x_{\R{R}}, \beta^x_{\B{B}})=&\sum_{L_{\B{B}}, L_{\R{R}}}\frac{\tanh(\beta_{\R{R}}^x)^{|L_{\R{R}}|}\tanh(\beta_{\B{B}}^x)^{|L_{\B{B}}|}}{\sqrt{2}^{|L_{\R{R}}\cup L_{\B{B}}|}}\\
     &\times 2^{C_{L_{\R{R}}\cup L_{\B{B}}}},
\end{aligned}
 \end{equation}
 for the deformed pure wavefunction, and 
  \begin{equation}
  \begin{aligned}
     \textrm{tr}(\rho^2) \;
    \overset{\rm decohere}{\underset{  m_\R{R}\, , \, m_\B{B}}{\propto}}
    \;  &\sum_{L_{\B{B}}, L_{\R{R}}}\frac{r_{\R{R}}^{|L_{\R{R}}|}r_{\B{B}}^{|L_{\B{B}}|}}{2^{|L_{\R{R}}|+| L_{\B{B}}|}} 4^{C_{L_{\R{R}}\cup L_{\B{B}}}},
\end{aligned}
 \end{equation}
 when computing the purity of the resulting decohered density matrix. Here, we have used the fact that only closed loop configurations  $L_{\R{R}}, L_{\B{B}}$  on the red $\R{\mathcal{R}_R}$ and blue $\B{\mathcal{R}_B}$ sublattices, respectively, lead to non-vanishing contributions. Moreover, $C_{L_{\R{R}}\cup L_{\B{B}}}$ corresponds to the number of independent loops in the (colorless) union ${L_{\R{R}}\cup L_{\B{B}}}$, or equivalently and perhaps more clearly, the number of faces enclosed (see Fig.~\ref{fig:two_NA}}a for an example with $C_{L_{\R{R}}\cup L_{\B{B}}}=4$). This coupling effectively works as an attractive potential between loops of different color that increases their probability of crossing. This enhancement is directly related to their non-trivial braiding. If the models were decoupled, we would have instead found $2^{C_{L_{\R{R}}}+C_{L_{\B{B}}}}$. Hence, we find that the combination of two non-Abelian errors leads to two coupled O$(2)$ loop models. Unfortunately, we are not aware of previous works studying the phase diagram of the resulting coupled loop model, so we leave their study to future work. Finally, including the action of an $X$ error on the remaining green sublattice $\G{\mathcal{R}_G}$ follows a similar logic. However, in this case three (instead of two) different colored loops can intersect in two different ways, and results in vanishing contributions. What would be the fate of $D_4$ TO under the simultaneous action of these errors? Does the braiding have an impact on the robustness of $D_4$ TO under decoherence? 

\begin{figure}
     \centering
     \includegraphics[width=0.7\linewidth]{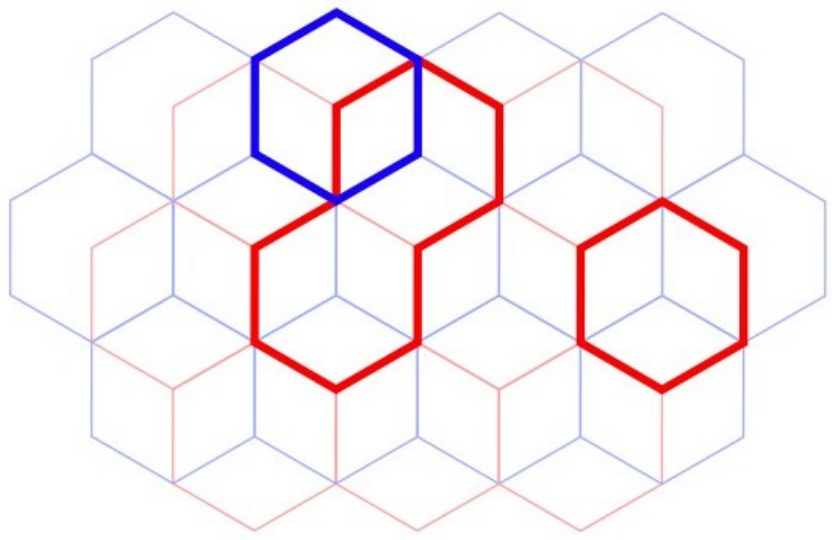}
     \caption{\textbf{Competing non-Abelian errors.} Resulting coupled O$(2)$ loop models when considering the proliferation of two non-Abelian errors acting on different sublattices.For the configuration shown in the figure, $C_{L_{\R{R}}\cup L_{\B{B}}}=4$.}
     \label{fig:two_NA}
 \end{figure}

\section{$D_4$ strong-to-weak spontaneous symmetry breaking} \label{sec:D4_SSB_X}

Previous sections have dealt with the loss of coherence arising from coupling to an environment, which can result in a topologically ordered state losing its ability to store quantum information.
However, the effect of a quantum channel does not necessarily lead to featureless stationary mixed states stripped of quantum correlations. 
There are various ways to avoid this fate---e.g., by fine-tuning the environment~\cite{ 2008_zoller_Lindblad_pre_state, 2009_verstraete_quantum_state_engineer, 2019_buca_jaksch_non_statioanry_dissipation, 2023_symmetry_induced_DFS, 2023_google_engineered_dissipation}, or restricting to strongly symmetric quantum channels, namely those for which every Kraus operator is symmetric~\cite{li_2023,li2024highlyentangledstationarystatesstrong,LeeYouXu2022, ma2024topological, Lessa_24, 2024_sala_SSSB,lessa2024mixedstatequantumanomalymultipartite}. As an example, the phenomenon of strong-to-weak spontaneous symmetry breaking (sw-SSB) has been shown to trigger long-range order in mixed states as measured by non-linear observables in the density matrix~ \cite{LeeYouXu2022, ma2024topological, Lessa_24, 2024_sala_SSSB}.

Following these works, let us introduce the concept by considering a $\mathbb{Z}_2$-symmetric short-range correlated initial state $|\psi_0\rangle$, under the action of a strongly $\mathbb{Z}_2$ -symmetric local quantum channel. For example, take $\ket{\psi_0}=\ket{+}^{\otimes N}$ in the $X$-basis and suppose that every Kraus operator commutes with the $\mathbb{Z}_2$ generator $G=\prod_j X_j$. Such a symmetric channel can be given by the composition of local quantum channels
\begin{equation}
    \rho_0 \to \mathcal{E}_{i,j}(\rho_0)=(1-p)\rho_0 + p Z_iZ_j \rho_0 Z_i Z_j
\end{equation}
acting on every bond $(i,j)$ of a square lattice. Notice that the resulting decohered density matrix $\rho$ is invariant under both left and right multiplication by the symmetry transformation $G$. When this is the case, $\rho$ is said to be strongly $G$-symmetric.

While a local quantum channel cannot modify the scaling of linear correlations of the form $\textrm{tr}(\rho O_xO_y)$ as imposed by Lieb-Robinson bounds, the phenomenon of sw-SSB is signalled by the generation of long-range correlations in non-linear quantities like the Rényi-2 correlators 
\begin{equation}
    R^{(2)}_{\mathbb{Z}_2}(x,y)\equiv \frac{\textrm{tr}\left(\rho O_xO_y \rho O_xO_y\right)}{\textrm{tr}(\rho^2)},
\end{equation}
with $O_x$ a charged operator under $G$, e.g., $Z_x$.
The definition of sw-SSB, up to subtleties that will be discussed later on, is as follows~\cite{Lessa_24}: A strongly $G$-symmetric mixed state $\rho$ has sw-SSB if the Rényi-$2$ correlator $ R^{(2)}_{\mathbb{Z}_2}(x,y)$ of some charged local operator (e.g., $Z_x$ in this case) is non-vanishing in the limit $|x-y|\to\infty$, while the conventional correlation function $\textrm{tr}(\rho Z_xZ_y)$ shows no long-range order.
When evaluating $R^{(2)}(x,y)$ for $\ket{\psi_0}$, one finds that this corresponds to the thermal correlation $\langle \sigma_x\sigma_y\rangle_{2\beta}$ on the $2$D Ising model at inverse temperature $2\beta$ with $\tanh(\beta)=p/(1-p)$. Hence, for error rates larger than a critical value $p_c^{(2)}$, this metric reveals that the density matrix $\rho$ displays sw-SSB.
It follows that for $p^{(2)}>p_c^{(2)}$ the decohered density matrix is not `symmetrically invertible', namely no mixed state $\tilde{\rho}$ exists such that $\rho \otimes \tilde{\rho}$ can be connected to a symmetric product state via a local quantum channel~\cite{Lessa_24}, showcasing the non-trivial structure of $\rho$.

While a priori this phenomenon appears to be unrelated to a decoherence transition for TO states, it has been shown that, in fact, these phenomena are intimately related. One can can connect the $\mathbb{Z}_2$ sw-SSB transition discussed above to decohering the TC wavefunction via an $Z$ quantum channel using Wegner's duality~\cite{wegner1971duality,LeeYouXu2022,chen2023separability}. This duality maps the toric code wavefunction to a trivial paramagnet $\ket{+}^{\otimes N}$ and vice versa. Similarly here, the combination of the gauging and entangling maps introduced in Sec.~\ref{sec:mapping_main} together with Eq.~\eqref{eq:map_X_red} provide a complementary dual picture of subjecting the $D_4$ topological order to decoherence as a $D_4$ sw-SSB transition.
To reveal this connection, we consider a bilayer system where each of the layers corresponds to a honeycomb lattice with qubits lying on the vertices as in Fig.~\ref{fig:bil_honey}a. The upper layer hosts red qubits initialized in the product state $\ket{\bm{\R{+}}}\equiv \ket{\R{+}}^{\otimes |\R{\mathcal{R}_R}|}$, while the lower layer hosts green $\ket{\bm{\G{\tilde{+}}}}\equiv \ket{\G{\tilde{+}}}^{\otimes |\G{\mathcal{R}_G}|}$ and blue  $\ket{\bm{\B{\tilde{+}}}}\equiv \ket{\B{\tilde{+}}}^{\otimes |\B{\mathcal{R}_B}|}$ qubits lying on the two $\G{\mathcal{R}_G}$ and $\B{\mathcal{R}_B}$ sublattices, respectively. The state of the system $\ket{\psi_0}=\ket{\bm{\R{+}}}\ket{\bm{\G{\tilde{+}}}} \ket{\bm{\B{\tilde{+}}}}$ is then invariant under the $D_4$ symmetry transformation generated by $R$ and $S$ as defined below Eq.~\eqref{eq:SR}. 
Following our discussion in the previous sections, one can readily find strongly $D_4$ symmetric local quantum channels corresponding to the ``ungauged'' version of the quantum channels in Eqs.~\eqref{eq:EX_D4_comb}. These are given by the composition of the commuting local quantum channels
\begin{equation} \label{eq:EX_D4}
\begin{aligned}
    &\mathcal{E}^{\R{X}}_{i,j}(\rho_0) \\
    &= (1-p_\R{x})\rho_0 + p_\R{x} \R{Z}_i\R{Z}_j \widetilde{\CZ}_{ij}\rho_0 \R{Z}_i\R{Z}_j \widetilde{\CZ}_{ij}, 
\end{aligned}
\end{equation}
 acting on every link $(i,j)$ of the bilayer system, and 
\begin{align} \label{eq:EZ_EX_D4}
    &\mathcal{E}^{\B{Z}}_{\B{b},\B{b'}}(\rho_0) = (1-p_{\B{z}})\rho_0 + p_{\B{z}} \B{\tilde{Z}}_{\B{b}}\B{\tilde{Z}}_{\B{b'}} \rho_0 \B{\tilde{Z}}_{\B{b}}\B{\tilde{Z}}_{\B{b'}},\\
     &\mathcal{E}^{\G{Z}}_{\G{g},\G{g'}}(\rho_0) = (1-p_\G{z})\rho_0 + p_\G{z} \G{\tilde{Z}}_{\G{g}}\G{\tilde{Z}}_{\G{g'}} \rho_0 \G{\tilde{Z}}_{\G{g}}\G{\tilde{Z}}_{\G{g'}},
\end{align}
with $\B{b},\B{b'}\in \B{\mathcal{R}_B}$ and $\G{g},\G{g'}\in \G{\mathcal{R}_G}$. Consequently, characterizing the possible spontaneous symmetry breaking patterns of $D_4$, is equivalent to the characterization of the possible ways on which the decoherence channels in Eqs.~\eqref{eq:EX_D4_comb} affect $D_4$ TO.

 In this case, one can diagnose sw-SSB by the non-vanishing of the Rényi-2 correlators
\begin{equation} \label{eq:W_R}
\begin{aligned}
    &R^{(2)}_{D_4}(\B{b},\G{g})=\frac{\mathrm{tr} (\rho \R{Z}_{\B{b}}\R{Z}_{\G{g}} \widetilde{\CZ}_{\B{b}\G{g}} \rho\R{Z}_{\B{b}}\R{Z}_{\G{g}} \widetilde{\CZ}_{\B{b}\G{g}} )}{\mathrm{tr} (\rho ^2)},
\end{aligned}
\end{equation}
and
\begin{align} \label{eq:W_BG}
    &R^{(2)}_\G{G}(\G{g},\G{g'})=\frac{\mathrm{tr} (\rho \G{\tilde{Z}}_{\G{g}}\G{\tilde{Z}}_{\G{g'}}  \rho\G{\tilde{Z}}_{\G{g}}\G{\tilde{Z}}_{\G{g'}} )}{\mathrm{tr} (\rho ^2)}, \\
    &R^{(2)}_\B{B}(\B{b},\B{b'})=\frac{\mathrm{tr} (\rho \B{\tilde{Z}}_{\B{b}}\B{\tilde{Z}}_{\B{b'}}\rho\B{\tilde{Z}}_{\B{b}}\B{\tilde{Z}}_{\B{b'}} )}{\mathrm{tr} (\rho ^2)},
\end{align}
in the limit $|{\G{g}}-{\G{g'}}|,|{\B{b}}-{\B{b'}}|\to \infty$. The former $R^{(2)}_{D_4}(\B{b},\G{g})$ corresponds to the thermal expectation value $\langle \R{Z}_{\B{b}}\R{Z}_{\G{g}}  \widetilde{\CZ}^{(2)}_{\B{b}\G{g}}  \widetilde{\CZ}^{(3)}_{\B{b}\G{g}} \rangle $ evaluated on the three-honeycomb-layer stat-mech model introduced in Eq.~\eqref{eq:Z_XR}. This quantity exactly matches the expectation value $\mathcal{W}_{m_{\R{R}}\times m_{\R{R}}}$ (when summed over all sites $x,y$) signalling the condensation of flux pairs $m_{\R{R}}\times m_{\R{R}}$. The latter two in Eq.~\eqref{eq:W_BG} correspond to the condensation of $e_\G{G}\times e_\G{G}$ and $e_\B{B}\times e_\B{B}$ pairs respectively, as signaled by the respective order parameters $\mathcal{W}_{e_{\G{G}}\times e_{\G{G}}}$ and $\mathcal{W}_{e_{\B{B}}\times e_{\B{B}}}$. From this perspective, these correlators detect different symmetry breaking patterns with the corresponding phase diagram resembling that in Fig.~\ref{fig:mixed_pG=PB}. A detailed analysis of this phase diagram will be included somewhere else.

Before closing this section we notice that an alternative, and inequivalent, characterization of sw-SSB is via the so-called fidelity correlator~\cite{Lessa_24} (corresponding to a generalization of the quantum fidelity involving the state $\rho$). Unlike for Rényi-$2$ correlators, this quantity acquiring a finite value permits to prove a stability theorem that establishes sw-SSB as a universal mixed-state property~\cite{Lessa_24}. Similar to the relation, through Kramers-Wannier duality, between the toric code ground state under bit-flip noise and the $\mathbb{Z}_2$ sw-SSB~\cite{LeeYouXu2022,chen2023separability,Lessa_24}, we notice that the fidelity correlator to diagnose $D_4$ sw-SSB, maps to the quantum fidelity defined in Eq.~\eqref{eq:fidelity}, after replacing the closed non-contractible loop $\R{\mathcal{X}}$ by the open string $\R{\mathcal{X}}_{t_i}^{t_f}$, and performing the ungauging map in Sec.~\ref{sec:mapping_main}. 
Hence, in combination to our findings in Section~\ref{sec:n_to_1} for $p_\G{G}=p_\B{B}=0$, it is plausible that the critical point for $D_4$ sw-SSB is attained at maximum error rate! 

\section{Conclusions and outlook} \label{sec:conclusions}

We have characterized the effect of decoherence and wavefunction deformation for a microscopic realization of $D_4$ TO. For the case of proliferating Abelian anyons, we recovered the phenomenology found in previous works for the toric code ground state, where the random bond Ising model governs the (intrinsic) quantum correction threshold $p_c$~\cite{Dennis_2002,fan2023diagnostics}. Our main focus is on the proliferation of non-Abelian anyons.
We found that even if we decohere \emph{all} possible non-Abelian anyons of a certain type (here, $m_\R{R}$), the $D_4$ TO is infinitely robust as characterized by the purity. However, a finite threshold exists for the largest moment $\textrm{tr}(\rho^\infty)$ of the decohered density matrix $\rho$, and we further argued that $p_c^{(\infty)}=1/2$ (i.e., even for $n \to \infty$ we only lose $D_4$ TO at the maximal decoherence rate). This scenario contrasts to the behavior of the toric code ground state, where the finite threshold $p_c^{(n)}$ monotonically \emph{increases} with the Rényi-index $n$~\cite{fan2023diagnostics}.  We argued that the underlying cause is the presence of a non-trivial loop weight appearing for non-Abelian anyons. Indeed, we showed how loop models provide a natural stat-mech model for these systems, physically corresponding to the worldlines of the `poisoning' anyons, whose quantum dimensions lead to topological loop weights.

Furthermore, by exactly diagonalizing $\rho$ at maximal error rate $p=1/2$ for the error proliferating non-Abelian anyons (namely, with quantum dimension $d_{m_{\R{R}}}=2$), we found that its spectrum is given by a random O$(2)$ loop model. We equivalently write this as a random $D_4$ symmetric four-state clock model, at a tuning parameter where the clean model is critical. This can be further pushed into the extended gapless phase by initially considering a deformed wavefunction with a finite correlation length. We argued that the presence of disorder likely lowers the ordering temperature compared to the clean model, implying that the non-Abelian quantum memory might be stable to arbitrary decoherence of $m_\R{R}$ anyons---although future work will have to clarify the effect of the frustration of the model. As a concrete diagnostic, we showed that the robustness of this non-Abelian memory (in terms of the quantum fidelity between two topologically distinct ground states) is encoded in the free energy cost of inserting a symmetry defect line in this disordered $D_4$ rotor model.

We then considered the action of various types of Pauli errors proliferating both Abelian and non-Abelian anyons, the former appearing as fusion channels of the latter. This gave rise to a rich phase diagram both for pure wavefunction deformations (Fig.~\ref{fig:Phase_diag_D4}) as well as for the effect of decoherence as characterized by the purity (Fig.~\ref{fig:mixed_pG=PB}). These were obtained by a combination of analytical results pertaining the analysis of the underlying stat-mech models in certain regimes, and numerical analysis via Monte-Carlo methods. Here, we found that different (condensation) transitions can be diagnosed by local order parameters of the underlying stat-mech model. An important finding for the deformed wavefunction was the appearance of an extended gapless phase showcasing power-law decaying correlations of the Wilson operators associated with the proliferated anyons. It would be interesting for future work to further explore the physical reason for this critical phase. It is suggestive that it arises due to an inability to condense $m_\R{R}$ anyons in isolation, which require a formalism beyond those of the recent Refs.~\cite{ellison2024classificationmixedstatetopologicalorders, sohal_24} which discussed the emergence of non-modular TOs in mixed states. Moreover, we notice that this gapless phase is reminiscent of that appearing in Ref.~\cite{chen2024unconventional}. It would be interesting to understand whether the tools discussed in this work can be used to investigate that set-up. For the decohered density matrix, we then found that even when non-Abelian errors are maximally proliferated on top of $D_4$ TO, the system is still robust to a finite proliferation of Abelian anyons.  The characterization of the resulting mixed state ``phases'' appearing beyond these thresholds is left as an open question. {It would also be worthwhile to analyze the setting where the initial wavefunction is not at the fixed-point of the topologically ordered phase, and whether this allows to tune the effective string tension of the resulting loop model.} Finally, we discussed the direct application of these results to the analysis of $D_4$ strong-to-weak spontaneous symmetry breaking (sw-SSB), exploiting a similar relation as the relating $\mathbb{Z}_2$ sw-SSB to the decoherence transition of the toric code \cite{LeeYouXu2022,chen2023separability}.

Our work also gives rise to other open questions both regarding the fate of $D_4$ TO to decoherence, as well as to that of other non-Abelian topological orders. First, it appears that the non-Abelian nature of the phase makes it robust against errors proliferating anyons of non-trivial quantum dimension. In a companion paper~\cite{short_paper} we indeed show that decohering quantum doubles with anyons with larger quantum dimension again produces loop models which prevent a straightforward proliferation. This motivates the question: \emph{Are non-Abelian topological orders with only large quantum dimensions more stable?} Second, while here we considered a combination of errors leading to anyons with trivial braiding, it would be interesting to explore whether the case with non-trivial braiding leads to distinct error thresholds. This scenario could be explored by the configuration discussed in Sec.~\ref{sec:var_NA}, where two or three different non-Abelian anyons with non-trivial braiding are produced. Here, the underlying stat-mech model corresponds to a two (or three) coupled O$(2)$ loop models where the braiding leads to a cancellation of certain configurations.

Our work shows that a simple generalization of the Ising model (the `$ZZ\CZ$' model) dictates much of what happens to deforming and decohering $D_4$ non-abelian TO. Here we focused on the ``intrinsic'' properties of the decohered density matrix as in Ref.~\onlinecite{fan2023diagnostics}, agnostic to a particular error correction scheme. The aforementioned work found that the threshold $p_c$ does indeed match the one obtained from an optimal error-correction protocol (when assuming perfect syndrome measurements). It would be relevant to perform a similar analysis for $D_4$ TO: Can one indeed correct for arbitrary strong decoherence proliferating (only) $m_\R{R}$ anyons? Moreover, how would the presence of faulty measurements affect this conclusion? Does a 3D version of the `$ZZ\CZ$' model then emerge? Moreover, our work shows that detailed studies for non-Abelian decohered models similar to those for the toric code are possible, including the formulation of relevant stat-mech models. With this new insight, it would be exciting to explore whether similar structures can be identified for other non-Abelian TOs. 

Finally, while we were able to harness known results about loop models on the honeycomb lattice, our work points the way to various new kinds of loop models (e.g., the coupled O$(2)$ loop models appearing for $\textrm{tr}(\rho^n)$ in Eq.~\eqref{eq:trrhon_local}, or the coupled O$(2)$ loop models appearing when including two or three non-Abelian anyons discussed in Sec.~\ref{sec:var_NA}). We believe these new stat-mech models can be of interest in their own right, and in addition these can in turn inform about error thresholds for non-Abelian TO. In particular, what is the fate of the corresponding random models even at the maximum error rate? It would be interesting to perform a numerical study of the fidelity as shown in Eq.~\eqref{eq:fidelity}, an analysis we leave for previous work.

\begin{acknowledgments}
    
\end{acknowledgments}
We are grateful to Ehud Altman, Henrik Dreyer, Ruihua Fan, Paul Fendley, Tarun Grover, Wenjie Ji, Yujie Liu, Roger Mong, Lesik Motrunich, Benedikt Placke, John Preskill, Daniel Ranard, Ramanjit Sohal, Nathanan Tantivasadakarn, Robijn Vanhove and Sagar Vijay for helpful discussions and feedback. We also thank Yizhi You for a previous collaboration on a project about strong-to-weak symmetry breaking.
P.S. acknowledges the Les Houches School ``Topological order: Anyons and Fractons" and its participants for their insightful lectures and discussions.
This work was partially conceived at the Aspen Center for Physics (P.S., R.V.), which is supported by National Science Foundation grant PHY-2210452 and Durand Fund. P.S.~and J.A.~acknowledge support from the Caltech Institute for Quantum Information and Matter, an NSF Physics Frontiers Center (NSF Grant PHY-1733907), and the Walter Burke Institute for Theoretical Physics at Caltech.  The U.S.~Department of Energy, Office of Science, National Quantum Information Science Research Centers, Quantum Science Center partially supported the field theory analysis of this work.


\begin{thebibliography}{146}%
\makeatletter
\providecommand \@ifxundefined [1]{%
 \@ifx{#1\undefined}
}%
\providecommand \@ifnum [1]{%
 \ifnum #1\expandafter \@firstoftwo
 \else \expandafter \@secondoftwo
 \fi
}%
\providecommand \@ifx [1]{%
 \ifx #1\expandafter \@firstoftwo
 \else \expandafter \@secondoftwo
 \fi
}%
\providecommand \natexlab [1]{#1}%
\providecommand \enquote  [1]{``#1''}%
\providecommand \bibnamefont  [1]{#1}%
\providecommand \bibfnamefont [1]{#1}%
\providecommand \citenamefont [1]{#1}%
\providecommand \href@noop [0]{\@secondoftwo}%
\providecommand \href [0]{\begingroup \@sanitize@url \@href}%
\providecommand \@href[1]{\@@startlink{#1}\@@href}%
\providecommand \@@href[1]{\endgroup#1\@@endlink}%
\providecommand \@sanitize@url [0]{\catcode `\\12\catcode `\$12\catcode `\&12\catcode `\#12\catcode `\^12\catcode `\_12\catcode `\%12\relax}%
\providecommand \@@startlink[1]{}%
\providecommand \@@endlink[0]{}%
\providecommand \url  [0]{\begingroup\@sanitize@url \@url }%
\providecommand \@url [1]{\endgroup\@href {#1}{\urlprefix }}%
\providecommand \urlprefix  [0]{URL }%
\providecommand \Eprint [0]{\href }%
\providecommand \doibase [0]{https://doi.org/}%
\providecommand \selectlanguage [0]{\@gobble}%
\providecommand \bibinfo  [0]{\@secondoftwo}%
\providecommand \bibfield  [0]{\@secondoftwo}%
\providecommand \translation [1]{[#1]}%
\providecommand \BibitemOpen [0]{}%
\providecommand \bibitemStop [0]{}%
\providecommand \bibitemNoStop [0]{.\EOS\space}%
\providecommand \EOS [0]{\spacefactor3000\relax}%
\providecommand \BibitemShut  [1]{\csname bibitem#1\endcsname}%
\let\auto@bib@innerbib\@empty
\bibitem [{\citenamefont {Hastings}(2011)}]{Hastings_11}%
  \BibitemOpen
  \bibfield  {author} {\bibinfo {author} {\bibfnamefont {M.~B.}\ \bibnamefont {Hastings}},\ }\bibfield  {title} {\bibinfo {title} {Topological order at nonzero temperature},\ }\href {https://doi.org/10.1103/PhysRevLett.107.210501} {\bibfield  {journal} {\bibinfo  {journal} {Phys. Rev. Lett.}\ }\textbf {\bibinfo {volume} {107}},\ \bibinfo {pages} {210501} (\bibinfo {year} {2011})}\BibitemShut {NoStop}%
\bibitem [{\citenamefont {Coser}\ and\ \citenamefont {Pérez-García}(2019)}]{Coser_2019}%
  \BibitemOpen
  \bibfield  {author} {\bibinfo {author} {\bibfnamefont {A.}~\bibnamefont {Coser}}\ and\ \bibinfo {author} {\bibfnamefont {D.}~\bibnamefont {Pérez-García}},\ }\bibfield  {title} {\bibinfo {title} {Classification of phases for mixed states via fast dissipative evolution},\ }\href {https://doi.org/10.22331/q-2019-08-12-174} {\bibfield  {journal} {\bibinfo  {journal} {Quantum}\ }\textbf {\bibinfo {volume} {3}},\ \bibinfo {pages} {174} (\bibinfo {year} {2019})}\BibitemShut {NoStop}%
\bibitem [{\citenamefont {Lu}\ \emph {et~al.}(2020)\citenamefont {Lu}, \citenamefont {Hsieh},\ and\ \citenamefont {Grover}}]{Tarun_finiteT}%
  \BibitemOpen
  \bibfield  {author} {\bibinfo {author} {\bibfnamefont {T.-C.}\ \bibnamefont {Lu}}, \bibinfo {author} {\bibfnamefont {T.~H.}\ \bibnamefont {Hsieh}},\ and\ \bibinfo {author} {\bibfnamefont {T.}~\bibnamefont {Grover}},\ }\bibfield  {title} {\bibinfo {title} {Detecting topological order at finite temperature using entanglement negativity},\ }\href {https://doi.org/10.1103/PhysRevLett.125.116801} {\bibfield  {journal} {\bibinfo  {journal} {Phys. Rev. Lett.}\ }\textbf {\bibinfo {volume} {125}},\ \bibinfo {pages} {116801} (\bibinfo {year} {2020})}\BibitemShut {NoStop}%
\bibitem [{\citenamefont {de~Groot}\ \emph {et~al.}(2022)\citenamefont {de~Groot}, \citenamefont {Turzillo},\ and\ \citenamefont {Schuch}}]{de_Groot_2022}%
  \BibitemOpen
  \bibfield  {author} {\bibinfo {author} {\bibfnamefont {C.}~\bibnamefont {de~Groot}}, \bibinfo {author} {\bibfnamefont {A.}~\bibnamefont {Turzillo}},\ and\ \bibinfo {author} {\bibfnamefont {N.}~\bibnamefont {Schuch}},\ }\bibfield  {title} {\bibinfo {title} {Symmetry protected topological order in open quantum systems},\ }\href {https://doi.org/10.22331/q-2022-11-10-856} {\bibfield  {journal} {\bibinfo  {journal} {Quantum}\ }\textbf {\bibinfo {volume} {6}},\ \bibinfo {pages} {856} (\bibinfo {year} {2022})}\BibitemShut {NoStop}%
\bibitem [{\citenamefont {Bao}\ \emph {et~al.}(2023)\citenamefont {Bao}, \citenamefont {Fan}, \citenamefont {Vishwanath},\ and\ \citenamefont {Altman}}]{bao2023mixedstate}%
  \BibitemOpen
  \bibfield  {author} {\bibinfo {author} {\bibfnamefont {Y.}~\bibnamefont {Bao}}, \bibinfo {author} {\bibfnamefont {R.}~\bibnamefont {Fan}}, \bibinfo {author} {\bibfnamefont {A.}~\bibnamefont {Vishwanath}},\ and\ \bibinfo {author} {\bibfnamefont {E.}~\bibnamefont {Altman}},\ }\href@noop {} {\bibinfo {title} {Mixed-state topological order and the errorfield double formulation of decoherence-induced transitions}} (\bibinfo {year} {2023}),\ \Eprint {https://arxiv.org/abs/2301.05687} {arXiv:2301.05687 [quant-ph]} \BibitemShut {NoStop}%
\bibitem [{\citenamefont {Fan}\ \emph {et~al.}(2024)\citenamefont {Fan}, \citenamefont {Bao}, \citenamefont {Altman},\ and\ \citenamefont {Vishwanath}}]{fan2023diagnostics}%
  \BibitemOpen
  \bibfield  {author} {\bibinfo {author} {\bibfnamefont {R.}~\bibnamefont {Fan}}, \bibinfo {author} {\bibfnamefont {Y.}~\bibnamefont {Bao}}, \bibinfo {author} {\bibfnamefont {E.}~\bibnamefont {Altman}},\ and\ \bibinfo {author} {\bibfnamefont {A.}~\bibnamefont {Vishwanath}},\ }\bibfield  {title} {\bibinfo {title} {Diagnostics of mixed-state topological order and breakdown of quantum memory},\ }\href {https://doi.org/10.1103/PRXQuantum.5.020343} {\bibfield  {journal} {\bibinfo  {journal} {PRX Quantum}\ }\textbf {\bibinfo {volume} {5}},\ \bibinfo {pages} {020343} (\bibinfo {year} {2024})}\BibitemShut {NoStop}%
\bibitem [{\citenamefont {Lee}\ \emph {et~al.}(2023)\citenamefont {Lee}, \citenamefont {Jian},\ and\ \citenamefont {Xu}}]{LeeYouXu2022}%
  \BibitemOpen
  \bibfield  {author} {\bibinfo {author} {\bibfnamefont {J.~Y.}\ \bibnamefont {Lee}}, \bibinfo {author} {\bibfnamefont {C.-M.}\ \bibnamefont {Jian}},\ and\ \bibinfo {author} {\bibfnamefont {C.}~\bibnamefont {Xu}},\ }\bibfield  {title} {\bibinfo {title} {Quantum criticality under decoherence or weak measurement},\ }\href {https://doi.org/10.1103/PRXQuantum.4.030317} {\bibfield  {journal} {\bibinfo  {journal} {PRX Quantum}\ }\textbf {\bibinfo {volume} {4}},\ \bibinfo {pages} {030317} (\bibinfo {year} {2023})}\BibitemShut {NoStop}%
\bibitem [{\citenamefont {{Sang}}\ \emph {et~al.}(2023)\citenamefont {{Sang}}, \citenamefont {{Zou}},\ and\ \citenamefont {{Hsieh}}}]{Renorm_QECC_23}%
  \BibitemOpen
  \bibfield  {author} {\bibinfo {author} {\bibfnamefont {S.}~\bibnamefont {{Sang}}}, \bibinfo {author} {\bibfnamefont {Y.}~\bibnamefont {{Zou}}},\ and\ \bibinfo {author} {\bibfnamefont {T.~H.}\ \bibnamefont {{Hsieh}}},\ }\bibfield  {title} {\bibinfo {title} {{Mixed-state Quantum Phases: Renormalization and Quantum Error Correction}},\ }\href {https://doi.org/10.48550/arXiv.2310.08639} {\bibfield  {journal} {\bibinfo  {journal} {arXiv e-prints}\ ,\ \bibinfo {eid} {arXiv:2310.08639}} (\bibinfo {year} {2023})},\ \Eprint {https://arxiv.org/abs/2310.08639} {arXiv:2310.08639 [quant-ph]} \BibitemShut {NoStop}%
\bibitem [{\citenamefont {Chen}\ and\ \citenamefont {Grover}(2024{\natexlab{a}})}]{chen2023separability}%
  \BibitemOpen
  \bibfield  {author} {\bibinfo {author} {\bibfnamefont {Y.-H.}\ \bibnamefont {Chen}}\ and\ \bibinfo {author} {\bibfnamefont {T.}~\bibnamefont {Grover}},\ }\bibfield  {title} {\bibinfo {title} {Separability transitions in topological states induced by local decoherence},\ }\href {https://doi.org/10.1103/PhysRevLett.132.170602} {\bibfield  {journal} {\bibinfo  {journal} {Phys. Rev. Lett.}\ }\textbf {\bibinfo {volume} {132}},\ \bibinfo {pages} {170602} (\bibinfo {year} {2024}{\natexlab{a}})}\BibitemShut {NoStop}%
\bibitem [{\citenamefont {Wang}\ \emph {et~al.}(2023)\citenamefont {Wang}, \citenamefont {Wu},\ and\ \citenamefont {Wang}}]{wang2023intrinsic}%
  \BibitemOpen
  \bibfield  {author} {\bibinfo {author} {\bibfnamefont {Z.}~\bibnamefont {Wang}}, \bibinfo {author} {\bibfnamefont {Z.}~\bibnamefont {Wu}},\ and\ \bibinfo {author} {\bibfnamefont {Z.}~\bibnamefont {Wang}},\ }\href@noop {} {\bibinfo {title} {Intrinsic mixed-state topological order without quantum memory}} (\bibinfo {year} {2023}),\ \Eprint {https://arxiv.org/abs/2307.13758} {arXiv:2307.13758 [quant-ph]} \BibitemShut {NoStop}%
\bibitem [{\citenamefont {Ma}\ and\ \citenamefont {Wang}(2023)}]{Ma_23}%
  \BibitemOpen
  \bibfield  {author} {\bibinfo {author} {\bibfnamefont {R.}~\bibnamefont {Ma}}\ and\ \bibinfo {author} {\bibfnamefont {C.}~\bibnamefont {Wang}},\ }\bibfield  {title} {\bibinfo {title} {Average symmetry-protected topological phases},\ }\href {https://doi.org/10.1103/PhysRevX.13.031016} {\bibfield  {journal} {\bibinfo  {journal} {Phys. Rev. X}\ }\textbf {\bibinfo {volume} {13}},\ \bibinfo {pages} {031016} (\bibinfo {year} {2023})}\BibitemShut {NoStop}%
\bibitem [{\citenamefont {Lu}\ \emph {et~al.}(2023)\citenamefont {Lu}, \citenamefont {Zhang}, \citenamefont {Vijay},\ and\ \citenamefont {Hsieh}}]{Lu23}%
  \BibitemOpen
  \bibfield  {author} {\bibinfo {author} {\bibfnamefont {T.-C.}\ \bibnamefont {Lu}}, \bibinfo {author} {\bibfnamefont {Z.}~\bibnamefont {Zhang}}, \bibinfo {author} {\bibfnamefont {S.}~\bibnamefont {Vijay}},\ and\ \bibinfo {author} {\bibfnamefont {T.~H.}\ \bibnamefont {Hsieh}},\ }\bibfield  {title} {\bibinfo {title} {Mixed-state long-range order and criticality from measurement and feedback},\ }\href {https://doi.org/10.1103/PRXQuantum.4.030318} {\bibfield  {journal} {\bibinfo  {journal} {PRX Quantum}\ }\textbf {\bibinfo {volume} {4}},\ \bibinfo {pages} {030318} (\bibinfo {year} {2023})}\BibitemShut {NoStop}%
\bibitem [{\citenamefont {Lee}\ \emph {et~al.}(2022)\citenamefont {Lee}, \citenamefont {You},\ and\ \citenamefont {Xu}}]{lee2022symmetry}%
  \BibitemOpen
  \bibfield  {author} {\bibinfo {author} {\bibfnamefont {J.~Y.}\ \bibnamefont {Lee}}, \bibinfo {author} {\bibfnamefont {Y.-Z.}\ \bibnamefont {You}},\ and\ \bibinfo {author} {\bibfnamefont {C.}~\bibnamefont {Xu}},\ }\bibfield  {title} {\bibinfo {title} {Symmetry protected topological phases under decoherence},\ }\href@noop {} {\bibfield  {journal} {\bibinfo  {journal} {arXiv preprint arXiv:2210.16323}\ } (\bibinfo {year} {2022})}\BibitemShut {NoStop}%
\bibitem [{\citenamefont {Zhu}\ \emph {et~al.}(2023)\citenamefont {Zhu}, \citenamefont {Tantivasadakarn}, \citenamefont {Vishwanath}, \citenamefont {Trebst},\ and\ \citenamefont {Verresen}}]{Zhu23}%
  \BibitemOpen
  \bibfield  {author} {\bibinfo {author} {\bibfnamefont {G.-Y.}\ \bibnamefont {Zhu}}, \bibinfo {author} {\bibfnamefont {N.}~\bibnamefont {Tantivasadakarn}}, \bibinfo {author} {\bibfnamefont {A.}~\bibnamefont {Vishwanath}}, \bibinfo {author} {\bibfnamefont {S.}~\bibnamefont {Trebst}},\ and\ \bibinfo {author} {\bibfnamefont {R.}~\bibnamefont {Verresen}},\ }\bibfield  {title} {\bibinfo {title} {Nishimori's cat: Stable long-range entanglement from finite-depth unitaries and weak measurements},\ }\href {https://doi.org/10.1103/PhysRevLett.131.200201} {\bibfield  {journal} {\bibinfo  {journal} {Phys. Rev. Lett.}\ }\textbf {\bibinfo {volume} {131}},\ \bibinfo {pages} {200201} (\bibinfo {year} {2023})}\BibitemShut {NoStop}%
\bibitem [{\citenamefont {Lyons}(2024)}]{lyons24}%
  \BibitemOpen
  \bibfield  {author} {\bibinfo {author} {\bibfnamefont {A.}~\bibnamefont {Lyons}},\ }\href {https://arxiv.org/abs/2403.03955} {\bibinfo {title} {Understanding stabilizer codes under local decoherence through a general statistical mechanics mapping}} (\bibinfo {year} {2024}),\ \Eprint {https://arxiv.org/abs/2403.03955} {arXiv:2403.03955 [quant-ph]} \BibitemShut {NoStop}%
\bibitem [{\citenamefont {{Li}}\ and\ \citenamefont {{Mong}}(2024)}]{Mong_24}%
  \BibitemOpen
  \bibfield  {author} {\bibinfo {author} {\bibfnamefont {Z.}~\bibnamefont {{Li}}}\ and\ \bibinfo {author} {\bibfnamefont {R.~S.~K.}\ \bibnamefont {{Mong}}},\ }\bibfield  {title} {\bibinfo {title} {{Replica topological order in quantum mixed states and quantum error correction}},\ }\href {https://doi.org/10.48550/arXiv.2402.09516} {\bibfield  {journal} {\bibinfo  {journal} {arXiv e-prints}\ ,\ \bibinfo {eid} {arXiv:2402.09516}} (\bibinfo {year} {2024})},\ \Eprint {https://arxiv.org/abs/2402.09516} {arXiv:2402.09516 [quant-ph]} \BibitemShut {NoStop}%
\bibitem [{\citenamefont {{Sohal}}\ and\ \citenamefont {{Prem}}(2024)}]{sohal_24}%
  \BibitemOpen
  \bibfield  {author} {\bibinfo {author} {\bibfnamefont {R.}~\bibnamefont {{Sohal}}}\ and\ \bibinfo {author} {\bibfnamefont {A.}~\bibnamefont {{Prem}}},\ }\bibfield  {title} {\bibinfo {title} {{A Noisy Approach to Intrinsically Mixed-State Topological Order}},\ }\href {https://doi.org/10.48550/arXiv.2403.13879} {\bibfield  {journal} {\bibinfo  {journal} {arXiv e-prints}\ ,\ \bibinfo {eid} {arXiv:2403.13879}} (\bibinfo {year} {2024})},\ \Eprint {https://arxiv.org/abs/2403.13879} {arXiv:2403.13879 [cond-mat.str-el]} \BibitemShut {NoStop}%
\bibitem [{\citenamefont {Ellison}\ and\ \citenamefont {Cheng}(2024)}]{ellison2024classificationmixedstatetopologicalorders}%
  \BibitemOpen
  \bibfield  {author} {\bibinfo {author} {\bibfnamefont {T.}~\bibnamefont {Ellison}}\ and\ \bibinfo {author} {\bibfnamefont {M.}~\bibnamefont {Cheng}},\ }\href {https://arxiv.org/abs/2405.02390} {\bibinfo {title} {Towards a classification of mixed-state topological orders in two dimensions}} (\bibinfo {year} {2024}),\ \Eprint {https://arxiv.org/abs/2405.02390} {arXiv:2405.02390 [cond-mat.str-el]} \BibitemShut {NoStop}%
\bibitem [{\citenamefont {Su}\ \emph {et~al.}(2024{\natexlab{a}})\citenamefont {Su}, \citenamefont {Yang},\ and\ \citenamefont {Jian}}]{tapestry_24}%
  \BibitemOpen
  \bibfield  {author} {\bibinfo {author} {\bibfnamefont {K.}~\bibnamefont {Su}}, \bibinfo {author} {\bibfnamefont {Z.}~\bibnamefont {Yang}},\ and\ \bibinfo {author} {\bibfnamefont {C.-M.}\ \bibnamefont {Jian}},\ }\bibfield  {title} {\bibinfo {title} {Tapestry of dualities in decohered quantum error correction codes},\ }\href {https://doi.org/10.1103/PhysRevB.110.085158} {\bibfield  {journal} {\bibinfo  {journal} {Phys. Rev. B}\ }\textbf {\bibinfo {volume} {110}},\ \bibinfo {pages} {085158} (\bibinfo {year} {2024}{\natexlab{a}})}\BibitemShut {NoStop}%
\bibitem [{\citenamefont {Chen}\ and\ \citenamefont {Grover}(2024{\natexlab{b}})}]{chen2024unconventional}%
  \BibitemOpen
  \bibfield  {author} {\bibinfo {author} {\bibfnamefont {Y.-H.}\ \bibnamefont {Chen}}\ and\ \bibinfo {author} {\bibfnamefont {T.}~\bibnamefont {Grover}},\ }\href@noop {} {\bibinfo {title} {Unconventional topological mixed-state transition and critical phase induced by self-dual coherent errors}} (\bibinfo {year} {2024}{\natexlab{b}}),\ \Eprint {https://arxiv.org/abs/2403.06553} {arXiv:2403.06553 [quant-ph]} \BibitemShut {NoStop}%
\bibitem [{\citenamefont {{Hauser}}\ \emph {et~al.}(2024)\citenamefont {{Hauser}}, \citenamefont {{Bao}}, \citenamefont {{Sang}}, \citenamefont {{Lavasani}}, \citenamefont {{Agrawal}},\ and\ \citenamefont {{Fisher}}}]{Hauser_24}%
  \BibitemOpen
  \bibfield  {author} {\bibinfo {author} {\bibfnamefont {J.}~\bibnamefont {{Hauser}}}, \bibinfo {author} {\bibfnamefont {Y.}~\bibnamefont {{Bao}}}, \bibinfo {author} {\bibfnamefont {S.}~\bibnamefont {{Sang}}}, \bibinfo {author} {\bibfnamefont {A.}~\bibnamefont {{Lavasani}}}, \bibinfo {author} {\bibfnamefont {U.}~\bibnamefont {{Agrawal}}},\ and\ \bibinfo {author} {\bibfnamefont {M.~P.~A.}\ \bibnamefont {{Fisher}}},\ }\bibfield  {title} {\bibinfo {title} {{Information dynamics in decohered quantum memory with repeated syndrome measurements: a dual approach}},\ }\href {https://doi.org/10.48550/arXiv.2407.07882} {\bibfield  {journal} {\bibinfo  {journal} {arXiv e-prints}\ ,\ \bibinfo {eid} {arXiv:2407.07882}} (\bibinfo {year} {2024})},\ \Eprint {https://arxiv.org/abs/2407.07882} {arXiv:2407.07882 [quant-ph]} \BibitemShut {NoStop}%
\bibitem [{\citenamefont {Sala}\ \emph {et~al.}(2024)\citenamefont {Sala}, \citenamefont {Gopalakrishnan}, \citenamefont {Oshikawa},\ and\ \citenamefont {You}}]{2024_sala_SSSB}%
  \BibitemOpen
  \bibfield  {author} {\bibinfo {author} {\bibfnamefont {P.}~\bibnamefont {Sala}}, \bibinfo {author} {\bibfnamefont {S.}~\bibnamefont {Gopalakrishnan}}, \bibinfo {author} {\bibfnamefont {M.}~\bibnamefont {Oshikawa}},\ and\ \bibinfo {author} {\bibfnamefont {Y.}~\bibnamefont {You}},\ }\href@noop {} {\bibinfo {title} {Spontaneous strong symmetry breaking in open systems: Purification perspective}} (\bibinfo {year} {2024}),\ \Eprint {https://arxiv.org/abs/2405.02402} {arXiv:2405.02402 [quant-ph]} \BibitemShut {NoStop}%
\bibitem [{\citenamefont {{Sang}}\ and\ \citenamefont {{Hsieh}}(2024)}]{Markov_length_24}%
  \BibitemOpen
  \bibfield  {author} {\bibinfo {author} {\bibfnamefont {S.}~\bibnamefont {{Sang}}}\ and\ \bibinfo {author} {\bibfnamefont {T.~H.}\ \bibnamefont {{Hsieh}}},\ }\bibfield  {title} {\bibinfo {title} {{Stability of mixed-state quantum phases via finite Markov length}},\ }\href {https://doi.org/10.48550/arXiv.2404.07251} {\bibfield  {journal} {\bibinfo  {journal} {arXiv e-prints}\ ,\ \bibinfo {eid} {arXiv:2404.07251}} (\bibinfo {year} {2024})},\ \Eprint {https://arxiv.org/abs/2404.07251} {arXiv:2404.07251 [quant-ph]} \BibitemShut {NoStop}%
\bibitem [{\citenamefont {{Lu}}(2024)}]{TshungCheng_24}%
  \BibitemOpen
  \bibfield  {author} {\bibinfo {author} {\bibfnamefont {T.-C.}\ \bibnamefont {{Lu}}},\ }\bibfield  {title} {\bibinfo {title} {{Disentangling transitions in topological order induced by boundary decoherence}},\ }\href {https://doi.org/10.48550/arXiv.2404.06514} {\bibfield  {journal} {\bibinfo  {journal} {arXiv e-prints}\ ,\ \bibinfo {eid} {arXiv:2404.06514}} (\bibinfo {year} {2024})},\ \Eprint {https://arxiv.org/abs/2404.06514} {arXiv:2404.06514 [quant-ph]} \BibitemShut {NoStop}%
\bibitem [{\citenamefont {Lee}(2024)}]{lee2024exactcalculationscoherentinformation}%
  \BibitemOpen
  \bibfield  {author} {\bibinfo {author} {\bibfnamefont {J.~Y.}\ \bibnamefont {Lee}},\ }\href {https://arxiv.org/abs/2402.16937} {\bibinfo {title} {Exact calculations of coherent information for toric codes under decoherence: Identifying the fundamental error threshold}} (\bibinfo {year} {2024}),\ \Eprint {https://arxiv.org/abs/2402.16937} {arXiv:2402.16937 [cond-mat.stat-mech]} \BibitemShut {NoStop}%
\bibitem [{\citenamefont {Ma}\ \emph {et~al.}(2024)\citenamefont {Ma}, \citenamefont {Zhang}, \citenamefont {Bi}, \citenamefont {Cheng},\ and\ \citenamefont {Wang}}]{ma2024topological}%
  \BibitemOpen
  \bibfield  {author} {\bibinfo {author} {\bibfnamefont {R.}~\bibnamefont {Ma}}, \bibinfo {author} {\bibfnamefont {J.-H.}\ \bibnamefont {Zhang}}, \bibinfo {author} {\bibfnamefont {Z.}~\bibnamefont {Bi}}, \bibinfo {author} {\bibfnamefont {M.}~\bibnamefont {Cheng}},\ and\ \bibinfo {author} {\bibfnamefont {C.}~\bibnamefont {Wang}},\ }\href@noop {} {\bibinfo {title} {Topological phases with average symmetries: the decohered, the disordered, and the intrinsic}} (\bibinfo {year} {2024}),\ \Eprint {https://arxiv.org/abs/2305.16399} {arXiv:2305.16399 [cond-mat.str-el]} \BibitemShut {NoStop}%
\bibitem [{\citenamefont {Rakovszky}\ \emph {et~al.}(2024)\citenamefont {Rakovszky}, \citenamefont {Gopalakrishnan},\ and\ \citenamefont {von Keyserlingk}}]{rakovszky2024definingstablephasesopen}%
  \BibitemOpen
  \bibfield  {author} {\bibinfo {author} {\bibfnamefont {T.}~\bibnamefont {Rakovszky}}, \bibinfo {author} {\bibfnamefont {S.}~\bibnamefont {Gopalakrishnan}},\ and\ \bibinfo {author} {\bibfnamefont {C.}~\bibnamefont {von Keyserlingk}},\ }\href {https://arxiv.org/abs/2308.15495} {\bibinfo {title} {Defining stable phases of open quantum systems}} (\bibinfo {year} {2024}),\ \Eprint {https://arxiv.org/abs/2308.15495} {arXiv:2308.15495 [quant-ph]} \BibitemShut {NoStop}%
\bibitem [{\citenamefont {Ma}\ and\ \citenamefont {Turzillo}(2024)}]{ma2024symmetry}%
  \BibitemOpen
  \bibfield  {author} {\bibinfo {author} {\bibfnamefont {R.}~\bibnamefont {Ma}}\ and\ \bibinfo {author} {\bibfnamefont {A.}~\bibnamefont {Turzillo}},\ }\bibfield  {title} {\bibinfo {title} {Symmetry protected topological phases of mixed states in the doubled space},\ }\href@noop {} {\bibfield  {journal} {\bibinfo  {journal} {arXiv preprint arXiv:2403.13280}\ } (\bibinfo {year} {2024})}\BibitemShut {NoStop}%
\bibitem [{\citenamefont {Hsin}\ \emph {et~al.}(2023)\citenamefont {Hsin}, \citenamefont {Luo},\ and\ \citenamefont {Sun}}]{hsin2023anomaliesaveragesymmetriesentanglement}%
  \BibitemOpen
  \bibfield  {author} {\bibinfo {author} {\bibfnamefont {P.-S.}\ \bibnamefont {Hsin}}, \bibinfo {author} {\bibfnamefont {Z.-X.}\ \bibnamefont {Luo}},\ and\ \bibinfo {author} {\bibfnamefont {H.-Y.}\ \bibnamefont {Sun}},\ }\href {https://arxiv.org/abs/2312.09074} {\bibinfo {title} {Anomalies of average symmetries: Entanglement and open quantum systems}} (\bibinfo {year} {2023}),\ \Eprint {https://arxiv.org/abs/2312.09074} {arXiv:2312.09074 [cond-mat.str-el]} \BibitemShut {NoStop}%
\bibitem [{\citenamefont {Su}\ \emph {et~al.}(2024{\natexlab{b}})\citenamefont {Su}, \citenamefont {Myerson-Jain},\ and\ \citenamefont {Xu}}]{su_2024}%
  \BibitemOpen
  \bibfield  {author} {\bibinfo {author} {\bibfnamefont {K.}~\bibnamefont {Su}}, \bibinfo {author} {\bibfnamefont {N.}~\bibnamefont {Myerson-Jain}},\ and\ \bibinfo {author} {\bibfnamefont {C.}~\bibnamefont {Xu}},\ }\bibfield  {title} {\bibinfo {title} {Conformal field theories generated by chern insulators under decoherence or measurement},\ }\href {https://doi.org/10.1103/PhysRevB.109.035146} {\bibfield  {journal} {\bibinfo  {journal} {Phys. Rev. B}\ }\textbf {\bibinfo {volume} {109}},\ \bibinfo {pages} {035146} (\bibinfo {year} {2024}{\natexlab{b}})}\BibitemShut {NoStop}%
\bibitem [{\citenamefont {Wang}\ and\ \citenamefont {Li}(2024)}]{wang2024anomalyopenquantumsystems}%
  \BibitemOpen
  \bibfield  {author} {\bibinfo {author} {\bibfnamefont {Z.}~\bibnamefont {Wang}}\ and\ \bibinfo {author} {\bibfnamefont {L.}~\bibnamefont {Li}},\ }\href {https://arxiv.org/abs/2403.14533} {\bibinfo {title} {Anomaly in open quantum systems and its implications on mixed-state quantum phases}} (\bibinfo {year} {2024}),\ \Eprint {https://arxiv.org/abs/2403.14533} {arXiv:2403.14533 [quant-ph]} \BibitemShut {NoStop}%
\bibitem [{\citenamefont {Kawabata}\ \emph {et~al.}(2024)\citenamefont {Kawabata}, \citenamefont {Sohal},\ and\ \citenamefont {Ryu}}]{Kawabata_24}%
  \BibitemOpen
  \bibfield  {author} {\bibinfo {author} {\bibfnamefont {K.}~\bibnamefont {Kawabata}}, \bibinfo {author} {\bibfnamefont {R.}~\bibnamefont {Sohal}},\ and\ \bibinfo {author} {\bibfnamefont {S.}~\bibnamefont {Ryu}},\ }\bibfield  {title} {\bibinfo {title} {Lieb-schultz-mattis theorem in open quantum systems},\ }\href {https://doi.org/10.1103/PhysRevLett.132.070402} {\bibfield  {journal} {\bibinfo  {journal} {Phys. Rev. Lett.}\ }\textbf {\bibinfo {volume} {132}},\ \bibinfo {pages} {070402} (\bibinfo {year} {2024})}\BibitemShut {NoStop}%
\bibitem [{\citenamefont {Guo}\ and\ \citenamefont {Ashida}(2024)}]{guo_24}%
  \BibitemOpen
  \bibfield  {author} {\bibinfo {author} {\bibfnamefont {Y.}~\bibnamefont {Guo}}\ and\ \bibinfo {author} {\bibfnamefont {Y.}~\bibnamefont {Ashida}},\ }\bibfield  {title} {\bibinfo {title} {Two-dimensional symmetry-protected topological phases and transitions in open quantum systems},\ }\href {https://doi.org/10.1103/PhysRevB.109.195420} {\bibfield  {journal} {\bibinfo  {journal} {Phys. Rev. B}\ }\textbf {\bibinfo {volume} {109}},\ \bibinfo {pages} {195420} (\bibinfo {year} {2024})}\BibitemShut {NoStop}%
\bibitem [{\citenamefont {Zhang}\ \emph {et~al.}(2024{\natexlab{a}})\citenamefont {Zhang}, \citenamefont {Agrawal},\ and\ \citenamefont {Vijay}}]{zhang2024quantumcommunicationmixedstateorder}%
  \BibitemOpen
  \bibfield  {author} {\bibinfo {author} {\bibfnamefont {Z.}~\bibnamefont {Zhang}}, \bibinfo {author} {\bibfnamefont {U.}~\bibnamefont {Agrawal}},\ and\ \bibinfo {author} {\bibfnamefont {S.}~\bibnamefont {Vijay}},\ }\href {https://arxiv.org/abs/2405.05965} {\bibinfo {title} {Quantum communication and mixed-state order in decohered symmetry-protected topological states}} (\bibinfo {year} {2024}{\natexlab{a}}),\ \Eprint {https://arxiv.org/abs/2405.05965} {arXiv:2405.05965 [quant-ph]} \BibitemShut {NoStop}%
\bibitem [{\citenamefont {Lavasani}\ and\ \citenamefont {Vijay}(2024)}]{lavasani2024stabilitygappedquantummatter}%
  \BibitemOpen
  \bibfield  {author} {\bibinfo {author} {\bibfnamefont {A.}~\bibnamefont {Lavasani}}\ and\ \bibinfo {author} {\bibfnamefont {S.}~\bibnamefont {Vijay}},\ }\href {https://arxiv.org/abs/2402.14906} {\bibinfo {title} {The stability of gapped quantum matter and error-correction with adiabatic noise}} (\bibinfo {year} {2024}),\ \Eprint {https://arxiv.org/abs/2402.14906} {arXiv:2402.14906 [cond-mat.str-el]} \BibitemShut {NoStop}%
\bibitem [{\citenamefont {{Lessa}}\ \emph {et~al.}(2024)\citenamefont {{Lessa}}, \citenamefont {{Ma}}, \citenamefont {{Zhang}}, \citenamefont {{Bi}}, \citenamefont {{Cheng}},\ and\ \citenamefont {{Wang}}}]{Lessa_24}%
  \BibitemOpen
  \bibfield  {author} {\bibinfo {author} {\bibfnamefont {L.~A.}\ \bibnamefont {{Lessa}}}, \bibinfo {author} {\bibfnamefont {R.}~\bibnamefont {{Ma}}}, \bibinfo {author} {\bibfnamefont {J.-H.}\ \bibnamefont {{Zhang}}}, \bibinfo {author} {\bibfnamefont {Z.}~\bibnamefont {{Bi}}}, \bibinfo {author} {\bibfnamefont {M.}~\bibnamefont {{Cheng}}},\ and\ \bibinfo {author} {\bibfnamefont {C.}~\bibnamefont {{Wang}}},\ }\bibfield  {title} {\bibinfo {title} {{Strong-to-Weak Spontaneous Symmetry Breaking in Mixed Quantum States}},\ }\href {https://doi.org/10.48550/arXiv.2405.03639} {\bibfield  {journal} {\bibinfo  {journal} {arXiv e-prints}\ ,\ \bibinfo {eid} {arXiv:2405.03639}} (\bibinfo {year} {2024})},\ \Eprint {https://arxiv.org/abs/2405.03639} {arXiv:2405.03639 [quant-ph]} \BibitemShut {NoStop}%
\bibitem [{\citenamefont {Lessa}\ \emph {et~al.}(2024)\citenamefont {Lessa}, \citenamefont {Cheng},\ and\ \citenamefont {Wang}}]{lessa2024mixedstatequantumanomalymultipartite}%
  \BibitemOpen
  \bibfield  {author} {\bibinfo {author} {\bibfnamefont {L.~A.}\ \bibnamefont {Lessa}}, \bibinfo {author} {\bibfnamefont {M.}~\bibnamefont {Cheng}},\ and\ \bibinfo {author} {\bibfnamefont {C.}~\bibnamefont {Wang}},\ }\href {https://arxiv.org/abs/2401.17357} {\bibinfo {title} {Mixed-state quantum anomaly and multipartite entanglement}} (\bibinfo {year} {2024}),\ \Eprint {https://arxiv.org/abs/2401.17357} {arXiv:2401.17357 [cond-mat.str-el]} \BibitemShut {NoStop}%
\bibitem [{\citenamefont {Li}\ \emph {et~al.}(2024{\natexlab{a}})\citenamefont {Li}, \citenamefont {Lee},\ and\ \citenamefont {Yoshida}}]{li_lee_Yoshida_24}%
  \BibitemOpen
  \bibfield  {author} {\bibinfo {author} {\bibfnamefont {Z.}~\bibnamefont {Li}}, \bibinfo {author} {\bibfnamefont {D.}~\bibnamefont {Lee}},\ and\ \bibinfo {author} {\bibfnamefont {B.}~\bibnamefont {Yoshida}},\ }\href {https://arxiv.org/abs/2405.07970} {\bibinfo {title} {How much entanglement is needed for emergent anyons and fermions?}} (\bibinfo {year} {2024}{\natexlab{a}}),\ \Eprint {https://arxiv.org/abs/2405.07970} {arXiv:2405.07970 [quant-ph]} \BibitemShut {NoStop}%
\bibitem [{\citenamefont {Liu}\ and\ \citenamefont {Lieu}(2024)}]{Leo_24}%
  \BibitemOpen
  \bibfield  {author} {\bibinfo {author} {\bibfnamefont {Y.-J.}\ \bibnamefont {Liu}}\ and\ \bibinfo {author} {\bibfnamefont {S.}~\bibnamefont {Lieu}},\ }\bibfield  {title} {\bibinfo {title} {Dissipative phase transitions and passive error correction},\ }\href {https://doi.org/10.1103/PhysRevA.109.022422} {\bibfield  {journal} {\bibinfo  {journal} {Phys. Rev. A}\ }\textbf {\bibinfo {volume} {109}},\ \bibinfo {pages} {022422} (\bibinfo {year} {2024})}\BibitemShut {NoStop}%
\bibitem [{\citenamefont {Lieu}\ \emph {et~al.}(2024)\citenamefont {Lieu}, \citenamefont {Liu},\ and\ \citenamefont {Gorshkov}}]{Lieu_24}%
  \BibitemOpen
  \bibfield  {author} {\bibinfo {author} {\bibfnamefont {S.}~\bibnamefont {Lieu}}, \bibinfo {author} {\bibfnamefont {Y.-J.}\ \bibnamefont {Liu}},\ and\ \bibinfo {author} {\bibfnamefont {A.~V.}\ \bibnamefont {Gorshkov}},\ }\bibfield  {title} {\bibinfo {title} {Candidate for a passively protected quantum memory in two dimensions},\ }\href {https://doi.org/10.1103/PhysRevLett.133.030601} {\bibfield  {journal} {\bibinfo  {journal} {Phys. Rev. Lett.}\ }\textbf {\bibinfo {volume} {133}},\ \bibinfo {pages} {030601} (\bibinfo {year} {2024})}\BibitemShut {NoStop}%
\bibitem [{\citenamefont {Nielsen}\ and\ \citenamefont {Chuang}(2010)}]{Nielsen_Chuang_2010}%
  \BibitemOpen
  \bibfield  {author} {\bibinfo {author} {\bibfnamefont {M.~A.}\ \bibnamefont {Nielsen}}\ and\ \bibinfo {author} {\bibfnamefont {I.~L.}\ \bibnamefont {Chuang}},\ }\href@noop {} {\emph {\bibinfo {title} {Quantum Computation and Quantum Information: 10th Anniversary Edition}}}\ (\bibinfo  {publisher} {Cambridge University Press},\ \bibinfo {year} {2010})\BibitemShut {NoStop}%
\bibitem [{\citenamefont {Wen}(2007)}]{Wen_book}%
  \BibitemOpen
  \bibfield  {author} {\bibinfo {author} {\bibfnamefont {X.-G.}\ \bibnamefont {Wen}},\ }\href {https://doi.org/10.1093/acprof:oso/9780199227259.001.0001} {\emph {\bibinfo {title} {{Quantum Field Theory of Many-Body Systems: From the Origin of Sound to an Origin of Light and Electrons}}}}\ (\bibinfo  {publisher} {Oxford University Press},\ \bibinfo {year} {2007})\BibitemShut {NoStop}%
\bibitem [{\citenamefont {Sachdev}(2023)}]{Sachdev_2023}%
  \BibitemOpen
  \bibfield  {author} {\bibinfo {author} {\bibfnamefont {S.}~\bibnamefont {Sachdev}},\ }\href@noop {} {\emph {\bibinfo {title} {Quantum Phases of Matter}}}\ (\bibinfo  {publisher} {Cambridge University Press},\ \bibinfo {year} {2023})\BibitemShut {NoStop}%
\bibitem [{\citenamefont {Leinaas}\ and\ \citenamefont {Myrheim}(1977)}]{Leinaas_77}%
  \BibitemOpen
  \bibfield  {author} {\bibinfo {author} {\bibfnamefont {J.~M.}\ \bibnamefont {Leinaas}}\ and\ \bibinfo {author} {\bibfnamefont {J.}~\bibnamefont {Myrheim}},\ }\bibfield  {title} {\bibinfo {title} {On the theory of identical particles},\ }\href {https://doi.org/10.1007/BF02727953} {\bibfield  {journal} {\bibinfo  {journal} {Il Nuovo Cimento B (1971-1996)}\ }\textbf {\bibinfo {volume} {37}},\ \bibinfo {pages} {1} (\bibinfo {year} {1977})}\BibitemShut {NoStop}%
\bibitem [{\citenamefont {Goldin}\ \emph {et~al.}(1981)\citenamefont {Goldin}, \citenamefont {Menikoff},\ and\ \citenamefont {Sharp}}]{Goldin_81}%
  \BibitemOpen
  \bibfield  {author} {\bibinfo {author} {\bibfnamefont {G.~A.}\ \bibnamefont {Goldin}}, \bibinfo {author} {\bibfnamefont {R.}~\bibnamefont {Menikoff}},\ and\ \bibinfo {author} {\bibfnamefont {D.~H.}\ \bibnamefont {Sharp}},\ }\bibfield  {title} {\bibinfo {title} {{Representations of a local current algebra in nonsimply connected space and the Aharonov–Bohm effect}},\ }\href {https://doi.org/10.1063/1.525110} {\bibfield  {journal} {\bibinfo  {journal} {Journal of Mathematical Physics}\ }\textbf {\bibinfo {volume} {22}},\ \bibinfo {pages} {1664} (\bibinfo {year} {1981})},\ \Eprint {https://arxiv.org/abs/https://pubs.aip.org/aip/jmp/article-pdf/22/8/1664/19078813/1664\_1\_online.pdf} {https://pubs.aip.org/aip/jmp/article-pdf/22/8/1664/19078813/1664\_1\_online.pdf} \BibitemShut {NoStop}%
\bibitem [{\citenamefont {Wilczek}(1982)}]{Wilczek_82}%
  \BibitemOpen
  \bibfield  {author} {\bibinfo {author} {\bibfnamefont {F.}~\bibnamefont {Wilczek}},\ }\bibfield  {title} {\bibinfo {title} {Quantum mechanics of fractional-spin particles},\ }\href {https://doi.org/10.1103/PhysRevLett.49.957} {\bibfield  {journal} {\bibinfo  {journal} {Phys. Rev. Lett.}\ }\textbf {\bibinfo {volume} {49}},\ \bibinfo {pages} {957} (\bibinfo {year} {1982})}\BibitemShut {NoStop}%
\bibitem [{\citenamefont {Einarsson}(1990)}]{Einarsson90}%
  \BibitemOpen
  \bibfield  {author} {\bibinfo {author} {\bibfnamefont {T.}~\bibnamefont {Einarsson}},\ }\bibfield  {title} {\bibinfo {title} {Fractional statistics on a torus},\ }\href {https://doi.org/10.1103/PhysRevLett.64.1995} {\bibfield  {journal} {\bibinfo  {journal} {Phys. Rev. Lett.}\ }\textbf {\bibinfo {volume} {64}},\ \bibinfo {pages} {1995} (\bibinfo {year} {1990})}\BibitemShut {NoStop}%
\bibitem [{\citenamefont {Kitaev}(2003)}]{Kitaev_2003}%
  \BibitemOpen
  \bibfield  {author} {\bibinfo {author} {\bibfnamefont {A.}~\bibnamefont {Kitaev}},\ }\bibfield  {title} {\bibinfo {title} {Fault-tolerant quantum computation by anyons},\ }\bibfield  {journal} {\bibinfo  {journal} {Annals of Physics}\ }\textbf {\bibinfo {volume} {303}},\ \href {https://doi.org/10.1016/s0003-4916(02)00018-0} {10.1016/s0003-4916(02)00018-0} (\bibinfo {year} {2003})\BibitemShut {NoStop}%
\bibitem [{\citenamefont {Freedman}\ \emph {et~al.}(2002)\citenamefont {Freedman}, \citenamefont {Larsen},\ and\ \citenamefont {Wang}}]{Freedman_2000gwh}%
  \BibitemOpen
  \bibfield  {author} {\bibinfo {author} {\bibfnamefont {M.~H.}\ \bibnamefont {Freedman}}, \bibinfo {author} {\bibfnamefont {M.}~\bibnamefont {Larsen}},\ and\ \bibinfo {author} {\bibfnamefont {Z.}~\bibnamefont {Wang}},\ }\bibfield  {title} {\bibinfo {title} {{A Modular Functor Which is Universal for Quantum Computation}},\ }\href {https://doi.org/10.1007/s002200200645} {\bibfield  {journal} {\bibinfo  {journal} {Commun. Math. Phys.}\ }\textbf {\bibinfo {volume} {227}},\ \bibinfo {pages} {605} (\bibinfo {year} {2002})},\ \Eprint {https://arxiv.org/abs/quant-ph/0001108} {arXiv:quant-ph/0001108} \BibitemShut {NoStop}%
\bibitem [{\citenamefont {Freedman}\ \emph {et~al.}(2006)\citenamefont {Freedman}, \citenamefont {Nayak},\ and\ \citenamefont {Walker}}]{Freedman_2006}%
  \BibitemOpen
  \bibfield  {author} {\bibinfo {author} {\bibfnamefont {M.}~\bibnamefont {Freedman}}, \bibinfo {author} {\bibfnamefont {C.}~\bibnamefont {Nayak}},\ and\ \bibinfo {author} {\bibfnamefont {K.}~\bibnamefont {Walker}},\ }\bibfield  {title} {\bibinfo {title} {Towards universal topological quantum computation in the $\nu=5/2$ fractional quantum hall state},\ }\bibfield  {journal} {\bibinfo  {journal} {Physical Review B}\ }\textbf {\bibinfo {volume} {73}},\ \href {https://doi.org/10.1103/physrevb.73.245307} {10.1103/physrevb.73.245307} (\bibinfo {year} {2006})\BibitemShut {NoStop}%
\bibitem [{\citenamefont {Nayak}\ \emph {et~al.}(2008)\citenamefont {Nayak}, \citenamefont {Simon}, \citenamefont {Stern}, \citenamefont {Freedman},\ and\ \citenamefont {Das~Sarma}}]{Nayak_08}%
  \BibitemOpen
  \bibfield  {author} {\bibinfo {author} {\bibfnamefont {C.}~\bibnamefont {Nayak}}, \bibinfo {author} {\bibfnamefont {S.~H.}\ \bibnamefont {Simon}}, \bibinfo {author} {\bibfnamefont {A.}~\bibnamefont {Stern}}, \bibinfo {author} {\bibfnamefont {M.}~\bibnamefont {Freedman}},\ and\ \bibinfo {author} {\bibfnamefont {S.}~\bibnamefont {Das~Sarma}},\ }\bibfield  {title} {\bibinfo {title} {Non-abelian anyons and topological quantum computation},\ }\href {https://doi.org/10.1103/RevModPhys.80.1083} {\bibfield  {journal} {\bibinfo  {journal} {Rev. Mod. Phys.}\ }\textbf {\bibinfo {volume} {80}},\ \bibinfo {pages} {1083} (\bibinfo {year} {2008})}\BibitemShut {NoStop}%
\bibitem [{\citenamefont {Terhal}(2015)}]{Terhal_15}%
  \BibitemOpen
  \bibfield  {author} {\bibinfo {author} {\bibfnamefont {B.~M.}\ \bibnamefont {Terhal}},\ }\bibfield  {title} {\bibinfo {title} {Quantum error correction for quantum memories},\ }\href {https://doi.org/10.1103/RevModPhys.87.307} {\bibfield  {journal} {\bibinfo  {journal} {Rev. Mod. Phys.}\ }\textbf {\bibinfo {volume} {87}},\ \bibinfo {pages} {307} (\bibinfo {year} {2015})}\BibitemShut {NoStop}%
\bibitem [{\citenamefont {Read}\ and\ \citenamefont {Sachdev}(1991{\natexlab{a}})}]{read_91}%
  \BibitemOpen
  \bibfield  {author} {\bibinfo {author} {\bibfnamefont {N.}~\bibnamefont {Read}}\ and\ \bibinfo {author} {\bibfnamefont {S.}~\bibnamefont {Sachdev}},\ }\bibfield  {title} {\bibinfo {title} {Large-n expansion for frustrated quantum antiferromagnets},\ }\href {https://doi.org/10.1103/PhysRevLett.66.1773} {\bibfield  {journal} {\bibinfo  {journal} {Phys. Rev. Lett.}\ }\textbf {\bibinfo {volume} {66}},\ \bibinfo {pages} {1773} (\bibinfo {year} {1991}{\natexlab{a}})}\BibitemShut {NoStop}%
\bibitem [{\citenamefont {Wen}(1991{\natexlab{a}})}]{WEN1991}%
  \BibitemOpen
  \bibfield  {author} {\bibinfo {author} {\bibfnamefont {X.~G.}\ \bibnamefont {Wen}},\ }\bibfield  {title} {\bibinfo {title} {Mean-field theory of spin-liquid states with finite energy gap and topological orders},\ }\href {https://doi.org/10.1103/PhysRevB.44.2664} {\bibfield  {journal} {\bibinfo  {journal} {Phys. Rev. B}\ }\textbf {\bibinfo {volume} {44}},\ \bibinfo {pages} {2664} (\bibinfo {year} {1991}{\natexlab{a}})}\BibitemShut {NoStop}%
\bibitem [{\citenamefont {Wang}\ \emph {et~al.}(2003)\citenamefont {Wang}, \citenamefont {Harrington},\ and\ \citenamefont {Preskill}}]{WANG200331}%
  \BibitemOpen
  \bibfield  {author} {\bibinfo {author} {\bibfnamefont {C.}~\bibnamefont {Wang}}, \bibinfo {author} {\bibfnamefont {J.}~\bibnamefont {Harrington}},\ and\ \bibinfo {author} {\bibfnamefont {J.}~\bibnamefont {Preskill}},\ }\bibfield  {title} {\bibinfo {title} {Confinement-higgs transition in a disordered gauge theory and the accuracy threshold for quantum memory},\ }\href {https://doi.org/https://doi.org/10.1016/S0003-4916(02)00019-2} {\bibfield  {journal} {\bibinfo  {journal} {Annals of Physics}\ }\textbf {\bibinfo {volume} {303}},\ \bibinfo {pages} {31} (\bibinfo {year} {2003})}\BibitemShut {NoStop}%
\bibitem [{\citenamefont {Dennis}\ \emph {et~al.}(2002)\citenamefont {Dennis}, \citenamefont {Kitaev}, \citenamefont {Landahl},\ and\ \citenamefont {Preskill}}]{Dennis_2002}%
  \BibitemOpen
  \bibfield  {author} {\bibinfo {author} {\bibfnamefont {E.}~\bibnamefont {Dennis}}, \bibinfo {author} {\bibfnamefont {A.}~\bibnamefont {Kitaev}}, \bibinfo {author} {\bibfnamefont {A.}~\bibnamefont {Landahl}},\ and\ \bibinfo {author} {\bibfnamefont {J.}~\bibnamefont {Preskill}},\ }\bibfield  {title} {\bibinfo {title} {Topological quantum memory},\ }\href {https://doi.org/10.1063/1.1499754} {\bibfield  {journal} {\bibinfo  {journal} {Journal of Mathematical Physics}\ }\textbf {\bibinfo {volume} {43}},\ \bibinfo {pages} {4452–4505} (\bibinfo {year} {2002})}\BibitemShut {NoStop}%
\bibitem [{\citenamefont {Nussinov}\ and\ \citenamefont {Ortiz}(2008)}]{Nussinov_08}%
  \BibitemOpen
  \bibfield  {author} {\bibinfo {author} {\bibfnamefont {Z.}~\bibnamefont {Nussinov}}\ and\ \bibinfo {author} {\bibfnamefont {G.}~\bibnamefont {Ortiz}},\ }\bibfield  {title} {\bibinfo {title} {Autocorrelations and thermal fragility of anyonic loops in topologically quantum ordered systems},\ }\href {https://doi.org/10.1103/PhysRevB.77.064302} {\bibfield  {journal} {\bibinfo  {journal} {Phys. Rev. B}\ }\textbf {\bibinfo {volume} {77}},\ \bibinfo {pages} {064302} (\bibinfo {year} {2008})}\BibitemShut {NoStop}%
\bibitem [{\citenamefont {Bravyi}\ and\ \citenamefont {Terhal}(2009)}]{Bravyi_2009}%
  \BibitemOpen
  \bibfield  {author} {\bibinfo {author} {\bibfnamefont {S.}~\bibnamefont {Bravyi}}\ and\ \bibinfo {author} {\bibfnamefont {B.}~\bibnamefont {Terhal}},\ }\bibfield  {title} {\bibinfo {title} {A no-go theorem for a two-dimensional self-correcting quantum memory based on stabilizer codes},\ }\href {https://doi.org/10.1088/1367-2630/11/4/043029} {\bibfield  {journal} {\bibinfo  {journal} {New Journal of Physics}\ }\textbf {\bibinfo {volume} {11}},\ \bibinfo {pages} {043029} (\bibinfo {year} {2009})}\BibitemShut {NoStop}%
\bibitem [{\citenamefont {Landon-Cardinal}\ and\ \citenamefont {Poulin}(2013)}]{Poulin_13}%
  \BibitemOpen
  \bibfield  {author} {\bibinfo {author} {\bibfnamefont {O.}~\bibnamefont {Landon-Cardinal}}\ and\ \bibinfo {author} {\bibfnamefont {D.}~\bibnamefont {Poulin}},\ }\bibfield  {title} {\bibinfo {title} {Local topological order inhibits thermal stability in 2d},\ }\href {https://doi.org/10.1103/PhysRevLett.110.090502} {\bibfield  {journal} {\bibinfo  {journal} {Phys. Rev. Lett.}\ }\textbf {\bibinfo {volume} {110}},\ \bibinfo {pages} {090502} (\bibinfo {year} {2013})}\BibitemShut {NoStop}%
\bibitem [{\citenamefont {Brown}\ \emph {et~al.}(2016)\citenamefont {Brown}, \citenamefont {Loss}, \citenamefont {Pachos}, \citenamefont {Self},\ and\ \citenamefont {Wootton}}]{Brown_16}%
  \BibitemOpen
  \bibfield  {author} {\bibinfo {author} {\bibfnamefont {B.~J.}\ \bibnamefont {Brown}}, \bibinfo {author} {\bibfnamefont {D.}~\bibnamefont {Loss}}, \bibinfo {author} {\bibfnamefont {J.~K.}\ \bibnamefont {Pachos}}, \bibinfo {author} {\bibfnamefont {C.~N.}\ \bibnamefont {Self}},\ and\ \bibinfo {author} {\bibfnamefont {J.~R.}\ \bibnamefont {Wootton}},\ }\bibfield  {title} {\bibinfo {title} {Quantum memories at finite temperature},\ }\href {https://doi.org/10.1103/RevModPhys.88.045005} {\bibfield  {journal} {\bibinfo  {journal} {Rev. Mod. Phys.}\ }\textbf {\bibinfo {volume} {88}},\ \bibinfo {pages} {045005} (\bibinfo {year} {2016})}\BibitemShut {NoStop}%
\bibitem [{\citenamefont {Goldin}\ \emph {et~al.}(1985)\citenamefont {Goldin}, \citenamefont {Menikoff},\ and\ \citenamefont {Sharp}}]{Goldin_85}%
  \BibitemOpen
  \bibfield  {author} {\bibinfo {author} {\bibfnamefont {G.~A.}\ \bibnamefont {Goldin}}, \bibinfo {author} {\bibfnamefont {R.}~\bibnamefont {Menikoff}},\ and\ \bibinfo {author} {\bibfnamefont {D.~H.}\ \bibnamefont {Sharp}},\ }\bibfield  {title} {\bibinfo {title} {Comments on "general theory for quantum statistics in two dimensions"},\ }\href {https://doi.org/10.1103/PhysRevLett.54.603} {\bibfield  {journal} {\bibinfo  {journal} {Phys. Rev. Lett.}\ }\textbf {\bibinfo {volume} {54}},\ \bibinfo {pages} {603} (\bibinfo {year} {1985})}\BibitemShut {NoStop}%
\bibitem [{\citenamefont {Wen}(1991{\natexlab{b}})}]{Wen_91_FQH}%
  \BibitemOpen
  \bibfield  {author} {\bibinfo {author} {\bibfnamefont {X.~G.}\ \bibnamefont {Wen}},\ }\bibfield  {title} {\bibinfo {title} {Non-abelian statistics in the fractional quantum hall states},\ }\href {https://doi.org/10.1103/PhysRevLett.66.802} {\bibfield  {journal} {\bibinfo  {journal} {Phys. Rev. Lett.}\ }\textbf {\bibinfo {volume} {66}},\ \bibinfo {pages} {802} (\bibinfo {year} {1991}{\natexlab{b}})}\BibitemShut {NoStop}%
\bibitem [{\citenamefont {Moore}\ and\ \citenamefont {Read}(1991)}]{MOORE1991362}%
  \BibitemOpen
  \bibfield  {author} {\bibinfo {author} {\bibfnamefont {G.}~\bibnamefont {Moore}}\ and\ \bibinfo {author} {\bibfnamefont {N.}~\bibnamefont {Read}},\ }\bibfield  {title} {\bibinfo {title} {Nonabelions in the fractional quantum hall effect},\ }\href {https://doi.org/https://doi.org/10.1016/0550-3213(91)90407-O} {\bibfield  {journal} {\bibinfo  {journal} {Nuclear Physics B}\ }\textbf {\bibinfo {volume} {360}},\ \bibinfo {pages} {362} (\bibinfo {year} {1991})}\BibitemShut {NoStop}%
\bibitem [{\citenamefont {Moore}\ and\ \citenamefont {Seiberg}(1989)}]{Moore1989}%
  \BibitemOpen
  \bibfield  {author} {\bibinfo {author} {\bibfnamefont {G.}~\bibnamefont {Moore}}\ and\ \bibinfo {author} {\bibfnamefont {N.}~\bibnamefont {Seiberg}},\ }\bibfield  {title} {\bibinfo {title} {Classical and quantum conformal field theory},\ }\href {https://doi.org/10.1007/BF01238857} {\bibfield  {journal} {\bibinfo  {journal} {Communications in Mathematical Physics}\ }\textbf {\bibinfo {volume} {123}},\ \bibinfo {pages} {177} (\bibinfo {year} {1989})}\BibitemShut {NoStop}%
\bibitem [{\citenamefont {Wootton}\ \emph {et~al.}(2014)\citenamefont {Wootton}, \citenamefont {Burri}, \citenamefont {Iblisdir},\ and\ \citenamefont {Loss}}]{Wootton_14}%
  \BibitemOpen
  \bibfield  {author} {\bibinfo {author} {\bibfnamefont {J.~R.}\ \bibnamefont {Wootton}}, \bibinfo {author} {\bibfnamefont {J.}~\bibnamefont {Burri}}, \bibinfo {author} {\bibfnamefont {S.}~\bibnamefont {Iblisdir}},\ and\ \bibinfo {author} {\bibfnamefont {D.}~\bibnamefont {Loss}},\ }\bibfield  {title} {\bibinfo {title} {Error correction for non-abelian topological quantum computation},\ }\href {https://doi.org/10.1103/PhysRevX.4.011051} {\bibfield  {journal} {\bibinfo  {journal} {Phys. Rev. X}\ }\textbf {\bibinfo {volume} {4}},\ \bibinfo {pages} {011051} (\bibinfo {year} {2014})}\BibitemShut {NoStop}%
\bibitem [{\citenamefont {Brell}\ \emph {et~al.}(2014)\citenamefont {Brell}, \citenamefont {Burton}, \citenamefont {Dauphinais}, \citenamefont {Flammia},\ and\ \citenamefont {Poulin}}]{Brell_14}%
  \BibitemOpen
  \bibfield  {author} {\bibinfo {author} {\bibfnamefont {C.~G.}\ \bibnamefont {Brell}}, \bibinfo {author} {\bibfnamefont {S.}~\bibnamefont {Burton}}, \bibinfo {author} {\bibfnamefont {G.}~\bibnamefont {Dauphinais}}, \bibinfo {author} {\bibfnamefont {S.~T.}\ \bibnamefont {Flammia}},\ and\ \bibinfo {author} {\bibfnamefont {D.}~\bibnamefont {Poulin}},\ }\bibfield  {title} {\bibinfo {title} {Thermalization, error correction, and memory lifetime for ising anyon systems},\ }\href {https://doi.org/10.1103/PhysRevX.4.031058} {\bibfield  {journal} {\bibinfo  {journal} {Phys. Rev. X}\ }\textbf {\bibinfo {volume} {4}},\ \bibinfo {pages} {031058} (\bibinfo {year} {2014})}\BibitemShut {NoStop}%
\bibitem [{\citenamefont {Wootton}\ and\ \citenamefont {Hutter}(2016)}]{Wootton_16}%
  \BibitemOpen
  \bibfield  {author} {\bibinfo {author} {\bibfnamefont {J.~R.}\ \bibnamefont {Wootton}}\ and\ \bibinfo {author} {\bibfnamefont {A.}~\bibnamefont {Hutter}},\ }\bibfield  {title} {\bibinfo {title} {Active error correction for abelian and non-abelian anyons},\ }\href {https://doi.org/10.1103/PhysRevA.93.022318} {\bibfield  {journal} {\bibinfo  {journal} {Phys. Rev. A}\ }\textbf {\bibinfo {volume} {93}},\ \bibinfo {pages} {022318} (\bibinfo {year} {2016})}\BibitemShut {NoStop}%
\bibitem [{\citenamefont {Burton}\ \emph {et~al.}(2017)\citenamefont {Burton}, \citenamefont {Brell},\ and\ \citenamefont {Flammia}}]{Burton_2017}%
  \BibitemOpen
  \bibfield  {author} {\bibinfo {author} {\bibfnamefont {S.}~\bibnamefont {Burton}}, \bibinfo {author} {\bibfnamefont {C.~G.}\ \bibnamefont {Brell}},\ and\ \bibinfo {author} {\bibfnamefont {S.~T.}\ \bibnamefont {Flammia}},\ }\bibfield  {title} {\bibinfo {title} {Classical simulation of quantum error correction in a fibonacci anyon code},\ }\bibfield  {journal} {\bibinfo  {journal} {Physical Review A}\ }\textbf {\bibinfo {volume} {95}},\ \href {https://doi.org/10.1103/physreva.95.022309} {10.1103/physreva.95.022309} (\bibinfo {year} {2017})\BibitemShut {NoStop}%
\bibitem [{\citenamefont {Dauphinais}\ and\ \citenamefont {Poulin}(2017)}]{Dauphinais_2017}%
  \BibitemOpen
  \bibfield  {author} {\bibinfo {author} {\bibfnamefont {G.}~\bibnamefont {Dauphinais}}\ and\ \bibinfo {author} {\bibfnamefont {D.}~\bibnamefont {Poulin}},\ }\bibfield  {title} {\bibinfo {title} {Fault-tolerant quantum error correction for non-abelian anyons},\ }\href {https://doi.org/10.1007/s00220-017-2923-9} {\bibfield  {journal} {\bibinfo  {journal} {Communications in Mathematical Physics}\ }\textbf {\bibinfo {volume} {355}},\ \bibinfo {pages} {519–560} (\bibinfo {year} {2017})}\BibitemShut {NoStop}%
\bibitem [{\citenamefont {{Schotte}}\ \emph {et~al.}(2022)\citenamefont {{Schotte}}, \citenamefont {{Burgelman}},\ and\ \citenamefont {{Zhu}}}]{Schotte_22a}%
  \BibitemOpen
  \bibfield  {author} {\bibinfo {author} {\bibfnamefont {A.}~\bibnamefont {{Schotte}}}, \bibinfo {author} {\bibfnamefont {L.}~\bibnamefont {{Burgelman}}},\ and\ \bibinfo {author} {\bibfnamefont {G.}~\bibnamefont {{Zhu}}},\ }\bibfield  {title} {\bibinfo {title} {{Fault-tolerant error correction for a universal non-Abelian topological quantum computer at finite temperature}},\ }\href {https://doi.org/10.48550/arXiv.2301.00054} {\bibfield  {journal} {\bibinfo  {journal} {arXiv e-prints}\ ,\ \bibinfo {eid} {arXiv:2301.00054}} (\bibinfo {year} {2022})},\ \Eprint {https://arxiv.org/abs/2301.00054} {arXiv:2301.00054 [quant-ph]} \BibitemShut {NoStop}%
\bibitem [{\citenamefont {Schotte}\ \emph {et~al.}(2022)\citenamefont {Schotte}, \citenamefont {Zhu}, \citenamefont {Burgelman},\ and\ \citenamefont {Verstraete}}]{Schotte_22b}%
  \BibitemOpen
  \bibfield  {author} {\bibinfo {author} {\bibfnamefont {A.}~\bibnamefont {Schotte}}, \bibinfo {author} {\bibfnamefont {G.}~\bibnamefont {Zhu}}, \bibinfo {author} {\bibfnamefont {L.}~\bibnamefont {Burgelman}},\ and\ \bibinfo {author} {\bibfnamefont {F.}~\bibnamefont {Verstraete}},\ }\bibfield  {title} {\bibinfo {title} {Quantum error correction thresholds for the universal fibonacci turaev-viro code},\ }\href {https://doi.org/10.1103/PhysRevX.12.021012} {\bibfield  {journal} {\bibinfo  {journal} {Phys. Rev. X}\ }\textbf {\bibinfo {volume} {12}},\ \bibinfo {pages} {021012} (\bibinfo {year} {2022})}\BibitemShut {NoStop}%
\bibitem [{\citenamefont {Iqbal}\ \emph {et~al.}(2024)\citenamefont {Iqbal}, \citenamefont {Tantivasadakarn}, \citenamefont {Verresen}, \citenamefont {Campbell}, \citenamefont {Dreiling}, \citenamefont {Figgatt}, \citenamefont {Gaebler}, \citenamefont {Johansen}, \citenamefont {Mills}, \citenamefont {Moses}, \citenamefont {Pino}, \citenamefont {Ransford}, \citenamefont {Rowe}, \citenamefont {Siegfried}, \citenamefont {Stutz}, \citenamefont {Foss-Feig}, \citenamefont {Vishwanath},\ and\ \citenamefont {Dreyer}}]{iqbal2023creation}%
  \BibitemOpen
  \bibfield  {author} {\bibinfo {author} {\bibfnamefont {M.}~\bibnamefont {Iqbal}}, \bibinfo {author} {\bibfnamefont {N.}~\bibnamefont {Tantivasadakarn}}, \bibinfo {author} {\bibfnamefont {R.}~\bibnamefont {Verresen}}, \bibinfo {author} {\bibfnamefont {S.~L.}\ \bibnamefont {Campbell}}, \bibinfo {author} {\bibfnamefont {J.~M.}\ \bibnamefont {Dreiling}}, \bibinfo {author} {\bibfnamefont {C.}~\bibnamefont {Figgatt}}, \bibinfo {author} {\bibfnamefont {J.~P.}\ \bibnamefont {Gaebler}}, \bibinfo {author} {\bibfnamefont {J.}~\bibnamefont {Johansen}}, \bibinfo {author} {\bibfnamefont {M.}~\bibnamefont {Mills}}, \bibinfo {author} {\bibfnamefont {S.~A.}\ \bibnamefont {Moses}}, \bibinfo {author} {\bibfnamefont {J.~M.}\ \bibnamefont {Pino}}, \bibinfo {author} {\bibfnamefont {A.}~\bibnamefont {Ransford}}, \bibinfo {author} {\bibfnamefont {M.}~\bibnamefont {Rowe}}, \bibinfo {author} {\bibfnamefont {P.}~\bibnamefont {Siegfried}}, \bibinfo {author} {\bibfnamefont {R.~P.}\ \bibnamefont {Stutz}}, \bibinfo {author}
  {\bibfnamefont {M.}~\bibnamefont {Foss-Feig}}, \bibinfo {author} {\bibfnamefont {A.}~\bibnamefont {Vishwanath}},\ and\ \bibinfo {author} {\bibfnamefont {H.}~\bibnamefont {Dreyer}},\ }\bibfield  {title} {\bibinfo {title} {Non-abelian topological order and anyons on a trapped-ion processor},\ }\href {https://doi.org/10.1038/s41586-023-06934-4} {\bibfield  {journal} {\bibinfo  {journal} {Nature}\ }\textbf {\bibinfo {volume} {626}},\ \bibinfo {pages} {505} (\bibinfo {year} {2024})}\BibitemShut {NoStop}%
\bibitem [{\citenamefont {Levin}\ and\ \citenamefont {Wen}(2005)}]{Levin_Wen_05}%
  \BibitemOpen
  \bibfield  {author} {\bibinfo {author} {\bibfnamefont {M.~A.}\ \bibnamefont {Levin}}\ and\ \bibinfo {author} {\bibfnamefont {X.-G.}\ \bibnamefont {Wen}},\ }\bibfield  {title} {\bibinfo {title} {String-net condensation: A physical mechanism for topological phases},\ }\href {https://doi.org/10.1103/PhysRevB.71.045110} {\bibfield  {journal} {\bibinfo  {journal} {Phys. Rev. B}\ }\textbf {\bibinfo {volume} {71}},\ \bibinfo {pages} {045110} (\bibinfo {year} {2005})}\BibitemShut {NoStop}%
\bibitem [{\citenamefont {Tantivasadakarn}\ \emph {et~al.}(2023{\natexlab{a}})\citenamefont {Tantivasadakarn}, \citenamefont {Vishwanath},\ and\ \citenamefont {Verresen}}]{Tantivasadakarn_2023}%
  \BibitemOpen
  \bibfield  {author} {\bibinfo {author} {\bibfnamefont {N.}~\bibnamefont {Tantivasadakarn}}, \bibinfo {author} {\bibfnamefont {A.}~\bibnamefont {Vishwanath}},\ and\ \bibinfo {author} {\bibfnamefont {R.}~\bibnamefont {Verresen}},\ }\bibfield  {title} {\bibinfo {title} {Hierarchy of topological order from finite-depth unitaries, measurement, and feedforward},\ }\bibfield  {journal} {\bibinfo  {journal} {PRX Quantum}\ }\textbf {\bibinfo {volume} {4}},\ \href {https://doi.org/10.1103/prxquantum.4.020339} {10.1103/prxquantum.4.020339} (\bibinfo {year} {2023}{\natexlab{a}})\BibitemShut {NoStop}%
\bibitem [{\citenamefont {Yoshida}(2016)}]{Yoshida_2016}%
  \BibitemOpen
  \bibfield  {author} {\bibinfo {author} {\bibfnamefont {B.}~\bibnamefont {Yoshida}},\ }\bibfield  {title} {\bibinfo {title} {Topological phases with generalized global symmetries},\ }\bibfield  {journal} {\bibinfo  {journal} {Physical Review B}\ }\textbf {\bibinfo {volume} {93}},\ \href {https://doi.org/10.1103/physrevb.93.155131} {10.1103/physrevb.93.155131} (\bibinfo {year} {2016})\BibitemShut {NoStop}%
\bibitem [{\citenamefont {Dijkgraaf}\ and\ \citenamefont {Witten}(1990)}]{Dijkgraaf1990}%
  \BibitemOpen
  \bibfield  {author} {\bibinfo {author} {\bibfnamefont {R.}~\bibnamefont {Dijkgraaf}}\ and\ \bibinfo {author} {\bibfnamefont {E.}~\bibnamefont {Witten}},\ }\bibfield  {title} {\bibinfo {title} {Topological gauge theories and group cohomology},\ }\href {https://doi.org/10.1007/BF02096988} {\bibfield  {journal} {\bibinfo  {journal} {Communications in Mathematical Physics}\ }\textbf {\bibinfo {volume} {129}},\ \bibinfo {pages} {393} (\bibinfo {year} {1990})}\BibitemShut {NoStop}%
\bibitem [{\citenamefont {Dijkgraaf}\ \emph {et~al.}(1991)\citenamefont {Dijkgraaf}, \citenamefont {Pasquier},\ and\ \citenamefont {Roche}}]{Dijkgraaf1991}%
  \BibitemOpen
  \bibfield  {author} {\bibinfo {author} {\bibfnamefont {R.}~\bibnamefont {Dijkgraaf}}, \bibinfo {author} {\bibfnamefont {V.}~\bibnamefont {Pasquier}},\ and\ \bibinfo {author} {\bibfnamefont {P.}~\bibnamefont {Roche}},\ }\bibfield  {title} {\bibinfo {title} {Quasi hope algebras, group cohomology and orbifold models},\ }\href {https://doi.org/https://doi.org/10.1016/0920-5632(91)90123-V} {\bibfield  {journal} {\bibinfo  {journal} {Nuclear Physics B - Proceedings Supplements}\ }\textbf {\bibinfo {volume} {18}},\ \bibinfo {pages} {60} (\bibinfo {year} {1991})}\BibitemShut {NoStop}%
\bibitem [{\citenamefont {Hu}\ \emph {et~al.}(2013)\citenamefont {Hu}, \citenamefont {Wan},\ and\ \citenamefont {Wu}}]{Hu13}%
  \BibitemOpen
  \bibfield  {author} {\bibinfo {author} {\bibfnamefont {Y.}~\bibnamefont {Hu}}, \bibinfo {author} {\bibfnamefont {Y.}~\bibnamefont {Wan}},\ and\ \bibinfo {author} {\bibfnamefont {Y.-S.}\ \bibnamefont {Wu}},\ }\bibfield  {title} {\bibinfo {title} {Twisted quantum double model of topological phases in two dimensions},\ }\href {https://doi.org/10.1103/PhysRevB.87.125114} {\bibfield  {journal} {\bibinfo  {journal} {Phys. Rev. B}\ }\textbf {\bibinfo {volume} {87}},\ \bibinfo {pages} {125114} (\bibinfo {year} {2013})}\BibitemShut {NoStop}%
\bibitem [{\citenamefont {Haegeman}\ \emph {et~al.}(2015)\citenamefont {Haegeman}, \citenamefont {Van~Acoleyen}, \citenamefont {Schuch}, \citenamefont {Cirac},\ and\ \citenamefont {Verstraete}}]{Haegeman_15}%
  \BibitemOpen
  \bibfield  {author} {\bibinfo {author} {\bibfnamefont {J.}~\bibnamefont {Haegeman}}, \bibinfo {author} {\bibfnamefont {K.}~\bibnamefont {Van~Acoleyen}}, \bibinfo {author} {\bibfnamefont {N.}~\bibnamefont {Schuch}}, \bibinfo {author} {\bibfnamefont {J.~I.}\ \bibnamefont {Cirac}},\ and\ \bibinfo {author} {\bibfnamefont {F.}~\bibnamefont {Verstraete}},\ }\bibfield  {title} {\bibinfo {title} {Gauging quantum states: From global to local symmetries in many-body systems},\ }\href {https://doi.org/10.1103/PhysRevX.5.011024} {\bibfield  {journal} {\bibinfo  {journal} {Phys. Rev. X}\ }\textbf {\bibinfo {volume} {5}},\ \bibinfo {pages} {011024} (\bibinfo {year} {2015})}\BibitemShut {NoStop}%
\bibitem [{\citenamefont {Zhu}\ and\ \citenamefont {Zhang}(2019)}]{GuoYi_19}%
  \BibitemOpen
  \bibfield  {author} {\bibinfo {author} {\bibfnamefont {G.-Y.}\ \bibnamefont {Zhu}}\ and\ \bibinfo {author} {\bibfnamefont {G.-M.}\ \bibnamefont {Zhang}},\ }\bibfield  {title} {\bibinfo {title} {Gapless coulomb state emerging from a self-dual topological tensor-network state},\ }\href {https://doi.org/10.1103/PhysRevLett.122.176401} {\bibfield  {journal} {\bibinfo  {journal} {Phys. Rev. Lett.}\ }\textbf {\bibinfo {volume} {122}},\ \bibinfo {pages} {176401} (\bibinfo {year} {2019})}\BibitemShut {NoStop}%
\bibitem [{\citenamefont {Mariën}\ \emph {et~al.}(2017)\citenamefont {Mariën}, \citenamefont {Haegeman}, \citenamefont {Fendley},\ and\ \citenamefont {Verstraete}}]{Mari_n_2017}%
  \BibitemOpen
  \bibfield  {author} {\bibinfo {author} {\bibfnamefont {M.}~\bibnamefont {Mariën}}, \bibinfo {author} {\bibfnamefont {J.}~\bibnamefont {Haegeman}}, \bibinfo {author} {\bibfnamefont {P.}~\bibnamefont {Fendley}},\ and\ \bibinfo {author} {\bibfnamefont {F.}~\bibnamefont {Verstraete}},\ }\bibfield  {title} {\bibinfo {title} {Condensation-driven phase transitions in perturbed string nets},\ }\bibfield  {journal} {\bibinfo  {journal} {Physical Review B}\ }\textbf {\bibinfo {volume} {96}},\ \href {https://doi.org/10.1103/physrevb.96.155127} {10.1103/physrevb.96.155127} (\bibinfo {year} {2017})\BibitemShut {NoStop}%
\bibitem [{\citenamefont {Xu}\ and\ \citenamefont {Schuch}(2021)}]{Xu_2021}%
  \BibitemOpen
  \bibfield  {author} {\bibinfo {author} {\bibfnamefont {W.-T.}\ \bibnamefont {Xu}}\ and\ \bibinfo {author} {\bibfnamefont {N.}~\bibnamefont {Schuch}},\ }\bibfield  {title} {\bibinfo {title} {Characterization of topological phase transitions from a non-abelian topological state and its galois conjugate through condensation and confinement order parameters},\ }\bibfield  {journal} {\bibinfo  {journal} {Physical Review B}\ }\textbf {\bibinfo {volume} {104}},\ \href {https://doi.org/10.1103/physrevb.104.155119} {10.1103/physrevb.104.155119} (\bibinfo {year} {2021})\BibitemShut {NoStop}%
\bibitem [{\citenamefont {Xu}\ \emph {et~al.}(2022)\citenamefont {Xu}, \citenamefont {Garre-Rubio},\ and\ \citenamefont {Schuch}}]{Xu_2022}%
  \BibitemOpen
  \bibfield  {author} {\bibinfo {author} {\bibfnamefont {W.-T.}\ \bibnamefont {Xu}}, \bibinfo {author} {\bibfnamefont {J.}~\bibnamefont {Garre-Rubio}},\ and\ \bibinfo {author} {\bibfnamefont {N.}~\bibnamefont {Schuch}},\ }\bibfield  {title} {\bibinfo {title} {Complete characterization of non-abelian topological phase transitions and detection of anyon splitting with projected entangled pair states},\ }\bibfield  {journal} {\bibinfo  {journal} {Physical Review B}\ }\textbf {\bibinfo {volume} {106}},\ \href {https://doi.org/10.1103/physrevb.106.205139} {10.1103/physrevb.106.205139} (\bibinfo {year} {2022})\BibitemShut {NoStop}%
\bibitem [{\citenamefont {Schotte}\ \emph {et~al.}(2019)\citenamefont {Schotte}, \citenamefont {Carrasco}, \citenamefont {Vanhecke}, \citenamefont {Vanderstraeten}, \citenamefont {Haegeman}, \citenamefont {Verstraete},\ and\ \citenamefont {Vidal}}]{Schotte_2019}%
  \BibitemOpen
  \bibfield  {author} {\bibinfo {author} {\bibfnamefont {A.}~\bibnamefont {Schotte}}, \bibinfo {author} {\bibfnamefont {J.}~\bibnamefont {Carrasco}}, \bibinfo {author} {\bibfnamefont {B.}~\bibnamefont {Vanhecke}}, \bibinfo {author} {\bibfnamefont {L.}~\bibnamefont {Vanderstraeten}}, \bibinfo {author} {\bibfnamefont {J.}~\bibnamefont {Haegeman}}, \bibinfo {author} {\bibfnamefont {F.}~\bibnamefont {Verstraete}},\ and\ \bibinfo {author} {\bibfnamefont {J.}~\bibnamefont {Vidal}},\ }\bibfield  {title} {\bibinfo {title} {Tensor-network approach to phase transitions in string-net models},\ }\bibfield  {journal} {\bibinfo  {journal} {Physical Review B}\ }\textbf {\bibinfo {volume} {100}},\ \href {https://doi.org/10.1103/physrevb.100.245125} {10.1103/physrevb.100.245125} (\bibinfo {year} {2019})\BibitemShut {NoStop}%
\bibitem [{\citenamefont {Fendley}(2008)}]{Fendley_2008}%
  \BibitemOpen
  \bibfield  {author} {\bibinfo {author} {\bibfnamefont {P.}~\bibnamefont {Fendley}},\ }\bibfield  {title} {\bibinfo {title} {Topological order from quantum loops and nets},\ }\href {https://doi.org/10.1016/j.aop.2008.04.011} {\bibfield  {journal} {\bibinfo  {journal} {Annals of Physics}\ }\textbf {\bibinfo {volume} {323}},\ \bibinfo {pages} {3113–3136} (\bibinfo {year} {2008})}\BibitemShut {NoStop}%
\bibitem [{\citenamefont {Domany}\ \emph {et~al.}(1981)\citenamefont {Domany}, \citenamefont {Mukamel}, \citenamefont {Nienhuis},\ and\ \citenamefont {Schwimmer}}]{Nienhuis_81}%
  \BibitemOpen
  \bibfield  {author} {\bibinfo {author} {\bibfnamefont {E.}~\bibnamefont {Domany}}, \bibinfo {author} {\bibfnamefont {D.}~\bibnamefont {Mukamel}}, \bibinfo {author} {\bibfnamefont {B.}~\bibnamefont {Nienhuis}},\ and\ \bibinfo {author} {\bibfnamefont {A.}~\bibnamefont {Schwimmer}},\ }\bibfield  {title} {\bibinfo {title} {Duality relations and equivalences for models with o(n) and cubic symmetry},\ }\href {https://doi.org/https://doi.org/10.1016/0550-3213(81)90559-9} {\bibfield  {journal} {\bibinfo  {journal} {Nuclear Physics B}\ }\textbf {\bibinfo {volume} {190}},\ \bibinfo {pages} {279} (\bibinfo {year} {1981})}\BibitemShut {NoStop}%
\bibitem [{\citenamefont {Nienhuis}(1982)}]{Nienhuis_82}%
  \BibitemOpen
  \bibfield  {author} {\bibinfo {author} {\bibfnamefont {B.}~\bibnamefont {Nienhuis}},\ }\bibfield  {title} {\bibinfo {title} {Exact critical point and critical exponents of $\mathrm{O}(n)$ models in two dimensions},\ }\href {https://doi.org/10.1103/PhysRevLett.49.1062} {\bibfield  {journal} {\bibinfo  {journal} {Phys. Rev. Lett.}\ }\textbf {\bibinfo {volume} {49}},\ \bibinfo {pages} {1062} (\bibinfo {year} {1982})}\BibitemShut {NoStop}%
\bibitem [{\citenamefont {Peled}\ and\ \citenamefont {Spinka}(2019)}]{peled2019lectures}%
  \BibitemOpen
  \bibfield  {author} {\bibinfo {author} {\bibfnamefont {R.}~\bibnamefont {Peled}}\ and\ \bibinfo {author} {\bibfnamefont {Y.}~\bibnamefont {Spinka}},\ }\href@noop {} {\bibinfo {title} {Lectures on the spin and loop $o(n)$ models}} (\bibinfo {year} {2019}),\ \Eprint {https://arxiv.org/abs/1708.00058} {arXiv:1708.00058 [math-ph]} \BibitemShut {NoStop}%
\bibitem [{\citenamefont {Sala}\ and\ \citenamefont {Verresen}(2024)}]{short_paper}%
  \BibitemOpen
  \bibfield  {author} {\bibinfo {author} {\bibfnamefont {P.}~\bibnamefont {Sala}}\ and\ \bibinfo {author} {\bibfnamefont {R.}~\bibnamefont {Verresen}},\ }\href {https://arxiv.org/abs/2409.12230} {\bibinfo {title} {{Stability and Loop Models from Decohering Non-Abelian Topological Order}}} (\bibinfo {year} {2024}),\ \Eprint {https://arxiv.org/abs/2409.12230} {arXiv:2409.12230 [quant-ph]} \BibitemShut {NoStop}%
\bibitem [{\citenamefont {de~Wild~Propitius}(1995)}]{propitius}%
  \BibitemOpen
  \bibfield  {author} {\bibinfo {author} {\bibfnamefont {M.}~\bibnamefont {de~Wild~Propitius}},\ }\href {https://arxiv.org/abs/hep-th/9511195} {\bibinfo {title} {Topological interactions in broken gauge theories}} (\bibinfo {year} {1995}),\ \Eprint {https://arxiv.org/abs/hep-th/9511195} {arXiv:hep-th/9511195 [hep-th]} \BibitemShut {NoStop}%
\bibitem [{\citenamefont {Kitaev}(2006)}]{Kitaev_2006}%
  \BibitemOpen
  \bibfield  {author} {\bibinfo {author} {\bibfnamefont {A.}~\bibnamefont {Kitaev}},\ }\bibfield  {title} {\bibinfo {title} {Anyons in an exactly solved model and beyond},\ }\href {https://doi.org/10.1016/j.aop.2005.10.005} {\bibfield  {journal} {\bibinfo  {journal} {Annals of Physics}\ }\textbf {\bibinfo {volume} {321}},\ \bibinfo {pages} {2–111} (\bibinfo {year} {2006})}\BibitemShut {NoStop}%
\bibitem [{\citenamefont {Lootens}\ \emph {et~al.}(2022)\citenamefont {Lootens}, \citenamefont {Vancraeynest-De~Cuiper}, \citenamefont {Schuch},\ and\ \citenamefont {Verstraete}}]{Lootens22}%
  \BibitemOpen
  \bibfield  {author} {\bibinfo {author} {\bibfnamefont {L.}~\bibnamefont {Lootens}}, \bibinfo {author} {\bibfnamefont {B.}~\bibnamefont {Vancraeynest-De~Cuiper}}, \bibinfo {author} {\bibfnamefont {N.}~\bibnamefont {Schuch}},\ and\ \bibinfo {author} {\bibfnamefont {F.}~\bibnamefont {Verstraete}},\ }\bibfield  {title} {\bibinfo {title} {Mapping between morita-equivalent string-net states with a constant depth quantum circuit},\ }\href {https://doi.org/10.1103/PhysRevB.105.085130} {\bibfield  {journal} {\bibinfo  {journal} {Phys. Rev. B}\ }\textbf {\bibinfo {volume} {105}},\ \bibinfo {pages} {085130} (\bibinfo {year} {2022})}\BibitemShut {NoStop}%
\bibitem [{\citenamefont {Barkeshli}\ \emph {et~al.}(2019)\citenamefont {Barkeshli}, \citenamefont {Bonderson}, \citenamefont {Cheng},\ and\ \citenamefont {Wang}}]{Barkeshli19}%
  \BibitemOpen
  \bibfield  {author} {\bibinfo {author} {\bibfnamefont {M.}~\bibnamefont {Barkeshli}}, \bibinfo {author} {\bibfnamefont {P.}~\bibnamefont {Bonderson}}, \bibinfo {author} {\bibfnamefont {M.}~\bibnamefont {Cheng}},\ and\ \bibinfo {author} {\bibfnamefont {Z.}~\bibnamefont {Wang}},\ }\bibfield  {title} {\bibinfo {title} {Symmetry fractionalization, defects, and gauging of topological phases},\ }\href {https://doi.org/10.1103/PhysRevB.100.115147} {\bibfield  {journal} {\bibinfo  {journal} {Phys. Rev. B}\ }\textbf {\bibinfo {volume} {100}},\ \bibinfo {pages} {115147} (\bibinfo {year} {2019})}\BibitemShut {NoStop}%
\bibitem [{\citenamefont {Chen}(2017)}]{Chen17}%
  \BibitemOpen
  \bibfield  {author} {\bibinfo {author} {\bibfnamefont {X.}~\bibnamefont {Chen}},\ }\bibfield  {title} {\bibinfo {title} {Symmetry fractionalization in two dimensional topological phases},\ }\href {https://doi.org/https://doi.org/10.1016/j.revip.2017.02.002} {\bibfield  {journal} {\bibinfo  {journal} {Reviews in Physics}\ }\textbf {\bibinfo {volume} {2}},\ \bibinfo {pages} {3} (\bibinfo {year} {2017})}\BibitemShut {NoStop}%
\bibitem [{\citenamefont {Ben-Zion}\ \emph {et~al.}(2016)\citenamefont {Ben-Zion}, \citenamefont {Das},\ and\ \citenamefont {McGreevy}}]{BenZion16}%
  \BibitemOpen
  \bibfield  {author} {\bibinfo {author} {\bibfnamefont {D.}~\bibnamefont {Ben-Zion}}, \bibinfo {author} {\bibfnamefont {D.}~\bibnamefont {Das}},\ and\ \bibinfo {author} {\bibfnamefont {J.}~\bibnamefont {McGreevy}},\ }\bibfield  {title} {\bibinfo {title} {Exactly solvable models of spin liquids with spinons, and of three-dimensional topological paramagnets},\ }\href {https://doi.org/10.1103/PhysRevB.93.155147} {\bibfield  {journal} {\bibinfo  {journal} {Phys. Rev. B}\ }\textbf {\bibinfo {volume} {93}},\ \bibinfo {pages} {155147} (\bibinfo {year} {2016})}\BibitemShut {NoStop}%
\bibitem [{\citenamefont {Stephen}\ \emph {et~al.}(2020)\citenamefont {Stephen}, \citenamefont {Garre-Rubio}, \citenamefont {Dua},\ and\ \citenamefont {Williamson}}]{Stephen20}%
  \BibitemOpen
  \bibfield  {author} {\bibinfo {author} {\bibfnamefont {D.~T.}\ \bibnamefont {Stephen}}, \bibinfo {author} {\bibfnamefont {J.}~\bibnamefont {Garre-Rubio}}, \bibinfo {author} {\bibfnamefont {A.}~\bibnamefont {Dua}},\ and\ \bibinfo {author} {\bibfnamefont {D.~J.}\ \bibnamefont {Williamson}},\ }\bibfield  {title} {\bibinfo {title} {Subsystem symmetry enriched topological order in three dimensions},\ }\href {https://doi.org/10.1103/PhysRevResearch.2.033331} {\bibfield  {journal} {\bibinfo  {journal} {Phys. Rev. Res.}\ }\textbf {\bibinfo {volume} {2}},\ \bibinfo {pages} {033331} (\bibinfo {year} {2020})}\BibitemShut {NoStop}%
\bibitem [{\citenamefont {Tantivasadakarn}\ \emph {et~al.}(2023{\natexlab{b}})\citenamefont {Tantivasadakarn}, \citenamefont {Verresen},\ and\ \citenamefont {Vishwanath}}]{Shortest_NA}%
  \BibitemOpen
  \bibfield  {author} {\bibinfo {author} {\bibfnamefont {N.}~\bibnamefont {Tantivasadakarn}}, \bibinfo {author} {\bibfnamefont {R.}~\bibnamefont {Verresen}},\ and\ \bibinfo {author} {\bibfnamefont {A.}~\bibnamefont {Vishwanath}},\ }\bibfield  {title} {\bibinfo {title} {Shortest route to non-abelian topological order on a quantum processor},\ }\href {https://doi.org/10.1103/PhysRevLett.131.060405} {\bibfield  {journal} {\bibinfo  {journal} {Phys. Rev. Lett.}\ }\textbf {\bibinfo {volume} {131}},\ \bibinfo {pages} {060405} (\bibinfo {year} {2023}{\natexlab{b}})}\BibitemShut {NoStop}%
\bibitem [{\citenamefont {Verstraete}\ \emph {et~al.}(2006)\citenamefont {Verstraete}, \citenamefont {Wolf}, \citenamefont {Perez-Garcia},\ and\ \citenamefont {Cirac}}]{Verstraete_06}%
  \BibitemOpen
  \bibfield  {author} {\bibinfo {author} {\bibfnamefont {F.}~\bibnamefont {Verstraete}}, \bibinfo {author} {\bibfnamefont {M.~M.}\ \bibnamefont {Wolf}}, \bibinfo {author} {\bibfnamefont {D.}~\bibnamefont {Perez-Garcia}},\ and\ \bibinfo {author} {\bibfnamefont {J.~I.}\ \bibnamefont {Cirac}},\ }\bibfield  {title} {\bibinfo {title} {Criticality, the area law, and the computational power of projected entangled pair states},\ }\href {https://doi.org/10.1103/PhysRevLett.96.220601} {\bibfield  {journal} {\bibinfo  {journal} {Phys. Rev. Lett.}\ }\textbf {\bibinfo {volume} {96}},\ \bibinfo {pages} {220601} (\bibinfo {year} {2006})}\BibitemShut {NoStop}%
\bibitem [{\citenamefont {Sahay}\ \emph {et~al.}(2025)\citenamefont {Sahay}, \citenamefont {von Keyserlingk}, \citenamefont {Verresen},\ and\ \citenamefont {Zhang}}]{sahay_2025}%
  \BibitemOpen
  \bibfield  {author} {\bibinfo {author} {\bibfnamefont {R.}~\bibnamefont {Sahay}}, \bibinfo {author} {\bibfnamefont {C.}~\bibnamefont {von Keyserlingk}}, \bibinfo {author} {\bibfnamefont {R.}~\bibnamefont {Verresen}},\ and\ \bibinfo {author} {\bibfnamefont {C.}~\bibnamefont {Zhang}},\ }\href {https://arxiv.org/abs/2503.01977} {\bibinfo {title} {Enforced gaplessness from states with exponentially decaying correlations}} (\bibinfo {year} {2025}),\ \Eprint {https://arxiv.org/abs/2503.01977} {arXiv:2503.01977 [cond-mat.str-el]} \BibitemShut {NoStop}%
\bibitem [{\citenamefont {Houtappel}(1950)}]{HOUTAPPEL1950425}%
  \BibitemOpen
  \bibfield  {author} {\bibinfo {author} {\bibfnamefont {R.}~\bibnamefont {Houtappel}},\ }\bibfield  {title} {\bibinfo {title} {Order-disorder in hexagonal lattices},\ }\href {https://doi.org/https://doi.org/10.1016/0031-8914(50)90130-3} {\bibfield  {journal} {\bibinfo  {journal} {Physica}\ }\textbf {\bibinfo {volume} {16}},\ \bibinfo {pages} {425} (\bibinfo {year} {1950})}\BibitemShut {NoStop}%
\bibitem [{\citenamefont {Verresen}\ and\ \citenamefont {Vishwanath}(2022)}]{unifyingQSL}%
  \BibitemOpen
  \bibfield  {author} {\bibinfo {author} {\bibfnamefont {R.}~\bibnamefont {Verresen}}\ and\ \bibinfo {author} {\bibfnamefont {A.}~\bibnamefont {Vishwanath}},\ }\bibfield  {title} {\bibinfo {title} {Unifying kitaev magnets, kagom\'e dimer models, and ruby rydberg spin liquids},\ }\href {https://doi.org/10.1103/PhysRevX.12.041029} {\bibfield  {journal} {\bibinfo  {journal} {Phys. Rev. X}\ }\textbf {\bibinfo {volume} {12}},\ \bibinfo {pages} {041029} (\bibinfo {year} {2022})}\BibitemShut {NoStop}%
\bibitem [{\citenamefont {Wannier}(1950)}]{triang_Ising}%
  \BibitemOpen
  \bibfield  {author} {\bibinfo {author} {\bibfnamefont {G.~H.}\ \bibnamefont {Wannier}},\ }\bibfield  {title} {\bibinfo {title} {Antiferromagnetism. the triangular ising net},\ }\href {https://doi.org/10.1103/PhysRev.79.357} {\bibfield  {journal} {\bibinfo  {journal} {Phys. Rev.}\ }\textbf {\bibinfo {volume} {79}},\ \bibinfo {pages} {357} (\bibinfo {year} {1950})}\BibitemShut {NoStop}%
\bibitem [{\citenamefont {Bl\"ote}\ and\ \citenamefont {Nightingale}(1993)}]{blote_93}%
  \BibitemOpen
  \bibfield  {author} {\bibinfo {author} {\bibfnamefont {H.~W.~J.}\ \bibnamefont {Bl\"ote}}\ and\ \bibinfo {author} {\bibfnamefont {M.~P.}\ \bibnamefont {Nightingale}},\ }\bibfield  {title} {\bibinfo {title} {Antiferromagnetic triangular ising model: Critical behavior of the ground state},\ }\href {https://doi.org/10.1103/PhysRevB.47.15046} {\bibfield  {journal} {\bibinfo  {journal} {Phys. Rev. B}\ }\textbf {\bibinfo {volume} {47}},\ \bibinfo {pages} {15046} (\bibinfo {year} {1993})}\BibitemShut {NoStop}%
\bibitem [{\citenamefont {Nienhuis}\ \emph {et~al.}(1984)\citenamefont {Nienhuis}, \citenamefont {Hilhorst},\ and\ \citenamefont {Blote}}]{Nienhuis84b}%
  \BibitemOpen
  \bibfield  {author} {\bibinfo {author} {\bibfnamefont {B.}~\bibnamefont {Nienhuis}}, \bibinfo {author} {\bibfnamefont {H.~J.}\ \bibnamefont {Hilhorst}},\ and\ \bibinfo {author} {\bibfnamefont {H.~W.~J.}\ \bibnamefont {Blote}},\ }\bibfield  {title} {\bibinfo {title} {Triangular sos models and cubic-crystal shapes},\ }\href {https://doi.org/10.1088/0305-4470/17/18/025} {\bibfield  {journal} {\bibinfo  {journal} {Journal of Physics A: Mathematical and General}\ }\textbf {\bibinfo {volume} {17}},\ \bibinfo {pages} {3559} (\bibinfo {year} {1984})}\BibitemShut {NoStop}%
\bibitem [{\citenamefont {Moessner}\ \emph {et~al.}(2000)\citenamefont {Moessner}, \citenamefont {Sondhi},\ and\ \citenamefont {Chandra}}]{Moessner00}%
  \BibitemOpen
  \bibfield  {author} {\bibinfo {author} {\bibfnamefont {R.}~\bibnamefont {Moessner}}, \bibinfo {author} {\bibfnamefont {S.~L.}\ \bibnamefont {Sondhi}},\ and\ \bibinfo {author} {\bibfnamefont {P.}~\bibnamefont {Chandra}},\ }\bibfield  {title} {\bibinfo {title} {Two-dimensional periodic frustrated ising models in a transverse field},\ }\href {https://doi.org/10.1103/physrevlett.84.4457} {\bibfield  {journal} {\bibinfo  {journal} {Physical Review Letters}\ }\textbf {\bibinfo {volume} {84}},\ \bibinfo {pages} {4457–4460} (\bibinfo {year} {2000})}\BibitemShut {NoStop}%
\bibitem [{\citenamefont {Rokhsar}\ and\ \citenamefont {Kivelson}(1988)}]{RK88}%
  \BibitemOpen
  \bibfield  {author} {\bibinfo {author} {\bibfnamefont {D.~S.}\ \bibnamefont {Rokhsar}}\ and\ \bibinfo {author} {\bibfnamefont {S.~A.}\ \bibnamefont {Kivelson}},\ }\bibfield  {title} {\bibinfo {title} {Superconductivity and the quantum hard-core dimer gas},\ }\href {https://doi.org/10.1103/PhysRevLett.61.2376} {\bibfield  {journal} {\bibinfo  {journal} {Phys. Rev. Lett.}\ }\textbf {\bibinfo {volume} {61}},\ \bibinfo {pages} {2376} (\bibinfo {year} {1988})}\BibitemShut {NoStop}%
\bibitem [{\citenamefont {Read}\ and\ \citenamefont {Sachdev}(1991{\natexlab{b}})}]{Read91}%
  \BibitemOpen
  \bibfield  {author} {\bibinfo {author} {\bibfnamefont {N.}~\bibnamefont {Read}}\ and\ \bibinfo {author} {\bibfnamefont {S.}~\bibnamefont {Sachdev}},\ }\bibfield  {title} {\bibinfo {title} {Large-n expansion for frustrated quantum antiferromagnets},\ }\href {https://doi.org/10.1103/PhysRevLett.66.1773} {\bibfield  {journal} {\bibinfo  {journal} {Phys. Rev. Lett.}\ }\textbf {\bibinfo {volume} {66}},\ \bibinfo {pages} {1773} (\bibinfo {year} {1991}{\natexlab{b}})}\BibitemShut {NoStop}%
\bibitem [{\citenamefont {Fisher}\ and\ \citenamefont {Stephenson}(1963)}]{Fisher63}%
  \BibitemOpen
  \bibfield  {author} {\bibinfo {author} {\bibfnamefont {M.~E.}\ \bibnamefont {Fisher}}\ and\ \bibinfo {author} {\bibfnamefont {J.}~\bibnamefont {Stephenson}},\ }\bibfield  {title} {\bibinfo {title} {Statistical mechanics of dimers on a plane lattice. ii. dimer correlations and monomers},\ }\href {https://doi.org/10.1103/PhysRev.132.1411} {\bibfield  {journal} {\bibinfo  {journal} {Phys. Rev.}\ }\textbf {\bibinfo {volume} {132}},\ \bibinfo {pages} {1411} (\bibinfo {year} {1963})}\BibitemShut {NoStop}%
\bibitem [{\citenamefont {Chen}\ and\ \citenamefont {Grover}(2024{\natexlab{c}})}]{Chen_2024}%
  \BibitemOpen
  \bibfield  {author} {\bibinfo {author} {\bibfnamefont {Y.-H.}\ \bibnamefont {Chen}}\ and\ \bibinfo {author} {\bibfnamefont {T.}~\bibnamefont {Grover}},\ }\bibfield  {title} {\bibinfo {title} {Symmetry-enforced many-body separability transitions},\ }\bibfield  {journal} {\bibinfo  {journal} {PRX Quantum}\ }\textbf {\bibinfo {volume} {5}},\ \href {https://doi.org/10.1103/prxquantum.5.030310} {10.1103/prxquantum.5.030310} (\bibinfo {year} {2024}{\natexlab{c}})\BibitemShut {NoStop}%
\bibitem [{\citenamefont {Ashida}\ \emph {et~al.}(2023)\citenamefont {Ashida}, \citenamefont {Furukawa},\ and\ \citenamefont {Oshikawa}}]{ashida2023systemenvironment}%
  \BibitemOpen
  \bibfield  {author} {\bibinfo {author} {\bibfnamefont {Y.}~\bibnamefont {Ashida}}, \bibinfo {author} {\bibfnamefont {S.}~\bibnamefont {Furukawa}},\ and\ \bibinfo {author} {\bibfnamefont {M.}~\bibnamefont {Oshikawa}},\ }\href@noop {} {\bibinfo {title} {System-environment entanglement phase transitions}} (\bibinfo {year} {2023}),\ \Eprint {https://arxiv.org/abs/2311.16343} {arXiv:2311.16343 [cond-mat.stat-mech]} \BibitemShut {NoStop}%
\bibitem [{\citenamefont {Schwartz}\ and\ \citenamefont {Fishman}(1980)}]{SCHWARTZ1980115}%
  \BibitemOpen
  \bibfield  {author} {\bibinfo {author} {\bibfnamefont {M.}~\bibnamefont {Schwartz}}\ and\ \bibinfo {author} {\bibfnamefont {S.}~\bibnamefont {Fishman}},\ }\bibfield  {title} {\bibinfo {title} {Real space renormalization group study of the random bond ising model},\ }\href {https://doi.org/https://doi.org/10.1016/0378-4371(80)90076-X} {\bibfield  {journal} {\bibinfo  {journal} {Physica A: Statistical Mechanics and its Applications}\ }\textbf {\bibinfo {volume} {104}},\ \bibinfo {pages} {115} (\bibinfo {year} {1980})}\BibitemShut {NoStop}%
\bibitem [{\citenamefont {Harris}(1974)}]{Harris_1974}%
  \BibitemOpen
  \bibfield  {author} {\bibinfo {author} {\bibfnamefont {A.~B.}\ \bibnamefont {Harris}},\ }\bibfield  {title} {\bibinfo {title} {Effect of random defects on the critical behaviour of ising models},\ }\href {https://doi.org/10.1088/0022-3719/7/9/009} {\bibfield  {journal} {\bibinfo  {journal} {Journal of Physics C: Solid State Physics}\ }\textbf {\bibinfo {volume} {7}},\ \bibinfo {pages} {1671} (\bibinfo {year} {1974})}\BibitemShut {NoStop}%
\bibitem [{\citenamefont {Nishimori}(1981)}]{Nishimori_81}%
  \BibitemOpen
  \bibfield  {author} {\bibinfo {author} {\bibfnamefont {H.}~\bibnamefont {Nishimori}},\ }\bibfield  {title} {\bibinfo {title} {{Internal Energy, Specific Heat and Correlation Function of the Bond-Random Ising Model}},\ }\href {https://doi.org/10.1143/PTP.66.1169} {\bibfield  {journal} {\bibinfo  {journal} {Progress of Theoretical Physics}\ }\textbf {\bibinfo {volume} {66}},\ \bibinfo {pages} {1169} (\bibinfo {year} {1981})},\ \Eprint {https://arxiv.org/abs/https://academic.oup.com/ptp/article-pdf/66/4/1169/5265369/66-4-1169.pdf} {https://academic.oup.com/ptp/article-pdf/66/4/1169/5265369/66-4-1169.pdf} \BibitemShut {NoStop}%
\bibitem [{\citenamefont {Ozeki}\ and\ \citenamefont {Nishimori}(1993)}]{Ozeki_1993}%
  \BibitemOpen
  \bibfield  {author} {\bibinfo {author} {\bibfnamefont {Y.}~\bibnamefont {Ozeki}}\ and\ \bibinfo {author} {\bibfnamefont {H.}~\bibnamefont {Nishimori}},\ }\bibfield  {title} {\bibinfo {title} {Phase diagram of gauge glasses},\ }\href {https://doi.org/10.1088/0305-4470/26/14/009} {\bibfield  {journal} {\bibinfo  {journal} {Journal of Physics A: Mathematical and General}\ }\textbf {\bibinfo {volume} {26}},\ \bibinfo {pages} {3399} (\bibinfo {year} {1993})}\BibitemShut {NoStop}%
\bibitem [{\citenamefont {Le~Doussal}\ and\ \citenamefont {Harris}(1988)}]{doussal_88}%
  \BibitemOpen
  \bibfield  {author} {\bibinfo {author} {\bibfnamefont {P.}~\bibnamefont {Le~Doussal}}\ and\ \bibinfo {author} {\bibfnamefont {A.~B.}\ \bibnamefont {Harris}},\ }\bibfield  {title} {\bibinfo {title} {Location of the ising spin-glass multicritical point on nishimori's line},\ }\href {https://doi.org/10.1103/PhysRevLett.61.625} {\bibfield  {journal} {\bibinfo  {journal} {Phys. Rev. Lett.}\ }\textbf {\bibinfo {volume} {61}},\ \bibinfo {pages} {625} (\bibinfo {year} {1988})}\BibitemShut {NoStop}%
\bibitem [{\citenamefont {Honecker}\ \emph {et~al.}(2001)\citenamefont {Honecker}, \citenamefont {Picco},\ and\ \citenamefont {Pujol}}]{Honecker_2001}%
  \BibitemOpen
  \bibfield  {author} {\bibinfo {author} {\bibfnamefont {A.}~\bibnamefont {Honecker}}, \bibinfo {author} {\bibfnamefont {M.}~\bibnamefont {Picco}},\ and\ \bibinfo {author} {\bibfnamefont {P.}~\bibnamefont {Pujol}},\ }\bibfield  {title} {\bibinfo {title} {Universality class of the nishimori point in the 2d<mml:math xmlns:mml="http://www.w3.org/1998/math/mathml" display="inline"><mml:mo>±</mml:mo><mml:mi mathvariant="italic">j</mml:mi></mml:math>random-bond ising model},\ }\bibfield  {journal} {\bibinfo  {journal} {Physical Review Letters}\ }\textbf {\bibinfo {volume} {87}},\ \href {https://doi.org/10.1103/physrevlett.87.047201} {10.1103/physrevlett.87.047201} (\bibinfo {year} {2001})\BibitemShut {NoStop}%
\bibitem [{\citenamefont {Gruzberg}\ \emph {et~al.}(2001)\citenamefont {Gruzberg}, \citenamefont {Read},\ and\ \citenamefont {Ludwig}}]{Gruzberg_2001}%
  \BibitemOpen
  \bibfield  {author} {\bibinfo {author} {\bibfnamefont {I.~A.}\ \bibnamefont {Gruzberg}}, \bibinfo {author} {\bibfnamefont {N.}~\bibnamefont {Read}},\ and\ \bibinfo {author} {\bibfnamefont {A.~W.~W.}\ \bibnamefont {Ludwig}},\ }\bibfield  {title} {\bibinfo {title} {Random-bond ising model in two dimensions: The nishimori line and supersymmetry},\ }\bibfield  {journal} {\bibinfo  {journal} {Physical Review B}\ }\textbf {\bibinfo {volume} {63}},\ \href {https://doi.org/10.1103/physrevb.63.104422} {10.1103/physrevb.63.104422} (\bibinfo {year} {2001})\BibitemShut {NoStop}%
\bibitem [{\citenamefont {de~Queiroz}(2006)}]{deQueiroz06}%
  \BibitemOpen
  \bibfield  {author} {\bibinfo {author} {\bibfnamefont {S.~L.~A.}\ \bibnamefont {de~Queiroz}},\ }\bibfield  {title} {\bibinfo {title} {Multicritical point of ising spin glasses on triangular and honeycomb lattices},\ }\href {https://doi.org/10.1103/PhysRevB.73.064410} {\bibfield  {journal} {\bibinfo  {journal} {Phys. Rev. B}\ }\textbf {\bibinfo {volume} {73}},\ \bibinfo {pages} {064410} (\bibinfo {year} {2006})}\BibitemShut {NoStop}%
\bibitem [{\citenamefont {Zhang}\ \emph {et~al.}(2024{\natexlab{b}})\citenamefont {Zhang}, \citenamefont {Qi},\ and\ \citenamefont {Bi}}]{zhang2024strangecorrelationfunctionaverage}%
  \BibitemOpen
  \bibfield  {author} {\bibinfo {author} {\bibfnamefont {J.-H.}\ \bibnamefont {Zhang}}, \bibinfo {author} {\bibfnamefont {Y.}~\bibnamefont {Qi}},\ and\ \bibinfo {author} {\bibfnamefont {Z.}~\bibnamefont {Bi}},\ }\href {https://arxiv.org/abs/2210.17485} {\bibinfo {title} {Strange correlation function for average symmetry-protected topological phases}} (\bibinfo {year} {2024}{\natexlab{b}}),\ \Eprint {https://arxiv.org/abs/2210.17485} {arXiv:2210.17485 [cond-mat.str-el]} \BibitemShut {NoStop}%
\bibitem [{\citenamefont {Shi}\ \emph {et~al.}(2020)\citenamefont {Shi}, \citenamefont {Kato},\ and\ \citenamefont {Kim}}]{Shi_2020}%
  \BibitemOpen
  \bibfield  {author} {\bibinfo {author} {\bibfnamefont {B.}~\bibnamefont {Shi}}, \bibinfo {author} {\bibfnamefont {K.}~\bibnamefont {Kato}},\ and\ \bibinfo {author} {\bibfnamefont {I.~H.}\ \bibnamefont {Kim}},\ }\bibfield  {title} {\bibinfo {title} {Fusion rules from entanglement},\ }\href {https://doi.org/10.1016/j.aop.2020.168164} {\bibfield  {journal} {\bibinfo  {journal} {Annals of Physics}\ }\textbf {\bibinfo {volume} {418}},\ \bibinfo {pages} {168164} (\bibinfo {year} {2020})}\BibitemShut {NoStop}%
\bibitem [{\citenamefont {Preskill}(2004)}]{Preskill_LN}%
  \BibitemOpen
  \bibfield  {author} {\bibinfo {author} {\bibfnamefont {J.}~\bibnamefont {Preskill}},\ }\href {http://theory.caltech.edu/~preskill/ph219/topological.pdf} {\bibinfo {title} {Chapter 9. topological quantum computation}} (\bibinfo {year} {2004})\BibitemShut {NoStop}%
\bibitem [{\citenamefont {Duminil-Copin}\ \emph {et~al.}(2020)\citenamefont {Duminil-Copin}, \citenamefont {Glazman}, \citenamefont {Peled},\ and\ \citenamefont {Spinka}}]{duminilcopin2020macroscopicloopsloopon}%
  \BibitemOpen
  \bibfield  {author} {\bibinfo {author} {\bibfnamefont {H.}~\bibnamefont {Duminil-Copin}}, \bibinfo {author} {\bibfnamefont {A.}~\bibnamefont {Glazman}}, \bibinfo {author} {\bibfnamefont {R.}~\bibnamefont {Peled}},\ and\ \bibinfo {author} {\bibfnamefont {Y.}~\bibnamefont {Spinka}},\ }\href {https://arxiv.org/abs/1707.09335} {\bibinfo {title} {Macroscopic loops in the loop $o(n)$ model at nienhuis' critical point}} (\bibinfo {year} {2020}),\ \Eprint {https://arxiv.org/abs/1707.09335} {arXiv:1707.09335 [math.PR]} \BibitemShut {NoStop}%
\bibitem [{\citenamefont {Chayes}\ \emph {et~al.}(2000)\citenamefont {Chayes}, \citenamefont {Pryadko},\ and\ \citenamefont {Shtengel}}]{Chayes_2000}%
  \BibitemOpen
  \bibfield  {author} {\bibinfo {author} {\bibfnamefont {L.}~\bibnamefont {Chayes}}, \bibinfo {author} {\bibfnamefont {L.~P.}\ \bibnamefont {Pryadko}},\ and\ \bibinfo {author} {\bibfnamefont {K.}~\bibnamefont {Shtengel}},\ }\bibfield  {title} {\bibinfo {title} {Intersecting loop models on $z^d$: rigorous results},\ }\href {https://doi.org/10.1016/s0550-3213(99)00780-4} {\bibfield  {journal} {\bibinfo  {journal} {Nuclear Physics B}\ }\textbf {\bibinfo {volume} {570}},\ \bibinfo {pages} {590–614} (\bibinfo {year} {2000})}\BibitemShut {NoStop}%
\bibitem [{\citenamefont {Guo}\ \emph {et~al.}(2000)\citenamefont {Guo}, \citenamefont {Bl\"ote},\ and\ \citenamefont {Wu}}]{Guo_2000}%
  \BibitemOpen
  \bibfield  {author} {\bibinfo {author} {\bibfnamefont {W.}~\bibnamefont {Guo}}, \bibinfo {author} {\bibfnamefont {H.~W.~J.}\ \bibnamefont {Bl\"ote}},\ and\ \bibinfo {author} {\bibfnamefont {F.~Y.}\ \bibnamefont {Wu}},\ }\bibfield  {title} {\bibinfo {title} {Phase transition in the $\mathit{n}>2$ honeycomb $o(\mathit{n})$ model},\ }\href {https://doi.org/10.1103/PhysRevLett.85.3874} {\bibfield  {journal} {\bibinfo  {journal} {Phys. Rev. Lett.}\ }\textbf {\bibinfo {volume} {85}},\ \bibinfo {pages} {3874} (\bibinfo {year} {2000})}\BibitemShut {NoStop}%
\bibitem [{\citenamefont {Kubica}\ \emph {et~al.}(2018)\citenamefont {Kubica}, \citenamefont {Beverland}, \citenamefont {Brandão}, \citenamefont {Preskill},\ and\ \citenamefont {Svore}}]{Kubica_2018}%
  \BibitemOpen
  \bibfield  {author} {\bibinfo {author} {\bibfnamefont {A.}~\bibnamefont {Kubica}}, \bibinfo {author} {\bibfnamefont {M.~E.}\ \bibnamefont {Beverland}}, \bibinfo {author} {\bibfnamefont {F.}~\bibnamefont {Brandão}}, \bibinfo {author} {\bibfnamefont {J.}~\bibnamefont {Preskill}},\ and\ \bibinfo {author} {\bibfnamefont {K.~M.}\ \bibnamefont {Svore}},\ }\bibfield  {title} {\bibinfo {title} {Three-dimensional color code thresholds via statistical-mechanical mapping},\ }\bibfield  {journal} {\bibinfo  {journal} {Physical Review Letters}\ }\textbf {\bibinfo {volume} {120}},\ \href {https://doi.org/10.1103/physrevlett.120.180501} {10.1103/physrevlett.120.180501} (\bibinfo {year} {2018})\BibitemShut {NoStop}%
\bibitem [{\citenamefont {Shimada}\ \emph {et~al.}(2014)\citenamefont {Shimada}, \citenamefont {Jacobsen},\ and\ \citenamefont {Kamiya}}]{Shimada_2014}%
  \BibitemOpen
  \bibfield  {author} {\bibinfo {author} {\bibfnamefont {H.}~\bibnamefont {Shimada}}, \bibinfo {author} {\bibfnamefont {J.~L.}\ \bibnamefont {Jacobsen}},\ and\ \bibinfo {author} {\bibfnamefont {Y.}~\bibnamefont {Kamiya}},\ }\bibfield  {title} {\bibinfo {title} {Phase diagram and strong-coupling fixed point in the disordered o(n) loop model},\ }\href {https://doi.org/10.1088/1751-8113/47/12/122001} {\bibfield  {journal} {\bibinfo  {journal} {Journal of Physics A: Mathematical and Theoretical}\ }\textbf {\bibinfo {volume} {47}},\ \bibinfo {pages} {122001} (\bibinfo {year} {2014})}\BibitemShut {NoStop}%
\bibitem [{\citenamefont {Alba}\ \emph {et~al.}(2009)\citenamefont {Alba}, \citenamefont {Pelissetto},\ and\ \citenamefont {Vicari}}]{Alba_2009}%
  \BibitemOpen
  \bibfield  {author} {\bibinfo {author} {\bibfnamefont {V.}~\bibnamefont {Alba}}, \bibinfo {author} {\bibfnamefont {A.}~\bibnamefont {Pelissetto}},\ and\ \bibinfo {author} {\bibfnamefont {E.}~\bibnamefont {Vicari}},\ }\bibfield  {title} {\bibinfo {title} {Quasi-long-range order in the 2d xy model with random phase shifts},\ }\href {https://doi.org/10.1088/1751-8113/42/29/295001} {\bibfield  {journal} {\bibinfo  {journal} {Journal of Physics A: Mathematical and Theoretical}\ }\textbf {\bibinfo {volume} {42}},\ \bibinfo {pages} {295001} (\bibinfo {year} {2009})}\BibitemShut {NoStop}%
\bibitem [{\citenamefont {Witten}(1982)}]{Witten982}%
  \BibitemOpen
  \bibfield  {author} {\bibinfo {author} {\bibfnamefont {E.}~\bibnamefont {Witten}},\ }\bibfield  {title} {\bibinfo {title} {Constraints on supersymmetry breaking},\ }\href {https://doi.org/https://doi.org/10.1016/0550-3213(82)90071-2} {\bibfield  {journal} {\bibinfo  {journal} {Nuclear Physics B}\ }\textbf {\bibinfo {volume} {202}},\ \bibinfo {pages} {253} (\bibinfo {year} {1982})}\BibitemShut {NoStop}%
\bibitem [{\citenamefont {Wouters}\ \emph {et~al.}(2021)\citenamefont {Wouters}, \citenamefont {Katsura},\ and\ \citenamefont {Schuricht}}]{Wouters_2021}%
  \BibitemOpen
  \bibfield  {author} {\bibinfo {author} {\bibfnamefont {J.}~\bibnamefont {Wouters}}, \bibinfo {author} {\bibfnamefont {H.}~\bibnamefont {Katsura}},\ and\ \bibinfo {author} {\bibfnamefont {D.}~\bibnamefont {Schuricht}},\ }\bibfield  {title} {\bibinfo {title} {Interrelations among frustration-free models via witten’s conjugation},\ }\bibfield  {journal} {\bibinfo  {journal} {SciPost Physics Core}\ }\textbf {\bibinfo {volume} {4}},\ \href {https://doi.org/10.21468/scipostphyscore.4.4.027} {10.21468/scipostphyscore.4.4.027} (\bibinfo {year} {2021})\BibitemShut {NoStop}%
\bibitem [{\citenamefont {Burnell}(2018)}]{Burnell_2018}%
  \BibitemOpen
  \bibfield  {author} {\bibinfo {author} {\bibfnamefont {F.}~\bibnamefont {Burnell}},\ }\bibfield  {title} {\bibinfo {title} {Anyon condensation and its applications},\ }\href {https://doi.org/10.1146/annurev-conmatphys-033117-054154} {\bibfield  {journal} {\bibinfo  {journal} {Annual Review of Condensed Matter Physics}\ }\textbf {\bibinfo {volume} {9}},\ \bibinfo {pages} {307–327} (\bibinfo {year} {2018})}\BibitemShut {NoStop}%
\bibitem [{\citenamefont {Chaikin}\ and\ \citenamefont {Lubensky}(1995)}]{chaikin_lubensky_1995}%
  \BibitemOpen
  \bibfield  {author} {\bibinfo {author} {\bibfnamefont {P.~M.}\ \bibnamefont {Chaikin}}\ and\ \bibinfo {author} {\bibfnamefont {T.~C.}\ \bibnamefont {Lubensky}},\ }\href {https://doi.org/10.1017/CBO9780511813467} {\emph {\bibinfo {title} {Principles of Condensed Matter Physics}}}\ (\bibinfo  {publisher} {Cambridge University Press},\ \bibinfo {year} {1995})\BibitemShut {NoStop}%
\bibitem [{\citenamefont {Halperin}(1981)}]{dasgupta1981phase}%
  \BibitemOpen
  \bibfield  {author} {\bibinfo {author} {\bibfnamefont {B.}~\bibnamefont {Halperin}},\ }\bibfield  {title} {\bibinfo {title} {Phase transition in a lattice model of superconductivity},\ }\href@noop {} {\bibfield  {journal} {\bibinfo  {journal} {Physical Review Letters}\ }\textbf {\bibinfo {volume} {47}},\ \bibinfo {pages} {1556} (\bibinfo {year} {1981})}\BibitemShut {NoStop}%
\bibitem [{\citenamefont {Fisher}\ and\ \citenamefont {Glazman}(1996)}]{fisher1996transportonedimensionalluttingerliquid}%
  \BibitemOpen
  \bibfield  {author} {\bibinfo {author} {\bibfnamefont {M.~P.~A.}\ \bibnamefont {Fisher}}\ and\ \bibinfo {author} {\bibfnamefont {L.~I.}\ \bibnamefont {Glazman}},\ }\href {https://arxiv.org/abs/cond-mat/9610037} {\bibinfo {title} {Transport in a one-dimensional luttinger liquid}} (\bibinfo {year} {1996}),\ \Eprint {https://arxiv.org/abs/cond-mat/9610037} {arXiv:cond-mat/9610037 [cond-mat.mes-hall]} \BibitemShut {NoStop}%
\bibitem [{\citenamefont {Nienhuis}(1984)}]{Nienhuis_84}%
  \BibitemOpen
  \bibfield  {author} {\bibinfo {author} {\bibfnamefont {B.}~\bibnamefont {Nienhuis}},\ }\bibfield  {title} {\bibinfo {title} {Critical behavior of two-dimensional spin models and charge asymmetry in the coulomb gas},\ }\href {https://doi.org/10.1007/BF01009437} {\bibfield  {journal} {\bibinfo  {journal} {Journal of Statistical Physics}\ }\textbf {\bibinfo {volume} {34}},\ \bibinfo {pages} {731 } (\bibinfo {year} {1984})}\BibitemShut {NoStop}%
\bibitem [{\citenamefont {Domb}\ and\ \citenamefont {Lebowitz}(2000)}]{domb2000phase}%
  \BibitemOpen
  \bibfield  {author} {\bibinfo {author} {\bibfnamefont {C.}~\bibnamefont {Domb}}\ and\ \bibinfo {author} {\bibfnamefont {J.}~\bibnamefont {Lebowitz}},\ }\href {https://books.google.es/books?id=FYvFm4xF7CgC} {\emph {\bibinfo {title} {Phase Transitions and Critical Phenomena}}},\ \bibinfo {series} {Phase Transitions and Critical Phenomena}\ No.\ \bibinfo {number} {v. 18}\ (\bibinfo  {publisher} {Elsevier Science},\ \bibinfo {year} {2000})\BibitemShut {NoStop}%
\bibitem [{\citenamefont {Affleck}(1991)}]{Affleck_O2_K2}%
  \BibitemOpen
  \bibfield  {author} {\bibinfo {author} {\bibfnamefont {I.}~\bibnamefont {Affleck}},\ }\bibfield  {title} {\bibinfo {title} {Nonlinear \ensuremath{\sigma} model at \ensuremath{\theta}=\ensuremath{\pi}: Euclidean lattice formulation and solid-on-solid models},\ }\href {https://doi.org/10.1103/PhysRevLett.66.2429} {\bibfield  {journal} {\bibinfo  {journal} {Phys. Rev. Lett.}\ }\textbf {\bibinfo {volume} {66}},\ \bibinfo {pages} {2429} (\bibinfo {year} {1991})}\BibitemShut {NoStop}%
\bibitem [{\citenamefont {Kraus}\ \emph {et~al.}(2008)\citenamefont {Kraus}, \citenamefont {B\"uchler}, \citenamefont {Diehl}, \citenamefont {Kantian}, \citenamefont {Micheli},\ and\ \citenamefont {Zoller}}]{2008_zoller_Lindblad_pre_state}%
  \BibitemOpen
  \bibfield  {author} {\bibinfo {author} {\bibfnamefont {B.}~\bibnamefont {Kraus}}, \bibinfo {author} {\bibfnamefont {H.~P.}\ \bibnamefont {B\"uchler}}, \bibinfo {author} {\bibfnamefont {S.}~\bibnamefont {Diehl}}, \bibinfo {author} {\bibfnamefont {A.}~\bibnamefont {Kantian}}, \bibinfo {author} {\bibfnamefont {A.}~\bibnamefont {Micheli}},\ and\ \bibinfo {author} {\bibfnamefont {P.}~\bibnamefont {Zoller}},\ }\bibfield  {title} {\bibinfo {title} {Preparation of entangled states by quantum {Markov} processes},\ }\href {https://doi.org/10.1103/PhysRevA.78.042307} {\bibfield  {journal} {\bibinfo  {journal} {Phys. Rev. A}\ }\textbf {\bibinfo {volume} {78}},\ \bibinfo {pages} {042307} (\bibinfo {year} {2008})}\BibitemShut {NoStop}%
\bibitem [{\citenamefont {Verstraete}\ \emph {et~al.}(2009)\citenamefont {Verstraete}, \citenamefont {Wolf},\ and\ \citenamefont {Ignacio~Cirac}}]{2009_verstraete_quantum_state_engineer}%
  \BibitemOpen
  \bibfield  {author} {\bibinfo {author} {\bibfnamefont {F.}~\bibnamefont {Verstraete}}, \bibinfo {author} {\bibfnamefont {M.~M.}\ \bibnamefont {Wolf}},\ and\ \bibinfo {author} {\bibfnamefont {J.}~\bibnamefont {Ignacio~Cirac}},\ }\bibfield  {title} {\bibinfo {title} {Quantum computation and quantum-state engineering driven by dissipation},\ }\href {https://doi.org/10.1038/nphys1342} {\bibfield  {journal} {\bibinfo  {journal} {Nature Physics}\ }\textbf {\bibinfo {volume} {5}},\ \bibinfo {pages} {633} (\bibinfo {year} {2009})}\BibitemShut {NoStop}%
\bibitem [{\citenamefont {Buča}\ \emph {et~al.}(2019)\citenamefont {Buča}, \citenamefont {Tindall},\ and\ \citenamefont {Jaksch}}]{2019_buca_jaksch_non_statioanry_dissipation}%
  \BibitemOpen
  \bibfield  {author} {\bibinfo {author} {\bibfnamefont {B.}~\bibnamefont {Buča}}, \bibinfo {author} {\bibfnamefont {J.}~\bibnamefont {Tindall}},\ and\ \bibinfo {author} {\bibfnamefont {D.}~\bibnamefont {Jaksch}},\ }\bibfield  {title} {\bibinfo {title} {Non-stationary coherent quantum many-body dynamics through dissipation},\ }\href {https://doi.org/10.1038/s41467-019-09757-y} {\bibfield  {journal} {\bibinfo  {journal} {Nature Communications}\ }\textbf {\bibinfo {volume} {10}},\ \bibinfo {pages} {1730} (\bibinfo {year} {2019})}\BibitemShut {NoStop}%
\bibitem [{\citenamefont {Dubois}\ \emph {et~al.}(2023)\citenamefont {Dubois}, \citenamefont {Saalmann},\ and\ \citenamefont {Rost}}]{2023_symmetry_induced_DFS}%
  \BibitemOpen
  \bibfield  {author} {\bibinfo {author} {\bibfnamefont {J.}~\bibnamefont {Dubois}}, \bibinfo {author} {\bibfnamefont {U.}~\bibnamefont {Saalmann}},\ and\ \bibinfo {author} {\bibfnamefont {J.~M.}\ \bibnamefont {Rost}},\ }\bibfield  {title} {\bibinfo {title} {Symmetry-induced decoherence-free subspaces},\ }\href {https://doi.org/10.1103/PhysRevResearch.5.L012003} {\bibfield  {journal} {\bibinfo  {journal} {Phys. Rev. Res.}\ }\textbf {\bibinfo {volume} {5}},\ \bibinfo {pages} {L012003} (\bibinfo {year} {2023})}\BibitemShut {NoStop}%
\bibitem [{\citenamefont {Mi}\ \emph {et~al.}(2023)\citenamefont {Mi} \emph {et~al.}}]{2023_google_engineered_dissipation}%
  \BibitemOpen
  \bibfield  {author} {\bibinfo {author} {\bibfnamefont {X.}~\bibnamefont {Mi}} \emph {et~al.},\ }\href@noop {} {\bibinfo {title} {Stable quantum-correlated many body states via engineered dissipation}} (\bibinfo {year} {2023}),\ \Eprint {https://arxiv.org/abs/2304.13878} {arXiv:2304.13878} \BibitemShut {NoStop}%
\bibitem [{\citenamefont {Li}\ \emph {et~al.}(2023)\citenamefont {Li}, \citenamefont {Sala},\ and\ \citenamefont {Pollmann}}]{li_2023}%
  \BibitemOpen
  \bibfield  {author} {\bibinfo {author} {\bibfnamefont {Y.}~\bibnamefont {Li}}, \bibinfo {author} {\bibfnamefont {P.}~\bibnamefont {Sala}},\ and\ \bibinfo {author} {\bibfnamefont {F.}~\bibnamefont {Pollmann}},\ }\bibfield  {title} {\bibinfo {title} {Hilbert space fragmentation in open quantum systems},\ }\href {https://doi.org/10.1103/PhysRevResearch.5.043239} {\bibfield  {journal} {\bibinfo  {journal} {Phys. Rev. Res.}\ }\textbf {\bibinfo {volume} {5}},\ \bibinfo {pages} {043239} (\bibinfo {year} {2023})}\BibitemShut {NoStop}%
\bibitem [{\citenamefont {Li}\ \emph {et~al.}(2024{\natexlab{b}})\citenamefont {Li}, \citenamefont {Pollmann}, \citenamefont {Read},\ and\ \citenamefont {Sala}}]{li2024highlyentangledstationarystatesstrong}%
  \BibitemOpen
  \bibfield  {author} {\bibinfo {author} {\bibfnamefont {Y.}~\bibnamefont {Li}}, \bibinfo {author} {\bibfnamefont {F.}~\bibnamefont {Pollmann}}, \bibinfo {author} {\bibfnamefont {N.}~\bibnamefont {Read}},\ and\ \bibinfo {author} {\bibfnamefont {P.}~\bibnamefont {Sala}},\ }\href {https://arxiv.org/abs/2406.08567} {\bibinfo {title} {Highly-entangled stationary states from strong symmetries}} (\bibinfo {year} {2024}{\natexlab{b}}),\ \Eprint {https://arxiv.org/abs/2406.08567} {arXiv:2406.08567 [quant-ph]} \BibitemShut {NoStop}%
\bibitem [{\citenamefont {Wegner}(1971)}]{wegner1971duality}%
  \BibitemOpen
  \bibfield  {author} {\bibinfo {author} {\bibfnamefont {F.~J.}\ \bibnamefont {Wegner}},\ }\bibfield  {title} {\bibinfo {title} {Duality in generalized ising models and phase transitions without local order parameters},\ }\href@noop {} {\bibfield  {journal} {\bibinfo  {journal} {Journal of Mathematical Physics}\ }\textbf {\bibinfo {volume} {12}},\ \bibinfo {pages} {2259} (\bibinfo {year} {1971})}\BibitemShut {NoStop}%
\bibitem [{\citenamefont {Ginsparg}(1988)}]{ginsparg1988appliedconformalfieldtheory}%
  \BibitemOpen
  \bibfield  {author} {\bibinfo {author} {\bibfnamefont {P.}~\bibnamefont {Ginsparg}},\ }\href {https://arxiv.org/abs/hep-th/9108028} {\bibinfo {title} {Applied conformal field theory}} (\bibinfo {year} {1988}),\ \Eprint {https://arxiv.org/abs/hep-th/9108028} {arXiv:hep-th/9108028 [hep-th]} \BibitemShut {NoStop}%
\bibitem [{\citenamefont {Francesco}\ \emph {et~al.}(1997)\citenamefont {Francesco}, \citenamefont {Mathieu},\ and\ \citenamefont {S{\'e}n{\'e}chal}}]{francesco1997conformal}%
  \BibitemOpen
  \bibfield  {author} {\bibinfo {author} {\bibfnamefont {P.}~\bibnamefont {Francesco}}, \bibinfo {author} {\bibfnamefont {P.}~\bibnamefont {Mathieu}},\ and\ \bibinfo {author} {\bibfnamefont {D.}~\bibnamefont {S{\'e}n{\'e}chal}},\ }\href {https://books.google.fr/books?id=keUrdME5rhIC} {\emph {\bibinfo {title} {Conformal Field Theory}}},\ Graduate Texts in Contemporary Physics\ (\bibinfo  {publisher} {Springer},\ \bibinfo {year} {1997})\BibitemShut {NoStop}%
\end{thebibliography}

%

\appendix

\onecolumngrid

\section{Pure wavefunction deformations and ground-state phase transitions}
\label{app:pure_def_stat}
Let us consider pure wavefunctions deformations of the form 
\begin{equation} \label{eq:unn_psi}
    \ket{\psi(\beta)}= e^{ \frac{\beta}{2} \sum_{j\in {\mathcal{R}}} \hat{O}_j}\ket{\textrm{TO}}
\end{equation}
with $\mathcal{R}$ a subset of the lattice, $\hat{O}_j$ a Pauli or combination of Pauli operators of different strength and $\beta>0$. In the main text, we focus on the norm of $\ket{\psi(\beta)}$ as a function of $\beta$, which can be interpreted as the partition function of a classical stat-mech model 
\begin{equation}
    \mathcal{Z}(\beta)\equiv \avg{\psi(\beta)|\psi(\beta)}.
\end{equation}
One may wonder whether singularities in $\mathcal{Z}(\beta)$, i.e., finite temperature phase transitions in the $(2+0)$-dimensional classical stat-mech model, necessarily correspond to ground-state phase transitions in the corresponding parent Hamiltonians. We claim that these are equivalent as long as the corresponding stat-mech model only involves local interactions as we argue in the following. Let us assume that the partition function can be written as 
\begin{equation}
    \mathcal{Z}(\beta)=\langle \psi(\beta)|\psi(\beta) \rangle = \sum_{\{\sigma\}} e^{-\beta H(\{\sigma\})}
\end{equation}
with $H(\{\sigma\})=\sum_{\bm{r}}h_{\bm{r}}$ a local spin Hamiltonian with $h_{\bm{r}}=-\frac{1}{2}\sum_{\bm{r}} \sigma_{\bm{r}} \sum_{j\in \mathcal{N}_{\bm{r}}} \sigma_j$  in terms of ``Ising''-like variables $\sigma_{\bm{r}}$. Then diagonal correlations (i.e., those only involving $\hat{Z}$ Paulis) are given by thermal correlations evaluated on the corresponding stat-mech model
\begin{equation}
    \langle \psi(\beta)|Z_{\bm{r}} Z_{\bm{r}^\prime}|\psi(\beta) \rangle = \langle \sigma_{\bm{r}} \sigma_{\bm{r}^\prime} \rangle_{\beta}.
\end{equation}
Moreover, off-diagonal correlations can also be computed as a thermal average of \emph{local} observables~\cite{Verstraete_06}
\begin{equation}
\begin{aligned}
    \langle X_{\bm{r}}X_{\bm{r}^\prime}\rangle &= \frac{1}{\mathcal{Z}}\sum_{\{\sigma\}}e^{-\beta H(\{\sigma\})}e^{\beta h_{\bm{r}}}e^{\beta h_{\bm{r}^\prime}} \\ &=\langle e^{\beta h_{\bm{r}}}e^{\beta h_{\bm{r}^\prime}} \rangle_{\beta}.
\end{aligned}
\end{equation}
and
\begin{equation}
\begin{aligned}
    \langle Y_{\bm{r}}Y_{\bm{r}^\prime}\rangle &= \frac{1}{\mathcal{Z}}\sum_{\{\sigma\}}  e^{-\beta H(\{\sigma\})}\left(i\sigma_{\bm{r}}e^{\beta h_{\bm{r}}}\right) \left(i\sigma_{\bm{r}^\prime}e^{\beta h_{\bm{r}^\prime}}\right)\\ &=-\langle \sigma_{\bm{r}}e^{\beta h_{\bm{r}}}\sigma_{\bm{r}^\prime}e^{\beta h_{\bm{r}^\prime}} \rangle_{\beta} .
\end{aligned}
\end{equation}

On the one hand, we recall that a ground state phase transition corresponds to closing the gap, and this affects the scaling of certain correlation functions, e.g., modifying their scaling from exponential, to constant or to power-law decay. On the other, it is clear that phase transitions in $\mathcal{Z}(\beta)$ (as a function of $\beta$), are captured by the scaling of diagonal correlations, but as we showed before the stat-mech model can also be used to evaluated off-diagonal ones!

\section{(Un)gauging and (dis)entangling mappings}
\label{app:mappings}

The ground state of the $D_4$ topological order model in Eq.~\eqref{eq:H_D4} is specified by the following stabilizers (including mirrored version) equaling unity

\begin{center}
\includegraphics[scale=0.45]{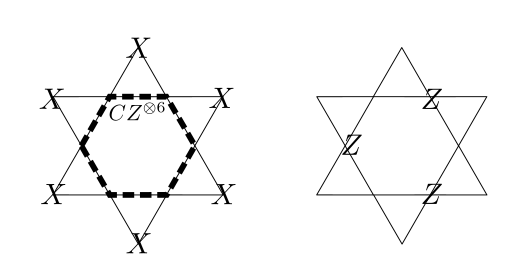}
\end{center}

\begin{figure}
    \centering
    \includegraphics[scale=0.35]{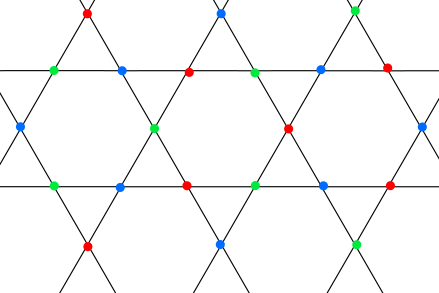}
    \caption{\textbf{kagome lattice.} Three color sublattices red, green and blue.}
    \label{fig:Kagome}
\end{figure}

It will be useful to keep track of the three $\R{\mathcal{R}_R}$, $\G{\mathcal{R}_G}$ and $\B{\mathcal{R}_B}$ sublattices of the model (on the kagome lattice) as shown in Fig.~\ref{fig:Kagome}. A key property we use is that if e.g., $X$ only appears on red, then blue and green have conserved $B_t=Z^{\otimes 3}$, which we can use to dualize (i.e., gauge the 1-form, or ungauge the $\mathbb Z_2$ symmetry).    

\subsection{`Ungauging' the green sublattice}

Doing this first for green: we use \color{DarkGreen}$Z^{\otimes 3}=1$ \color{black} to map the problem to the following lattice
\begin{center}
\includegraphics[scale=0.35]{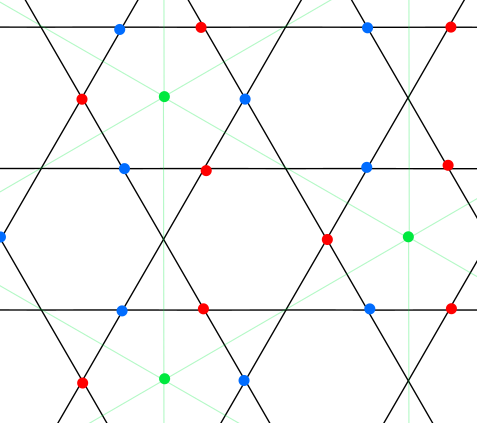}
\end{center}
where a green $Z$ on the original lattice maps to $ZZ$ on the new lattice. Now our state has mapped to $\ket{D_4}\to   \ket{R\tilde GB}.$
We now work out the structure of $\ket{R \tilde G B}$ by specifying its stabilizers on this new lattice.

The stabilizers for this state are as follows: the simplest ones are the $Z^{\otimes 3}$ stabilizers on the red and blue sublattices, which have not changed at all. Secondly, the 12-body stabilizer whose $X$'s were living on the green sublattice now map to the following:
\begin{center}
\includegraphics[scale=0.3]{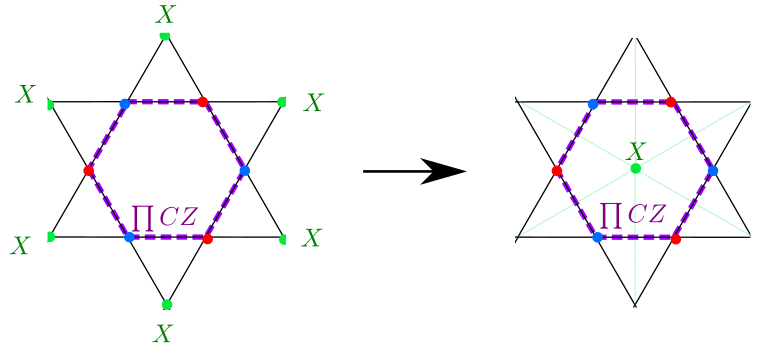}
\end{center}
Finally, the other 12-body stabilizers (i.e., where the $X$'s are on blue or red) map as follows:
\begin{center}
\includegraphics[scale=0.5]{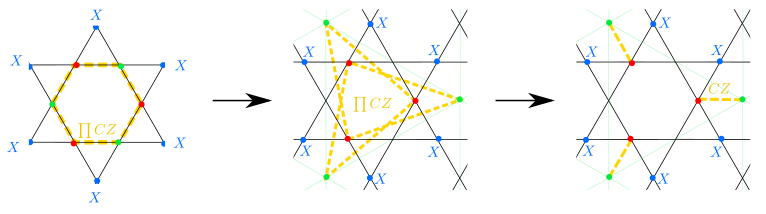}
\end{center}
where in the last step we used the \color{red}$Z^{\otimes 3} = 1$ \color{black} stabilizer to simplify the $CZ$'s. A similar stabilizer of course holds upon replacing blue and red.

\subsection{`Ungauging' the blue sublattice}

We now use the $Z^{\otimes 3}$ stabilizer on the blue sublattice to similarly map to the following lattice, where a single blue $Z$ maps to two $ZZ$'s on the new blue sublattice:
\begin{center}
\includegraphics[scale=0.5]{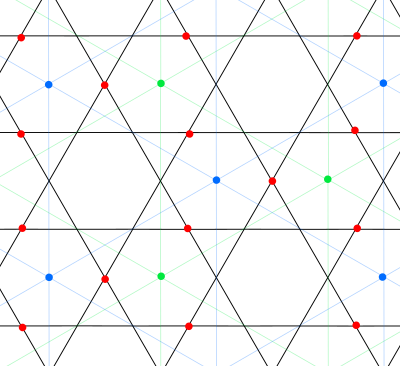}
\end{center}
Since this is starting to get a bit messy, note that we can also draw these points as lying on the following Lieb (`heavy hex') honeycomb lattice:
\begin{center}
\includegraphics[scale=0.3]{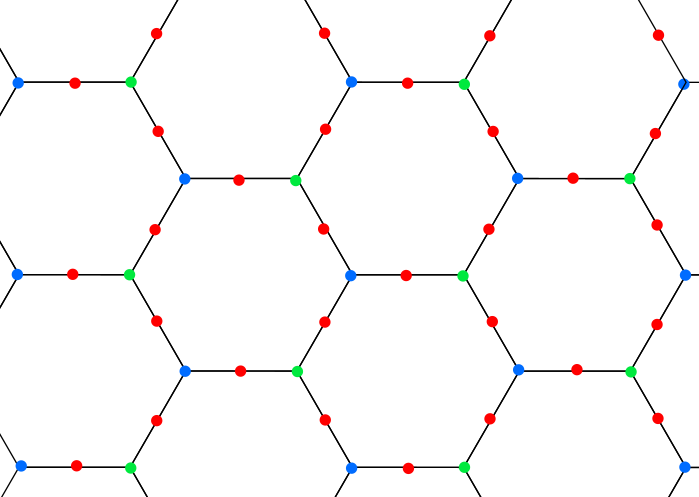}
\end{center}

Our wavefunction on this lattice then becomes $ \ket{R\tilde G \tilde B}$, with stabilizers:
\begin{enumerate}
    \item The red sublattice still has the usual $Z^{\otimes 3}$ stabilizer.
    \item For every blue vertex we have: \begin{center}
\includegraphics[scale=0.45]{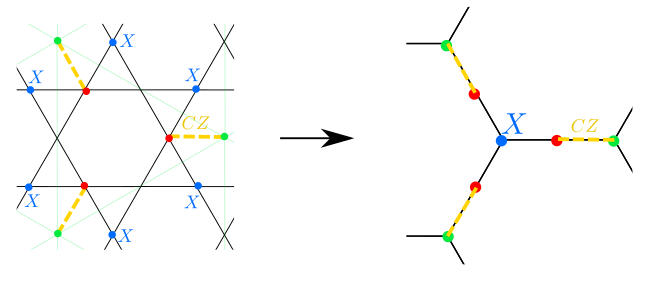}
\end{center}
\item For every green vertex we have:
\begin{center}
\includegraphics[scale=0.5]{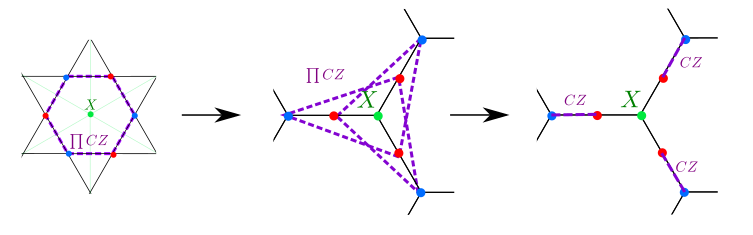}
\end{center}
where the last step again involves using the \color{red}$Z^{\otimes 3}$ \color{black} stabilizer. The resulting stabilizer is equivalent to the blue one (after exchanging blue and green), which confirms that blue and green are playing equivalent roles in this formulation.
\item Lastly, there is an $X$-type stabilizer associated to the red sublattice:
\begin{center}
\includegraphics[scale=0.5]{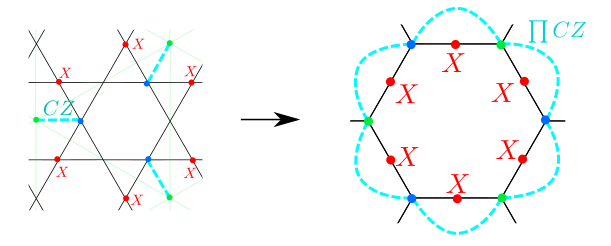}
\end{center}
    
\end{enumerate}

\subsection{(Dis)entangling}
\label{sec:disent}
Thus far, we got  $\ket{R\tilde G \tilde B}$ on a honeycomb Lieb lattice (i.e., vertices and bonds of the honeycomb lattice), where $\ket{R \tilde G \tilde B}$ is specified by the above list of stabilizers.

We now further simplify this state by acting with a (dis)entangler $U$ consisting of a three-body $CCZ$ gate on every bond. Note that every bond has three qubits: blue and green vertex qubits, and a red bond qubit. We can thus write
\begin{equation}
U = \prod_{\langle \color{blue} v \color{black} , \color{DarkGreen} v' \color{black} \rangle} \color{blue} C \color{red} C \color{DarkGreen} Z_{\color{blue} v, \color{red} \langle v,v'\rangle, \color{DarkGreen} v'}.
\end{equation}
The action of this unitary on our state is quite simple, by just repeatedly using the fact that conjugating $X_i$ (or $X_j$) by $CCZ_{i,j,k}$ gives $X_i CZ_{j,k}$ (or $X_j CZ_{i,k}$). Thus:
\begin{equation}
U \ket{\psi(\beta)} \propto e^{\beta \sum_{r \in R} X_r CZ_{b,g}}  \left( U \ket{R\tilde G \tilde B}\right), \label{eq:temp}
\end{equation}
where $X_r CZ_{b,g}$ is to be understood as a three-body operator on a single bond, acting with $X$ on the red bond qubit and acting with $CZ$ on the blue and green vertices of that bond, and where $U \ket{R\tilde G \tilde B}$ is defined by the following list of stabilizers:
\begin{enumerate}
    \item The red sublattice still has the usual $Z^{\otimes 3}$ stabilizer.
    \item For every blue and green vertex we just have the stablizers \color{blue} $X_b = 1$ \color{black} and \color{DarkGreen} $X_g = 1$ \color{black}. Explicitly for the green sublattice:
    \begin{center}
\includegraphics[scale=0.45]{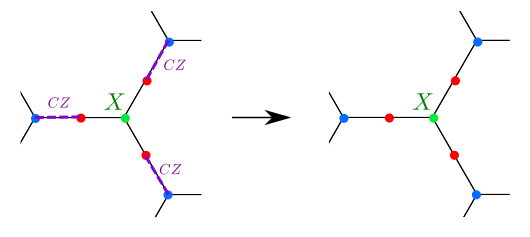}
\end{center}
    \item Lastly, the red $X$-stabilizer also dis-entangles:
    \begin{center}
\includegraphics[scale=0.45]{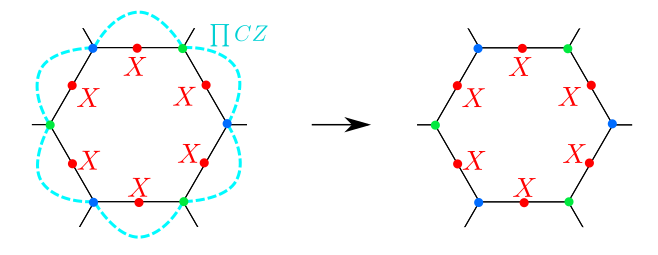}
\end{center}
\end{enumerate}

In conclusion, at this point, $U \ket{R\tilde G \tilde B}$ consists of \emph{trivial product states} on the blue and green sublattices, and a usual toric code state on the red sublattice.

\subsection{Dualizing and then `ungauging' the red sublattice} \label{app:dual}
Since our imaginary time-evolution does not commute with the red $Z^{\otimes 3}$ stabilizer of our initial state, we cannot use the latter to ungauge the model, unlike what we did for the green and blue sublattices. However, we can do so if we first perform an $e-m$ duality, i.e., using the $X$- instead of $Z$-stabilizer. In particular, note that the red six-body $X^{\otimes 6}$ stabilizer \emph{does} commute with our imaginary time-evolution. This means we can perform a Kramers-Wannier transformation such that the red qubits now live on the sites of the honeycomb lattice. For convenience, we choose the convention that a bond \color{red}$X$ \color{black} qubit maps to a \color{red} $ZZ$ \color{black} interaction on the new lattice. This maps the toric code on the red sublattice into a $\ket{+}^{\otimes N}$ product state. Note that the effective lattice is a honeycomb lattice with \emph{two} qubits per site! If we denote the Pauli operators as $X,Z$ and $\tilde X, \tilde Z$, then we can write our resulting state as follows:
\begin{equation}
\ket{\psi(\beta)} \propto e^{\beta \sum_{\langle v,v'\rangle } \tilde Z_v \tilde Z_{v'} CZ_{v,v'}} |+\rangle^{\otimes N} \otimes |\tilde +\rangle^{\otimes N}
\end{equation}
We can think of this as a \emph{honeycomb bilayer}, where the coupling consists of $\tilde Z \tilde Z$ on the bond of one layer coupled to $CZ$ on the bond of the other layer. 

\begin{figure}
    \centering
    \includegraphics[width=0.75\linewidth]{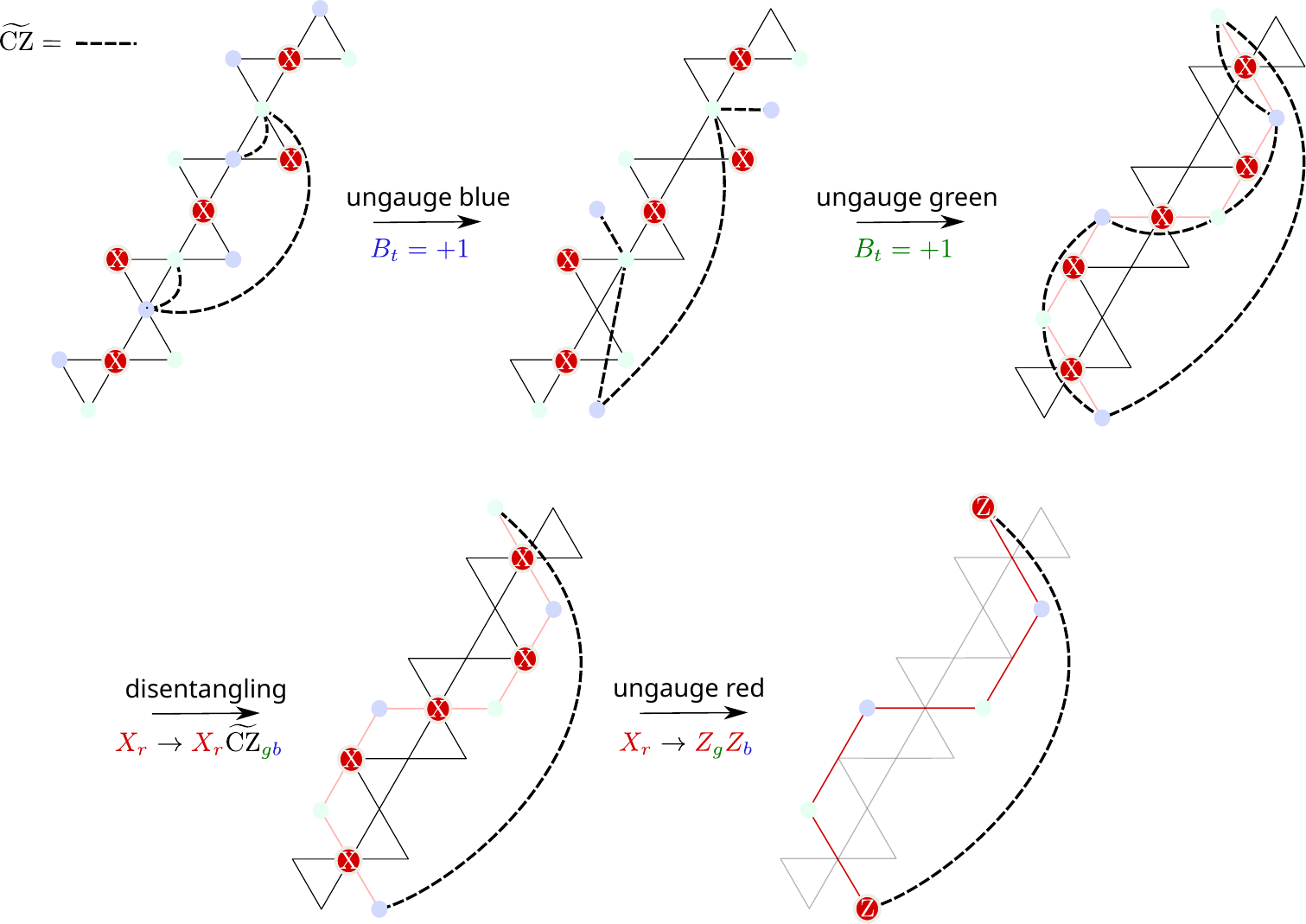}
    \caption{\textbf{Ungauging $\mathcal{\R{X}}_{t_i}^{t_f}$ in Eq.~\eqref{eq:X_if} }}
    \label{fig:ungauging_X}
\end{figure}

The result of applying these mappings to the Wilson operator $\R{X}_{t_i}^{t_f}$ in Eq.~\eqref{eq:X_if} is provided in Fig.~\ref{fig:ungauging_X} in App.~\ref{app:mappings}.

\section{From $D_4$ TO to $O(N)$ loop models: Analytical derivations}
\label{app:Analy_der}

This appendix contains the analytical derivations showing that the action of pure wavefunction deformations and quantum channels acting on the $D_4$ topological ordered ground state of Hamiltonian in Eq.~\eqref{eq:H_D4} with all logical operators $\mathcal{Z}=+1$, can be exactly mapped to $O(N)$ loop models. We show that $N$ is given by the quantum dimension $d_a$ (respectively $d_a^2$) of the anyon the said deformation (quantum channel) can proliferate. We provide additional details to the derivations presented in the main text.

\subsection{Non-Abelian error $X$}
\label{app:X_error}

In this section we compute the expectation value $\langle D_4|\prod_{j\in L_{\R{R}}} X_{j} |D_4\rangle$. To do so, we make use of the ungauging and disentangling maps introduced in the previous Sec.~\ref{app:mappings}, leading to
\begin{equation}
    \langle D_4|\prod_{j\in L_{\R{R}}} X_{j} |D_4\rangle=\langle \tilde{+} |\langle\R{\textrm{TC}} | \prod_{e\in L_{\R{R}}} {\color{red} X}_{e} \widetilde{{\color{blue} C}{\color{DarkGreen} Z}}_{e_+,e_-}| \tilde{+} \rangle | \R{\textrm{T.C.}} \rangle ,
\end{equation}
where in the first equality, $e$ corresponds to a bond of the honeycomb lattice as in Fig.~\ref{fig:D4_TO} with vertex boundaries denoted by $e_{\pm}$, the latter hosting spin-$1/2$'s on the state $\ket{\tilde{+}}$ along $X$, and with $\widetilde{{\color{blue} C}{\color{DarkGreen} Z}}_{i,j}=\frac{1}{2}(\mathds{1}+\tilde{Z}_i + \tilde{Z}_j - \tilde{Z}_i\tilde{Z}_j)$ coupling {\color{blue} B} and {\color{DarkGreen} G} sublattices. Notice that this operator is left invariant under the exchange $\tilde{Z}_i \leftrightarrow \tilde{Z}_j$. Using
\begin{equation} \label{eq:W_R_exp}
    \langle \tilde{+} |\langle\R{\textrm{T.C.}} | \prod_{e\in L_{\R{R}}} {\color{red}X}_{e} \widetilde{{\color{blue} C}{\color{DarkGreen} Z}}_{e_+,e_-}| \tilde{+} \rangle |\R{\textrm{T.C.}} \rangle=  \langle \R{\textrm{T.C.}} | \prod_{e\in L_{\R{R}}} {\color{red} X}_e| \R{\textrm{T.C.}} \rangle \langle \tilde{+} |\prod_{e\in L_{\R{R}}} \widetilde{{\color{blue} C}{\color{DarkGreen} Z}}_{e_+,e_-}| \tilde{+} \rangle
\end{equation}
we again find that only closed loop configurations $L_{\R{R}}$ (now on the honeycomb lattice) lead to non-vanishing $\langle\R{\textrm{T.C.}} | \prod_{e\in L_{\R{R}}} {\color{red}X}_e| \R{\textrm{T.C.}}\rangle=1$. Everything that remains is to evaluate $W_{L_{\R{R}}}\equiv \langle \tilde{+} |\prod_{e\in L_{\R{R}}} \widetilde{CZ}_{e_+,e_-}| \tilde{+} \rangle$ on such closed loop configurations. In the following we prove that $W_{L_{\R{R}}}=2^{C_{L_{\R{R}}}}/\sqrt{2}^{|L_{\R{R}}|}$, where $C_{L_{\R{R}}}$ corresponds to the number of disconnected components in $L_{\R{R}}$. Let us denote by $\ell\in L_{\R{R}}$ a connected component of $L_{\R{R}}$, and define $CZ_{\ell}=\prod_{e\in \ell} CZ_{e_+,e_-}$. Then one finds
\begin{equation} \label{eq:W_gammaR}
    W_{L_{\R{R}}}=\prod_{\ell\in L_{\R{R}}}  \frac{1}{2^{|\ell|}}\textrm{tr}\left( CZ_{\ell}\right).
\end{equation}
Hence we only need to evaluate $\textrm{tr}\left( CZ_{\ell}\right)$ on every connected component $\ell$. First of all, being $\CZ$ a diagonal (in the $Z$ basis) $2$-body gate it can be written as
\begin{equation} 
\vcenter{\hbox{\begin{tikzpicture}
      \draw[-] (0,0) -- (0.9,0.);
      \draw[-] (0,0) -- (0,0.4);
      \draw[-] (.9,0) -- (.9,0.4);
      \draw[-] (0,0.4) -- (0.9,0.4);
      \node at (0.45,0.2) {$\CZ$};
      \draw[-] (0.1,0.0) -- (0.1,-0.1);
      \node at (0.1,-0.3) {$\sigma_i$};
      \draw[-] (0.8,0.0) -- (0.8,-0.1);
      \node at (0.8,-0.3) {$\sigma_j$};
      \draw[-] (0.1,0.4) -- (0.1,0.5);
      \node at (0.1,0.6) {$\sigma_i$};
      \draw[-] (0.8,0.4) -- (0.8,0.5);
      \node at (0.8,.6) {$\sigma_j$};

      \node at (1.4, 0.2) {$=\sqrt{2}$};

      \draw[-] (2.,0.5) -- (2.,-0.1);
      \draw[-] (2.9,0.5) -- (2.9,-0.1);
      \draw[-] (2.,0.2) -- (2.15,0.2);
      \draw[-] (2.65,0.2) -- (2.9,0.2);
      \node at (2.,-0.3) {$\sigma_i$};
      \node at (2.,0.6) {$\sigma_i$};
      \node at (2.9,-0.3) {$\sigma_j$};
    \node at (2.9,.6) {$\sigma_j$};
    \draw[opacity=0.5] (2.4,0.2) circle (0.25);
    \node at (2.4,0.2) {$H$};
\end{tikzpicture} }}
\end{equation}
since when $\sigma_i=\sigma_j=-1$ the result is the same configurations multiply by an overall $-1$ phase, and by $+1$ in any other configuration. Here the symbol
\begin{equation}
\vcenter{\hbox{\begin{tikzpicture}
      \draw[-] (2.,0.5) -- (2.,-0.1);
      \draw[-] (2.,0.2) -- (2.4,0.2);
      \node at (2.,0.6) {$\sigma_i$};
      \node at (2.6,0.2) {$\sigma_j$};
      \node at (2.,-0.3) {$\sigma_k$};
       \node at (3., 0.2) {$=$};
       \node at (3.7, 0.2) {$\delta_{\sigma_i \sigma_j \sigma_j}$};
\end{tikzpicture} }}
\end{equation}
corresponds to the identity tensor $\delta_{\sigma_i \sigma_j \sigma_j}$ which is non-vanishing and equals $1$ if only if $\sigma_i =\sigma_j =\sigma_j$. For example, multiplying two overlaping $\CZ$ gates $\CZ_{ij}\CZ_{jk}$ corresponds to
\begin{equation} 
\vcenter{\hbox{\begin{tikzpicture}
       \node at (1.5,0.2) {$\sqrt{2}^2$};
      \draw[-] (2.,0.5) -- (2.,-0.1);
      \draw[-] (2.9,0.5) -- (2.9,-0.1);
      \draw[-] (2.,0.2) -- (2.15,0.2);
      \draw[-] (2.65,0.2) -- (2.9,0.2);
    \draw[opacity=0.5] (2.4,0.2) circle (0.25);
    \node at (2.4,0.2) {$H$};

     \draw[-] (2.9,0.5) -- (2.9,-0.1);
      \draw[-] (3.8,0.5) -- (3.8,-0.1);
      \draw[-] (2.9,0.2) -- (3.05,0.2);
      \draw[-] (3.55,0.2) -- (3.8,0.2);
    \draw[opacity=0.5] (3.3,0.2) circle (0.25);
    \node at (3.3,0.2) {$H$};

    \node at (2.,-0.3) {$\sigma_i$};
      \node at (2.9,-0.3) {$\sigma_j$};
      \node at (3.8,-0.3) {$\sigma_k$};
\end{tikzpicture} }}
\end{equation}
Therefore, we just found that  $\textrm{tr}\left( CZ_{\ell_\R{R}}\right)=\sqrt{2}^{|L_{\R{R}}|}\textrm{tr}\left( H^{|\ell_\R{R}|}\right)$ with $H=(X + Z)/\sqrt{2}$ the Hadamard matrix. Since the Hadamard transformation $H$ has eigenvalues $\lambda_{\pm}=\pm 1$, we then find that $\textrm{tr}\left( CZ_{\ell}\right)=2\sqrt{2}^{|\ell_\R{R}|}$ if $\ell_\R{R}$ has even length $|\ell|$, but vanishing otherwise. However, being $\ell_\R{R}$ a closed loop on the honeycomb lattice, its length is always even. Notice that this also implies $(\pm 1)^{|L_{\R{R}}|}=+1$ when summing over closed loop configurations. Hence, 
\begin{equation}
    \begin{aligned}
        W_{L_{\R{R}}}= \prod_{\ell_\R{R}\in L_{\R{R}}}  \frac{1}{2^{|\ell_\R{R}|}}2\sqrt{2}^{|\ell_\R{R}|}= 2^{C_{L_{\R{R}}}}{2^{-\frac{|L_{\R{R}}|}{2}}}.
    \end{aligned}
\end{equation}

An alternative way to obtain the value of $\langle D_4|\prod_{j\in L_{\R{R}}} X_{j} |D_4\rangle$, is by mapping the $D_4$ TO to a $\mathbb{Z}_2^3$ symmetry protected topological (SPT)  phase following Ref.~\cite{Yoshida_2016}. This corresponds to applying the ``ungauging map'' not only to operators commuting with \G{green} and \B{blue} plaquettes terms $B_t=Z^{\otimes 3}$, but also on the \R{red} sublattice, instead of following subsection~\ref{app:dual}. Let us consider a single loop configuration $L_{\R{R}}$. First, notice that the product of $X_j \in \R{\mathcal{R}_R}$ along the boundary of $L_{\R{R}}$ can be written as the product of the $6$-body operators $X^{\otimes 6}$ with $X$ acting on the tips of a star \textrm{\ding{65}} lying on $ \R{\mathcal{R}_R}$
\begin{equation}
    \prod_{j\in L_{\R{R}}} \R{X}_{j} = \prod_{\textrm{\ding{65}} \in \textrm{int}(L_{\R{R}})} \R{X}^{\otimes 6}.
\end{equation}
Then, following Ref.~\cite{Yoshida_2016} we map $\ket{D_4}\to \ket{\text{SPT}}$, and 
\begin{equation}
    \prod_{\textrm{\ding{65}} \in \textrm{int}(L_{\R{R}})}\R{X}^{\otimes 6} \longrightarrow  \prod_{\textrm{\ding{65}} \in \textrm{int}(L_{\R{R}})} \R{X}_{\textrm{\ding{65}}}
\end{equation}
with $X_{\textrm{\ding{65}}}$ lying at the center of ${\textrm{\ding{65}}}$. In fact, this SPT is defined on a triangular lattice (which is also a trivalent lattice), with the stabilizers on the red sublattice given by
\begin{equation} 
\begin{tikzpicture}
   \newdimen\Rad
   \Rad=1.cm
   \draw[dashed] (0:\Rad) \foreach \x in {60,120,...,360} {  -- (\x:\Rad) };
    \draw[dashed] (1,0) -- (-1,0);
    \draw[dashed] (0.5,-0.867) -- (-0.5,0.867);
    \draw[dashed] (0.5,0.867) -- (-0.5,-0.867);
    
   \node at (0,0) {$\color{red}{X}$};
   \node[cm={cos(60) ,-sin(60) ,sin(60) ,cos(60) ,(0.89,0.55)}]  {$\CZ$};
   \node[cm={cos(60) ,sin(60) ,-sin(60) ,cos(60) ,(0.89,-0.55)}]  {$\CZ$};
   \node[cm={cos(60) ,sin(60) ,-sin(60) ,cos(60) ,(-0.89,0.55)}]  {$\CZ$};
   \node[cm={cos(60) ,-sin(60) ,sin(60) ,cos(60) ,(-0.89,-0.55)}]  {$\CZ$};
   \node at (0., -1.1) {$\CZ$};
   \node at (0., 1.1) {$\CZ$};
   \node at (2.3, 0.) {$\ket{\text{SPT}}=\ket{\text{SPT}}$,};
\end{tikzpicture}
\end{equation}
and hence
\begin{equation} 
\begin{tikzpicture}
   \newdimen\Rad
   \Rad=1.cm
   \draw[dashed] (0:\Rad) \foreach \x in {60,120,...,360} {  -- (\x:\Rad) };
   \node at (0,0) {$\color{red}{X}$};
   \draw[dashed] (1,0) -- (-1,0);
    \draw[dashed] (0.5,-0.867) -- (-0.5,0.867);
    \draw[dashed] (0.5,0.867) -- (-0.5,-0.867);
   \node at (1.9, 0.) {$\ket{\text{SPT}}=$};
   \draw[dashed, cm={1,0,0,1,(4.,0.)}] (0:\Rad) \foreach \x in {60,120,...,360} {  -- (\x:\Rad) };
   \draw[dashed] (5,0) -- (3,0);
    \draw[dashed] (4.5,-0.867) -- (3.5,0.867);
    \draw[dashed] (4.5,0.867) -- (3.5,-0.867);
   \node[cm={cos(60) ,-sin(60) ,sin(60) ,cos(60) ,(4.89,0.55)}]  {$\CZ$};
   \node[cm={cos(60) ,sin(60) ,-sin(60) ,cos(60) ,(4.89,-0.55)}]  {$\CZ$};
   \node[cm={cos(60) ,sin(60) ,-sin(60) ,cos(60) ,(4.-0.89,0.55)}]  {$\CZ$};
   \node[cm={cos(60) ,-sin(60) ,sin(60) ,cos(60) ,(4.-0.89,-0.55)}]  {$\CZ$};
   \node at (4., -1.1) {$\CZ$};
   \node at (4., 1.1) {$\CZ$};
   \node at (5.6, 0.) {$\ket{\text{SPT}}$.};
\end{tikzpicture}
\end{equation}

Then, we find that
\begin{equation}
        \langle D_4|\prod_{j\in L_{\R{R}}} \R{X}_j |D_4\rangle = \langle \text{SPT} |\prod_{ b \in  L_{\R{R}}} \CZ_b |\text{SPT} \rangle
\end{equation}
where we have used the fact that $\CZ^2_b=1$, and with $b \in L_{\R{R}} $ a bond of the honeycomb lattice lying on $L_{\R{R}}$. Finally, the SPT can be mapped to a trivial product state $\ket{+}$ on every site of the triangular lattice by applying a $\textrm{CCZ}$ unitary transformation 
\begin{equation}
        \langle D_4|\prod_{j\in L_{\R{R}}} \R{X}_j |D_4\rangle = \langle + |\prod_{ b \in  L_{\R{R}}} \CZ_b |+ \rangle
\end{equation}
then obtaining the same expression for $W_{L_{\R{R}}}$ as in Eq.~\eqref{eq:W_R_exp}.

\subsection{Combined errors}
\label{app:comb_ZpX}
In this section we combine both Abelian $Z$ and non-Abelian errors $X$ acting on one or several sublattices. As the previous section has made explicit, the difference will come from the evaluation of expectation of the corresponding errors on the $\ket{D_4}$ TO ground state. 

\subsubsection{Abelian plus non-Abelian errors on different sublattices}
\label{app:ab_plus_nonab}
\begin{figure}
    \centering
    \includegraphics[width=0.85\linewidth]{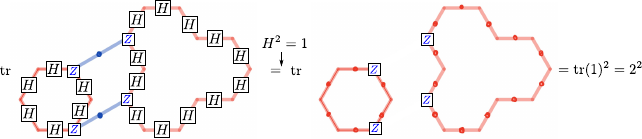}
    \caption{\textbf{Example of the computation of the weight in Eq.~\eqref{eq:Had_weight}.}}
    \label{fig:grap_not}
\end{figure}

 We start by combining an $X$ deformation acting on the red sublattice, and $Z$ acting on blue and green sublattices. If $\beta^z_{\G{G}}=0$ one finds that the norm of the deformed wave function 
 \begin{equation}
     \ket{\psi(\beta^x_{\R{R}}, \beta^z_{\B{B}})}=e^{\frac{ \beta_{\R{R}}^x}{2} \sum_{j\in \R{\mathcal{R}_R}} X_j} e^{\frac{ \beta_B^z}{2} \sum_{j\in \B{\mathcal{R}_B}} Z_j}\ket{D_4},
 \end{equation}
 is given by
 \begin{equation}
     \mathcal{Z}_{\ket{\psi}}(\beta^x_{\R{R}}, \beta^z_{\B{B}})=\mathcal{Z}_{0,R}\mathcal{Z}_{0,B}\sum_{\textrm{closed}\,\,L_{\R{R}}}\sum_{\gamma_{\B{B}}}\tanh(\beta_{\R{R}}^x)^{|L_{\R{R}}|}\tanh(\beta_{\B{B}}^z)^{|\gamma_{\B{B}}|}\langle{D_4|\prod_{j\in L_{\R{R}}}X_j \prod_{i\in \gamma_{\B{B}}} Z_i|D_4}\rangle,
 \end{equation}
 where now the errors are coupled through the last factor. As in Sec.~\ref{app:X_error} we know that only closed red loop configurations $L_{\R{R}}$ on the honeycomb lattice lead to a finite contribution, and hence we need to evaluate
\begin{equation}
    W_{L_{\R{R}}}(\gamma_{\B{B}})\equiv \langle{D_4|\prod_{j\in L_{\R{R}}}X_j \prod_{i\in \gamma_{\B{B}}} Z_i|D_4}\rangle=  \langle \tilde{+}|\prod_{e\in L_{\R{R}}} \widetilde{CZ}_{e_+,e_-} \prod_{b\in \gamma_{\B{B}}} {\color{blue}\tilde{Z}}_{b_+}{\color{blue}\tilde{Z}}_{b_-}|\tilde{+}\rangle,
\end{equation}
where once again $\gamma_{\B{B}}$ collects the coordinates of the bonds on the (blue) triangular lattice ---or equivalently, blue sites of the original kagome lattice--- which corresponds to a sublattice of the honeycomb. Now we need to distinguish two possible scenarios. The case where $L_{\R{R}}$ and $\gamma_{\B{B}}$ do not intersect reduces to the discussion in the previous sections, with $\gamma_{\B{B}}$ required to be a closed loop configuration. On the other hand, one can wonder whether $\gamma_{\B{B}}$ needs to be closed if they intersect, given that a $L_{\R{R}}$ loop contains $e_{\B{B}}$ Abelian anyons as coming from the fusion channel $m_{\R{R}}\times m_{\R{R}}=1+e_{\B{B}} + e_{\G{G}} + e_{\B{B}}e_{\G{G}}$. To compute $W_{L_{\R{R}}}(\gamma_{\B{B}})$ we recall that  $\textrm{tr}\left( CZ_{L_{\R{R}}}\right)=\sqrt{2}^{|L_{\R{R}}|}\textrm{tr}\left( H^{|L_{\R{R}}|}\right)$, and hence the weight requires calculating quantities of the form
\begin{equation} \label{eq:tr_HZ}
    \textrm{tr}(HH..HZH...HZH...HH...H)
\end{equation}
with an even number of $H$ for any closed loop configuration $L_{\R{R}}$ on the honeycomb lattice.
Using the cyclicity of the trace, that $H^2=1$ and finally that $ZH=HX$, one finds that Eq.~\eqref{eq:tr_HZ} is only non-vanishing if there are an even number of $Z$'s. Hence, $\gamma_{\B{B}}$ needs to intersect (on the blue sublattice) each single connected component $\ell_{\R{R}}$ of $L_{\R{R}}=\cup\ell_{\R{R}}$ an even number of times. Let's assume the coordinates of these points along one of these components are given by $\{b_j\}_{j=1}^{2n}$ with $b_1<b_2<\cdots <b_{2n}$. Then 
\begin{equation} \label{eq:Had_weight}
      \langle \tilde{+}|\prod_{e\in \ell_\R{R}} \widetilde{CZ}_{e_+,e_-} \prod_{j=1}^{2n} {\color{blue}\tilde{Z}}_{b_j}|\tilde{+}\rangle=\frac{\sqrt{2}^{|\ell_\R{R}|}}{2^{|\ell_\R{R}|}}\textrm{tr}\left({\color{blue}{Z}}H^{b_2-b_1}{\color{blue}{Z}}H^{b_3-b_2}\cdots {\color{blue}{Z}}H^{|\ell_\R{R}|-(b_{2n}-b_1)}\right),
\end{equation}
where $H$ lies on the links of the honeycomb lattice. This calculation can be efficiently performed introducing a graphical notation shown in Fig.~\ref{fig:grap_not}.
Notice that there are always an even number of $H$ in between two consecutive ${\color{blue}Z}$'s, and using $H^2=\mathds{1}$ we find that $ \langle \tilde{+}|\prod_{e\in \ell_\R{R}} \widetilde{CZ}_{e_+,e_-} \prod_{j=1}^{2n} {\color{blue}\tilde{Z}}_{b_j}|\tilde{+}\rangle=2\sqrt{2}^{-|\ell_\R{R}|}$. Hence we find the weight
\begin{equation}
     W_{L_{\R{R}}}(\gamma_{\B{B}})=\frac{2^{C_{L_{\R{R}}}}}{\sqrt{2}^{|L_{\R{R}}|}},
\end{equation}
whenever either closed loops in $\gamma_{\B{B}}$ do not intersect with $L_{\R{R}}$, or if components of $\gamma_{\B{B}}$ end on $L_{\R{R}}$. See Fig.~\ref{fig:D4_TO}. This result can be understood from the fact that indeed $e_{\B{B}}$ anyons are contained in the fusion channel of $m_{\R{R}}\times m_{\R{R}}$, and hence a loop $\gamma_{\B{B}}$ can close on a $L_{\R{R}}$ loop. Putting it all together we find
 \begin{equation} 
     \mathcal{Z}_{\ket{\psi}}(\beta^x_{\R{R}}, \beta^z_{\B{B}})=\mathcal{Z}_{0,R}\mathcal{Z}_{0,B}\sum_{L_{\R{R}}}\sum_{\gamma_{\B{B}}}\tanh(\beta_{\B{B}}^z)^{|\gamma_{\B{B}}|} \left(\frac{\tanh(\beta_{\R{R}}^x)}{\sqrt{2}}\right)^{|L_{\R{R}}|}2^{C_{L_{\R{R}}}},
 \end{equation}
 where $\sum_{\textrm{``closed"} \,\,\gamma_{\B{B}}}$ is a sum over all loop configurations in the triangular blue sublattice that are either closed, or end up on a $L_{\R{R}}$ loop. These correspond to $e_{\B{B}}$ anyons lying on a closed $L_{\R{R}}$ loop. As a next step, we can add Abelian ${\color{DarkGreen} Z}$ errors on the remaining (green) sublattice. A similar calculation to the one above leads to the classical partition function
\begin{equation} \label{eq:X_R_Z_BG}
\begin{aligned}
     \mathcal{Z}_{\ket{\psi}}(\beta^x_{\R{R}}, \beta^z_{\B{B}},\beta^z_{\G{G}})&=\mathcal{Z}_{0,R}\mathcal{Z}_{0,B}\mathcal{Z}_{0,G}\sum_{L_{\R{R}}}\sum_{\gamma_{\B{B}}}\sum_{\gamma_{\G{G}}}{\sigma_{L_{\R{R}}}(\gamma_{\B{B}},\gamma_{\G{G}})}\tanh(\beta_{\B{B}}^z)^{|\gamma_{\B{B}}|}\tanh(\beta_{\G{G}}^z)^{|\gamma_{\G{G}}|} \left(\frac{\tanh(\beta_{\R{R}}^x)}{\sqrt{2}}\right)^{|L_{\R{R}}|}2^{C_{L_{\R{R}}}},
\end{aligned}
 \end{equation}
 where now both $\gamma_{\B{B}}$ and $\gamma_{\G{G}}$ are loop configurations that can be either closed or end up on a $L_{\R{R}}$ loop configuration on the blue and green sublattices respectively. Moreover, unlike for vanishing $\beta_{\G{G}}^z$, there is an additional sign contribution $ {\sigma_{L_{\R{R}}}(\gamma_{\B{B}},\gamma_{\G{G}})}$ which comes from evaluating a weight analogous to that in Eq.~\eqref{eq:Had_weight} for every connected component $\gamma$ of $L_{\R{R}}$, which now will include both ${\color{blue}Z}$ as well as ${\color{DarkGreen}Z}$ insertions. For a given $\gamma_{\B{B}},\gamma_{\G{G}},L_{\R{R}}$, this sign is then given by
 \begin{equation}
      {\sigma_{L_{\R{R}}}(\gamma_{\B{B}},\gamma_{\G{G}})}=\textrm{sign}\left[\prod_{\ell_{\R{R}} \in L_{\R{R}}}\textrm{tr}\left({\color{blue}{Z}}H {\color{DarkGreen}{Z}} H{\color{blue}{Z}}{\color{blue}{Z}}\cdots H{\color{blue}{Z}}H{\color{DarkGreen}{Z}}\right)\right]
 \end{equation}
 where every non-vanishing factor contains in total an even number of ${\color{blue}Z}$ and ${\color{DarkGreen}Z}$ insertions. However, as explained in the main text, a closed expression for this sign is not necessary to characterize the phase diagram of the deformed wavefunction. Moreover notice that by construction an $H$ needs to appear in between insertions of different color as they lie on different sublattices. The analogous result for the decohered mixed state reads
 \begin{equation} 
\begin{aligned}
    \textrm{tr}\left[\rho(r^x_{\R{R}}, r^z_{\B{B}},r^z_{\G{G}})^2\right]\propto\sum_{L_{\R{R}}}\sum_{\gamma_{\B{B}}}\sum_{\gamma_{\G{G}}} (r_{\B{B}}^z)^{|\gamma_{\B{B}}|}(r_{\G{G}}^z)^{|\gamma_{\G{G}}|} \left(\frac{r_{\R{R}}^x}{2}\right)^{|L_{\R{R}}|}4^{C_{L_{\R{R}}}},
\end{aligned}
 \end{equation}
 i.e., three coupled O$(N)$ loop models with $N=1,4$.

\subsubsection{Only non-Abelian errors on different sublattices} \label{app:2_NA}

We now consider the scenario of two non-Abelian errors acting on different sublattices (e.g., red and blue) 
 \begin{equation}
     \ket{\psi(\beta^x_{\R{R}}, \beta^x_B)}=e^{\frac{ \beta_{\R{R}}^x}{2} \sum_{r\in \R{\mathcal{R}_R}} {\color{red} X}_r} e^{\frac{ \beta_B^x}{2} \sum_{b\in \B{\mathcal{R}_B}} {\color{blue} X}_b}\ket{D_4}.
 \end{equation}
 In this case the classical partition function reads
  \begin{equation}
     \mathcal{Z}_{\ket{\psi}}(\beta^x_{\R{R}}, \beta^z_{\B{B}})=\mathcal{Z}_{0,R}\mathcal{Z}_{0,B}\sum_{L_{\R{R}}}\sum_{L_{\B{B}}}\tanh(\beta_{\R{R}}^x)^{|L_{\R{R}}|}\tanh(\beta_B^x)^{|L_{\B{B}}|}\langle{D_4|\prod_{r\in L_{\R{R}}}{\color{red}X}_r \prod_{b\in L_{\B{B}}} {\color{blue}X}_b|D_4}\rangle,
 \end{equation}
 where we have used the fact that only closed loop configurations  $L_{\R{R}},L_{\B{B}}$  in either the red or blue sublattices respectively, have non-vanishing contributions. Nonetheless, we still need to evaluate 
 \begin{equation}
     W(L_{\R{R}},L_{\B{B}})=\langle{D_4|\prod_{r\in L_{\R{R}}}{\color{red}X}_r \prod_{b\in L_{\B{B}}} {\color{blue}X}_b|D_4}\rangle.
 \end{equation}
To do so we notice that products of $X$'s around a closed red (blue) loop can be written as a product of $6$-body operators $X^{\otimes 6}$ acting on the vertices of stars \ding{65} and lying within the loop. Hence, we can follow the same approach shown at the end of Sec.~\ref{app:X_error}, and map $\ket{D_4}$ to a $\mathbb{Z}_2^3$ SPT defined on a triangular lattice, and finally apply the $\textrm{CCZ}$ disentangling transformation in Sec.~\ref{sec:disent}. One then finds  
\begin{equation}
W(L_{\R{R}},L_{\B{B}})=\langle{+|\CZ_{L_{\R{R}}}\CZ_{{L_{\B{B}}}}|+}\rangle
\end{equation}
similar to the result in Sec.~\ref{app:X_error}, where $\CZ_{{L_{\R{R}}}}$ is defined as the product of two-qubits $\CZ$ gates along the loop configuration $L_\R{R}$. However, unlike in that section, the weight includes two different loops ${L_{\R{R}}},{L_{\B{B}}}$ which can now overlap as shown in Fig.~\ref{fig:two_NA}. Hence, we should rather consider the global loop configuration $L_{\R{R}}\oplus L_{\B{B}}$ given by the union. Let's denote by $\ell$ a connected component of $L_{\R{R}}\oplus L_{\B{B}}$ lying on the intertwined red and blue honeycomb lattices as in Fig.~\ref{fig:two_NA}. Then, $W(L_{\R{R}},L_{\B{B}})=\prod_{\ell \in L_{\R{R}}\oplus L_{\B{B}}}W_\ell$, with $W_\ell$ given by
\begin{equation}
     W_\ell=\frac{1}{2^{|\ell|-|V|}}\textrm{tr}\left(\prod_{e\in\ell}\CZ_e\right)=\frac{2^{|V|}}{\sqrt{2}^{|\ell|}}\textrm{tr}\left(\prod_{e\in\ell}H_e\right)
\end{equation}
where $|V(\ell)|$ is the number of vertices of the resulting planar graph drawn by $\ell$ (namely, vertices where $L_\R{R}$ and $L_\B{B}$ intersect). The difference to the calculation for a single type of error is that now $\ell$ is a closed loop configuration on the intersection of two honeycomb lattices. However, we can then use the fact that on every connected component $\ell$, only an even number of Hadamards can appear between intersections of two loops and hence can be paired up resulting in a factor of $2$ for every (colorless) connected component. Hence, we find 
\begin{equation}
     W_\ell=\frac{2^{|V(\ell)|+1}}{\sqrt{2}^{|\ell|}}=\frac{2^{|V(\ell)|+1}}{\sqrt{2}^{|\ell|}},
\end{equation}
where we have used that for a planar graph Euler's theorem applies together with the fact that the number of vertices is twice the number of edges, leading to the relation $|V(\ell)|+1=|F(\ell)|-1\equiv C_\ell$, with $C_\ell$ the number of loops or also called cyclomatic number. Namely, the minimum number of edges that must be removed from $\ell$ to break all its cycles, making it into a tree. In the example of Fig.~\ref{fig:two_NA}, $C_{L_{\R{R}}\oplus L_{\B{B}}}=4$.

\section{Stat-mech formulation of $\textrm{tr}(\rho^n)$ for non-Abelian noise} \label{app:trrhon_local}

\subsection{Coupled O(2) loop model representation}

Let us consider the decohered density matrix 
\begin{equation} \label{eq:rho_E}
    \rho=(1-p)^{|\R{\mathcal{R}_R}|}\sum_E t^{|E|}\left(\prod_{\R{r}\in E}\R{X_r}\right)\ket{D_4}\bra{D_4}\left(\prod_{\R{r}\in E}\R{X_r}\right),
\end{equation}
appearing as a result of applying an $\R{X}$ Pauli channel on the sublattice $\R{\mathcal{R}_R}$. The goal of this appendix is to compute $\textrm{tr}(\rho^n)$ for any $n$, and provide representation in terms of a local stat-mech model. Here, $t\equiv\frac{p_\R{R}}{1-p_\R{R}}$, and $E$ is a collection of sites in $\R{\mathcal{R}_R}$. To compute $\textrm{tr}(\rho^n)$ we need to evaluate the overlaps  $\langle D_4| \prod_{\R{r}\in E\oplus E'} \R{X_r}|D_4\rangle$ that we already encountered in the main text to give 
\begin{equation}
 \langle D_4| \prod_{\R{r}\in E\oplus E'} |D_4\rangle=f(E\oplus E')\equiv \frac{2^{C(E\oplus E')}}{\sqrt{2}^{|E\oplus E '|}},
 \end{equation}
when $E\oplus E'$ is a contractible closed loop configuration and zero otherwise. To lighten the notation let us denote $\ket{E}=\prod_{\R{r}\in E}\R{X_r}\ket{D_4}$. Then, the $n$th power of $\rho$ reads
\begin{equation}
    \rho^n = (1-p)^{n|\R{\mathcal{R}_R}|}\sum_{\{E^{(s)}\}_{s=1}^n} \prod_{s=1}^n t^{|E^{(s)}|} |E^{(s)}\rangle\langle E^{(s)}|.
\end{equation}

From here, we can compute any moment $\textrm{tr}(\rho^n)$, which requires to evaluate the overlaps $\langle E^{(s)}|E^{(s+1)}\rangle$ with $E^{(n+1)}=E^{(1)}$.  This explicitly shows that each pair of layers contribute with a topological weight that relates to the quantum dimension $d_{m_\R{R}}=2$ of $m_{\R{R}}$. From here we find
\begin{equation} \label{eq:trrhon_symm}
\begin{aligned}
    &\textrm{tr}(\rho^n)= (1-p)^{n|\R{\mathcal{R}_R}|} \sum^{\prime}_{\{E^{(s)}\}_{s=1}^n}\prod_{s=1}^n t^{|E^{(s)}|}
    \prod_{s=1}^{n}f(E^{(s)}\oplus E^{(s+1)}) ,
\end{aligned}
\end{equation}
where the sum over $E^{(s)}$ is constrained to those for which $E^{(s)}\oplus E^{(s+1)}$ forms a closed loop configuration. Notice that this condition implies that all combinations $E^{(s)}\oplus E^{(s')}$ for any $s,s'=1,\dots, n$ correspond to a contractible loop. This also implies that any $E^{(s)}$ with $s>1$, is related to $E^{(1)}$ via a contractible closed loop configuration $L^{(s-1)}$ as $E^{(s)}=E^{(1)}\oplus L^{(s-1)}$. Hence, Eq.~\eqref{eq:trrhon_symm} admits the equivalent representation
\begin{equation} \label{eq:trrhon_1}
\begin{aligned}
    &\textrm{tr}(\rho^n)=(1-p)^{n|\R{\mathcal{R}_R}|} \sum_{E^{(1)}} t^{|E^{(1)}|}\sum_{\{L^{(s)}\}_{s=1}^{n-1}}\prod_{s=1}^{n-1}t^{|E^{(1)}\oplus L^{(s)}|}
    f(L^{(1)}) \left(\prod_{s=1}^{n-2}f(L^{(s)}\oplus L^{(s+1)}) \right) f(L^{(n-1)}).
\end{aligned}
\end{equation}

\subsection{Local spin model representation}

For convenience, we start from the `ungauged' representation of the channel as in Sec.~\ref{sec:D4_SSB_X}:
\begin{equation}
\mathcal{E}^{\R{X}}_{i,j}(\rho_0) = (1-p_\R{R})\rho_0 + p_\R{R} \;  \hat h_{ij} \; \rho_0 \;  \hat h_{ij} \qquad \textrm{where } \hat h_{ij} = \R{Z}_i\R{Z}_j \widetilde{\CZ}_{ij},
\end{equation}
and where the initial state is a trivial product state $\ket{\psi_0}\bra{\psi_0}$ with $\ket{\psi_0}=\ket{\bm{\R{+}}}\ket{\bm{\G{\tilde{+}}}} \ket{\bm{\B{\tilde{+}}}}$.

Hence, up to an irrelevant prefactor, we have:
\begin{equation}
\textrm{tr}\left( \rho^n \right) = \sum_{\{ E^{(s)}\}_{s=1}^n} t^{|E^{(1)}| + |E^{(2)}| + \cdots + |E^{(n)}|}
\bra{\psi_0} \prod_{\langle ij \rangle \in E^{(1)} \oplus E^{(2)}} \hat h_{ij} \ket{\psi_0} \;
\bra{\psi_0} \prod_{\langle ij \rangle \in E^{(2)} \oplus E^{(3)}} \hat h_{ij} \ket{\psi_0} \;
\cdots \;
\bra{\psi_0} \prod_{\langle ij \rangle \in E^{(n)} \oplus E^{(1)}} \hat h_{ij} \ket{\psi_0}
\end{equation}
with $t=\frac{p_\R{R}}{1-p_\R{R}}$. Since $\ket{\psi_0}$ is a superposition of all classical states, we can replace the expectation values by a sum over all classical $
h_{ij}^{(s)} = \sigma_i^{(s)} \sigma_j^{(s)} \widetilde{\CZ}_{ij}^{(s)}$ for each of the $s=1,2,\cdots,n$ expectation values, namely
\begin{equation}
    \bra{\psi_0} \prod_{\langle ij \rangle \in E_{s} \oplus E_{s+1}} \hat h_{ij} \ket{\psi_0} \to \sum_{\{\sigma_j^{(s)}\}}\sum_{\{\tilde{\sigma}_j^{(s)}\}} \prod_{\langle ij \rangle \in E_{s} \oplus E_{s+1}}  h^{(s)}_{ij}
\end{equation}
with $E_{n+1}=E^{(1)}$. This gives the following classical stat-mech model:
\begin{align}
\textrm{tr}\left( \rho^n \right) &= \sum_{\{\sigma^{(s)}_j\}_{s=1}^{n}}\sum_{\{\tilde{\sigma}_j^{(s)}\}_{s=1}^n} \sum_{\{ E^{(s)}\}_{s=1}^n} t^{|E^{(1)}| + |E^{(2)}| + \cdots + |E^{(n)}|}
\prod_{\langle ij \rangle \in E^{(1)} \oplus E^{(2)}} h_{ij}^{(1)} \times
\prod_{\langle ij \rangle \in E^{(2)} \oplus E^{(3)}} h_{ij}^{(2)} \times \cdots \times
\prod_{\langle ij \rangle \in E^{(n)} \oplus E^{(1)} } h_{ij}^{(n)} \label{eq:intermediate} \\
&= \sum_{\{\sigma^{(s)}_j\}_{s=1}^{n}}\sum_{\{\tilde{\sigma}_j^{(s)}\}_{s=1}^n} \sum_{\{ E^{(s)}\}_{s=1}^n} 
\prod_{\langle ij \rangle \in E^{(1)} } t\; h_{ij}^{(n)}h_{ij}^{(1)} \times
\prod_{\langle ij \rangle \in E^{(2)} } t\; h_{ij}^{(1)} h_{ij}^{(2)} \times \cdots \times
\prod_{\langle ij \rangle \in E^{(n)}} t\; h_{ij}^{(n-1)} h_{ij}^{(n)} \\
&= \sum_{\{\sigma^{(s)}_j\}_{s=1}^{n}}\sum_{\{\tilde{\sigma}_j^{(s)}\}_{s=1}^n} \prod_{\langle i ,j\rangle_{\hexagon}} \prod_{s=1}^n \left( 1 + t \; h_{ij}^{(s)} h_{ij}^{(s+1)} \right) \propto \sum_{\{\sigma^{(s)}_j\}_{s=1}^{n}}\sum_{\{\tilde{\sigma}_j^{(s)}\}_{s=1}^n}
\exp\left( \beta  \sum_{\langle i ,j\rangle_{\hexagon}} \sum_{s=1}^n h_{ij}^{(s)} h_{ij}^{(s+1)} \right),
\end{align}
where $\tanh(\beta) = t$. In the second step we have used that 
\begin{equation}
\prod_{\langle ij \rangle \in E^{(s)} \oplus E^{(s+1)}} h_{ij}^{(s)} =   \prod_{\langle ij \rangle \in E^{(s)} } h_{ij}^{(s)}  \prod_{\langle ij \rangle \in E^{(s+1)} } h_{ij}^{(s)}.
\end{equation}
This derives Eq.~\eqref{eq:trrhon_local} in the main text. We note that tracing out the $\sigma$ variables in Eq.~\eqref{eq:intermediate} would directly re-derive  the loop model representation in Eq.~\eqref{eq:trrhon_symm}, thereby establishing a direct link between these two equivalent formulations.

\subsubsection*{Alternative spin model}

One can reduce the number of Ising spins $\sigma_i^{(s)}$. Note that if we flip all replicas $\sigma_i^{(s)} \to - \sigma_i^{(s)}$ for a \emph{fixed} index $i$, the model is symmetric. This means that without loss of generality, we can fix the $\sigma$-spins on the last layer: $\sigma_i^{(n)} = 1$ (up to an irrelevant prefactor of the partition function). There are now $n-1$ $\sigma$-spins remaining. It is convenient to introduce a change of variables:
\begin{equation}
\tau_i^{(1)} = \sigma_i^{(1)}\sigma_i^{(2)}, \quad \tau_i^{(2)} = \sigma_i^{(2)} \sigma_i^{(3)}, \quad \dots, \quad \tau_i^{(n-2)} = \sigma_i^{(n-2)} \sigma_i^{(n-1)}, \quad \tau_i^{(n-1)} = \sigma_i^{(n-1)}.
\end{equation}
Note that this implies $\prod_{s=1}^{n-1} \tau_i^{(s)} = \sigma_i^{(1)} =  \sigma_i^{(n)} \sigma_i^{(1)}$.
In conclusion, we arrive at:
\begin{equation} \label{eq:trrhon_local_app}
\begin{aligned}
\textrm{tr}(\rho^n) =&\sum_{\{\tau^{(s)}_j\}_{s=1}^{n-1}}\sum_{\{\tilde{\sigma}_j^{(s-1/2)}\}_{s=1}^n}  \exp\left({\beta \sum_{\langle i,j\rangle_{\hexagon}}\left[\sum_{s=1}^{n-1} \mathrm h_{i,j}^{(s)} + \prod_{s=1}^{n-1} \mathrm h_{i,j}^{(s)}\right]}\right)
\end{aligned}
\end{equation}
where each local term for $s=1,\dots, n-1$ is given by
\begin{equation} \label{eq.h_ij_app}
    \mathrm h_{i,j}^{(s)}=\tau_i^{(s)}\tau_j^{(s)}\widetilde{\CZ}_{ij}^{(s-1/2)}\widetilde{\CZ}_{ij}^{(s+1/2)};
\end{equation}
here we have labeled the $\tilde{\cdot}$-spins by a half-integer index to emphasize that we can think of them as living in between the $\tau$-spin layers. Note that this formulation closely resembles the structure found for the toric code in Ref.~\cite{fan2023diagnostics}: $n-1$ terms and a last term involving all replicas. In facat, one can  recover the same result by setting $\CZ\to 1$.

\section{Quantum fidelity} \label{ref:fidelity}
The quantum fidelity between two density matrices $\rho$ and $\sigma$ is defined as
\begin{equation}
    F(\rho, \sigma)\equiv \left(\textrm{tr}\sqrt{\sqrt{\rho}\sigma \sqrt{\rho}}\right)^2,
\end{equation}
and quantifies how distinguishable these density matrices are. Sometimes $F\to \sqrt{F}$ is also used as a definition of quantum fidelity. For example, when $\rho=\ket{\psi_\rho}\bra{\psi_\rho}$ and $\sigma = \ket{\psi_\sigma}\bra{\psi_\sigma}$ are projectors on normalized pure states, then the quantum fidelity agrees with the overlap
\begin{equation}
    F(\rho, \sigma) = |\langle  \psi_\rho |\psi_\sigma \rangle |^2.
\end{equation}
This vanishes when $\ket{\psi_\rho}$ is orthogonal to $\ket{\psi_\sigma}$.  

 Let us now consider the two initial states $\ket{D_4}$ and $\R{\mathcal{X}}\ket{D_4}$ and apply $\mathcal{E}^{\R{X}}$ at maximum error rate $p_\R{R}=1/2$, such that $\rho=\mathcal{E}^{\R{X}}(\ket{D_4}\bra{D_4})$ and $\sigma=\mathcal{E}^{\R{X}}(\R{\mathcal{X}}\ket{D_4}\bra{D_4}\R{\mathcal{X}})$. Using the fact that $\mathcal{E}^{\R{X}}$ can be written as a random projector channel as in Eq.~\eqref{eq:E_proj}, we denote by
 \begin{align}
     &\ket{\eta}=\prod_{\R{r} \in \R{\mathcal{R}_R}} \frac{1}{2}(1+\eta_\R{r} X_\R{r})\ket{D_4},\\
     &\ket{\R{\mathcal{X}},\eta}=\prod_{\R{r} \in \R{\mathcal{R}_R}} \frac{1}{2}(1+\eta_\R{r} X_\R{r})\R{\mathcal{X}}\ket{D_4}.
 \end{align}
 Then
\begin{equation}
    \rho=\sum_{\{\eta\}} \ket{\eta}\bra{\eta},\hspace{15pt} \sigma=\sum_{\{\eta\}} \ket{\R{\mathcal{X}},\eta}\bra{\R{\mathcal{X}},\eta}
\end{equation}
where $\avg{\eta|\eta'}=\delta_{\eta,\eta'}\avg{\eta|\eta}$ due to the orthogonality of the projectors $P_{\pm, j}$. Hence,
\begin{equation}
    \sqrt{\rho} = \sum_{\{\eta\}} \frac{\ket{\eta}\bra{\eta}}{\sqrt{\avg{\eta|\eta}}}
\end{equation}
and we thus have
\begin{equation}
    \sqrt{\rho} \sigma \sqrt{\rho}= \sum_{\{\eta\}} \frac{|\avg{\eta|\R{\mathcal{X}},\eta}|^2}{{\avg{\eta|\eta}}}\ket{\eta}\bra{\eta},
\end{equation}
which implies 
\begin{equation}
    \sqrt{\sqrt{\rho} \sigma \sqrt{\rho}}= \sum_{\{\eta\}} \frac{|\avg{\eta|\R{\mathcal{X}},\eta}|}{\avg{\eta|\eta}}\ket{\eta}\bra{\eta}.
\end{equation}
One then finds that the (square-root) fidelity equals the average overlap
\begin{equation} \label{eq:F_gen}
    F'(\rho, \sigma) = \sum_{\{\eta\}}  |\avg{\eta|\R{\mathcal{X}},\eta}| =  \sum_{\{\eta\}}  P(\eta) \frac{|\avg{\eta|\R{\mathcal{X}},\eta}|}{\avg{\eta|\eta}}
\end{equation}
with 
\begin{equation}
    P(\eta)=\avg{\eta|\eta}=\frac{1}{4^{|\R{\mathcal{R}_R}|}}\sum_{ L_{\R{R}}} \left(\prod_{\R{e} \in L_{\R{R}}}\frac{\eta_\R{e}}{\sqrt{2}} \right) 2^{C_{L_{\R{R}}}}\propto{\frac{1}{2}(\mathcal{Z}(\eta)-\mathcal{Z}_-(\eta))}
\end{equation}
with 
\begin{equation}
    \mathcal{Z}(\eta)=\sum_{\{\sigma, \tilde{\sigma}\}} \prod_{\langle i,j\rangle}\left(1+\eta_{ij} \sigma_i\sigma_j\widetilde{\CZ}_{ij}\right),
\end{equation}
agreeing with the partition function of the disorder O$(2)$ loop model, and $\mathcal Z_-(\eta)$ a partition function taking the same form but with a line of flipped antiferromagnetic bonds along a non-contractible loop $\mathcal{C}$ perpendicular to $\R{\mathcal{X}}$.
On the other hand, the overlap $ \avg{\eta|\R{\mathcal{X}},\eta} = \avg{\eta|\R{\mathcal{X}}|\eta}$ corresponds to a similar partition function but constrained to include an odd number of non-contractible loops on each configuration $L_\R{R}$. This constraint can be naturally accounted for on the local-stat mech model in terms of Ising variables by writing
\begin{equation}
     |\avg{\eta|\R{\mathcal{X}}|\eta}|= \frac{1}{2}|\mathcal{Z}(\eta)-\mathcal{Z}_-(\eta)|
\end{equation}
Indeed, the difference $\mathcal{Z}(\eta)-\mathcal{Z}_-(\eta)$ vanishes whenever an even number of non-contractible closed loops intersects this defect line. Hence, the fidelity becomes
\begin{equation}
 F'(\rho, \sigma)  =  {\frac{1}{2}\sum_{\{\eta\}}  P(\eta)\frac{\left|\mathcal{Z}(\eta)-\mathcal{Z}_-(\eta)\right|}{\mathcal{Z}(\eta)+\mathcal{Z}_-(\eta)}= \frac{1}{2}\sum_{\{\eta\}}  P(\eta)\frac{\left|1-e^{-\Delta F_\mathcal{C}}\right|}{1+e^{-\Delta F_\mathcal{C}}}}
\end{equation}
as stated in the main text, with $e^{-\Delta F_\mathcal{C}}=\frac{\mathcal{Z}_-(\eta)}{\mathcal{Z}(\eta)}$.

\section{Consistent field theory} \label{app:field_theory}

A consistent field theory description characterizing the phase diagrams in Fig.~\ref{fig:Phase_diag_D4}b is given by 
\begin{equation} \label{eq:H_complete}
\begin{aligned}
     \mathcal{H} &= \frac{1}{2\pi}\left(K(\partial_x \theta)^2 +\frac{1}{4K}(\partial_x \varphi)^2 \right) -g\cos(8\theta) - \lambda \cos(\varphi) - J\G{\sigma}_g \sin(2\theta)-J\B{\sigma}_b \cos(2\theta) - m\G{\varepsilon}_g - m\B{\varepsilon}_b,
\end{aligned}
\end{equation}
when $\beta^z_\G{G}= \beta^z_\B{B}$. Here, $J\geq 0$; $\lambda, m$ can take both positive and negative values; and finally  $\G{\sigma}_g, \B{\sigma}_b$ and $\G{\varepsilon}_{g},\B{\varepsilon}_{b}$ respectively correspond to the order and energy fields for the Ising CFT~\cite{ginsparg1988appliedconformalfieldtheory,francesco1997conformal}. The motivation to include these new Ising variables is the following: If $\beta^z_{\B{B}}$ is large, $\cos(2 \B{\theta_b} - 2 \B{\theta_{b'}})$ tries to develop long-range order in $\cos(2 \B{\theta_b})=\pm 1$.
If we think of the four values that $\B{\theta_b}$ can take (namely, $\B{\theta_b}\in \{0, \pi/2, \pi, 3\pi/2\}$), this implies the following two patterns:\\

    \begin{center}
    \includegraphics[width=0.4\linewidth]
    {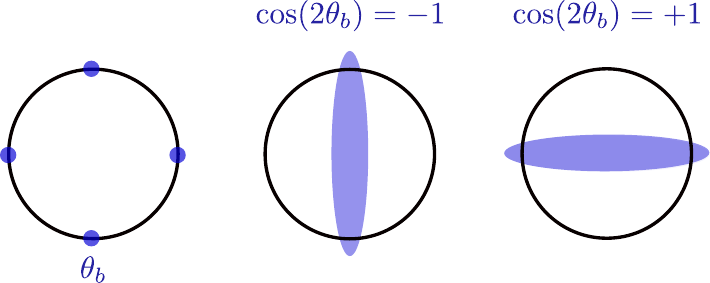}
    \end{center}
    
Hence, we expect a type of `nematic' order for large $\beta^z_{\B{B}}$, where either the system spontanoeusly chooses either the horizontal or vertical axis.
Note that this nematic ordering would indeed be Ising-like: it would break $D_4$ down to a remaining $\mathbb{Z}_2\times \mathbb{Z}_2$, which means only one $\mathbb{Z}_2$ is spontaneously broken. That is consistent with the solvable limit where $\beta^x_{\R{R}} = \beta^z_{\G{G}} = 0$, because we already know that $\beta^z_{\B{B}}$ by itself leads to an Ising transition. A similar reasoning applies when $\beta^z_{\G{G}}$ is large.

For the green sublattice the $\mathbb{Z}_2$ Ising symmetry is given by $S$, which acts as $S:\G{\sigma}_g \to -\G{\sigma}_g $ and $\sin(2\theta)\to - \sin(2\theta)$. For the blue sublattice this corresponds to $RS$ which similarly acts on $\B{\sigma}_b$ and $\cos(2\theta)$. The corresopnding phase diagram for $\beta^z_{\G{G}}=0$ is shown in Fig.~\ref{fig:Phase_diag_D4}a. Moreover, under $M$ $\G{\sigma}_g$ and $\B{\sigma}_b$ are exchanged, and $\sin(2\theta)\leftrightarrow \cos(2\theta)$. 
When instead $\beta^z_\G{G} \neq \beta^z_\B{B}$, this field theory includes an additional potential term $\cos(4\theta)$ and the parameters $J$ and $m$ can take a different value on each sublattice, namely $J\to J_\G{G}, J_\B{B}$ and similarly $m\to m_\G{G},m_\B{B}$.
Moreover, since the condensation of $\cos(2\theta)$ and $\sin(2\theta)$ gives rise to the Ising transition that we found in the previous section when tuning $\beta^z_\G{G}= \beta^z_\B{B}$, we expect that the mass of each species of Ising variables is controlled by $ m_\G{G}\sim \beta^z_\G{G}$, $ m_\B{B}\sim \beta^z_\B{B}$. Once the system undergoes the Ising transition indicated with a horizontal black line in the lower right corner of Fig.~\ref{fig:Phase_diag_D4}b, one can replace $\B{\sigma}_b,\G{\sigma}_g \to $constant, and hence the terms $\cos(2\theta), \sin(2\theta)$ are effectively added to the Hamiltonian for the compact boson. Then, $\mathcal{W}_{e_\B{B}}$ (and also $\mathcal{W}_{e_\G{G}}$ when $\beta^z_\G{G}= \beta^z_\B{B}$) picks up a finite value, leading to the trivial (toric code) phase in the upper right of panel a (b) in Fig.~\ref{fig:Phase_diag_D4}. At this point $\cos (\theta)$ has also acquired a finite value. 

Hence, once the system undergoes the Ising transition $\B{\sigma}_b,\G{\sigma}_g \to $constant, the terms $\cos(2\theta), \sin(2\theta)$ become relevant perturbations for the compact boson Hamiltonian, pinning the value of $\theta$. Upon minimizing $- \sin(2\theta)- \cos(2\theta)$, one finds that the minimum is attained for $\theta=\pi/8$, $\pi+\pi/8$, right in between the possible values that $\G{\theta}_g$ and $\B{\theta}_b$ can take, and along the symmetry axis on which the $M$ symmetry is defined below Eq.~\eqref{eq:SR}. Given these constraints, one of the symmetry-breaking patterns consistent with this minimization is given by
\begin{center}
\begin{tikzpicture}
\node at (3,0){
    \begin{tikzpicture}
    \draw[opacity=0.5] (0,0) circle (0.5);
    \draw[-,dashed,opacity=0.25] (-0.7,-0.2899) -- (0.7,0.2899);
    \draw[->,blue] (-0.0,0) -- (-0.5,0);
    \draw[->,blue] (-0.0,0) -- (0.5,0);
    \draw[->,green] (-0.0,0) -- (-0.353,-0.353);
    \draw[->,green] (-0.0,0) -- (0.353,0.353);
    \node at (0.8,0.31) {$M$};
    \end{tikzpicture}.
};
\end{tikzpicture}
\end{center}

\section{Numerical simulation using Monte-Carlo } \label{app:MC}

In this appendix we provide details about the Monte-Carlo simulations that we use to obtain the numerical results shown in Figs.~\ref{fig:Purewf_phaseDiag_bG=0}, \ref{fig:Purewf_phaseDiag_bG=bB}, and \ref{fig:Correlations_gapless} when considering pure wavefunction deformations, and Fig.~\ref{fig:mixed_pG=PB} when dealing with the decohered mixed states characterized by the purity. To simplify the numerical implementation, we coordinate each honeycomb layer with a brick wall structure shown below

\begin{center}
    \includegraphics[width=0.3\linewidth]{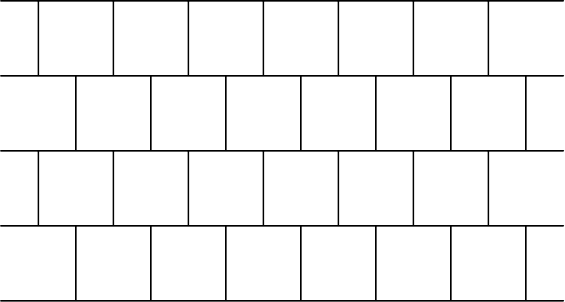}
\end{center}

We take even linear system sizes such that we can fix periodic boundary conditions, and take the number of sites in the horizontal direction $\mathcal{N}_x$ doubled of that in the vertical one $\mathcal{N}_y$, i.e., $\mathcal{N}_x=2\mathcal{N}_y$. For the characterization of the phase diagrams in Figs.~\ref{fig:Purewf_phaseDiag_bG=0}, \ref{fig:Purewf_phaseDiag_bG=bB} and Fig.~\ref{fig:mixed_pG=PB} we take $\mathcal{N}_y=40$, while $\mathcal{N}_y=260$ in the simulations of Fig.~\ref{fig:Correlations_gapless}.

To address the deformed pure wavefunction scenario, we consider a bilayer system composed of two stacked honeycomb layers as shown in Fig.~\ref{fig:bil_honey}a. No tilde $\sigma_j=\pm 1$ spins lie on the upper layer while tilde spins $\tilde{\sigma}_j$ lie on the lower one. Moreover, we refer to the two honeycomb sublattices on each layer as green $\G{G}$ and $\B{B}$. To numerically tackle this problem we consider two equivalent formulations. On the one hand we utilize the Hamiltonian as presented in Eq.~\eqref{eq:main_H_ZZCZ_ZZ} in the main text, to evaluate the order parameter $\mathcal{W}_{m_\R{R}}$. Second, after performing the unitary transformation $\tilde{\sigma}_j\to \sigma_j\tilde{\sigma}_j, \sigma_j\to \sigma_j$ explained above Eq.~\eqref{eq:main_H_X_II}, an equivalent presentation of the Hamiltonian reads (including the temperatures) 
\begin{equation} \label{eq:H_app_1}
    H_{\ket{\psi}}=-\frac{\beta_{\R{R}}^x}{2} \sum_{\langle i,j \rangle_{\hexagon}}({\sigma}_i {\sigma}_j+{\sigma}_i \tilde{\sigma}_j + \tilde{\sigma}_i {\sigma}_j -\tilde{\sigma}_i \tilde{\sigma}_j) - \beta_{\G{G}}^z \sum_{\langle \langle g, g' \rangle \rangle} \G{{\sigma}}_g\G{\tilde{\sigma}}_g\G{{\sigma}}_{g'}\G{\tilde{\sigma}}_{g'}- \beta_{\B{B}}^z \sum_{\langle \langle b, b' \rangle \rangle} \B{{\sigma}}_b\B{\tilde{\sigma}}_b\B{{\sigma}}_{b'}\B{\tilde{\sigma}}_{b'},
\end{equation}
that we observed lead to less fluctuation numerical results for the order parameters $\mathcal{W}_{e_\G{G}}$ and $\mathcal{W}_{e_\B{B}}$.

The algorithm then runs as follows:
\begin{enumerate}
    \item Fix parameters $\beta_{\R{R}}^x,\beta_{\G{G}}^z $ and $\beta_{\B{B}}^z $.
    \item Initialize random configuration of $\pm 1$ per site on the bilayer system. 
    \item Perform $\textrm{eqSteps}=7\cdot 10^4$ number of Metropolis steps --each of them involving $2\times \mathcal{N}_x\times \mathcal{N}_y$ single site updates--- with acceptance ratio given by the energy difference with respect to the energy function Eq.~\eqref{eq:H_app_1}. Notice that we assume that single-site updates lead to non-reducible dnamics with the uniqu stationary distribution given by $e^{-H_{\ket{\psi}}}$.
    \item Compute average quantities by performing additional $\textrm{mcSteps}=7\cdot 10^4$ Metropolis-steps.
\end{enumerate}

When dealing with the decohered density matrix, as characterized by the purity, we instead use the representation of the stat-mech model given by Eq.~\eqref{eq:main_rho_H_XpZ}. Here we used $\textrm{eqSteps}=6\cdot 10^4$ and $\textrm{mcSteps}=7\cdot 10^4$.\\

\end{document}